\documentclass[10pt,english]{article}
\usepackage[T1]{fontenc}
\usepackage[latin9]{inputenc}
\usepackage{babel}
\usepackage{latexsym}
\usepackage{amsmath, amsthm, amssymb, amsfonts, bm}
\usepackage{cancel}
\usepackage[normalem]{ulem}
\usepackage{graphicx}
\usepackage{placeins}
\usepackage{sectsty}
\usepackage[font=small]{caption}
\usepackage{mathrsfs}
\usepackage{subcaption}
\usepackage{xcolor}
\usepackage{multirow}
\usepackage{relsize}
\usepackage{mathpazo}
\usepackage{lettrine}

\usepackage[a4paper]{geometry}
\usepackage{setspace}
\definecolor{page_backgroundcolor}{rgb}{1, 1, 1}
\pagecolor{page_backgroundcolor}
\definecolor{sangre}{rgb}{0.6,0.18,0.19}
\definecolor{dullmagenta}{rgb}{0.4,0,0.4}
\definecolor{darkblueroyale}{rgb}{0,0,0.6}
\definecolor{verdeprofundo}{rgb}{0.02,0.376,0.031}
\definecolor{aqua}{rgb}{0.0, 1.0, 1.0}

\usepackage[authoryear]{natbib}
\usepackage[pdfusetitle,
 bookmarks=true,bookmarksnumbered=false,bookmarksopen=false,
 breaklinks=true,pdfborder={0 0 0},pdfborderstyle={},backref=false,colorlinks=true]
 {hyperref}
\hypersetup{citecolor=sangre,urlcolor=darkblueroyale,linkcolor=darkblueroyale}

    \newcommand{\pperp}{\perp\!\!\!\!\perp} 
    \newcommand{\ind}[2]{\mathbf{1}_{#1}\!\left\{#2\right\}}
    \newcommand{\prob}[1]{ \mathbf{Pr}\big( #1\big) }
    \newcommand{\Prob}[1]{ \mathbf{Pr}\left( #1\right) }
    \newcommand{\expect}[1]{ \mathbb{E}\big[ #1\big] }
    \newcommand{\Expect}[1]{ \mathbb{E}\left[ #1\right] }
    
    \newcommand{\condProb}[2]{\mathbf{Pr}\left(#1\,\middle|\,#2\right)}
    \newcommand{\condexpect}[2]{ \mathbb{E}\big[ #1 \,\big| #2 \big] }
    \newcommand{\condExpect}[2]{\mathbb E\left[#1\,\middle|\,#2\right]}
    \newcommand{\expectF}[2]{ \mathbb{E}_{#1}\big[ #2\big] }
    \newcommand{\ExpectF}[2]{ \mathbb{E}_{#1}\left[ #2\right] }
    \newcommand{\condexpectF}[3]{ \mathbb{E}_{#1}\big[ #2 \,\big|\, #3 \big] }
    \newcommand{\condExpectF}[3]{\mathbb E_{#1}\left[#2\,\middle|\,#3\right]}
    
    \newcommand{\argmin}[2]{ \underset{#1}{\textnormal{argmin\,}}{#2} }
    
    \newcommand{\norm}[1]{ \big\| #1 \big\| }
    \newcommand{\Norm}[1]{ \left\| #1 \right\| }

    \newcommand{\vecdel}{ {\pmb{\delta}} }

    \newcommand{\widebar}[2][0.9em]{\mathord{\vphantom{#2}\smash[t]{\overset{\rule{#1}{0.4pt}}{#2}}} }
    
\numberwithin{equation}{section}
\providecommand{\U}[1]{\protect\rule{.1in}{.1in}}

    \theoremstyle{plain}
        \newtheorem{theorem}{Theorem}
        \newtheorem{lemma}{Lemma}
        \newtheorem{proposition}{Proposition}
        \newtheorem{corollary}[theorem]{Corollary}
        \newtheorem{definition}{Definition}
        
    \theoremstyle{definition}
        \newtheorem{example}{Example}
        \newtheorem{remark}{Remark}

\makeatletter
\newtheoremstyle{assumptionstyle}%
  {10pt}% Space above
  {10pt}% Space below
  {\normalfont}% Body font
  {}% Indent amount
  {\bfseries}% Head font
  {.}% Punctuation after head
  {\newline}% Space after head
  {}% Head specification
\theoremstyle{assumptionstyle}
\newtheorem{assumption}{Assumption}
\AtBeginDocument{

}
\makeatother

\begin{document}

\title{
    \textbf{Identification Preserving Estimation with Nonparametric/Machine Learning Generated Conditioning Variables}
%    \thanks{We are grateful (...)}
}
\author{$%
    \begin{array}[c]{cc}%
    \text{Taegyu Yang}\footnote{E-mail : tzy5176@psu.edu, taegyuyang94@gmail.com}&
    \text{Ryo Kimoto}\footnote{E-mail : rqk5514@psu.edu, Webpage : \url{https://ryokimoto.github.io/} }\\
    \text{Department of Economics} & \text{Department of Economics}\\
    \text{Pennsylvania State University}& \text{Pennsylvania State University}
    \end{array} $}
    %\date{First Version: August, 1 2025\\
    \date{November 27, 2026}
%    Revised: \today\vspace{0.25in}}
\maketitle

\begin{abstract}
    Many economic models, including those used in proxy-variable approaches to production-function estimation, imply conditional moment restrictions. In these models, latent variables can enter both the residual and the conditioning set. These variables can be generated in a first stage using nonparametric or machine learning methods. We develop a double/debiased machine learning estimator of the finite-dimensional parameter identified by the conditional restriction, allowing the generated variables to enter the conditioning set. To preserve identification by the conditional restriction, we use a criterion based on an integrally strictly positive definite (ISPD) kernel, as in \citet{dml_escanciano_terschuur2026}. We derive an orthogonal score to account for first-stage estimation error in both the moment function and the variables used to evaluate the ISPD kernel. Under suitable conditions, the cross-fitted estimator is asymptotically normal at the root-$N$ rate. We propose a data-dependent rule that selects the ISPD kernel bandwidth by minimizing an estimated measure of asymptotic variance over a fixed finite grid. We establish asymptotically valid frequentist inference using the selected bandwidth. We also develop a Hausman test to assess whether the observed first-stage regressors can replace the generated conditioning variables. In a production-function application using Chilean manufacturing data, conventional instruments yield four solutions to the sample moment equations. In cereal demand, the Hausman test rejects conditioning on inventory purchase costs in place of predicted retail prices for four of five first-stage learners.

    \noindent\textbf{Keywords:} Conditional Moment Restrictions, Generated Regressor, Neyman-Orthogonal Score, Double/Debiased Machine Learning, U-Statistic, Production Function Estimation, Demand Estimation \\

    \noindent\textbf{JEL classification:} C13, C14, C21, C25, C26, D22, D24, L11, L66 \\
\end{abstract}

%%%%%%%%%%%%%%%%%%%%%%%%%%%%%%%%%%%%%%%%%%%%%%%%%%%%%%%%%%%%%%%%%%%%%%%%%%%%%%%%%%%%%%%%%%%%%%%%%%%%%%%%%%%%%%%%%%%%%%%%%%%%%%%%%%%%%%%%%%%%%%%%%%%%%%%%%%%%%%%%%%%%%%%%%%%%%%%%%%%%%%%%%%%%%%%%%%%%%%%%%%%%%%%%%%%%%%%%%%%%%%%%%%%%%%%%%%%%%%%%%%%%%%%%%%%%%%%%%%%%%%%%%%%%%%%%%%%%%%%%%%%%%%%%%%%%%%%%%%%%%%%%%%%%%%%%%%%%%%%%%%%%%%%%%%%%%%%%%%%%%%%%%%%%%%%%%%%%%%%%%%%%%%%%%%%%%%%%%%%%%%%%%%%%%%%%%%%%%%%%%%%%%%%%%%%%%%%%%%%%%%%%%%%%%%%%%%%%%%%%%%%%%%%%%%%%%%%%%%%%%%%%%%%%%%%%%%%%%%%%%%%%%%%%%%%%%%%%%%%%%%%%%%%%%%%%%%%%%%%%%%%%%%%%%%%%%%%%%%%%%%%%%%%%%%%%%%%%%%%%%%%%%%%%%%%%%%%%
\pagebreak \newpage
\section{Introduction}
Many economic models imply a conditional moment restriction (CMR) of the form
\begin{equation} \label{eq:CMR}
    \condexpect{\rho(W,\theta_0)}{Z}=0 \quad\text{almost surely},
\end{equation}
where $W$ collects the variables entering the moment function, $Z$ is a vector of conditioning variables, and $\theta_0$ is a finite-dimensional parameter of interest. The economic model specifies the moment function $\rho$ and the conditioning variables $Z$. It is often assumed that $\theta_0$ is the unique parameter value satisfying the CMR.

In many applications, some variables entering $\rho(W,\theta)$ or the conditioning set $Z$ are unobserved and must be generated in a preliminary estimation step. Prominent examples include proxy-variable approaches to production-function estimation, which use a nonparametric first stage to control for unobserved productivity \citep{olley_pakes1996,levinsohn_petrin2003,ackerberg_caves_frazer2015}, and control-function approaches to demand estimation, where an estimated control enters both the choice probabilities and the conditioning set \citep{petrin_train2010}.

Researchers typically substitute first-stage estimates for unobserved variables, use the resulting conditioning vector $\widehat Z$ to construct a finite set of instruments $\Gamma(\widehat Z)$, and estimate $\theta_0$ from the sample analogue of the moment condition $\expect{\Gamma(Z)\rho(W,\theta_0)}=0$. This procedure raises two issues. (1) Estimation error in the generated variables affects the properties of the resulting estimator. (2) The selected unconditional moments may have multiple roots even when the CMR uniquely identifies $\theta_0$, as emphasized by \citet{cmr_DL2004}. Estimation and inference must therefore account for first-stage estimation error while preserving the identifying information in the conditional restriction.

\citet{dml_escanciano_terschuur2026} address these two challenges in CMR models where unknown nuisance functions only enter the moment function $\rho$ and must be estimated in a preliminary step, while the conditioning variables $Z$ are observed directly. Their approach combines an identification-preserving quadratic criterion based on an integrally strictly positive definite (ISPD) kernel with double/debiased machine learning (DDML). The criterion retains the identifying information in the CMR, following the integrated-moment and kernel minimum-distance approaches of \citet{cmr_DL2004} and \citet{kernel_lavergne_patilea2013}. DDML then accounts for estimation error in these nuisance functions through Neyman orthogonalization and cross-fitting \citep{dml_chernozhukov_et_al2018,dml_chernozhukov_et_al2022}.

We extend this framework to allow nuisance functions to enter the conditioning vector $Z$ as well as the moment function $\rho$. First-stage estimation error affects both the moment function $\rho$ and the ISPD kernel evaluations, because the conditioning variables entering the kernel are generated using machine learning. We derive a Neyman-orthogonal score that accounts for both effects and use it to construct a cross-fitted DDML estimator. Under suitable rate conditions on the nuisance estimators and other regularity conditions, we establish root-$N$ asymptotic normality of the proposed estimator. We generalize the treatment of generated conditioning variables in \citet{cmr_kimoto_otsu2022} by allowing a nonparametric/machine-learning first stage. Their analysis assumes a parametric first stage.

Allowing generated conditioning variables also permits a comparison with restrictions based on the first-stage regressors. Conditioning directly on these regressors generally imposes a stronger CMR, which the original economic model need not justify. We construct a Hausman-type test comparing the parameter estimates under these two conditioning specifications. 

Our procedure preserves identification by the CMR without requiring the researcher to select a finite collection of instruments and verify that the resulting unconditional moments identify $\theta_0$. Implementation still requires choosing the bandwidth of the ISPD kernel. Every admissible fixed positive bandwidth preserves identification, although its choice affects asymptotic variance. We therefore provide a data-dependent rule that selects the bandwidth by minimizing an estimated measure of asymptotic variance over a fixed finite grid, and establish the asymptotic validity of inference using the chosen bandwidth.

The simulations in Section~\ref{sec06} first vary the rate of first-stage convergence to assess its effect on inference, then examine performance when the first stage is estimated using several nonparametric learners. Under a correctly specified CMR and a sufficiently fast first stage, the DDML estimator has little bias and its confidence intervals achieve near-nominal coverage. When the rate conditions hold, the Hausman test's empirical size approaches its nominal level and its power against the simulated alternative rises toward one as the sample grows, as our large-sample theory predicts. For correctly specified CMRs, our data-dependent bandwidth rule maintains near-nominal coverage across the designs and generally yields shorter intervals on average than fixing the bandwidth at either endpoint of the grid.

In the Chilean manufacturing data, instruments adapted from \citet{ackerberg_caves_frazer2015} yield four roots of the sample moment equations. The estimated elasticities for skilled labor, unskilled labor, and capital each change sign across these roots (Section~\ref{sec02}). In Section~\ref{sec07:acf}, we use the same data to estimate the production function directly from the model's CMR, avoiding the need to choose a finite set of instruments or select among the resulting sample roots. The cereal-demand application uses Dominick's supermarket scanner data analyzed in \citet{fosgerau_monardo_depalma2024}. We predict retail prices using promotions and acquisition costs, which measure what the retailer paid for its inventory. The Hausman test rejects using acquisition costs in place of predicted retail prices in the conditioning set for four of the five first-stage learners (Section~\ref{sec07:cereal}).

Section~\ref{sec02} presents the motivating production-function example. Section~\ref{sec03} sets out the econometric framework, and Section~\ref{sec04} develops the DDML estimator and its asymptotic theory. Section~\ref{sec05} presents the conditioning-set test and the bandwidth selection rule. Sections~\ref{sec06} and~\ref{sec07:empirical} report the simulation results and empirical illustrations, respectively. Section~\ref{sec08:conclusion} concludes. The appendix contains the proofs and numerical implementation details.
%%%%%%%%%%%%%%%%%%%%%%%%%%%%%%%%%%%%%%%%%%%%%%%%%%%%%%%%%%%%%%%%%%%%%%%%%%%%%%%%%%%%%%%%%%%%%%%%%%%%%%%%%%%%%%%%%%%%%%%%%%%%%%%%%%%%%%%%%%%%%%%%%%%%%%%%%%%%%%%%%%%%%%%%%%%%%%%%%%%%%%%%%%%%%%%%%%%%%%%%%%%%%%%%%%%%%%%%%%%%%%%%%%%%%%%%%%%%%%%%%%%%%%%%%%%%%%%%%%%%%%%%%%%%%%%%%%%%%%%%%%%%%%%%%%%%%%%%%%%%%%%%%%%%%%%%%%%%%%%%%%%%%%%%%%%%%%%%%%%%%%%%%%%%%%%%%%%%%%%%%%%%%%%%%%%%%%%%%%%%%%%%%%%%%%%%%%%%%%%%%%%%%%%%%%%%%%%%%%%%%%%%%%%%%%%%%%%%%%%%%%%%%%%%%%%%%%%%%%%%%%%%%%%%%%%%%%%%%%%%%%%%%%%%%%%%%%%%%%%%%%%%%%%%%%%%%%%%%%%%%%%%%%%%%%%%%%%%%%%%%%%%%%%%%%%%%%%%%%%%%%%%%%%%%%%%%%%%
\section{Motivating Example}\label{sec02}
We use Chilean manufacturing data to estimate a Cobb--Douglas value-added production function with AR(1) productivity. We adapt the instruments in equation~(28) of \citet[p.~2430]{ackerberg_caves_frazer2015}, hereafter ACF, to form five unconditional moment equations for the model's five parameters. Four distinct parameter vectors satisfy all five sample equations, and the signs of the estimated production elasticities depend on which solution we choose.

\paragraph{Data and production model.}
Chilean manufacturing data have an established place in the production-function literature; see, for example, \citet{levinsohn_petrin2003}. We use the publicly distributed \texttt{chilean} sample from the \texttt{prodest} package \citep{rovigatti2017}, containing 2,544 plant-year observations on 497 plants over 1996--2006. The package supplies a sectoral subsample covering a later period than the panel in \citet{levinsohn_petrin2003}. It provides value added, skilled and unskilled labor, capital, and materials. Requiring a one-year lag leaves an estimation sample of 1,944 plant-year observations from 401 plants. All production variables used below are in logs.

Let $Y_{it}$ denote log value added and let $X_{it}=(\ell^s_{it},\ell^u_{it},k_{it})^\top$ collect log skilled labor, unskilled labor, and capital. We maintain
\begin{align}
 Y_{it}&=X_{it}^\top\beta_0+\omega_{it}+\varepsilon_{it},
 &\omega_{it}&=\alpha_0+\gamma_0\omega_{i,t-1}+\xi_{it},
 \label{sec02:eq:production}
\end{align}
where productivity $\omega_{it}$ is observed by the plant when it chooses inputs. Let $\mathcal I_{it}$ contain the plant's information after input choices and before the transitory shock $\varepsilon_{it}$. Assume $\mathbb E[\varepsilon_{it}\mid\mathcal I_{it}]=0$ and $\mathbb E[\xi_{it}\mid\mathcal I_{i,t-1}]=0$. Inputs may therefore respond to current productivity, making ordinary least squares inappropriate. Under the first-stage outcome-invertibility assumption, materials $M_{it}$ permit the representation $\omega_{it}=f_0(X_{it},M_{it})$. Consequently,
\begin{equation}
 \eta_0(x,m)=\mathbb E[Y_{it}\mid X_{it}=x,M_{it}=m]
             =x^\top\beta_0+f_0(x,m).
 \label{sec02:eq:firststage}
\end{equation}
These are maintained restrictions for the empirical application; the data are not simulated from a known parameter vector.

\paragraph{From the conditional restriction to five GMM equations.}
Write $\theta=(\beta_s,\beta_u,\beta_k,\alpha,\gamma)^\top$ and $\eta_{it}=\eta(X_{it},M_{it})$. Define
\begin{equation}
 \rho_{it}(\theta,\eta)
 =\eta_{it}-X_{it}^\top\beta-\alpha
  -\gamma\big(\eta_{i,t-1}-X_{i,t-1}^\top\beta\big).
 \label{sec02:eq:residual}
\end{equation}
At the true parameters and first-stage function, this residual equals $\xi_{it}$. The productivity timing assumption implies the conditional moment restriction
\begin{equation}
 \mathbb E\!\left[\rho_{it}(\theta_0,\eta_0)
       \mid X_{i,t-1},\eta_{0,i,t-1}\right]=0.
 \label{sec02:eq:cmr}
\end{equation}
We adapt the illustrative second-stage moment conditions in \citet[p.~2430, equation~(28)]{ackerberg_caves_frazer2015}, using the five instruments
\begin{equation}
 Z_{i,t-1}(\eta)
 =\big(1,\ell^s_{i,t-1},\ell^u_{i,t-1},k_{i,t-1},\eta_{i,t-1}\big)^\top.
 \label{sec02:eq:instruments}
\end{equation}
The components $1$ and $X_{i,t-1}$ are observed, whereas $\eta_{0,i,t-1}$ is unknown and must be estimated.\footnote{ACF's illustration uses a constant, current capital, lagged labor, and the lagged first-stage conditional mean. Our specification distinguishes skilled and unskilled labor and uses lagged capital.}

Following convention, we estimate $\eta_0$ by regressing $Y_{it}$ on a quadratic polynomial in $(X_{it},M_{it})$, including all pairwise interaction terms. We then evaluate the fitted function at current and lagged inputs to obtain $\widehat\eta_{it}=\widehat\eta(X_{it},M_{it})$ and $\widehat\eta_{i,t-1}=\widehat\eta(X_{i,t-1},M_{i,t-1})$, respectively. Holding $\widehat\eta$ and the instruments fixed, we estimate all five parameters simultaneously by minimizing $Q_{N}(\theta)$, where
\begin{align}
    \widehat{g}_N(\theta)
 & =\frac{1}{N}\sum_{(i,t)\in\mathcal T_N} Z_{i,t-1}(\widehat\eta)\rho_{it}(\theta,\widehat\eta),
 & Q_{N}(\theta)&=\widehat{g}_N(\theta)^\top\widehat{W}_{N}\widehat{g}_N(\theta), \label{sec02:eq:gmm}
\end{align}
where $\mathcal T_N$ indexes the estimation sample of plant-year observations with data available in the preceding year, $N=1,944$, and
\begin{equation}\label{sec02:eq:weight}
    \widehat{W}_{N} = \left\{\frac{1}{N}\sum_{(i,t)\in\mathcal T_N}  Z_{i,t-1}(\widehat\eta)Z_{i,t-1}(\widehat\eta)^\top\right\}^{-1}.
\end{equation}
We search over $\Theta=[-3,3]^3\times[-2,2]\times[0.1,0.99]$, with identical symmetric bounds for the three production elasticities.

\paragraph{Four distinct solutions.}
Table~\ref{sec02:tab:roots} reports four distinct interior solutions. One has positive production elasticities of approximately $(0.637,0.601,0.310)$; the others include a negative labor or capital elasticity and substantially different productivity persistence. All four attain a numerically zero objective despite implying different production elasticities.

\begin{table}[htbp]
 
 \centering
 \caption{Multiple roots of five ACF-type GMM equations}
 \label{sec02:tab:roots}
 \begin{tabular}{l|rrrr}
 \hline\hline
 & Solution 1 & Solution 2 & Solution 3 & Solution 4\\
 \hline
 Skilled labor, $\beta_s$ & 0.637071 & 1.765550 & $-0.093579$ & 0.000049\\
 Unskilled labor, $\beta_u$ & 0.600547 & $-2.738719$ & 0.283731 & 0.056381\\
 Capital, $\beta_k$ & 0.310070 & 0.816349 & 0.757987 & $-0.175379$\\
 Intercept, $\alpha$ & 1.848467 & 0.776753 & 0.635149 & 0.315284\\
 Persistence, $\gamma$ & 0.743655 & 0.794770 & 0.822883 & 0.981275\\
\hline
 $Q_{N}(\widehat{\theta})$ & $4.00\!\times\!10^{-26}$ & $6.25\!\times\!10^{-26}$ & $5.33\!\times\!10^{-26}$ & $1.66\!\times\!10^{-28}$\\
 Moment-Jacobian rank & 5 & 5 & 5 & 5\\
 \hline\hline
 \end{tabular}
 \par\smallskip
 \begin{minipage}{0.96\textwidth}
 \footnotesize\textit{Notes:} Both stages use 1,944 current-period observations from 401 plants, with lagged inputs used for evaluation. The first stage is a full quadratic in the three inputs and materials. Objective values use full-precision parameters and the weight in equation~\eqref{sec02:eq:weight}; displayed coefficients are rounded. All solutions are strictly inside $\Theta$. The largest absolute direct sample moment is below $1.7\times10^{-13}$ at each solution. Small differences in objective values reflect numerical precision.
 \end{minipage}
\end{table}

Optimization from each of 66,339 grid-based starting values converges to one of the four roots displayed in Table~\ref{sec02:tab:roots}. Multiple roots also arise when the quadratic estimate of $\eta_0$ is replaced by local-quadratic regression or random forests, which yield four and three interior roots, respectively. Multiplicity persists when the first stage is fitted using all 2,544 plant-year observations, while retaining the same 1,944 plant-year observations for GMM. Related sensitivity of ACF estimates to optimization choices is documented by \citet{rovigatti_mollisi2018}. Appendix~\ref{app:empiricaldetails} documents our numerical checks and robustness results.

Root multiplicity extends beyond our baseline choice of instruments. Holding the sample, the quadratic first stage, and the parameter space $\Theta$ fixed, we combine a constant with any four of the levels, squares, and pairwise products of the four standardized lagged variables in equation~\eqref{sec02:eq:instruments}. Of the resulting 1,001 systems, 493 have multiple interior roots, 445 have one, and 63 have none. For example, replacing lagged fitted output with the interaction of lagged unskilled labor and lagged capital reduces the number of interior roots from four to one. The economic model implies the CMR, so each of these instrument sets yields valid population moments. Optimal instruments provide a theoretical benchmark for efficiency, but they depend on unknown conditional moments and are rarely used directly in applications. Practitioners must still choose a finite set of instruments and establish that it preserves global identification, even when the instruments are optimal \citep[Example~2]{cmr_DL2004}.

The signs of the estimated elasticities for skilled labor, unskilled labor, and capital depend on which of the four baseline roots is selected. Having as many moments as parameters, with a full-rank sample Jacobian at each solution, does not resolve this ambiguity. \citet{komunjer2012} gives sufficient conditions for global identification from a finite set of unconditional moments. Applying these conditions involves verifying properties of the population moment map over the parameter space, which can be difficult in complex structural models. This motivates an estimator that preserves identification by the model's CMR and avoids a separate global identification argument for a chosen set of instruments. In Sections~\ref{sec03}--\ref{sec04}, we develop such an approach and account for the estimation of the first-stage function through an orthogonal correction. We return to the Chilean data in Section~\ref{sec07:empirical} to examine its empirical performance.

%%%%%%%%%%%%%%%%%%%%%%%%%%%%%%%%%%%%%%%%%%%%%%%%%%%%%%%%%%%%%%%%%%%%%%%%%%%%%%%%%%%%%%%%%%%%%%%%%%%%%%%%%%%%%%%%%%%%%%%%%%%%%%%%%%%%%%%%%%%%%%%%%%%%%%%%%%%%%%%%%%%%%%%%%%%%%%%%%%%%%%%%%%%%%%%%%%%%%%%%%%%%%%%%%%%%%%%%%%%%%%%%%%%%%%%%%%%%%%%%%%%%%%%%%%%%%%%%%%%%%%%%%%%%%%%%%%%%%%%%%%%%%%%%%%%%%%%%%%%%%%%%%%%%%%%%%%%%%%%%%%%%%%%%%%%%%%%%%%%%%%%%%%%%%%%%%%%%%%%%%%%%%%%%%%%%%%%%%%%%%%%%%%%%%%%%%%%%%%%%%%%%%%%%%%%%%%%%%%%%%%%%%%%%%%%%%%%%%%%%%%%%%%%%%%%%%%%%%%%%%%%%%%%%%%%%%%%%%%%%%%%%%%%%%%%%%%%%%%%%%%%%%%%%%%%%%%%%%%%%%%%%%%%%%%%%%%%%%%%%%%%%%%%%%%%%%%%%%%%%%%%%%%%%%%%%%%%%
\section{Econometric framework}\label{sec03}
\subsection*{Basic Notation}
We introduce some basic notation which we will use throughout the paper. We denote the true probability measure and joint distribution of all underlying random variables by $P_0$ and $F_0$, respectively. The support of a random variable $X$ is denoted by $\text{supp}(X)$. For a probability measure $P$, $P\otimes P$ denotes its product measure with itself. We write $\mathbb{E}_{F}[X]$ for the expectation of $X$ under distribution $F$, and use $\mathbb{E}_{F_0}[X]$ and $\mathbb{E}[X]$ interchangeably for expectations under $F_0$.

For a positive integer $N>0$, $[N] \equiv \{1,....N\}$ denotes an index set. We will use $\text{int}(A)$ to mean the interior of a set $A$, where $A$ is a subset of $\mathbb{R}^{d}$ with $(\mathbb{R}^{d},\lVert\cdot\rVert)$ being a $d$--dimensional metric space, where $d\geq 1$, with a Euclidean metric $\lVert\cdot\rVert$. 

For a generic measure $Q$ and $1\leq p<\infty$, the $L^p(Q)$-norm of a measurable function $f$ is
\[ \lVert f\rVert_{Q,p}
 \equiv\left(\int\lVert f(x)\rVert^p\,dQ(x)\right)^{1/p}, \]
where the norm inside the integral is Euclidean when $f$ is vector-valued. If $Q=P_0$ or $Q=P_0\otimes P_0$, we omit $Q$ and write $\lVert f\rVert_p$. We use
\[ \lVert f\rVert_\infty\equiv\sup_{x\in\operatorname{dom}(f)}\lVert f(x)\rVert\]
for the uniform norm, with the supremum taken over the domain of $f$, rather than an essential supremum with respect to a measure.

For a generic function $(\theta,\eta)\mapsto f(\theta,\eta)$, we denote its directional derivative with respect to $\eta$ in an arbitrary direction $h$ by $D_{\eta}f(\theta,\eta)[h]$. If a function $\theta\mapsto f(\theta)$ is differentiable at $\theta_0$, we denote its Jacobian and Hessian matrices by $\nabla_{\theta}f(\theta_0)$ and $\nabla_{\theta}^2 f(\theta_0)$, respectively.

\subsection{Data, conditional moment restrictions, and instrument variables}
For each observation $i\in[N]$, let $Y_i\in\mathbb{R}$ denote an outcome of interest and let denote a vector of covariates, where $X_{1i}\in\mathbb{R}^{d_{X_1}}$ may be endogenous, whereas $X_{2i}\in\mathbb{R}^{d_{X_2}}$ is restricted to be exogenous. Let $S_i\in\mathbb{R}^{d_S}$ and $T_i\in \mathbb{R}^{d_V}$ denote a a vector of first-stage regressors and the first-stage outcome, respectively. The researcher observes $N$ independent and identically distributed (i.i.d.) copies $\{W_i\}_{i\in[N]}$ of $W\sim F_0$, where $W=(Y,X^\top,T^\top,S^\top)^\top\in\operatorname{supp}(W)$. Our asymptotic experiment has $N\to\infty$.

Let $\theta\in\Theta$ denote a candidate value of the parameter of interest, where $\Theta\subset\mathbb{R}^{d_\theta}$ is a finite-dimensional parameter space. To identify the true parameter $\theta_0$, the researcher imposes a known restriction $\rho : \text{supp}(W) \times \Theta \times \mathcal{H} \mapsto \mathbb{R}$ implied by the economic model. For simplicity, we focus on a single scalar-valued restriction. The extension to multiple restrictions and the construction of the corresponding DDML estimator are discussed in Corollary \ref{sec04 - corollary01}. The true parameter $\theta_0\in\Theta$ is defined as a solution to the following conditional moment equations:
\begin{equation}\label{sec03 - framework - CMR}
    \expect{ \rho(W_i, \theta, \eta_0 ) \,\mid\, Z_{i}^{c}(\eta_0) } = 0  \quad[P_0]-\text{a.s,}
\end{equation}
where the conditioning variable $Z_i^{c}(\eta_0)$ is defined below.

Let $V_i\in\mathbb{R}^{d_V}$ denote a vector of valid but latent instruments. The researcher's economic theory implies that the relevant conditioning variables, denoted by $Z_i^{c}$, take the form :
\begin{equation}\label{sec03 - framework - instrument}
    Z_{i}^{c} = \begin{cases}
                    (X_{2i}^\top, 0^\top)^\top      &\text{ if }c = 0\\
                    (X_{2i}^\top, V_i^\top)^\top    &\text{ if }c = 1.
                \end{cases}
\end{equation}
This notation provides a unified framework accommodating both the case in which $V_i$ is excluded from the conditioning set and the case in which it is included. Let $d_{Z,c}\in\mathbb{N}$ denote the dimension of the instrument. Then $d_{Z,c} = d_{X_2}$ if If $c=0$, and $d_{Z,c} = d_{X_2} + d_{V}$ otherwise.

Although $V_i$ is unobserved, we suppose that there exists a function $\eta_0 \in \mathcal{H} \equiv \{ \eta : \text{supp}(S)\to \mathbb{R}^{d_{V}}\}$ such that $V_i = \eta_0(S_i)$ almost surely. We refer to $\mathcal{H}$ as the nuisance function space. We further assume that the unknown function $\eta_0$ is nonparametrically identified by:
\begin{equation}\label{sec03 - framework - first stage}
    \expect{ \varsigma(S_i)\{ T_i -  \eta(S_i)\} }  = 0 \text{ for every }\varsigma(\cdot)\in\mathcal{H}\quad\Longleftrightarrow\quad \eta(\cdot) = \eta_0(\cdot),
\end{equation}
where $T_i\in \mathbb{R}^{d_V}$ denotes the first-stage outcome to recover $\eta(S_i)$, and $\varsigma(S_i)$ denotes the admissible generic test function. Accordingly, the conditioning variables can be viewed as a functional of $\eta$. For any $\eta\in\mathcal{H}$, define:
\begin{equation}\label{sec03 - framework - instrument fuctional}
    Z_{i}^{c}(\eta) = \begin{cases}
                    (X_{2i}^\top, 0^\top)^\top      &\text{ if }c = 0\\
                    (X_{2i}^\top, \eta(S_i)^\top)^\top    &\text{ if }c = 1.
                \end{cases}
\end{equation}
In particular, the true conditioning variables satisfy $Z_i^{c}=Z_i^{c}(\eta_0)$.

The following three examples illustrate how the models in \citet{ackerberg_caves_frazer2015}, \citet{petrin_train2010}, and \citet{imbens_newey2009} fit into our framework.

\begin{example}[Production Function Estimation, \citet{ackerberg_caves_frazer2015}]\label{sec03 - example01 - ACF}

We maintain the model~\eqref{sec02:eq:production} in ACF:
\begin{align*}
    Y_{i,t} = & \beta_0 X_{i,t}+\omega_{i,t}+\varepsilon_{i,t}, & \condexpect{\varepsilon_{i,t}}{\mathcal{I}_{i,t}}&= 0,\quad
    t\in\{1,2\},\\
    \omega_{i,2} &= \alpha_0\omega_{i,1}+\xi_{i,2}, &\condexpect{\xi_{i,2}}{\mathcal{I}_{i,1}}&= 0.
\end{align*}
Let $M_{i,1}$ denote an observed first-stage outcome, such as materials. Under first-stage outcome invertibility, $\omega_{i,1}=f_0(X_{i,1},M_{i,1})$. To express the first stage in our framework, set $T_i=Y_{i,1}$ and $S_i=(X_{i,1},M_{i,1})^\top$. Since $S_i$ is measurable with respect to $\mathcal{I}_{i,1}$, the first-stage regression identifies:
\[    V_i\equiv\eta_0(S_i)=\condexpect{T_i}{S_i}    =   \beta_0 X_{i,1}+\omega_{i,1}, \]
and $\eta(S_i)-\beta X_{i,1}$ is the implied period-1 productivity at a candidate $(\beta,\eta)$. Let $W_i = (Y_{i,2}, X_{i,1}, X_{i,2}, M_{i,1})$ and $\theta=(\beta,\alpha)^\top$. Define:
\[    \rho(W_i,\theta,\eta)
        =Y_{i,2} - \beta X_{i,2}-\alpha\{\eta(S_i)-\beta X_{i,1}\}, \qquad
        Z_i^1(\eta)=\left(X_{i,1},\eta(S_i)\right)^\top.\]
At the truth, $\eta_0(S_i)-\beta_0X_{2i}=\omega_{i,1}$ and $\rho(W_i,\theta_0,\eta_0)=\xi_{i,2}+\varepsilon_{i,2}$. Since $Z_i^1(\eta_0)$ is measurable with respect to $\mathcal{I}_{i,1}$, the restricton $\condexpect{\varepsilon_{i,t}}{\mathcal{I}_{i,t}} = 0$ a.s. implies $\condexpect{\rho(W_i,\theta_0,\eta_0)}{Z_i^1(\eta_0)}=0$ a.s.
\end{example}

\begin{example}[Demand Estimation, \citet{petrin_train2010}]\label{sec03 - example02 - PT}

Consider the control-function approach to demand estimation in \citet{petrin_train2010}. Index consumers by $i\in[N]$ and products by $p\in[P]$. Let $X_{1i,p}$ denote an endogenous product characteristic, such as the price of product $p$ faced by consumer $i$, and let $X_{2i,p}$ collect exogenous product characteristics. Stack these variables in $X_{1i}$ and $X_{2i}$, respectively. Let $S_i$ collect the exogenous variables and excluded instruments used in the first-stage regression.

Write the first-stage equation as
\begin{equation*}
    X_{1i}=\eta_0(S_i)+V_i,
    \qquad
    \eta_0(S_i)=\condexpect{X_{1i}}{S_i}.
\end{equation*}
Here, $T_i=X_{1i}$ is the observed first-stage dependent variable. For any candidate $\eta$, define:
\begin{equation*}
    V_i(\eta)=X_{1i}-\eta(S_i),
    \qquad
    V_i=V_i(\eta_0).
\end{equation*}

The component $V_{i,p}$ serves as a control variable for the dependence between the endogenous product characteristic and the unobserved component of utility. In particular, utility is specified as
\begin{equation*}
    \mathcal{U}_{i,p}
    =u(X_{1i,p},X_{2i,p},\beta_0)+CF(V_{i,p},\gamma_0)+\widetilde{\varepsilon}_{i,p},
\end{equation*}
where $u(X_{1i,p}, X_{2i,p}, \beta)$ and $CF(V_{i,p},\gamma)$ are known up to $\beta$ and $\gamma$, respectively. Let $\theta=(\beta^\top,\gamma^\top)^\top$.

Let $Y_{i,p}$ indicate that consumer $i$ chooses product $p$, and let $\Lambda_p(X_{1i},X_{2i},V_i,\theta_0)$ denote its conditional choice probability given $(X_{1i},X_{2i},V_i)$. Define:
\begin{equation*}
    \rho_p(W_i,\theta,\eta)=Y_{i,p}-\Lambda_p\big(X_{1i},X_{2i},X_{1i}-\eta(S_i),\theta\big)\qquad Z_i^1(\eta) = \left( X_{1i}^\top, X_{2i}^\top, \eta(S_i)^\top \right)^\top.
\end{equation*}
Since $V_i(\eta) = X_{1i}-\eta(S_i)$, the model implies $\condexpect{\rho_p(W_i,\theta_0,\eta_0)}{Z_i^1(\eta_0)}=0$ a.s. for each $p\in[P]$.
\end{example}

\begin{example}[Triangular System, \citet{imbens_newey2009}]\label{sec03 - example03 - NI}

Consider the triangular system discussed by \citet{imbens_newey2009}. The researcher observes $(Y_i,X_i,S_i)$, where $X_i$ is endogenous and $S_i$ is an excluded instrument, and the model is:
\begin{align*}
  Y_i &=m_0(X_i,V_i)+\varepsilon_i,
        &\mathbb{E}[\varepsilon_i\mid S_i,V_i]&=0,\\
  X_i &=\eta_0(S_i)+V_i,
        &\mathbb{E}[V_i\mid S_i]&=0.
\end{align*}
The additive first stage is a simplified version of their triangular control function framework. It identifies:
\begin{equation*}
    \eta_0(S_i)=\mathbb{E}[X_i\mid S_i],
        \qquad
    V_i=X_i - \eta_0(S_i).
\end{equation*}
The control $V_i$ is therefore generated by a nonparametric first-stage regression. Suppose further that $m_0(x, v) = m(x, v, \theta_0)$, where $m$ is known up to the finite-dimensional parameter $\theta_{0}\in \mathbb{R}^{d_{\theta}}$. Define
\[ \rho(Y_i,X_i,V_i,\theta) = Y_i - m(X_i, V_i, \theta),\qquad Z_{i}^{1}(\eta) = \left(S_i^\top, X_i - \eta(S_i) \right)^\top. \]
The relevant conditional moment restriction then takes the form $\condexpect{\rho(Y_i,X_i,V_i,\theta_0)}{Z_{i}^{1}(\eta_0)}=0$ a.s.
\end{example}

\subsection{Assumptions}
In this subsection, we introduce two basic assumptions about the first stage and the CMRs.

\hypertarget{shrinking-neighborhood-property}{}Let $\{\mathcal{H}_{N}\}_{N\geq1}$ be a sequence of convex neighborhoods of $\eta_0$ contained in $\mathcal H$, shrinking in the uniform norm so that
\[ \sup_{\eta\in\mathcal{H}_{N}}\lVert\eta-\eta_0\rVert_\infty\to0 \qquad\text{as }N\to\infty, \]
at a rate specified below. We denote the nonparametric estimator of $\eta_0$ by $\widehat\eta$ and require $\Pr(\widehat\eta\in\mathcal{H}_{N})$ to converge to one.

We impose the following conditions on the nuisance space, the first stage learner and its nonparametric estimator:

\begin{assumption}[Nuisance space and the first stage]\leavevmode\vspace{-\baselineskip}\label{sec03 - assumption01 - first stage}
\begin{enumerate}
    \item[(a)] The nuisance class $\mathcal H$ is a convex subset of a normed linear space equipped with the uniform norm $\lVert\cdot\rVert_\infty$.
    \item[(b)] $V_i = \eta_0(S_i)$ almost surely and $\eta_{0}(S)\in\mathcal{H}$ is a solution to equation~\eqref{sec03 - framework - first stage}.
    \item[(c)] There is a positive constant $C_{T}<\infty$ such that 
    \[  \condexpectF{F_0}{ \norm{T - \eta_0(S)}^2 }{S = s} < C_{T}\quad\textnormal{for almost every $s\in\textnormal{supp}(S)$.} \]
    \item[(d)] For each $N>0$ and for any $\eta\in \mathcal{H}_N$, there is a deterministic sequence $a_N\downarrow 0$ such that $\norm{ \eta - \eta_0}_{\infty} = o(a_N)$.
\end{enumerate}
\end{assumption}

Assumption \ref{sec03 - assumption01 - first stage} imposes standard conditions on the nuisance space and the first-stage nuisance function in the DML literature; see, for example, \citet{dml_chernozhukov_et_al2018}. Assumption \ref{sec03 - assumption01 - first stage}(b) requires that the first-stage model non-parametrically identify $V_i = \eta_0(S_i)$. Assumption \ref{sec03 - assumption01 - first stage}(c) requires that the first-stage error evaluated at the truth have finite and positive definite variance. Assumption \ref{sec03 - assumption01 - first stage}(d) specifies the required convergence rate of the nuisance estimator.

Existence and uniqueness of the first-stage solution are population properties. For example, if $\mathbb E[\lVert T \rVert] < \infty$, the conditional mean $\mathbb E[T\mid S]$ exists and is unique almost surely. When this conditional mean belongs to $\mathcal H$ and the test functions in equation~\eqref{sec03 - framework - first stage} are sufficiently rich to characterize the conditional mean restriction, it identifies $\eta_0$ up to equality almost surely under the distribution of $S$. Pointwise identification on the entire support requires additional restrictions, such as continuity of the admissible functions on that support. The identification of the first-stage solution is distinct from the convergence rate of its estimator. Smoothness of $\eta_0$ and compactness of $\operatorname{supp}(S)$ help verify uniform convergence rates for a specified learner, together with the appropriate sampling and tuning conditions.

We next impose the following conditions on the CMR.

\begin{assumption}[Regularity conditions on the restriction $\rho(W,\theta,\eta)$]\leavevmode\vspace{-\baselineskip}\label{sec03 - assumption02 - CMR}
\begin{enumerate}
    \item[(a)] Parameter space $\Theta \subset \mathbb{R}^{d_{\theta}}$ with $d_{\theta}<\infty$ is compact, and $\theta_0\in \textnormal{int}(\Theta)$ is the unique solution to the CMR in equation~\eqref{sec03 - framework - CMR}.
    \item[(b)] $(\theta,\eta) \mapsto \rho(\cdot,\theta,\eta)$ is continuous in $\theta\in\Theta$ with respect to Euclidean norm $\norm{\cdot}$ and $\eta\in\mathcal{H}$ with respect to sup-norm $\norm{\cdot}_\infty$. %There exists an open neighborhood $\textnormal{int}(\Theta)$ of $\theta_0$ satisfying $\textnormal{int}(\Theta)\subset\operatorname{int}(\Theta)$ such that, 
    For each $N>0$ and any $\eta\in\mathcal{H}_{N}$, the map $\theta\mapsto\rho(\cdot,\theta,\eta)$ is twice continuously differentiable on $\textnormal{int}(\Theta)$ almost surely. There exists a positive constant $\widebar{R}$ such that:
    \begin{equation}\label{sec03 - assumption02 - diff - theta}
        \max\left\{ 
            \sup_{\theta\in\Theta,\,\eta\in\mathcal{H}_N}\norm{\rho(\cdot,\theta,\eta)}_{2}, 
            \sup_{\theta\in\textnormal{int}(\Theta_0),\,\eta\in\mathcal{H}_N}\norm{ \nabla_{\theta}\rho(\cdot,\theta,\eta) }_{2},
            \sup_{\theta\in\textnormal{int}(\Theta_0),\,\eta\in\mathcal{H}_N}\norm{ \nabla_{\theta}^2\rho(\cdot,\theta,\eta) }_{2}
            \right\} \leq \widebar{R} < \infty   
    \end{equation}
    \item[(c)] For each conditioning specification $c\in\{0,1\}$ under consideration, 
    \[  \textnormal{\textbf{Var}}\big(\rho(W_i,\theta_0,\eta_0)\mid Z_i^{c}(\eta_0),S_i,T_i\big) >   0   \quad   \textnormal{a.s.} \]
    Furthermore, the matrix $\mathbb E[g_c(Z_i^{c}(\eta_0))g_c(Z_i^{c}(\eta_0))^\top]$ is positive definite, where
    \[  g_c(z)  \equiv  \condexpect{\nabla_{\theta}\rho(W_i, \theta_0, \eta_0)}{Z_{i}^{c}(\eta_0) = z}. \]
    
    \item[(d)] For $\theta\in\textnormal{int}(\Theta)$, $\eta \mapsto \rho(\cdot,\theta,\eta)$ is twice continuously Hadamard-differentiable \footnote{{see, for example, \citet[section~20.2]{vdv1998}.}} in $\eta\in\mathcal{H}_N$ almost surely. For $\theta\in\textnormal{int}(\Theta_0)$, $\eta\mapsto \nabla_\theta\rho(\cdot,\theta,\eta)$ is twice continuously Hadamard-differentiable in $\eta\in\mathcal{H}_N$. There exist nonnegative measurable functions $\widebar{J}(W_i),\widebar{H}(W_i)$, independent of $N$, such that the following bounds hold for every direction $h_1=\eta_1'-\eta_1$, $h_2 = \eta_2' - \eta_2$ with $\eta_1, \eta_1',\eta_2,\eta_2'\in\mathcal{H}_N$:
    \begin{equation}\label{sec03 - assumption02 - diff - eta}
        \begin{aligned}
            \max\left\{ 
                \sup_{\theta\in\Theta,\,\eta\in\mathcal{H}_N} \norm{ D_{\eta}\rho(W_i,\theta,\eta)[h_1] }\, ,\,\,
                \sup_{\theta\in\textnormal{int}(\Theta_0),\,\eta\in\mathcal{H}_N} \norm{ D_{\eta}\big(\nabla_\theta\rho(W_i,\theta,\eta)\big)[h_1] }
            \right\}
                \leq    &\widebar{J}(W_i)  \norm{h_1}_{\infty}&\\
            \max\left\{ 
                \sup_{\theta\in\Theta,\,\eta\in\mathcal{H}_N}\norm{D_\eta^2\rho(W_i,\theta,\eta)[h_1,h_2]}\, ,\,\,
                \sup_{\theta\in\textnormal{int}(\Theta_0),\,\eta\in\mathcal{H}_N}\norm{D_\eta^2\big(\nabla_\theta\rho (W_i,\theta,\eta) \big) [h_1,h_2]}
            \right\}
                \leq    &\widebar{H}(W_i) \norm{h_1}_\infty \norm{h_2}_\infty&
        \end{aligned}       
    \end{equation}
    Furthermore, there exist positive constants $C_{J}$ and $C_{H}$, independent of $N$, such that 
    \[ \Norm{ \widebar{J} }_{2} \leq C_{J} < \infty \quad\text{and}\quad \Norm{ \widebar{H} }_{2} \leq C_{H} < \infty. \]
    \item[(e)] There is a compact subset $\widebar{\Theta}_{o}\subset \textnormal{int}(\Theta)$ including $\theta_0$, such that for each $N>0$, for any $\eta\in\mathcal{H}_N$, the function classes $\mathcal{R}_{\eta}^{0} \equiv \{ \rho(\cdot, \theta, \eta) : \theta\in \Theta \}$ and $\mathcal{R}_{\eta}^{1} \equiv \{ \nabla_{\theta}\rho(\cdot, \theta, \eta) : \theta\in \widebar{\Theta}_{o} \}$ satisfy the following uniform entropy bound for $j = 0, 1$\textnormal{:}
        \begin{equation}\label{sec03 - assumption02 - entropy}
            \sup_{Q}\log{ N( \epsilon \norm{\widebar{\rho}^{j}}_{Q,2}, \mathcal{R}_{\eta}^{j}, \norm{\cdot}_{Q,2} ) } \leq A_0\log(A_1/\epsilon)\quad\text{for all }0 < \epsilon \leq 1,
        \end{equation}
        where $\widebar{\rho}^{j}$ is a measurable envelope for $\mathcal{R}_{\eta}^{j}$ with $\norm{\widebar{\rho}^{j}}_2 < \infty$, and $A_0,A_1$ are finite positive constants. The quantities $\widebar{\rho}^{0}$, $\widebar{\rho}^{1}$, $A_0$, and $A_1$ are independent of $\theta$, $\eta\in \mathcal{H}_N$ and $N$.
\end{enumerate}
\end{assumption}

Assumption \ref{sec03 - assumption02 - CMR} states the identification and regularity conditions for the CMR. Assumption \ref{sec03 - assumption02 - CMR}(a) requires a compact parameter space and a unique interior solution $\theta_0$ to \eqref{sec03 - framework - CMR}. Assumption \ref{sec03 - assumption02 - CMR}(b) imposes continuity in both $\theta$ and $\eta$, as well as twice continuous differentiability in $\theta$ with uniform $L_{2}(P)$ bounds on the derivatives. Assumption \ref{sec03 - assumption02 - CMR}(b) can be replaced with the differentiability locally on an open neighborhood, say $\mathcal{N}_0 \subset \textnormal{int}(\Theta)$ of $\theta_0$. We mainly impose the global differentiability on $\textnormal{int}(\Theta)$ to simplify the proofs. Under the local condition, the proofs of the main results require an additional step to verify that the evaluation points or the relevant estimators lie in $\mathcal{N}_0$ with probability approaching one.

Assumption \ref{sec03 - assumption02 - CMR}(c) requires the generalized residual at the truth to have positive variance conditional on the first-stage variables and the conditioning variable. It also imposes a rank condition on the conditional mean of the gradient of the generalized residual with respect to $\theta$. Together with the other maintained assumptions, these conditions ensure that the asymptotic variance matrix of the DDML estimator, defined below, is finite and positive definite. Assumption \ref{sec03 - assumption02 - CMR}(d) requires that the generalized residual and its parameter gradient change smoothly with perturbations of the first-stage nuisance function. The first and second nuisance derivatives are bounded by square-integrable envelopes uniformly over $\textnormal{int}(\theta)$ and the shrinking nuisance set $\mathcal{H}_N$. Assumption \ref{sec03 - assumption02 - CMR}(e) is  standard conditions for extremum estimator theory with nonlinear restrictions, ensuring the uniform weak law of large number and the uniform central limit theorem; see, for example, \citet{pakespollard1989, vw2023}. When $\rho(W_i,\theta,\eta)$ is affine in $\theta$, both classes are finite-dimensional for each fixed $\eta$, and the entropy bound follows provided that the envelopes are square-integrable.

\subsection{Identification-preserving kernel}

For a kernel function $K : \mathbb{R}^{d_{Z,c}}\times \mathbb{R}^{d_{Z,c}} \to [0,\infty)$, we define the population criterion $Q:\Theta\times\mathcal{H}\to[0,\infty)$ as:
\begin{equation}\label{sec04 - DL criterion}
    Q(\theta,\eta)
    \equiv \expectF{F_0}{
        \rho(W_i,\theta,\eta(S_i))
        \rho(W_j,\theta,\eta(S_j))
        K\big(Z_i^{c}(\eta),Z_j^{c}(\eta)\big) }.
\end{equation}

The key restriction on $K$ is that it preserves the identification provided by the CMR in \eqref{sec03 - framework - CMR}. In particular, at the true nuisance function $\eta_0$, we want the unique solution $\theta_0$ to the CMR to be the unique minimizer of $Q(\theta,\eta_0)$. An integrally strictly positive definite (ISPD) kernel ensures this property; see, for example, \citet{kernel_sriperumbudur_et_al2011,kernel_lavergne_patilea2013,kernel_muandet_et_al2020}. We impose the following conditions on the kernel.

\begin{assumption}[Restrictions on Kernel]\leavevmode\vspace{-\baselineskip}\label{sec03 - assumption03 - kernel}
\begin{enumerate}
    \item[(a)] A function $K : \mathbb{R}^{d_{Z,c}}\times \mathbb{R}^{d_{Z,c}} \to [0,\infty)$ is \textit{integrally strictly positive definite} (ISPD): $K(z_1, z_2) = K(z_2, z_1)$ for any $z_1, z_2 \in \mathbb{R}^{d_{Z,c}}$ and for a generic measure $P$ and any nonzero function $g\in L_2(P)$:
    \[ \iint g(z)K(z,z')g(z')dP(z)dP(z') > 0. \]
    \item[(b)] $K$ is twice continuously differentiable on $\mathbb{R}^{d_{Z,c}}\times \mathbb{R}^{d_{Z,c}}$. There exist finite positive constants $\widebar{K}_0, \widebar{K}_1, \widebar{K}_2$ such that 
    \[ \sup_{z, z'}\norm{ K(z,z') } < \widebar{K}_0 ,\quad \sup_{z, z'}\norm{ \nabla K(z,z') }  <  \widebar{K}_1 \quad\text{and}\quad \sup_{z, z'}\norm{ \nabla^2 K(z,z') }  < \widebar{K}_2. \]
\end{enumerate}
\end{assumption}

The role of the ISPD kernel is to aggregate the infinitely many restrictions in the CMR into a single criterion while preserving identification. In particular, for an ISPD kernel $K$ and a function $g\in L_2(P)$,
\[  \iint g(z)K(z,z')g(z')\,dP(z)\,dP(z')=0     \quad\Longleftrightarrow\quad
    g(z)=0 \quad P\text{-almost surely}.    \]
Taking $P$ to be the distribution of $Z_i^{c}(\eta_0)$ and $g(z)$ to be the conditional mean of $\rho(W_i,\theta,\eta_0(S_i))$ given $Z_i^{c}(\eta_0)=z$, the double integral equals $Q(\theta,\eta_0)$. Thus, $Q(\theta,\eta_0)=0$ if and only if $\theta$ satisfies the CMR in \eqref{sec03 - framework - CMR}. Since $\theta_0$ is the unique solution to the CMR, it is also the unique minimizer of $Q(\theta,\eta_0)$. Together with the rank condition in Assumption \ref{sec03 - assumption02 - CMR}(c), the ISPD condition in Assumption \ref{sec03 - assumption03 - kernel} also ensures that the population score Jacobian is nonsingular at $\theta_0$.

Since the researcher can choose the kernel, the boundedness and smoothness conditions are natural restrictions. In particular, when $c=1$, perturbations of $\eta$ change the kernel through the generated component of $Z_i^1(\eta)$. The smoothness and the boundedness control this change.

For example, the researcher may choose the Gaussian kernel
\[  K_b(z,z')=\exp\!\left(-\frac{\lVert z-z'\rVert^2}{2b^2}\right),\qquad b>0,  \]
with a fixed bandwidth $b$. Its derivatives through order two are bounded, and it is ISPD for probability measures on Euclidean space. We note that the fixed bandwidth $b$ does not affect the convergence rate of the proposed estimator, although it affects its asymptotic variance and may influence its finite-sample performance. Other ISPD kernels can be chosen, see, for example, \citet[p.~2393]{kernel_sriperumbudur_et_al2011}.

\begin{remark}
The original criterion of \citet{cmr_DL2004} also admits a pairwise kernel representation. To display the connection, fix $\eta=\eta_0$, write $Z_i=Z_i^{c}(\eta_0)$ and $\rho_i(\theta)=\rho(W_i,\theta,\eta_0)$, and let $P_Z$ be the distribution of $Z_i$. Define
\[  Q_{\mathrm{DL}}(\theta)=\int \mathbb E[\rho_i(\theta)\mathbf 1\{Z_i\leq u\}]^2\,dP_Z(u),    \]
where inequalities between vectors are componentwise. Expanding the square gives
\begin{align*}
 Q_{\mathrm{DL}}(\theta)    &=\mathbb E[\rho_i(\theta)\rho_j(\theta)K_{\mathrm{DL}}(Z_i,Z_j)],\\
 K_{\mathrm{DL}}(z,z')      &=\int\mathbf 1\{z\leq u\}\mathbf 1\{z'\leq u\}\,dP_Z(u).
\end{align*}
The indicator-based kernel recovers the original DL criterion. Even when scalar $Z$ has a smooth density, $K_{\mathrm{DL}}(z,z')=\Pr(Z\geq\max\{z,z'\})$ has a kink at $z=z'$ wherever the density is positive, violating Assumption~\ref{sec03 - assumption03 - kernel}(b).        
\end{remark}

%%%%%%%%%%%%%%%%%%%%%%%%%%%%%%%%%%%%%%%%%%%%%%%%%%%%%%%%%%%%%%%%%%%%%%%%%%%%%%%%%%%%%%%%%%%%%%%%%%%%%%%%%%%%%%%%%%%%%%%%%%%%%%%%%%%%%%%%%%%%%%%%%%%%%%%%%%%%%%%%%%%%%%%%%%%%%%%%%%%%%%%%%%%%%%%%%%%%%%%%%%%%%%%%%%%%%%%%%%%%%%%%%%%%%%%%%%%%%%%%%%%%%%%%%%%%%%%%%%%%%%%%%%%%%%%%%%%%%%%%%%%%%%%%%%%%%%%%%%%%%%%%%%%%%%%%%%%%%%%%%%%%%%%%%%%%%%%%%%%%%%%%%%%%%%%%%%%%%%%%%%%%%%%%%%%%%%%%%%%%%%%%%%%%%%%%%%%%%%%%%%%%%%%%%%%%%%%%%%%%%%%%%%%%%%%%%%%%%%%%%%%%%%%%%%%%%%%%%%%%%%%%%%%%%%%%%%%%%%%%%%%%%%%%%%%%%%%%%%%%%%%%%%%%%%%%%%%%%%%%%%%%%%%%%%%%%%%%%%%%%%%%%%%%%%%%%%%%%%%%%%%%%%%%%%%%%%%%
\section{DDML estimator and its large-sample properties}\label{sec04}
In this section, we study the convergence speed of the plug-in estimator. Next, we show how to construct the Neyman-orthogonal score accordingly, and provide the specific form of the orthogonal score for several examples. Then we define the DDML estimator, analyze the asymptotic properties of the DDML estimator such as the asymptotic normality, and provide sufficient conditions for the convergence rate of the first-stage nuisance estimates for valid inference.

\subsection{Plug-in Estimator} \newcommand{\plugin}{ \widehat{\theta}_{\textnormal{plugin}}^{c} }
To construct the plug-in estimator, we first define the sample criterion $(\theta,\eta)\mapsto Q_N(\theta,\eta)$ as the following symmetric U-statistic:
\[ Q_N(\theta,\eta) \equiv \binom{N}{2}^{-1}\sum_{i<j}\rho(W_i,\theta,\eta(S_i))\rho(W_j,\theta,\eta(S_j))K(Z_i^{c}(\eta),Z_j^{c}(\eta)), \]
which is a sample analogue of the population criterion function given in equation~\eqref{sec04 - DL criterion}. Then, the plug-in estimator $\plugin$ is defined as a minimizer of the criterion $Q_N(\theta,\widehat\eta)$, where $\widehat\eta$ denotes the nonparametric first-stage estimate; or, formally, we have:
\[\plugin \equiv \argmin{\theta\in \Theta }{Q_N(\theta,\widehat{\eta})}.\]

By Assumption \ref{sec03 - assumption02 - CMR}(b), the map $\theta\mapsto\rho(\cdot,\theta,\eta)$ is smooth on $\textnormal{int}(\Theta)$, and hence so is the sample criterion $Q_N(\theta,\eta)$. Therefore, if the plug-in estimator lies in $\textnormal{int}(\Theta)$, it satisfies the following sample first-order condition:
\[ \binom{N}{2}^{-1}\sum_{i<j} r^{c}\big(W_i, W_j, \theta, \widehat{\eta}(\cdot)\big) = 0\quad(\textnormal{or } o_{P}(1)\,\, ) \]
Here, the score function is given by:
\begin{equation}
 r^{c}(W_i,W_j,\theta,\eta)
 \equiv \nabla_\theta\big\{
    \rho\big(W_i,\theta,\eta(S_i)\big)
    \rho\big(W_j,\theta,\eta(S_j)\big)
 \big\}\,K\big(Z_i^{c}(\eta),Z_j^{c}(\eta)\big).
\end{equation}

The naive plug-in estimator $\plugin$ need not attain the parametric $N^{-1/2}$ convergence rate, because it can inherit the slower convergence rate of the nuisance estimator $\widehat\eta$. The following lemma formalizes this dependence.

\begin{lemma}\label{sec04 - lemma01}
    Suppose Assumptions \ref{sec03 - assumption01 - first stage}, \ref{sec03 - assumption02 - CMR}, and \ref{sec03 - assumption03 - kernel}. Then,
    \[ \norm{\plugin - \theta_0} = O_{P}(N^{-1/2}) + O_{P}(\norm{\hat{\eta} - \eta_0}_{\infty}).  \]    
\end{lemma}

\textbf{Proof of Lemma \ref{sec04 - lemma01} : }See Appendix \ref{sec04 - lemma01 - proof}.\\

As Lemma \ref{sec04 - lemma01} shows, a sufficient condition for root-$N$ consistency of the plug-in estimator is $\sqrt{N}\lVert\widehat\eta-\eta_0\rVert_\infty=O_P(1)$. Such a uniform root-$N$ rate is generally unavailable for many flexible machine-learning first-stage estimators. Subsection \ref{sec04 - orthogonal score construction} derives Neyman-orthogonal scores, and Subsection \ref{sec04 - DDML construction} proposes an estimator based on these scores that accommodates slower first-stage convergence.

\subsection{Construction of Neyman-Orthogonal Score}\label{sec04 - orthogonal score construction} \newcommand{\dt}[1]{ \left.\frac{d}{d\tau}{#1}\right|_{\tau = 0^+} }
In this subsection, we construct a Neyman-orthogonal score from $r^{c}(W_i,W_j,\theta,\eta)$. For an index $\tau\in[0,1]$, consider the mixture distribution
\[  F_\tau=F_0+\tau(F-F_0), \]
where $F$ is an alternative distribution. Let $\eta_\tau$ be a solution to the following first-stage equation under $F_\tau$:
\[  \mathbb E_{F_\tau}\!\left[\varsigma(S_i)\{T_i-\eta_\tau(S_i)\}\right]=0 \qquad\text{for every }\varsigma\in\mathcal H. \]
At $\tau=0$, this solution is $\eta_0$. The argument of $\eta_\tau$ is the first-stage regressor $S_i$, not the conditioning variable $Z_i^{c}$. We consider alternative distributions $F$ for which the first-stage solution exists near $\tau=0$ and the differentiation and integrability conditions used below hold \citep{dml_chernozhukov_et_al2022,semi_ichimura_newey2022}. For functions of the independent pair $(W_i,W_j)$, expectation under $F_\tau$ means integration under $F_\tau\otimes F_\tau$.

Our objective is to find a bias-correction term $\phi(\cdot)$, involving a pair of additional nuisance functions
\[  \vecdel^{c}(\cdot,\theta)=\big(\vecdel_1^{c}(\cdot,\theta),\vecdel_2^{c}(\cdot,\theta)\big), \]
to be specified below. Along the distribution path $F_\tau$, write this pair as $\vecdel_\tau^{c}(\cdot,\theta)$. We seek a correction such that the function
\[  \psi^{c}\big(W_i,W_j,\theta,\eta_\tau,\vecdel_\tau^{c}(\cdot,\theta)\big)  = r^{c}(W_i,W_j,\theta,\eta_\tau) +\phi\big(W_i,W_j,\eta_\tau,\vecdel_\tau^{c}(\cdot,\theta)\big)  \]
satisfies the following notion of Neyman orthogonality \citep{dml_chernozhukov_et_al2018,dml_chernozhukov_et_al2022,dml_ahrens_et_al2026,dml_escanciano_terschuur2026}.

\begin{definition}[Neyman Orthogonality]\leavevmode\label{sec04 - definition - neyman orthogonality}
For every $\theta\in\textnormal{int}(\Theta)$ and every admissible distribution path $F_\tau=F_0+\tau(F-F_0)$, let $\eta_\tau$ and $\vecdel_\tau^{c}(\cdot,\theta)$ denote the associated nuisance functions. The following two conditions hold:
\begin{enumerate}
    \item[(i)] The correction term has mean zero along the path: for some $\widebar\tau\in(0,1]$,
        \[ \mathbb{E}_{F_\tau\otimes F_\tau}\!\left[\phi\big(W_i,W_j,\eta_\tau,\vecdel_\tau^{c}(\cdot,\theta)\big)\right]=0 \quad\text{for every }\tau\in[0,\widebar\tau).  \]
    \item[(ii)] Holding the distribution in the expectation fixed at the true distribution, the expected score is locally insensitive to nuisance perturbations:
        \begin{equation}\label{sec04 - neyman}
        \left.\frac{d}{d\tau}\mathbb{E}_{F_0\otimes F_0}\!\left[\psi^{c}\big(W_i,W_j,\theta,\eta_\tau,\vecdel_\tau^{c}(\cdot,\theta)\big)\right]\right|_{\tau=0^+}=0.
    \end{equation}
\end{enumerate}
\end{definition}

It turns out that the bias-correction term can be constructed as follows:
\begin{equation}\label{sec04  - def01 - phi}
    \phi(W_i, W_j, \eta, \vecdel^{c}) \equiv \vecdel_{1}^{c}(S_i, \theta)(T_{i} - \eta(S_i)) + \vecdel_{2}^{c}(S_j, \theta)(T_{j} - \eta(S_j)).
\end{equation}
We work with the following high-level condition to ensure the existence of the bias-correction term, commonly used in the literature \citep{dml_chernozhukov_et_al2022,dml_ahrens_et_al2026,dml_escanciano_terschuur2026}:

\paragraph*{Condition (Linear representation)} There exist two population $L^2$-functions $\vecdel_{0,1}^{c}(S_i, \theta)$, $\vecdel_{0,2}^{c}(S_j, \theta)$ and a finite positive constant $C_{\delta}$ such that:
    \begin{equation}\label{sec04 - assumption04 - a1 - eq02}
        \sup_{\theta\in\textnormal{int}(\Theta_0) }\big\{ \norm{\vecdel_{0,1}^{c}(S, \theta)}_{2} + \norm{\vecdel_{0,2}^{c}(S, \theta)}_{2}  \big\} < C_{\delta},
    \end{equation}
    \begin{equation}\label{sec04 - assumption04 - a1 - eq01}
        \dt{ \expectF{F_0\otimes F_0}{ r^{c}(W_i, W_j, \theta, \eta_{\tau}(\cdot)) } } = \dt{ \expectF{F_0\otimes F_0}{ \vecdel_{0,1}^{c}(S_i, \theta)\eta_{\tau}(S_i) + \vecdel_{0,2}^{c}(S_j, \theta)\eta_{\tau}(S_j)} },     
    \end{equation}
    for every $\theta\in\textnormal{int}(\Theta_0)$. \vspace{+0.5cm}

\citet{dml_chernozhukov_et_al2024} discuss estimating Riesz representers through Riesz regression. Here, we impose a sufficient condition that allows us to derive an explicit form of the bias-correction term.

\begin{assumption}[Pointwise Differentiability in the Nuisance]\label{sec04 - assumption04 - linear representation}

Two functions $v\mapsto \rho(\cdot, \theta, v)$ and $v\mapsto \nabla_{\theta}\rho(\cdot, \theta, v)$ are continuously differentiable in $v\in \mathbb{R}^{d_{V}}$ almost surely. For each $N>0$, for any direction $h = \eta' - \eta_0$ such that $\eta'\in\mathcal{H}_N$, the following equations hold:
    \begin{equation}\label{sec04 - assumption04 - a2 - eq01}
        \begin{aligned}
            D_{\eta} \rho(W_i, \theta, \eta_0)[h] 
                =   &\frac{\partial}{\partial v}\rho(W_i,\theta,v)\Big|_{v = \eta_0(S_i)}\cdot h(S_i),&\\
            D_{\eta} \big(\nabla_\theta\rho(W_i, \theta, \eta_0)\big)[h]
                =   &\frac{\partial}{\partial v}\big(\nabla_\theta \rho(W_i,\theta,v)\big)\Big|_{v = \eta_0(S_i)}\cdot h(S_i).&
        \end{aligned} 
    \end{equation}
\end{assumption}\vspace{+0.5cm}

The population correction functions need not be known to the researcher. Their conditional-expectation representations below depend on the unknown data distribution and first-stage function; the feasible estimator therefore estimates them from the training sample.

Assumption \ref{sec04 - assumption04 - linear representation} requires that the nuisance function $\eta$ enter the residual function only through its evaluation at $S$, namely, $\rho(W,\theta,\eta)=\rho(W,\theta,\eta(S))$.

If we suppose Assumptions \ref{sec03 - assumption02 - CMR}, \ref{sec03 - assumption03 - kernel} and \ref{sec04 - assumption04 - linear representation}, we can further construct the following functions to define the bias-correction term:
\begin{equation}\label{sec04 - def02 - delta}
    \begin{aligned}
        \vecdel_{0,1}^{c}(S_i, \theta)
            \equiv  &\delta_{0,11}^{c}(S_i,\theta) + \ind{}{c=1}\delta_{0,12}^{c}(S_i, \theta),&\\
        \vecdel_{0,2}^{c}(S_j, \theta)
            \equiv  &\delta_{0,21}^{c}(S_j,\theta) + \ind{}{c=1}\delta_{0,22}^{c}(S_j, \theta),&
    \end{aligned}
\end{equation}
where
\begin{equation}\label{sec04 - def02 - delta2}
    \begin{aligned}
        \delta_{0,11}^{c}(S_i,\theta)
            \equiv &\condExpectF{F_0\otimes F_0}{ \frac{\partial}{\partial v}\nabla_{\theta}\Big( \rho(W_i, \theta, v) \cdot \rho(W_j, \theta, \eta_0(S_j))\Big)\Big|_{v = \eta_0(S_i)} \cdot  K( Z_{i}^{c}(\eta_0), Z_{j}^{c}(\eta_0) ) }{S_i}&\\
        \delta_{0,21}^{c}(S_j, \theta)
            \equiv &\condExpectF{F_0\otimes F_0}{ \frac{\partial}{\partial v}\nabla_{\theta}\Big( \rho(W_i, \theta, \eta_0(S_i)) \cdot \rho(W_j, \theta, v)\Big)\Big|_{ v = \eta_0(S_j) }\cdot  K( Z_{i}^{c}(\eta_0), Z_{j}^{c}(\eta_0) ) }{S_j}&\\
        \delta_{0,12}^{c}(S_i, \theta)
            \equiv &\condExpectF{F_0\otimes F_0}{ \nabla_{\theta}\Big( \rho(W_i, \theta, \eta_0(S_i)) \cdot \rho(W_j, \theta, \eta_0(S_j))\Big) \cdot \frac{\partial}{\partial v}K( (X_{2i}, v), Z_{j}^{c}(\eta_0) )\Big|_{v = \eta_0(S_i)} }{S_i}&\\
        \delta_{0,22}^{c}(S_j, \theta)
            \equiv &\condExpectF{F_0\otimes F_0}{\nabla_{\theta}\Big( \rho(W_i, \theta, \eta_0(S_i)) \cdot \rho(W_j, \theta, \eta_0(S_j)) \Big) \cdot  \frac{\partial}{\partial v}K( Z_{i}^{c}(\eta_0), (X_{2j}, v) )\Big|_{v = \eta_0(S_j)}  }{S_j}&
    \end{aligned}
\end{equation}
The following proposition shows that a new score $\psi^{c}(W_i, W_j, \theta, \eta, \vecdel^{c}) \equiv r^{c}(W_i, W_j, \theta, \eta) + \phi(W_i, W_j, \eta, \vecdel^{c}(\cdot,\theta) )$ obeys the Neyman orthogonality.

\begin{proposition}\label{sec04 - proposition01}

Under Assumptions \ref{sec03 - assumption01 - first stage}, \ref{sec03 - assumption02 - CMR}, \ref{sec03 - assumption03 - kernel}, and \ref{sec04 - assumption04 - linear representation}, two functions $\vecdel_{0,1}^{c}(\cdot)$ and $\vecdel_{0,2}^{c}(\cdot)$ defined by equations~\eqref{sec04 - def02 - delta} and \eqref{sec04 - def02 - delta2} satisfy equations~\eqref{sec04 - assumption04 - a1 - eq01} and \eqref{sec04 - assumption04 - a1 - eq02}, and therefore, the score $\psi^{c}(W_i, W_j, \theta, \eta(\cdot), \vecdel_{0}^{c}(\cdot, \theta))$ obeys the Neyman orthogonality.
\end{proposition}

\textbf{Proof of Proposition \ref{sec04 - proposition01}} : See Appendix \ref{sec04 - proposition01 - proof}.\\

Our construction extends the conditional moment restriction setting of \citet{dml_escanciano_terschuur2026} by allowing the conditioning variables to depend on the first-stage nuisance. When $c=0$, the conditioning variables and the kernel do not depend on $\eta$, and the bias-correction term reduces to the form considered in their setting. When $c=1$, the generated conditioning variables introduce the additional terms $\delta_{12}^{c}$ and $\delta_{22}^{c}$ through the derivative of the kernel.

The explicit forms of $\vecdel_{0,1}^{c}$ and $\vecdel_{0,2}^{c}$ for the production function example and the demand function example with a control function approach are provided in Appendices~\ref{app:implementation_ACF} and~\ref{app:implementation_cereal}, respectively.

\subsection{DDML estimator and its large-sample properties}\label{sec04 - DDML construction}
We now construct a cross-fitted DDML estimator. Following \citet{dml_escanciano_terschuur2026}, we partition the observation pairs into evaluation folds. For each fold, the nuisance functions are estimated using observations that do not appear in any pair in that fold. We then pool the estimated scores across folds to construct a DDML2-type estimator.\footnote{In standard DDML settings, \citet{dml_velez2026} shows superior properties of DDML2 relative to DDML1. While it is unknown whether the same holds in the U-statistic setting, we use DDML2.}

\begin{definition}[DDML Estimator]\leavevmode\label{sec04 - definition01 - DML}
\begin{itemize}
    \item[(1)] Let $G\geq 2$ be a fixed integer, and suppose that $N$ is divisible by $G$. Partition $[N]$ into $G$ groups of equal size $N/G$. Following Algorithm 2 of \citet{dml_escanciano_terschuur2026}, partition the set of index pairs $\{(i,j)\in [N]^2 : i < j\}$ into $L=G(2G-1)$ folds. These folds satisfy $I_{1}\cup...\cup I_{L} = \{(i,j)\in [N]^2 : i < j\}$ and $I_{\ell}\cap I_{\ell'} = \emptyset$ for $\ell\ne\ell'$. Let $J_{\ell} \equiv \{i\in [N] : i$ appears in some pair in $I_{\ell}\}$ and $n_{\ell} \equiv |J_{\ell}|$. By construction, $n_{\ell}=N/G+O(1)$ and $N/n_{\ell}\to G$ for every $\ell\in[L]$. Let $I_{-\ell} \equiv \{(i,j)\in [N]^2 : i < j,\; i,j\notin J_{\ell}\}$. 
    \item[(2)] For each $\ell\in[L]$, use observations $W_i$ with $i\notin J_{\ell}$ to obtain a primitive plug-in estimate $\widetilde{\theta}_{\ell}$ and nuisance estimates $\widehat{\eta}_{\ell}(\cdot)$ and $\widehat{\vecdel}_{\ell}^{c}(\cdot, \theta)$.
    \item[(3)] For each $\ell\in [L]$ and for a pair of observations $(i,j)\in I_{\ell}$, we obtain the orthogonal score $\psi_{\ell}^{c}(W_i, W_j, \theta)$,
    \begin{equation}\label{sec04 - definition01 - DML - eq01}
        \psi_{\ell}^{c}(W_i, W_j, \theta) \equiv r^{c}\big( W_i, W_j, \theta, \widehat{\eta}_{\ell}\big) + \phi(W_i, W_j, \widehat{\eta}_{\ell}, \widehat{\vecdel}_{\ell}^{c}(\cdot, \widetilde{\theta}_{\ell})).
    \end{equation} 
    \item[(4)] Choose a sufficiently small radius $\epsilon>0$ and hold it fixed as $N$ increases. The DDML estimator $\widehat{\theta}_{\textnormal{DDML}}^{c}$ is a solution to the following minimization problem:
    \begin{equation}\label{sec04 - definition01 - DML - eq02}
        \widehat{\theta}_{\textnormal{DDML}}^{c} = \argmin{ \substack{\theta \in \widebar{\Theta}_{o},\,
                                            \| \theta - \plugin \|\leq \epsilon} }
                                            { \Norm{ \binom{ N }{2}^{-1}  \sum_{\ell=1}^{L} \sum_{(i,j) \in I_{\ell}}\psi_{\ell}^{c}(W_i, W_j, \theta)  }^2 }.
    \end{equation}
\end{itemize}
\end{definition}

\begin{remark}
In practice, one may start a local numerical search at the full-sample plug-in estimate $\plugin$ without specifying $\epsilon$. The fixed radius restricts the theoretical analysis to a neighborhood of the pilot. A numerical search implements the definition if it finds a minimizer of the criterion over that neighborhood.
\end{remark}

The correction in equation~\eqref{sec04 - definition01 - DML - eq01} is evaluated at the training-sample pilot and held fixed when optimizing over $\theta$. To distinguish this fitted correction from the estimated family $\widehat{\vecdel}_{\ell}^{c}(\cdot,\theta)$, define
\begin{equation}\label{sec04 - effective correction}
    \check{\vecdel}_{\ell}^{c}(\cdot)
    \equiv \widehat{\vecdel}_{\ell}^{c}(\cdot,\widetilde\theta_{\ell}),
    \qquad \vecdel_0^{c}(\cdot)\equiv\vecdel_0^{c}(\cdot,\theta_0).
\end{equation}
Once estimated from the training sample, the pilot and fitted correction are held fixed when optimizing over $\theta$. Consequently, $\nabla_\theta\psi_{\ell}^{c}(\theta)=\nabla_\theta r^{c}(\theta,\widehat{\eta}_{\ell})$.

We additionally impose the following conditions to obtain a Gaussian approximation result for the DDML estimator $\widehat{\theta}_{\textnormal{DDML}}^{c}$.

\begin{assumption}[Algorithm stability and restriction on convergence rate]\leavevmode\vspace{-\baselineskip}\label{sec04 - assumption05 - algorithm stability}
\begin{enumerate}
    \item[(a)] For each fold $\ell\in[L]$, $\prob{\widehat{\eta}_{\ell}\in\mathcal{H}_{N}}\longrightarrow1$. Furthermore, for each $\ell\in [L]$ and two vanishing sequences $b_N$ and $a_N$, which appears in Assumption \ref{sec03 - assumption01 - first stage}:
    \[  \norm{ \widehat{\eta}_{\ell} - \eta_0 }_{\infty} = o_{P}(a_N)\quad\text{and}\quad 
        \norm{ \check{\vecdel}_{\ell}^{c} - \vecdel_{0}^{c} }_{\infty} = o_{P}(b_N).\]
    \item[(b)] Two vanishing sequences $a_N$ and $b_N$ satisfy the following conditions:
    \[  \sqrt{N}a_N^2 = O(1) \quad\text{and}\quad \sqrt{N} a_N b_N = O(1).\]
\end{enumerate}
\end{assumption}

Assumption \ref{sec04 - assumption05 - algorithm stability}(a) controls the estimation errors of the first-stage function and the fitted correction. Because the correction is evaluated at the estimated pilot $\widetilde\theta_{\ell}$, its error includes the effect of estimating this pilot. Together, parts (a) and (b) ensure that the second-order term remaining in the estimated score is negligible at the $N^{-1/2}$ scale.

\paragraph{The role of the pilot.}For a preliminary estimate $\widetilde\theta_{\ell}\in\textnormal{int}(\Theta)$, the total correction error can be decomposed as
\[  \check{\vecdel}_{\ell}^{c}-\vecdel_0^{c}(\cdot,\theta_0)
        ={} \big\{\widehat{\vecdel}_{\ell}^{c}(\cdot,\widetilde\theta_{\ell})
              -\vecdel_0^{c}(\cdot,\widetilde\theta_{\ell})\big\}
            +\big\{\vecdel_0^{c}(\cdot,\widetilde\theta_{\ell})
              -\vecdel_0^{c}(\cdot,\theta_0)\big\}. \]
Now suppose that there exists a constant $C_\theta<\infty$ such that, for all $\theta,\theta'\in\textnormal{int}(\Theta)$,
\[  \sup_{s\in\operatorname{supp}(S)}   \big\|\vecdel_0^{c}(s,\theta)-\vecdel_0^{c}(s,\theta')\big\|  \leq C_\theta\|\theta-\theta'\|.\]
Writing the norm of the first term as $e_{\delta,\ell}$ and $e_{\theta,\ell}=\norm{\widetilde\theta_{\ell}-\theta_0}$, we then have
\[  \norm{\check{\vecdel}_{\ell}^{c}-\vecdel_0^{c}}_\infty  \leq  e_{\delta,\ell}+C_\theta e_{\theta,\ell}. \]
For example, if
\[  \max\left\{ \sup_{\theta\in\textnormal{int}(\Theta)}    \big\|\widehat{\vecdel}_{\ell}^{c}(\cdot,\theta)-\vecdel_0^{c}(\cdot,\theta)\big\|_\infty\,,\,  \|\widetilde\theta_{\ell}-\theta_0\| \right\}=o_P(b_N),   \]
then the preceding bound implies the correction rate in Assumption \ref{sec04 - assumption05 - algorithm stability}(a). A practitioner can therefore verify the correction rate in Assumption \ref{sec04 - assumption05 - algorithm stability}(a) by checking these sufficient conditions for the chosen correction estimator and preliminary parameter estimator. The proof shows that the conditional mean of the estimated score at $\theta_0$, given the training observations and multiplied by $\sqrt{N}$, has order
\[  O_P\!\left(\sqrt{N}\left[   e_{\eta,\ell}^2+e_{\eta,\ell}e_{\delta,\ell}+e_{\eta,\ell}e_{\theta,\ell}   \right]\right).\]
Here, $e_{\eta,\ell}=\|\widehat{\eta}_{\ell}-\eta_0\|_\infty$ is the first-stage error, $e_{\delta,\ell}$ is the correction regression error defined above, and $e_{\theta,\ell}=\|\widetilde\theta_{\ell}-\theta_0\|$ is the preliminary parameter error. Thus, requiring each of the three products to be $o_P(N^{-1/2})$ makes this conditional mean negligible at the $N^{-1/2}$ scale. We also require the fitted correction to converge to its population target.

Under Assumption \ref{sec04 - assumption05 - algorithm stability}, the plug-in rate in Lemma \ref{sec04 - lemma01} ensures that the preliminary parameter error contributes $o_P(1)$ to the root-$N$-scaled conditional mean. The correction regression error must still satisfy $\sqrt{N}\,e_{\eta,\ell}e_{\delta,\ell}=o_P(1)$.

\paragraph{$\sqrt{N}$-consistency of the DDML estimator}  To state the asymptotic normality result for the DDML estimator $\widehat{\theta}_{\textnormal{DDML}}^{c}$, we introduce the Jacobian matrix $\Gamma_c$ and the asymptotic variance matrix of the score average, $\Sigma_c$, defined as follows:
\begin{equation}\label{sec04:eq:scorejacobian}
    \Gamma_{c} = \Expect{ \nabla_{\theta} r^{c}\big(W_i, W_j,\theta_0,\eta_0\big) } ,
\end{equation}
\begin{equation}\label{sec04:eq:scoreprojectioncovariance}
    \Sigma_{c} = 4\Expect{ \, \condexpect{\psi^{c}(W_i, W_j, \theta_0, \eta_0, \vecdel_0^{c})}{W_i}\condexpect{\psi^{c}(W_i, W_j, \theta_0, \eta_0, \vecdel_0^{c})}{W_i}^\top\, }.
\end{equation}

Now we are ready to provide the main asymptotic result of the DDML estimator:

\begin{theorem}\label{sec04 - theorem01 - dml}
    Suppose Assumptions~\ref{sec03 - assumption01 - first stage}, \ref{sec03 - assumption02 - CMR}, \ref{sec03 - assumption03 - kernel}, \ref{sec04 - assumption04 - linear representation}, and \ref{sec04 - assumption05 - algorithm stability} hold. Then,
    \begin{itemize}
        \item[(i)] $\widehat{\theta}_{\textnormal{DDML}}^{c} \overset{p}{\longrightarrow} \theta_0$.
        \item[(ii)] For the variance matrix $\Omega_{c} \equiv \big\{ \Gamma_{c}^{\top}\Gamma_{c}\big\}^{-1}\Gamma_{c}^\top\Sigma_{c}\Gamma_{c}\big\{\Gamma_{c}^{\top}\Gamma_{c}\big\}^{-1}$,
        \begin{equation}
            \sqrt{N}(\widehat{\theta}_{\textnormal{DDML}}^{c} - \theta_0)
                \overset{d}{\longrightarrow}    \mathcal{N}\left(0\,,\,\, \Omega_{c} \right).
        \end{equation}
    \end{itemize}
\end{theorem}

\textbf{Proof of Theorem \ref{sec04 - theorem01 - dml}} : See Appendix \ref{sec04 - theorem01 - dml - proof}.\\

\paragraph{Consistent estimator for the asymptotic variance} For valid inference, the asymptotic variance should be consistently estimated. We define the estimators for the Jacobian $\Gamma_{c}$ and the variance matrix $\Sigma_{c}$ as follows:
\begin{equation}\label{sec04 - eq - variance}
    \begin{aligned}
    &\widehat{\Gamma}_{c}
    \equiv  \binom{ N }{2}^{-1}\sum_{\ell\in [L]}\sum_{(i,j)\in I_{\ell}}\nabla_\theta \psi^{c}(W_i, W_j, \widehat{\theta}_{\textnormal{DDML}}^{c}, \widehat{\eta}_{\ell}, \check{\vecdel}_{\ell}^{c}),&\\
    &\widehat{\Sigma}_{c}
    \equiv \frac{4}{N(2G-1)}\sum_{\ell\in[L]}\sum_{i\in J_{\ell}}  \widebar{\psi}_{\ell,i}^{c}\widebar{\psi}_{\ell,i}^{c\top}
    \quad\text{where}\quad
    \widebar{\psi}_{\ell,i}^{c} \equiv \frac{1}{n_{\ell}-1}\sum_{j\in J_{\ell}\backslash\{i\}}\psi^{c}(W_i, W_j, \widehat{\theta}_{\textnormal{DDML}}^{c}, \widehat{\eta}_{\ell}, \check{\vecdel}_{\ell}^{c}).&
    \end{aligned}    
\end{equation}
Then, the following proposition establishes that, for each $c\in\{0,1\}$, $\widehat{\Gamma}_{c}$ and $\widehat{\Sigma}_{c}$ are consistent estimators of $\Gamma_{c}$ and $\Sigma_{c}$, respectively, and so is the variance estimator $\widehat{\Omega}_{c}
    \equiv  \big\{ \widehat{\Gamma}_{c}^\top\widehat{\Gamma}_{c} \big\}^{-1} \widehat{\Gamma}_{c}^\top \widehat{\Sigma}_{c} \widehat{\Gamma}_{c} \big\{ \widehat{\Gamma}_{c}^\top \widehat{\Gamma}_{c} \big\}^{-1}.$

\begin{proposition}\label{sec04 - proposition02}
    Under Assumptions~\ref{sec03 - assumption01 - first stage}, \ref{sec03 - assumption02 - CMR}, \ref{sec03 - assumption03 - kernel}, \ref{sec04 - assumption04 - linear representation}, and \ref{sec04 - assumption05 - algorithm stability},  
    $\widehat{\Gamma}_{c}\overset{p}{\longrightarrow} \Gamma_{c}$ and $\widehat{\Sigma}_{c}\overset{p}{\longrightarrow} \Sigma_{c}$. Therefore, $\widehat{\Omega}_{c}\overset{p}{\longrightarrow} \Omega_{c}$.
\end{proposition}

\textbf{Proof of Proposition \ref{sec04 - proposition02}} : See Appendix \ref{sec04 - proposition02 - proof}.\\

\paragraph{DDML estimator with vector-valued restrictions}
Throughout this section, we mainly analyze the DDML estimator for the CMRs with a scalar-valued restriction. Next, we provide the DDML estimator and its limiting distribution when the researcher has $M < \infty$ valid restrictions, each of which uniquely identifies $\theta_0$ through the following CMRs:
\[ \condexpect{\rho_{m}(W_i,\theta,\eta_0)}{Z_i^{c}(\eta_0)} = 0\quad\text{almost surely if and only if}\quad\theta=\theta_0,\qquad m \in [M]. \]
For each $m\in[M]$, let $\psi_m^{c}(W_i,W_j,\theta,\eta,\vecdel_{m}^{c})$ and $\widehat{\theta}_{N,m}^{c}$ denote the corresponding Neyman-orthogonal score and DDML estimator. We combine these estimators using a minimum-distance criterion. Let $\widehat{\pmb{\theta}}_N^{c}\equiv\big[ (\widehat{\theta}_{N,1}^{c})^\top,\ldots,(\widehat{\theta}_{N,M}^{c})^\top\big]^\top$. We define a block matrix $\pmb{\Omega}_c=\{\pmb{\Omega}_{c,mm'}\}_{m,m'\in[M]}$, where for $m,m'\in [M]$,
\begin{align*}
    \Gamma_{c,m}
        \equiv  &\expect{\nabla_\theta\psi_m^{c}(W_i,W_j,\theta_0,\eta_0,\vecdel_{0,m}^{c})}&\\
    \Sigma_{c,mm'}
        \equiv  &4\Expect{\condexpect{\psi_m^{c}(W_i,W_j,\theta_0,\eta_0,\vecdel_{0,m}^{c})}{W_i}\condexpect{\psi_{m'}^{c}(W_i,W_j,\theta_0,\eta_0,\vecdel_{0,m'}^{c})}{W_i}^\top}&\\
    \pmb{\Omega}_{c,mm'}
        \equiv  &\big( \Gamma_{c,m}^\top\Gamma_{c,m}\big)^{-1} \Gamma_{c,m}^\top\Sigma_{c,mm'}\Gamma_{c,m'} \big(\Gamma_{c,m'}^\top\Gamma_{c,m'}\big)^{-1}.&
\end{align*}
A minimum-distance estimator combining the $M$ estimates is:
\[ \widehat{\theta}_{\textnormal{MD}}^{c}\equiv\argmin{\theta\in\Theta}{\big(\widehat{\pmb{\theta}}_N^{c}-\pmb{A}\theta\big)^\top\widehat{\pmb{\Omega}}_c^{-1}\big(\widehat{\pmb{\theta}}_N^{c}-\pmb{A}\theta\big)}, \]
where $\pmb{A}\equiv\pmb{1}_M\otimes\mathbb{I}_{d_\theta}$, $\otimes$ denotes the Kronecker product, $\pmb{1}_M$ the $M$-dimensional column vector of $1$'s, and $\mathbb{I}_{d_\theta}$ the $d_{\theta}$-dimensional identity matrix. The following corollary characterizes the limit distribution of $\sqrt{N} (\widehat{\theta}_{\textnormal{MD}}^{c}-\theta_0)$:

\begin{corollary}\label{sec04 - corollary01}
Suppose Assumptions~\ref{sec03 - assumption01 - first stage}, \ref{sec03 - assumption02 - CMR}, \ref{sec03 - assumption03 - kernel}, \ref{sec04 - assumption04 - linear representation}, and \ref{sec04 - assumption05 - algorithm stability} hold for each restriction $\rho_m$, $m\in[M]$ with fixed $M<\infty$, with the same true parameter $\theta_0$. Suppose that $\pmb{\Omega}_c$ is positive definite and $\widehat{\pmb{\Omega}}_c\overset{p}{\longrightarrow}\pmb{\Omega}_c$. Then the minimum-distance estimator satisfies:
\[ \sqrt{N}(\widehat{\theta}_{\mathrm{MD}}^{c}-\theta_0)\overset{d}{\longrightarrow}\mathcal{N}\left(0,\,(\pmb{A}^\top\pmb{\Omega}_c^{-1}\pmb{A})^{-1}\right), \]
and $(\pmb{A}^\top\widehat{\pmb{\Omega}}_c^{-1}\pmb{A})^{-1}\overset{p}{\longrightarrow}(\pmb{A}^\top\pmb{\Omega}_c^{-1}\pmb{A})^{-1}$.
\end{corollary}

\textbf{Proof of Corollary \ref{sec04 - corollary01} : }See Appendix \ref{sec04 - corollary01 - proof}.\\

%%%%%%%%%%%%%%%%%%%%%%%%%%%%%%%%%%%%%%%%%%%%%%%%%%%%%%%%%%%%%%%%%%%%%%%%%%%%%%%%%%%%%%%%%%%%%%%%%%%%%%%%%%%%%%%%%%%%%%%%%%%%%%%%%%%%%%%%%%%%%%%%%%%%%%%%%%%%%%%%%%%%%%%%%%%%%%%%%%%%%%%%%%%%%%%%%%%%%%%%%%%%%%%%%%%%%%%%%%%%%%%%%%%%%%%%%%%%%%%%%%%%%%%%%%%%%%%%%%%%%%%%%%%%%%%%%%%%%%%%%%%%%%%%%%%%%%%%%%%%%%%%%%%%%%%%%%%%%%%%%%%%%%%%%%%%%%%%%%%%%%%%%%%%%%%%%%%%%%%%%%%%%%%%%%%%%%%%%%%%%%%%%%%%%%%%%%%%%%%%%%%%%%%%%%%%%%%%%%%%%%%%%%%%%%%%%%%%%%%%%%%%%%%%%%%%%%%%%%%%%%%%%%%%%%%%%%%%%%%%%%%%%%%%%%%%%%%%%%%%%%%%%%%%%%%%%%%%%%%%%%%%%%%%%%%%%%%%%%%%%%%%%%%%%%%%%%%%%%%%%%%%%%%%%%%%%%%%
\section{Testing the conditioning set, double robustness, and bandwidth selection}\label{sec05}
Subsection~\ref{sec05:test} provide a specification test of whether the generated conditioning variable $\eta_0(S_i)$ can be replaced by the underlying first-stage regressors $S_i$ without invalidating the CMR. Subsection~\ref{sec05:bandwidth} then provides a practical selection rule choosing the ISPD-kernel bandwidth by minimizing the asymptotic variance criterion over a fixed finite grid. Throughout this section, the superscript $c$ in $Z^{c}$ indexes the model under consideration, extending the earlier convention in which $c\in\{0,1\}$ indicates whether the conditioning set includes generated instruments.

\subsection{Testing the conditioning set}\label{sec05:test}
In this subsection, we develop a specification test of whether the generated conditioning variable $\eta_0(S_i)$ can be replaced by the underlying first-stage regressors $S_i$ without invalidating the CMR. When this replacement is valid, the kernel can be constructed using observed variables alone, eliminating the correction terms $\delta_{12}(\cdot)$ and $\delta_{22}(\cdot)$.

We define a new conditioning variable $Z_{i}^S = (X_{2i}^\top,S_i^\top)^\top$, while retaining the original conditioning variable $Z_{i}^{1}(\eta) \equiv (X_{2i}^\top,\eta(S_i)^\top)^\top$. Throughout this section, we maintain that the restriction $\rho(\cdot)$ is scalar-valued and $\theta_0$ is the unique solution to the CMRs $\condexpect{\rho(W,\theta,\eta_0)}{ Z^{1}(\eta_0) } = 0$ almost surely. The tower property of conditional expectations gives:
\[ \condexpect{\rho(W,\theta,\eta_0)}{Z^{1}(\eta_0)} = \condexpect{\condexpect{\rho(W,\theta,\eta_0)}{Z^{S}}}{Z^{1}(\eta_0)}. \]
Thus, if the $Z^{S}$-CMR has a solution, that solution must also be $\theta_0$. Based on this observation, we test the following hypotheses:
\begin{align}
        H_0:\quad&
        \label{sec05 - eq01 - hypothesis - null}\condexpect{\rho(W,\theta_0,\eta_0)}{Z^{S}} = 0\quad\text{with probability one,}&\\
        H_1:\quad&
        \label{sec05 - eq01 - hypothesis - alt}\condexpect{\rho(W,\theta_0,\eta_0)}{Z^{S}}\ne0\quad\text{with positive probability}.&
\end{align}
Under $H_0$, both CMRs identify $\theta_0$. Under $H_1$, the $Z^{S}$-CMR has no solution, although the original $Z^{1}(\eta_0)$-CMR continues to identify $\theta_0$.

For the control-function choice model in Example~\ref{sec03 - example02 - PT}, write $C_i=(X_{1i}^{\top},X_{2i}^{\top})^{\top}$ and $V_i(\eta)=X_{1i}-\eta(S_i)$. The baseline and observed-signal conditioning variables can be written as $Z_i^V(\eta)=(C_i^{\top},V_i(\eta)^{\top})^{\top}$ and $Z_i^S=(C_i^{\top},S_i^{\top})^{\top}$. Because $V_i(\eta_0)$ is a
function of $(C_i,S_i)$, the latter restriction implies the former. A Hausman-type comparison tests whether conditioning additionally on the underlying first-stage regressors invalidates the maintained control-function CMR. It does not impose that the endogenous regressor be exogenous. An informative comparison requires that $S_i$ contain information not already recoverable from $(C_i,V_i)$; otherwise the two restrictions are equivalent.

\paragraph{Setup.} Now we introduce additional notation to construct the test statistic. For the $Z^{1}(\eta_0)$-CMR, we retain $r^1$, $\vecdel_{0}^{1}$, and $\psi^1$, defined in Section \ref{sec04}. For the $Z^{S}$-CMR, let $K^S(Z_i^S, Z_j^S) = K\big( (X_{2i}, S_i), (X_{2j}, S_j) \big)$ for an ISPD kernel $K(\cdot,\cdot)$ satisfying Assumption \ref{sec03 - assumption03 - kernel}, and define $Q^{S}$, $r^{S}$, $\vecdel_{0}^{S}$, and $\psi^{S}$ as in Section \ref{sec04}. We note that $\vecdel_{0,12}^S=\vecdel_{0,22}^S=0$, since $K^S(Z_i^S,Z_j^S)$ does not depend on $\eta$. For both constructions, use the same folds $I_1,\ldots,I_L$ from Definition \ref{sec04 - definition01 - DML}. For each $\ell\in[L]$ and $c\in\{1,S\}$, obtain $\widetilde{\theta}_{\ell}^{c}$, $\widehat{\eta}_{\ell}$, $\widehat{\vecdel}_{\ell}^{c}$, and $\check{\vecdel}_{\ell}^{c}$ from the training observations, and define
\begin{equation}\label{sec05 - eq02 - score}
    \psi_{\ell}^{c}(W_i,W_j,\theta)\equiv r^{c}(W_i,W_j,\theta,\widehat{\eta}_{\ell})+\phi(W_i,W_j,\widehat{\eta}_{\ell},\check{\vecdel}_{\ell}^{c})).    
\end{equation}
The two DDML estimators are:
\begin{equation}\label{sec05 - eq03 - DDML}
    \widehat{\theta}_{\textnormal{DDML}}^{c}\equiv\argmin{\theta\in\Theta}{\Norm{ \widebar{\psi}^{c}(W_i, W_j, \theta) }^2}
    ,\quad
    \widebar{\psi}^{c}(W_i, W_j, \theta) \equiv \binom{N}{2}^{-1}\sum_{\ell=1}^{L}\sum_{(i,j)\in I_{\ell}}\psi_{\ell}^{c}(W_i,W_j,\theta).
\end{equation}
For each model $c,c'\in\{1, S\}$, we define:
\begin{equation}\label{sec05 - eq04 - Variance}
    \begin{aligned}
        \Gamma_{c}
            \equiv  &\expect{\nabla_\theta\psi^{c}(W_i,W_j,\theta_0,\eta_0,\vecdel_{0}^{c})},&\\
        \Sigma_{cc'}
            \equiv  &4\Expect{\condexpect{\psi^{c}(W_i,W_j,\theta_0,\eta_0,\vecdel_{0}^{c})}{W_i}\condexpect{\psi^{c'}(W_i,W_j,\theta_0,\eta_0,\vecdel_{0}^{c'})}{W_i}^\top},&\\
        {\Omega}_{cc'}
            \equiv  &\big(\Gamma_{c}^\top\Gamma_{c}\big)^{-1} \Gamma_{c}^\top\Sigma_{cc'}\Gamma_{c'} \big(\Gamma_{c'}^\top\Gamma_{c'}\big)^{-1},&
    \end{aligned}
\end{equation}
and let $\widehat{{\Omega}}_{cc'}$ be the corresponding estimator.

\paragraph{Test Statistic.} The testable implication is as follows : Let $\theta^{S}$ be the minimizer of the population DL-criterion $Q^{S}(\theta, \eta_0)$. Since $\Theta$ is compact and $\theta \mapsto Q^{S}(\theta, \eta_0)$ is continuous in $\theta$, the pseudo-true parameter $\theta^{S}$ is well-defined under both hypotheses. If the pseudo-true value $\theta^{S}$ lies in the interior of $\Theta$, it solves a first-order condition $\expect{r^{S}(W_i,W_j,\theta,\eta_0)} = 0$, and therefore Theorem \ref{sec04 - theorem01 - dml} implies $\sqrt{N}\big(\widehat{\theta}_{\textnormal{DDML}}^{S} - \theta^{S}\big) \overset{d}{\longrightarrow} \mathcal{N}(0, \Omega_{SS})$. Under the null hypothesis, $\theta^{S} = \theta_0$ holds and therefore the discrepancy between $\widehat{\theta}_{\textnormal{DDML}}^{1}$ and $\widehat{\theta}_{\textnormal{DDML}}^{S}$ converges to zero. Under the alternative hypothesis, however, it implies $\sqrt{N}\big( \theta^{S} - \theta_0\big) \to \infty$ when $\theta^{S}\ne \theta_0$.

Based on this observation, we define a Hausman-type test statistic \citep{hausman1978}:
\begin{equation}\label{sec05 - eq05 - Hausman}
    \mathcal{T}_{N} \equiv N \big( \widehat{\theta}_{\textnormal{DDML}}^{1} - \widehat{\theta}_{\textnormal{DDML}}^{S} \big)^{\top}\big(\widehat{{\Omega}}_{11}+\widehat{{\Omega}}_{SS}-\widehat{{\Omega}}_{1S}-\widehat{{\Omega}}_{S1}\big)^{-1} \big( \widehat{\theta}_{\textnormal{DDML}}^{1} - \widehat{\theta}_{\textnormal{DDML}}^{S} \big).    
\end{equation}

\begin{assumption}[Additional regularities for the specification test]\leavevmode\vspace{-\baselineskip}\label{sec05 - assumption06}
    \begin{enumerate}
        \item[(a)] $\theta_0$ is the unique solution to $Z^{1}(\eta_0)$-CMR.
        \item[(b)] The minimizer $\theta^{S}$ lies in $\textnormal{int}(\Theta_0)$, and for any $\epsilon>0$,
        \[ \inf_{ \|\theta - \theta^{S}\| \geq \epsilon } Q^{S}(\theta, \eta_0) > Q^{S}(\theta^{S}, \eta_0). \]
        \item[(c)] ${\Omega}_{11}+{\Omega}_{SS}-{\Omega}_{1S}-{\Omega}_{S1}$ is finite and positive definite.
        \item[(d)] For each $c,c'\in\{1,S\}$, $\widehat{{\Omega}}_{cc'}\overset{p}{\longrightarrow}{\Omega}_{cc'}$.
    \end{enumerate}
\end{assumption}

Assumption \ref{sec05 - assumption06}(a) maintains identification of $\theta_0$ under the original $Z^{1}(\eta_0)$-CMR. Assumption \ref{sec05 - assumption06}(b) requires $\theta^{S}$ to be an interior, well-separated minimizer of $Q^{S}(\theta,\eta_0)$. Assumption \ref{sec05 - assumption06}(c) ensures that the asymptotic covariance matrix of the difference between the two estimators is invertible. Assumption \ref{sec05 - assumption06}(d) requires consistent estimation of the covariance blocks used in the test statistic.

The following proposition characterizes the asymptotic behavior of the Hausman-type test statistic under the null and alternative hypotheses.

\begin{proposition}\textnormal{}\label{sec05 - proposition03 - test}
    Suppose that the model $c=1$ satisfies Assumptions~\ref{sec03 - assumption01 - first stage}, \ref{sec03 - assumption02 - CMR}, \ref{sec03 - assumption03 - kernel}, \ref{sec04 - assumption04 - linear representation}, and \ref{sec04 - assumption05 - algorithm stability}, and that Assumption \ref{sec05 - assumption06} holds. Under the null hypothesis in equation~\eqref{sec05 - eq01 - hypothesis - null}, suppose that the model $c=S$ also satisfies these assumptions. Then, $\mathcal{T}_N \overset{d}{\longrightarrow} \chi^2_{d_\theta}$ under the null hypothesis, and $\mathcal{T}_N \overset{p}{\longrightarrow}+\infty$ under the alternative hypothesis {in} equation~\eqref{sec05 - eq01 - hypothesis - alt}.
\end{proposition}

\textbf{Proof of Proposition \ref{sec05 - proposition03 - test} : }See Appendix \ref{sec05 - proposition03 - proof}.\\
%%%%%%%%%%%%%%%%%%%%%%%%%%%%%%%%%%%%%%%%%%%%%%%%%%%%%%%%%%%%%%%%%%%%%%%%%%%%%%%%%%%%%%%%%%%%%%%%%%%%%%%%%%%%%%%%%%%%%%%%%%%%%%%%%%%%%%%%%%%%%%%%%%%%%%%%%%%%%%%%%%%%%%%%%%%%%%%%%%%%%%%%%%%%%%%%%%%%%%%%%%%%%%%%%%%%%%%%%%%%%%%%%%%%%%%%%%%%%%%%%%%%%%%%%%%%%%%%%%%%%%%%%%%%%%%%%%%%%%%%%%%%%%%%%%%%%%%%%%%%%%%%%%%%%%%%%%%%%%%%%%%%%%%%%%%%%%%%%%%%%%%%%%%%%%%%%%%%%%%%%%%%%%%%%%%%%%%%%%%%%%%%%%%%%%%%%%%%%%%%%%%%%%%%%%%%%%%%%%%%%%%%%%%%%%%%%%%%%%%%%%%%%%%%%%%%%%%%%%%%%%%%%%%%%%%%%%%%%%%%%%%%%%%%%%%%%%%%%%%%%%%%%%%%%%%%%%%%%%%%%%%%%%%%%%%%%%%%%%%%%%%%%%%%%%%%%%%%%%%%%%%%%%%%%%%%%%%%
\subsection{Variance-based bandwidth selection}\label{sec05:bandwidth}
This subsection provides a practical rule for choosing the bandwidth of the ISPD kernel. Even though every fixed bandwidth satisfying our assumptions permits valid asymptotic inference, different bandwidths can yield different asymptotic variances. In finite samples, bandwidth choice can also affect bias, variability, and confidence-interval coverage. We propose selecting a bandwidth from a fixed finite grid by minimizing an estimated measure of asymptotic variance for the parameters of interest.\footnote{For related bandwidth-indexed asymptotic theory for smooth minimum distance estimators, see \citet{kernel_lavergne_patilea2013}.}

Let $b$ denote the bandwidth of the ISPD kernel $K_b$. Suppose that the researcher supplies a fixed, nonrandom finite set of candidate bandwidths
\[ \mathcal{B}\equiv\{b_1,b_2,\ldots,b_B\}\subset[C_1,C_2], \qquad B<\infty,\qquad 0<C_1<C_2<\infty, \]
where neither the grid nor its cardinality depends on $N$. Let $\widehat{\theta}_{\textnormal{DDML}}^{c}(b)$ denote the estimator in Definition~\ref{sec04 - definition01 - DML} constructed using $K_b$, and let $\Omega_{c}(b)$ denote its asymptotic covariance matrix in Theorem~\ref{sec04 - theorem01 - dml}. The dependence of $\Omega_{c}(b)$ on $b$ is visible in the formulas for the correction functions $\vecdel^{c}$ in {equations}~\eqref{sec04 - def02 - delta}--\eqref{sec04 - def02 - delta2}. These formulas involve $K_b$ and, when the conditioning variables are generated, its derivatives, which generally depend on $b$. The score and Jacobian entering the asymptotic covariance also depend on the kernel.

For every fixed $b\in(0,\infty)$ for which the assumptions of Theorem~\ref{sec04 - theorem01 - dml} hold,
\[ \sqrt{N} \big( \widehat{\theta}_{\textnormal{DDML}}^{c}(b)-\theta_0 \big)
       \overset{d}{\longrightarrow}\mathcal{N}(0,\Omega_{c}(b)). \]
Together with the consistent covariance estimator in Proposition~\ref{sec04 - proposition02}, this result gives asymptotically valid confidence intervals and tests of the prescribed asymptotic size for each such fixed bandwidth, provided the relevant limiting variance is
positive. Nevertheless, bandwidth choice matters in two respects. First, $\Omega_{c}(b)$ generally varies with $b$, so the asymptotic variance can differ across bandwidths even though each permits valid asymptotic inference. Second,  $b$ can affect finite-sample bias, variability, and confidence-interval coverage. We propose choosing $b$ by minimizing an estimated scalar measure of asymptotic variance over $\mathcal{B}$. This criterion targets asymptotic variance; it does not directly minimize finite-sample mean squared error or coverage error.

Write $\widehat{\Omega}_{c}(b)$ for the covariance estimator in Proposition~\ref{sec04 - proposition02}. In the case in which there are multiple CMRs identifying the same parameter, it is understood that $\widehat{\theta}_{\textnormal{DDML}}^{c}(b)$ denotes the minimum-distance estimator in Corollary~\ref{sec04 - corollary01}, and $\Omega_{c}(b)$ and $\widehat{\Omega}_{c}(b)$ denote its asymptotic covariance matrix and the corresponding covariance estimator, respectively.

To ensure that bandwidth selection cannot drive the population asymptotic variance to zero, we impose the following sufficient condition. For a single CMR, let $\Gamma_{c}(b)$ denote the score Jacobian in equation~\eqref{sec04:eq:scorejacobian} evaluated using $K_b$, and let $\Sigma_{c}(b)$ denote the corresponding covariance of the first-order score projection in equation~\eqref{sec04:eq:scoreprojectioncovariance}, including its factor of four. In the case in which there are multiple CMRs, $\Gamma_{c,m}(b)$ denotes the Jacobian for restriction $\rho_m$, and $\pmb\Sigma_{c}(b)=\{\Sigma_{c,mm'}(b)\}_{m,m'\in[M]}$ denotes the covariance of the stacked projections defined before Corollary~\ref{sec04 - corollary01}.

\begin{assumption}[Nondegenerate score projections]\label{sec05 - assumption08 - bandwidth nondegeneracy}
For every $b\in\mathcal{B}$, $\Gamma_{c}(b)$ is nonsingular and $\Sigma_{c}(b)$ is positive definite. In the case in which there are multiple CMRs, each component Jacobian $\Gamma_{c,m}(b)$ is nonsingular and the joint covariance matrix $\pmb\Sigma_{c}(b)$ is positive definite.
\end{assumption}

This is an additional assumption that we impose, as it does not follow from the ISPD property in Assumption~\ref{sec03 - assumption03 - kernel}(a). The sandwich formula then implies that $\Omega_{c}(b)$ is positive definite for every $b\in\mathcal{B}$, also for the minimum-distance estimator. Since the grid is fixed and finite, the smallest eigenvalue of the matrix $\Omega_{c}(b)$ is strictly positive for every $b\in\mathcal{B}$. This ensures that, for any fixed nonzero vector $\pmb{a}$, the asymptotic variance for estimating $\pmb{a}^\top\theta_0$ is strictly positive throughout the grid.

We compare bandwidths using the asymptotic variances of the coefficients of interest. For a single coefficient, the criterion is its asymptotic variance. For several coefficients, a weighted sum of their variances allows the researcher to specify their relative importance.

Write $\pmb{R}\theta_0$ for the coefficients or linear combinations of interest, where $\pmb{R}$ is a fixed matrix. With a fixed positive-semidefinite weighting matrix $W$ satisfying $\pmb{R}^\top \pmb{W} R\ne0$, the population and estimated criteria are
\begin{equation}\label{sec05:eq:varianceloss}
    v^{c}(b)=\operatorname{tr}\{ \pmb{W} \pmb{R} \Omega_{c}(b) \pmb{R}^\top\},\qquad
     \widehat{v}_{N}^{c}(b)=\operatorname{tr} \{\widehat{\pmb{W}}_{N} \pmb{R} \widehat{\Omega}_{c}(b) \pmb{R}^\top\},
\end{equation}
where $\widehat{ \pmb{W} }_{N}\overset{p}{\longrightarrow} \pmb{W}$.

The matrix $\pmb{R}$ selects the economically relevant components of $\theta_0$, while $\pmb{W}$ determines their relative weights in the variance criterion. In the production-function model in Example~\ref{sec03 - example01 - ACF}, $\pmb{R}$ can select the input coefficients, which are production elasticities, leaving out the parameters of the productivity process. In the demand model in Example~\ref{sec03 - example02 - PT},
price elasticities may be of primary economic interest, so $\pmb{R}$ can select the coefficients governing price sensitivity, leaving out auxiliary parameters. Once $\pmb{R}$ specifies the components of interest, a diagonal $\pmb{W}$ gives a weighted sum of their asymptotic variances, with larger entries placing greater emphasis on the corresponding components.

We choose the bandwidth that minimizes the estimated variance criterion over the candidate set:
\begin{equation}\label{sec05:eq:bandwidthselector}
    \widehat{b}_{N}^{c}\in\argmin{b\in\mathcal{B}}{\widehat{v}_{N}^{c}(b)},
    \qquad
    \widehat{\theta}_{\textnormal{VS}}^{c}=\widehat{\theta}_{\textnormal{DDML}}^{c}(\widehat{b}_{N}^{c}).
\end{equation}
If several candidate bandwidths attain the minimum, we select the one with the smallest index in the prespecified list $(b_1,\ldots,b_B)$.

\begin{proposition}[Finite-grid variance selection]\label{sec05:prop:bandwidth}
Suppose Assumption~\ref{sec05 - assumption08 - bandwidth nondegeneracy} holds and that, for every $b\in\mathcal{B}$,
\[  \sqrt{N}\big( \widehat{\theta}_{\textnormal{DDML}}^{c}(b) - \theta_0 \big) \overset{d}{\longrightarrow}\mathcal{N}(0,\Omega_{c}(b)),  \qquad
    \widehat{\Omega}_{c}(b)\overset{p}{\longrightarrow}\Omega_{c}(b), \qquad
    \widehat{\pmb{W}}_{N}\overset{p}{\longrightarrow} \pmb{W}. \]
Suppose $v$ has a unique minimizer $b_*^{c}\in\mathcal{B}$. Then
\begin{equation}\label{sec05:eq:bandwidthoracle}
    \Pr(\widehat{b}_{N}^{c} = b_*^{c} )\longrightarrow 1,\qquad
    \sqrt{N} \big( \widehat{\theta}_{\textnormal{VS}}^{c}-\theta_0 \big) \overset{d}{\longrightarrow}\mathcal{N}(0,\Omega(b_*^{c})),\qquad
    \widehat{\Omega}_{N}(\widehat{b}_{N}^{c})\overset{p}{\longrightarrow}\Omega(b_*^{c}).
 \end{equation}
\end{proposition}
\textbf{Proof of Proposition \ref{sec05:prop:bandwidth} :} See Appendix~\ref{app:proof:bandwidth}.\\

The key step is to show that $\Pr(\widehat{b}_{N}^{c} = b_*^{c})\to1$. The fixed, finite grid makes this straightforward: pointwise consistency of the estimated variance criterion implies uniform consistency over $\mathcal{B}$, and the unique population minimizer is separated from every other candidate by a strictly positive gap. Consequently, the estimated criterion selects $b_*^{c}$ with probability approaching one. The limiting distribution and variance consistency then follow from the corresponding results at $b_*^{c}$.

We evaluate the performance of this procedure in the simulation study in Section~\ref{sec06} and the empirical applications in Section~\ref{sec07:empirical}.

%%%%%%%%%%%%%%%%%%%%%%%%%%%%%%%%%%%%%%%%%%%%%%%%%%%%%%%%%%%%%%%%%%%%%%%%%%%%%%%%%%%%%%%%%%%%%%%%%%%%%%%%%%%%%%%%%%%%%%%%%%%%%%%%%%%%%%%%%%%%%%%%%%%%%%%%%%%%%%%%%%%%%%%%%%%%%%%%%%%%%%%%%%%%%%%%%%%%%%%%%%%%%%%%%%%%%%%%%%%%%%%%%%%%%%%%%%%%%%%%%%%%%%%%%%%%%%%%%%%%%%%%%%%%%%%%%%%%%%%%%%%%%%%%%%%%%%%%%%%%%%%%%%%%%%%%%%%%%%%%%%%%%%%%%%%%%%%%%%%%%%%%%%%%%%%%%%%%%%%%%%%%%%%%%%%%%%%%%%%%%%%%%%%%%%%%%%%%%%%%%%%%%%%%%%%%%%%%%%%%%%%%%%%%%%%%%%%%%%%%%%%%%%%%%%%%%%%%%%%%%%%%%%%%%%%%%%%%%%%%%%%%%%%%%%%%%%%%%%%%%%%%%%%%%%%%%%%%%%%%%%%%%%%%%%%%%%%%%%%%%%%%%%%%%%%%%%%%%%%%%%%%%%%%%%%%%%%%
\section{Simulation Experiments}\label{sec06}
In this section, we run Monte Carlo experiments to evaluate the bias, root mean squared error (RMSE), and coverage probability of the plug-in, DDML, and the variance-selected estimators. The first experiment uses controlled perturbations of the nuisance functions to examine the effect of the first-stage convergence rate. The second estimates the first-stage function from the simulated data using several nonparametric learners. We also report rejection probabilities for the Hausman specification test and compare fixed and variance-selected kernel bandwidths.

\subsection{Setup.} We consider two data-generating processes (DGPs) that differ in whether the underlying first-stage regressor $S_i$ can replace the generated conditioning variable $\eta_0(S_i)$. In DGP 1, only the CMR conditioning on $\eta_0(S_i)$ is correctly specified. In DGP 2, both conditioning specifications identify the same parameter.

We generate independent observations across $i\in[N]$. For each $i$, draw $\epsilon_{Ti}$, $\epsilon_{Yi}$, and $S_i$ independently, with
\[      \epsilon_{Ti}\sim\mathcal{N}(0,1),  \qquad  \epsilon_{Yi}\sim\mathcal{N}(0,1),  \qquad  S_i\sim\operatorname{Unif}(-1,1).\]
The first-stage outcome $T$ and the main outcome variable $Y$ are generated as
\[      T_i=\eta_0(S_i)+\epsilon_{Ti},  \qquad  Y_i=\theta_0\eta_0(S_i)+u_i,    \qquad \theta_0=1.  \]
The observed data vector is $W_i=(Y_i,T_i,S_i)^\top$. The first-stage regression identifies $\eta_0(S_i)=\condexpect{T_i}{S_i}$, and the residual used in both CMRs is
\[      \rho(W_i,\theta,\eta)\equiv Y_i-\theta\eta(S_i).    \]
We compare the conditioning variables $Z_i^1(\eta)=\eta(S_i)$ and $Z_i^S=S_i$. In each DGP, the two specifications use the same first-stage target and differ only in the conditioning variable.

Define
\[      \eta_1(S_i)=1+2\sin(\pi S_i),       \qquad      \eta_2(S_i)=S_i+\frac{1}{2}S_i^3.   \]
The two DGPs are given by
\begin{equation*}
    \begin{array}{lll}
        \textnormal{DGP 1:}
            & \eta_0(S_i)=\eta_1(S_i),
            & u_i=\epsilon_{Yi}+3\cos(\pi S_i),\\
        \textnormal{DGP 2:}
            & \eta_0(S_i)=\eta_2(S_i),
            & u_i=\epsilon_{Yi}.
    \end{array}
\end{equation*}

Under DGP 1, $\condexpect{\cos(\pi S_i)}{\sin(\pi S_i)}=0$. Hence, 
\[      \condexpect{\rho(W_i,\theta,\eta_0)}{Z_i^1(\eta_0)} = (\theta_0-\theta)\eta_1(S_i),\]
meaning that $\theta_0$ is the unique solution. In contrast,
\[      \condexpect{\rho(W_i,\theta,\eta_0)}{Z_i^S}   = (\theta_0-\theta)\eta_1(S_i)+3\cos(\pi S_i). \]
Taking expectations shows that any solution to the $Z^S$-CMR must equal $\theta_0$, since $\mathbb{E}[\eta_1(S_i)]=1$ and $\mathbb{E}[\cos(\pi S_i)]=0$. At $\theta_0$, however, the conditional mean equals $3\cos(\pi S_i)$, so the $Z^S$-CMR has no solution.

Under DGP 2, $\eta_2$ is strictly increasing on $[-1,1]$, so $S_i$ and $\eta_2(S_i)$ generate the same sigma-field. Because $\epsilon_{Yi}$ is independent of $S_i$, both CMRs are correctly specified and uniquely identify $\theta_0$. The two DGPs therefore allow us to examine the power and size of the Hausman test, respectively.

\subsection{Experiment 1: Impact of the first-stage convergence rate.} We first examine how the convergence rate of the first-stage nuisance estimate affects bias and confidence interval coverage. To isolate the effect of this rate from the choice of learner, we construct nuisance estimates by perturbing the population functions. For $k\in\{1,2\}$, we set
\begin{align*}
    \widehat{\eta}_k(S_i)   &   \equiv \eta_k(S_i)     +   \left(\frac{1}{4}\eta_k(S_i)+e_{\eta,i}\right)a_N,&
    e_{\eta,i}      &   =\frac{S_i\xi_{\eta,i}-\widehat{\mathbb{E}}[S_i\xi_{\eta,i}]}{\widehat{\operatorname{SD}}(S_i\xi_{\eta,i})},\\
    \widehat{\delta}(S_i)
        &\equiv \delta_0(S_i)   +\left(\frac{1}{4}\delta_0(S_i)+e_{\delta,i}\right)b_N,&
    e_{\delta,i}    &   =\frac{S_i\xi_{\delta,i}-\widehat{\mathbb{E}}[S_i\xi_{\delta,i}]}{\widehat{\operatorname{SD}}(S_i\xi_{\delta,i})}.
\end{align*}
Here, $\xi_{\eta,i}$ and $\xi_{\delta,i}$ are independent standard normal draws, independent across observations and of the simulated data. The correction $\delta_0$ is evaluated at $\theta_0$ for the corresponding conditioning specification and kernel bandwidth.

We set $a_N=N^{-a}$ with $a\in\{0.1,0.3\}$ and $b_N=N^{-0.3}$. When $a=0.3$, the rate products $\sqrt{N}a_N^2$ and $\sqrt{N}a_Nb_N$ both vanish. When $a=0.1$, neither vanishes. The rate conditions in Theorem \ref{sec04 - theorem01 - dml} therefore support root-$N$ inference for the correctly specified DDML estimators in the former setting, whereas they do not guarantee nominal coverage in the latter.

We consider $N\in\{200, 500,1{,}000,2{,}000\}$ and use $10{,}000$ simulation replications for each design. We set $G=5$, so that the DDML estimators use $L=G(2G-1)=45$ pair folds. We use the Gaussian kernel
\[      K_b(z,z')=\exp\left(-\frac{|z-z'|^2}{2b^2}\right),  \]
with reference bandwidth $b_{\textnormal{ref}}=0.5$. We also evaluate the variance-based bandwidth selection rule in Section~\ref{sec05:bandwidth} over the fixed grid $\mathcal{B}=\{0.1,0.2,0.3,0.4,0.5,0.75, 1.0, 1.5, 2.0\}$. In each replication, we select the bandwidth that minimizes the estimated asymptotic variance of the $Z^1$-DDML estimator. We report the $Z^1$-DDML results at the minimum, reference, maximum, and selected bandwidths, along with the selection frequencies.

Table \ref{sec06 - table01 - bias - ref} reports the bias and RMSE of each estimate using the fixed reference bandwidth $b_\textnormal{ref} = 0.5$. For correctly specified CMRs, the plug-in estimates exhibit larger absolute biases than the DDML estimates under both values of $a$. The plug-in bias nevertheless declines with $N$, consistent with the convergence result in Lemma \ref{sec04 - lemma01}. The DDML estimates generally exhibit larger absolute biases for $a=0.1$ than for $a=0.3$, regardless of specification, consistent with the remainder bounds underlying Theorem \ref{sec04 - theorem01 - dml}. Under DGP 1, the $Z^{S}$-CMR is incorrectly specified, whereas the $Z^{1}(\eta_1)$-CMR is correctly specified. Panel A reflects this distinction. For $a=0.3$, the $Z^{S}$-DDML estimates reports positive bias, while the $Z^{1}(\widehat{\eta}_1)$-DDML biases are close to zero. Under DGP 2, the two conditioning variables $S$ and $\eta_2(S)$ generate equivalent information sets, and both CMRs are correctly specified. For $a=0.3$, Panel B shows small biases for both DDML estimates. 

Although the plug-in unscaled RMSEs generally decline with $N$, their scaled RMSEs increase under both values of $a$, showing that the plug-in errors do not vanish at the root-$N$ rate over the reported sample sizes. In contrast, for $a=0.3$, the scaled RMSEs of the correctly specified DDML estimates remain  stable. When $a=0.1$, however, the scaled RMSEs are not stable, consistent with the rate requirements of Assumption \ref{sec04 - assumption05 - algorithm stability} and prediction of Theorem \ref{sec04 - theorem01 - dml}.

Table \ref{sec06 - table02 - coverage - ref} reports the empirical coverage probabilities of nominal 90\%, 95\%, and 99\% confidence intervals for $a=0.1$ and $a=0.3$, using the fixed reference bandwidth $b_\textnormal{ref} = 0.5$. For correctly specified CMRs, the plug-in intervals exhibit extreme undercoverage getting worse as $N$ increases, consistent with the increasing scaled biases reported in Table \ref{sec06 - table01 - bias - ref}. For $a=0.3$, the correctly specified DDML intervals have coverage probabilities generally close to the nominal levels in both panels. In contrast, the misspecified $Z^{S}$-DDML intervals in Panel A exhibit undercoverage as $N$ increases.

Table~\ref{sec06 - table03 - hausman} reports the empirical rejection probabilities of the Hausman specification test. Under DGP 1, the $Z^1(\eta_0)$-CMR is correctly specified, whereas the $Z^S$-CMR is misspecified. The rejection probabilities in Panel A therefore measure empirical power and approach one as $N$ increases.

Under DGP 2, both CMRs identify the same parameter, so the null hypothesis is true and the rejection probabilities in Panel B measure empirical size. When $a=0.1$, the rejection probabilities exceed their nominal levels, consistent with the first-stage convergence rate falling short of the requirement for DDML inference. When $a=0.3$, the rejection probabilities are generally at or below their nominal levels and move closer to them as $N$ increases.

We next examine the performance of the variance-selected DDML estimator. We focus on $a=0.3$ because the slower first-stage convergence rate at $a=0.1$ does not satisfy the rate requirement for DDML inference. Table~\ref{sec06 - table04 - variance selection} reports Monte Carlo bias, unscaled RMSE, and the empirical coverage probabilities and mean lengths of nominal $90\%$, $95\%$, and $99\%$ confidence intervals. We compare the DDML estimates at $b_{\min}=0.1$ and $b_{\max}=2.0$ with the estimates at the variance-selected bandwidth $\widehat{\theta}_{\textnormal{VS}}$.

In Panel~A, the DDML estimates under the correctly specified conditioning set exhibit small biases and coverage close to the nominal levels at both fixed bandwidths and at the selected bandwidth. In Panel~B, both conditioning specifications, $Z^S$ and $Z^1(\eta)$, yield small biases and near-nominal coverage. These results are consistent with the theoretical justification for inference after selecting a bandwidth from a fixed finite grid. For the correctly specified CMRs, the intervals based on $\widehat{\theta}_{\textnormal{VS}}$ are generally shorter than those at the two fixed bandwidths, as expected from a rule that minimizes the estimated asymptotic variance. Table~\ref{sec06 - table05 - frequency} reports how often each candidate bandwidth is selected across the simulation replications.

\FloatBarrier
\begin{table}[htbp]\centering
\caption{Bias and RMSE for each estimate using the reference bandwidth ($b_\textnormal{ref} = 0.5$)}\label{sec06 - table01 - bias - ref}
\begin{footnotesize}
\begin{tabular}{c|c|c|c|cc|c|cc}
\hline\hline
\multicolumn{9}{l}{\textbf{Panel A: DGP 1 is True}} \\
\hline
\multirow{3}{*}{Estimator} & \multirow{3}{*}{$Z^{c}(\eta)$} & \multirow{3}{*}{$N$} & \multicolumn{3}{c|}{$a=0.1$} & \multicolumn{3}{c}{$a=0.3$} \\
\cline{4-9}
& & & \multirow{2}{*}{Bias} & \multicolumn{2}{c|}{RMSE} &  \multirow{2}{*}{Bias} & \multicolumn{2}{c}{RMSE} \\
& & & & (Unscaled) & (Scaled) & & (Unscaled) & (Scaled) \\
\hline
\multirow{8}{*}{Plug-in} & \multirow{4}{*}{$Z^{1}(\widehat{\eta}_{1})$} & 200 & -0.156 & 0.176 & 2.494  & -0.055 & 0.099 & 1.399  \\
 &  & 500 & -0.141 & 0.151 & 3.368  & -0.040 & 0.065 & 1.462  \\
 &  & 1,000 & -0.133 & 0.138 & 4.356  & -0.033 & 0.048 & 1.533  \\
 &  & 2,000 & -0.125 & 0.127 & 5.698  & -0.027 & 0.037 & 1.656  \\
\cline{2-9}
 & \multirow{4}{*}{$Z^{S}$} & 200 & 0.025 & 0.117 & 1.650  & 0.117 & 0.170 & 2.399  \\
 &  & 500 & 0.037 & 0.081 & 1.815  & 0.131 & 0.153 & 3.421  \\
 &  & 1,000 & 0.043 & 0.066 & 2.083  & 0.137 & 0.148 & 4.672  \\
 &  & 2,000 & 0.050 & 0.062 & 2.758  & 0.143 & 0.149 & 6.646  \\
\hline
\multirow{8}{*}{DDML} & \multirow{4}{*}{$Z^{1}(\widehat{\eta}_{1})$} & 200 & -0.002 & 0.111 & 1.571  & -0.003 & 0.111 & 1.564  \\
 &  & 500 & -0.003 & 0.067 & 1.509  & 0.000 & 0.068 & 1.529  \\
 &  & 1,000 & -0.006 & 0.047 & 1.482  & -0.001 & 0.048 & 1.518  \\
 &  & 2,000 & -0.007 & 0.033 & 1.480  & -0.001 & 0.034 & 1.518  \\
\cline{2-9}
 & \multirow{4}{*}{$Z^{S}$} & 200 & 0.122 & 0.169 & 2.389  & 0.158 & 0.203 & 2.871  \\
 &  & 500 & 0.127 & 0.147 & 3.279  & 0.163 & 0.182 & 4.067  \\
 &  & 1,000 & 0.127 & 0.137 & 4.344  & 0.163 & 0.172 & 5.450  \\
 &  & 2,000 & 0.130 & 0.135 & 6.028  & 0.164 & 0.169 & 7.562  \\
\hline
\multicolumn{9}{l}{\textbf{Panel B: DGP 2 is True}} \\
\hline
\multirow{3}{*}{Estimator} & \multirow{3}{*}{$Z^{c}(\eta)$} & \multirow{3}{*}{$N$} & \multicolumn{3}{c|}{$a=0.1$} & \multicolumn{3}{c}{$a=0.3$} \\
\cline{4-9}
& & & \multirow{2}{*}{Bias} & \multicolumn{2}{c|}{RMSE} &  \multirow{2}{*}{Bias} & \multicolumn{2}{c}{RMSE} \\
& & & & (Unscaled) & (Scaled) & & (Unscaled) & (Scaled) \\
\hline
\multirow{8}{*}{Plug-in} & \multirow{4}{*}{$Z^{1}(\widehat{\eta}_{2})$} & 200 & -0.297 & 0.311 & 4.405  & -0.056 & 0.115 & 1.623  \\
 &  & 500 & -0.256 & 0.262 & 5.865  & -0.038 & 0.073 & 1.638  \\
 &  & 1,000 & -0.228 & 0.232 & 7.333  & -0.031 & 0.054 & 1.692  \\
 &  & 2,000 & -0.203 & 0.205 & 9.185  & -0.025 & 0.040 & 1.783  \\
\cline{2-9}
 & \multirow{4}{*}{$Z^{S}$} & 200 & -0.121 & 0.160 & 2.263  & -0.049 & 0.111 & 1.565  \\
 &  & 500 & -0.115 & 0.131 & 2.940  & -0.037 & 0.072 & 1.601  \\
 &  & 1,000 & -0.110 & 0.119 & 3.755  & -0.031 & 0.053 & 1.677  \\
 &  & 2,000 & -0.104 & 0.109 & 4.878  & -0.025 & 0.040 & 1.778  \\
\hline
\multirow{8}{*}{DDML} & \multirow{4}{*}{$Z^{1}(\widehat{\eta}_{2})$} & 200 & -0.121 & 0.273 & 3.855  & 0.001 & 0.197 & 2.781  \\
 &  & 500 & -0.093 & 0.161 & 3.594  & 0.003 & 0.108 & 2.424  \\
 &  & 1,000 & -0.076 & 0.113 & 3.565  & 0.003 & 0.071 & 2.233  \\
 &  & 2,000 & -0.063 & 0.084 & 3.742  & 0.002 & 0.048 & 2.157  \\
\cline{2-9}
 & \multirow{4}{*}{$Z^{S}$} & 200 & -0.007 & 0.157 & 2.221  & -0.001 & 0.165 & 2.330  \\
 &  & 500 & -0.007 & 0.089 & 1.990  & 0.001 & 0.097 & 2.164  \\
 &  & 1,000 & -0.008 & 0.059 & 1.879  & 0.001 & 0.065 & 2.062  \\
 &  & 2,000 & -0.008 & 0.042 & 1.865  & 0.000 & 0.046 & 2.041  \\
\hline\hline
\end{tabular}
\end{footnotesize}
\par\smallskip\begin{minipage}{.97\textwidth}\footnotesize\textit{Notes:}
This table reports bias, RMSE, and the empirical coverage probabilities of nominal $90\%$, $95\%$, and $99\%$ confidence interval of the plug-in and DDML estimates for each first-stage rate $a\in\{0.1,0.3\}$. The number of replications is $10{,}000$, and the sample size is $N\in\{200, 500, 1{,}000, 2{,}000\}$. All estimates are obtained using the reference bandwidth $b_\textnormal{ref}=0.5$. The second column indicates the conditioning specification used for estimation. Specifically, $Z^{S}$ and $Z^{1}(\widehat{\eta}_{k})$ correspond to the kernels $K(S_i,S_j)$ and $K(\widehat{\eta}_{k}(S_i),\widehat{\eta}_{k}(S_j))$, respectively. Panels A and B report results under DGP 1 and DGP 2, respectively. Unscaled RMSE is the square root of the average squared estimation error across the $10,000$ replications, and scaled RMSE is multiplied by $\sqrt{N}$.
\end{minipage}
\end{table}

\begin{table}[htbp]
\centering
\caption{Coverage probabilities for each estimates using the reference bandwidth $(b_\textnormal{ref} = 0.5)$}\label{sec06 - table02 - coverage - ref}
\begin{footnotesize}
\begin{tabular}{c|c|c|ccc|ccc}
\hline\hline
\multicolumn{9}{l}{\textbf{Panel A: DGP 1 is True}} \\
\hline
\multirow{2}{*}{Estimator} & \multirow{2}{*}{$Z^{c}(\eta)$} & \multirow{2}{*}{$N$} & \multicolumn{3}{c|}{$a=0.1$} & \multicolumn{3}{c}{$a=0.3$} \\
\cline{4-9}
& & & 90\% & 95\% & 99\% & 90\% & 95\% & 99\% \\
\hline
\multirow{8}{*}{Plug-in} & \multirow{4}{*}{$Z^{1}(\widehat{\eta}_{1})$} & 200 & 0.42 & 0.54 & 0.76 & 0.81 & 0.89 & 0.96 \\
 &  & 500 & 0.15 & 0.24 & 0.46 & 0.80 & 0.87 & 0.96 \\
 &  & 1,000 & 0.03 & 0.05 & 0.15 & 0.77 & 0.86 & 0.96 \\
 &  & 2,000 & 0.00 & 0.00 & 0.01 & 0.73 & 0.82 & 0.94 \\
\cline{2-9}
 & \multirow{4}{*}{$Z^{S}$} & 200 & 0.89 & 0.94 & 0.99 & 0.75 & 0.84 & 0.95 \\
 &  & 500 & 0.86 & 0.93 & 0.98 & 0.48 & 0.60 & 0.82 \\
 &  & 1,000 & 0.80 & 0.88 & 0.97 & 0.19 & 0.29 & 0.53 \\
 &  & 2,000 & 0.62 & 0.73 & 0.89 & 0.02 & 0.04 & 0.13 \\
\hline
\multirow{8}{*}{DDML} & \multirow{4}{*}{$Z^{1}(\widehat{\eta}_{1})$} & 200 & 0.91 & 0.96 & 0.99 & 0.91 & 0.96 & 0.99 \\
 &  & 500 & 0.90 & 0.95 & 0.99 & 0.90 & 0.95 & 0.99 \\
 &  & 1,000 & 0.90 & 0.95 & 0.99 & 0.90 & 0.95 & 0.99 \\
 &  & 2,000 & 0.90 & 0.95 & 0.99 & 0.90 & 0.95 & 0.99 \\
\cline{2-9}
 & \multirow{4}{*}{$Z^{S}$} & 200 & 0.74 & 0.84 & 0.95 & 0.67 & 0.78 & 0.93 \\
 &  & 500 & 0.47 & 0.61 & 0.82 & 0.36 & 0.49 & 0.73 \\
 &  & 1,000 & 0.20 & 0.31 & 0.57 & 0.12 & 0.19 & 0.40 \\
 &  & 2,000 & 0.03 & 0.06 & 0.17 & 0.01 & 0.02 & 0.07 \\
\hline
\multicolumn{9}{l}{\textbf{Panel B: DGP 2 is True}} \\
\hline
\multirow{2}{*}{Estimator} & \multirow{2}{*}{$Z^{c}(\eta)$} & \multirow{2}{*}{$N$} & \multicolumn{3}{c|}{$a=0.1$} & \multicolumn{3}{c}{$a=0.3$} \\
\cline{4-9}
& & & 90\% & 95\% & 99\% & 90\% & 95\% & 99\% \\
\hline
\multirow{8}{*}{Plug-in} & \multirow{4}{*}{$Z^{1}(\widehat{\eta}_{2})$} & 200 & 0.07 & 0.13 & 0.28 & 0.83 & 0.90 & 0.97 \\
 &  & 500 & 0.00 & 0.01 & 0.05 & 0.83 & 0.90 & 0.97 \\
 &  & 1,000 & 0.00 & 0.00 & 0.00 & 0.82 & 0.89 & 0.97 \\
 &  & 2,000 & 0.00 & 0.00 & 0.00 & 0.79 & 0.87 & 0.96 \\
\cline{2-9}
 & \multirow{4}{*}{$Z^{S}$} & 200 & 0.65 & 0.75 & 0.88 & 0.84 & 0.91 & 0.97 \\
 &  & 500 & 0.43 & 0.55 & 0.76 & 0.83 & 0.91 & 0.97 \\
 &  & 1,000 & 0.22 & 0.32 & 0.54 & 0.82 & 0.89 & 0.97 \\
 &  & 2,000 & 0.05 & 0.09 & 0.23 & 0.79 & 0.87 & 0.96 \\
\hline
\multirow{8}{*}{DDML} & \multirow{4}{*}{$Z^{1}(\widehat{\eta}_{2})$} & 200 & 0.86 & 0.93 & 0.98 & 0.90 & 0.95 & 0.99 \\
 &  & 500 & 0.81 & 0.89 & 0.97 & 0.90 & 0.95 & 0.99 \\
 &  & 1,000 & 0.77 & 0.86 & 0.96 & 0.90 & 0.95 & 0.99 \\
 &  & 2,000 & 0.70 & 0.79 & 0.93 & 0.90 & 0.95 & 0.99 \\
\cline{2-9}
 & \multirow{4}{*}{$Z^{S}$} & 200 & 0.91 & 0.96 & 0.99 & 0.90 & 0.95 & 0.99 \\
 &  & 500 & 0.90 & 0.96 & 0.99 & 0.90 & 0.95 & 0.99 \\
 &  & 1,000 & 0.90 & 0.95 & 0.99 & 0.91 & 0.95 & 0.99 \\
 &  & 2,000 & 0.90 & 0.95 & 0.99 & 0.90 & 0.95 & 0.99 \\
\hline\hline
\end{tabular}
\end{footnotesize}
\par\smallskip\begin{minipage}{.97\textwidth}\footnotesize\textit{Notes:}
This table reports the empirical coverage probabilities of the nominal $90\%$, $95\%$, and $99\%$ confidence intervals of the plug-in and DDML estimates at each first-stage rate $a\in\{0.1,0.3\}$. The number of replications is $10{,}000$, and sample sizes are $N\in\{200, 500, 1{,}000, 2{,}000\}$.  The second column indicates the conditioning specification used for estimation. Specifically, $Z^{S}$ and $Z^{1}(\widehat{\eta}_{k})$ correspond to the kernels $K(S_i,S_j)$ and $K(\widehat{\eta}_{k}(S_i),\widehat{\eta}_{k}(S_j))$, respectively. Panels A and B report results under DGP 1 and DGP 2, respectively. Coverage probabilities uses two-sided standard normal critical values.
\end{minipage}
\end{table}

\begin{table}[htbp]
\centering
\caption{Rejection probabilities of the Hausman specification test using the reference bandwidth $(b_\textnormal{ref} = 0.5)$}\label{sec06 - table03 - hausman}
\begin{footnotesize}
\begin{tabular}{c|ccc|ccc}
\hline\hline
\multicolumn{7}{l}{\textbf{Panel A: DGP 1 is True ($H_0$ is False)}} \\
\hline
 & \multicolumn{3}{c|}{$a=0.1$} & \multicolumn{3}{c}{$a=0.3$} \\
\cline{2-4}\cline{5-7}
$N$ & 10\% & 5\% & 1\% & 10\% & 5\% & 1\% \\
\hline
  200 & 0.37 & 0.24 & 0.07 & 0.64 & 0.50 & 0.22 \\
  500 & 0.79 & 0.68 & 0.42 & 0.96 & 0.93 & 0.78 \\
1,000 & 0.99 & 0.96 & 0.87 & 1.00 & 1.00 & 0.99 \\
2,000 & 1.00 & 1.00 & 1.00 & 1.00 & 1.00 & 1.00 \\
\hline
\multicolumn{7}{l}{\textbf{Panel B: DGP 2 is True ($H_0$ is True)}} \\
\hline
 & \multicolumn{3}{c|}{$a=0.1$} & \multicolumn{3}{c}{$a=0.3$} \\
\cline{2-4}\cline{5-7}
$N$ & 10\% & 5\% & 1\% & 10\% & 5\% & 1\% \\
\hline
  200 & 0.19 & 0.11 & 0.02 & 0.06 & 0.03 & 0.00 \\
  500 & 0.33 & 0.22 & 0.07 & 0.08 & 0.04 & 0.01 \\
1,000 & 0.48 & 0.35 & 0.15 & 0.09 & 0.04 & 0.01 \\
2,000 & 0.67 & 0.54 & 0.29 & 0.10 & 0.05 & 0.01 \\
\hline\hline
\end{tabular}
\end{footnotesize}
\par\smallskip\begin{minipage}{.97\textwidth}\footnotesize\textit{Notes:}
This table reports empirical rejection probabilities for the Hausman specification test at the 10\%, 5\%, and 1\% significance levels for each first-stage rate $a\in\{0.1,0.3\}$. The test compares the $Z^{1}(\widehat{\eta})$- and $Z^{S}$-DDML estimators at the minimum, maximum, reference, and selected bandwidths. The number of replications is $10{,}000$, and sample sizes are $N\in\{200, 500, 1{,}000, 2{,}000\}$. The first-stage rate is fixed at $a=0.3$. The candidate bandwidth grid is $\mathcal{B}=\{0.1,0.2,0.3,0.4,0.5,0.75,1.0,1.5,2.0\}$, with $b_{\min}=0.1$ and $b_{\max}=2.0$. The second column indicates the conditioning specification used for estimation. Specifically, $Z^{S}$ and $Z^{1}(\widehat{\eta}_{k})$ correspond to the kernels $K(S_i,S_j)$ and $K(\widehat{\eta}_{k}(S_i),\widehat{\eta}_{k}(S_j))$, respectively. The rejection probabilities in Panel B corresponds to the empirical size of the test statistic.
\end{minipage}
\end{table}

\begin{table}[htbp]
\centering
\caption{Bias, RMSE, and confidence interval for the variance-selected DDML esitmator at $a=0.3$.} \label{sec06 - table04 - variance selection}
\begin{footnotesize}
\begin{tabular}{c|c|c|cc|cc|cc|cc}
\hline\hline
\multicolumn{11}{l}{\textbf{Panel A: DGP 1 is True}} \\
\hline
\multirow{2}{*}{Estimator} & \multirow{2}{*}{$Z^{c}(\eta)$} & \multirow{2}{*}{$N$} & \multirow{2}{*}{Bias} & RMSE & \multicolumn{2}{c|}{90\% CI} & \multicolumn{2}{c|}{95\% CI} & \multicolumn{2}{c}{99\% CI} \\
\cline{6-11}
& & & & (Unscaled) & Coverage & Length & Coverage & Length & Coverage & Length \\
\hline
\multirow{8}{*}{ $\widehat{\theta}_{\textnormal{DDML}}(b_{\min})$ } & \multirow{4}{*}{$Z^1(\widehat{\eta})$} & 200 & 0.014 & 0.179 & 0.92 & 0.610 & 0.96 & 0.727 & 1.00 & 0.956 \\
 &  & 500 & 0.009 & 0.101 & 0.90 & 0.336 & 0.95 & 0.400 & 0.99 & 0.526 \\
 &  & 1,000 & 0.004 & 0.067 & 0.90 & 0.223 & 0.95 & 0.265 & 0.99 & 0.348 \\
 &  & 2,000 & 0.002 & 0.045 & 0.90 & 0.150 & 0.95 & 0.179 & 0.99 & 0.235 \\
\cline{2-11}
 & \multirow{4}{*}{$Z^S$} & 200 & 0.036 & 0.157 & 0.90 & 0.521 & 0.95 & 0.621 & 0.99 & 0.817 \\
 &  & 500 & 0.039 & 0.104 & 0.88 & 0.320 & 0.93 & 0.381 & 0.98 & 0.501 \\
 &  & 1,000 & 0.038 & 0.077 & 0.85 & 0.225 & 0.92 & 0.268 & 0.98 & 0.352 \\
 &  & 2,000 & 0.038 & 0.061 & 0.80 & 0.158 & 0.88 & 0.189 & 0.96 & 0.248 \\
\hline
\multirow{8}{*}{ $\widehat{\theta}_{\textnormal{DDML}}(b_{\max})$ } & \multirow{4}{*}{$Z^1(\widehat{\eta})$} & 200 & -0.006 & 0.118 & 0.91 & 0.392 & 0.96 & 0.467 & 0.99 & 0.614 \\
 &  & 500 & -0.001 & 0.074 & 0.90 & 0.246 & 0.95 & 0.294 & 0.99 & 0.386 \\
 &  & 1,000 & -0.002 & 0.052 & 0.90 & 0.175 & 0.95 & 0.208 & 0.99 & 0.274 \\
 &  & 2,000 & -0.001 & 0.037 & 0.91 & 0.124 & 0.95 & 0.148 & 0.99 & 0.194 \\
\cline{2-11}
 & \multirow{4}{*}{$Z^S$} & 200 & 0.058 & 0.170 & 0.90 & 0.547 & 0.95 & 0.651 & 0.99 & 0.856 \\
 &  & 500 & 0.061 & 0.117 & 0.85 & 0.336 & 0.92 & 0.400 & 0.98 & 0.526 \\
 &  & 1,000 & 0.060 & 0.092 & 0.80 & 0.236 & 0.88 & 0.281 & 0.96 & 0.370 \\
 &  & 2,000 & 0.060 & 0.078 & 0.67 & 0.167 & 0.79 & 0.199 & 0.92 & 0.261 \\
\hline
\multirow{8}{*}{$\widehat{\theta}_{\textnormal{VS}}$} & \multirow{4}{*}{$Z^1(\widehat{\eta})$} & 200 & -0.005 & 0.109 & 0.91 & 0.361 & 0.95 & 0.430 & 0.99 & 0.565 \\
 &  & 500 & 0.000 & 0.068 & 0.90 & 0.225 & 0.95 & 0.268 & 0.99 & 0.352 \\
 &  & 1,000 & -0.001 & 0.048 & 0.90 & 0.158 & 0.95 & 0.189 & 0.99 & 0.248 \\
 &  & 2,000 & -0.001 & 0.034 & 0.90 & 0.112 & 0.95 & 0.133 & 0.99 & 0.175 \\
\cline{2-11}
 & \multirow{4}{*}{$Z^S$} & 200 & 0.145 & 0.199 & 0.71 & 0.449 & 0.81 & 0.535 & 0.94 & 0.703 \\
 &  & 500 & 0.155 & 0.177 & 0.40 & 0.273 & 0.53 & 0.325 & 0.76 & 0.427 \\
 &  & 1,000 & 0.154 & 0.165 & 0.16 & 0.192 & 0.25 & 0.228 & 0.47 & 0.300 \\
 &  & 2,000 & 0.152 & 0.158 & 0.02 & 0.136 & 0.05 & 0.162 & 0.14 & 0.213 \\
\hline
\multicolumn{11}{l}{\textbf{Panel B: DGP 2 is True}} \\
\hline
\multirow{2}{*}{Estimator} & \multirow{2}{*}{$Z^{c}(\eta)$} & \multirow{2}{*}{$N$} & \multirow{2}{*}{Bias} & RMSE & \multicolumn{2}{c|}{90\% CI} & \multicolumn{2}{c|}{95\% CI} & \multicolumn{2}{c}{99\% CI} \\
\cline{6-11}
& & & & (Unscaled) & Coverage & Length & Coverage & Length & Coverage & Length \\
\hline
\multirow{8}{*}{ $\widehat{\theta}_{\textnormal{DDML}}(b_{\min})$ } & \multirow{4}{*}{$Z^1(\widehat{\eta})$} & 200 & 0.022 & 0.532 & 0.91 & 1.749 & 0.96 & 2.084 & 0.99 & 2.739 \\
 &  & 500 & 0.015 & 0.245 & 0.90 & 0.806 & 0.95 & 0.961 & 0.99 & 1.263 \\
 &  & 1,000 & 0.011 & 0.140 & 0.90 & 0.463 & 0.95 & 0.552 & 0.99 & 0.725 \\
 &  & 2,000 & 0.006 & 0.083 & 0.90 & 0.273 & 0.95 & 0.325 & 0.99 & 0.428 \\
\cline{2-11}
 & \multirow{4}{*}{$Z^S$} & 200 & 0.001 & 0.267 & 0.91 & 0.879 & 0.96 & 1.048 & 0.99 & 1.377 \\
 &  & 500 & 0.001 & 0.138 & 0.90 & 0.455 & 0.95 & 0.542 & 0.99 & 0.713 \\
 &  & 1,000 & 0.001 & 0.085 & 0.91 & 0.285 & 0.95 & 0.339 & 0.99 & 0.446 \\
 &  & 2,000 & 0.000 & 0.055 & 0.90 & 0.182 & 0.95 & 0.217 & 0.99 & 0.286 \\
\hline
\multirow{8}{*}{ $\widehat{\theta}_{\textnormal{DDML}}(b_{\max})$ } & \multirow{4}{*}{$Z^1(\widehat{\eta})$} & 200 & -0.050 & 0.248 & 0.94 & 0.903 & 0.97 & 1.076 & 1.00 & 1.415 \\
 &  & 500 & -0.030 & 0.132 & 0.92 & 0.458 & 0.96 & 0.546 & 0.99 & 0.718 \\
 &  & 1,000 & -0.020 & 0.083 & 0.91 & 0.283 & 0.96 & 0.337 & 0.99 & 0.444 \\
 &  & 2,000 & -0.014 & 0.055 & 0.90 & 0.180 & 0.95 & 0.214 & 0.99 & 0.282 \\
\cline{2-11}
 & \multirow{4}{*}{$Z^S$} & 200 & 0.007 & 0.384 & 0.97 & 1.542 & 0.99 & 1.837 & 1.00 & 2.414 \\
 &  & 500 & 0.002 & 0.188 & 0.94 & 0.700 & 0.97 & 0.834 & 1.00 & 1.096 \\
 &  & 1,000 & 0.002 & 0.112 & 0.93 & 0.404 & 0.97 & 0.481 & 1.00 & 0.632 \\
 &  & 2,000 & 0.000 & 0.069 & 0.92 & 0.241 & 0.96 & 0.287 & 0.99 & 0.377 \\
\hline
\multirow{8}{*}{$\widehat{\theta}_{\textnormal{VS}}$} & \multirow{4}{*}{$Z^1(\widehat{\eta})$} & 200 & -0.020 & 0.179 & 0.90 & 0.593 & 0.95 & 0.707 & 0.99 & 0.929 \\
 &  & 500 & -0.012 & 0.101 & 0.90 & 0.333 & 0.95 & 0.397 & 0.99 & 0.521 \\
 &  & 1,000 & -0.008 & 0.066 & 0.90 & 0.221 & 0.95 & 0.264 & 0.99 & 0.346 \\
 &  & 2,000 & -0.006 & 0.046 & 0.90 & 0.150 & 0.95 & 0.179 & 0.99 & 0.235 \\
\cline{2-11}
 & \multirow{4}{*}{$Z^S$} & 200 & 0.000 & 0.178 & 0.92 & 0.618 & 0.96 & 0.736 & 0.99 & 0.967 \\
 &  & 500 & 0.001 & 0.103 & 0.91 & 0.348 & 0.96 & 0.415 & 0.99 & 0.545 \\
 &  & 1,000 & 0.001 & 0.068 & 0.91 & 0.231 & 0.96 & 0.275 & 0.99 & 0.361 \\
 &  & 2,000 & 0.000 & 0.047 & 0.90 & 0.155 & 0.95 & 0.185 & 0.99 & 0.243 \\
\hline\hline
\end{tabular}
\end{footnotesize}
\par\smallskip\begin{minipage}{.97\textwidth}\footnotesize\textit{Notes:}
This table reports bias, unscaled RMSE, and the empirical coverage probabilities and mean lengths of nominal $90\%$, $95\%$, and $99\%$ confidence intervals of the DDML estimators at the smallest, largest bandwidths and at the selected bandwidth $(\widehat{b}_N)$. The simulation selects the bandwidth by minimizing the estimated variance of the $Z^1(\hat{\eta})$-DDML estimator and uses that bandwidth for both conditioning specifications. The number of replications is $10{,}000$, and sample sizes are $N\in\{200, 500, 1{,}000, 2{,}000\}$. The first-stage rate is fixed at $a=0.3$. The candidate bandwidth grid is $\mathcal{B}=\{0.1,0.2,0.3,0.4,0.5,0.75,1.0,1.5,2.0\}$, with $b_{\min}=0.1$ and $b_{\max}=2.0$. The second column indicates the conditioning specification used for estimation. Specifically, $Z^{S}$ and $Z^{1}(\widehat{\eta}_{k})$ correspond to the kernels $K(S_i,S_j)$ and $K(\widehat{\eta}_{k}(S_i),\widehat{\eta}_{k}(S_j))$, respectively. Coverage uses two-sided standard normal critical values.
\end{minipage}
\end{table}

\begin{table}[htbp]
\centering
\caption{Selection frequencies for the Gaussian kernel bandwidth for $a=0.3$.}\label{sec06 - table05 - frequency}
\begin{footnotesize}
\begin{tabular}{c|ccccccccc}
\hline\hline
\multicolumn{10}{l}{\textbf{Panel A: DGP 1 is True}} \\
\hline
$N$ & $b=0.1$ & $b=0.2$ & $b=0.3$ & $b=0.4$ & $b=0.5$ & $b=0.75$ & $b=1$ & $b=1.5$ & $b=2$ \\
\hline
200 & 0.00 & 0.00 & 0.04 & 0.13 & 0.30 & 0.24 & 0.15 & 0.14 & 0.00 \\
500 & 0.00 & 0.00 & 0.05 & 0.22 & 0.45 & 0.19 & 0.05 & 0.03 & 0.00 \\
1,000 & 0.00 & 0.00 & 0.05 & 0.34 & 0.51 & 0.09 & 0.01 & 0.00 & 0.00 \\
2,000 & 0.00 & 0.00 & 0.05 & 0.46 & 0.47 & 0.01 & 0.00 & 0.00 & 0.00 \\
\hline
\multicolumn{10}{l}{\textbf{Panel B: DGP 2 is True}} \\
\hline
$N$ & $b=0.1$ & $b=0.2$ & $b=0.3$ & $b=0.4$ & $b=0.5$ & $b=0.75$ & $b=1$ & $b=1.5$ & $b=2$ \\
\hline
200 & 0.00 & 0.00 & 0.00 & 0.00 & 0.00 & 0.39 & 0.61 & 0.00 & 0.00 \\
500 & 0.00 & 0.00 & 0.00 & 0.00 & 0.00 & 0.17 & 0.83 & 0.00 & 0.00 \\
1,000 & 0.00 & 0.00 & 0.00 & 0.00 & 0.00 & 0.02 & 0.98 & 0.00 & 0.00 \\
2,000 & 0.00 & 0.00 & 0.00 & 0.00 & 0.00 & 0.00 & 1.00 & 0.00 & 0.00 \\
\hline\hline
\end{tabular}
\end{footnotesize}
\par\smallskip\begin{minipage}{.97\textwidth}\footnotesize\textit{Notes:}
This table reports the frequency for the Gaussian kernel bandwidth. The first-stage rate is fixed to be$a=0.3$. The number of replications is $10{,}000$, and sample sizes are $N\in\{200, 500, 1{,}000, 2{,}000\}$. The candidate bandwidth grid is $\mathcal{B}=\{0.1,0.2,0.3,0.4,0.5,0.75,1.0,1.5,2.0\}$, with $b_{\min}=0.1$ and $b_{\max}=2.0$. The simulation selects the bandwidth by minimizing the estimated variance of the $Z^1(\hat{\eta})$-DDML estimates.
\end{minipage}
\end{table} \FloatBarrier
%%%%%%%%%%%%%%%%%%%%%%%%%%%%%%%%%%%%%%%%%%%%%%%%%%%%%%%%%%%%%%%%%%%%%%%%%%%%%%%%%%%%%%%%%%%%%%%%%%%%%%%%%%%%%%%%%%%%%%%%%%%%%%%%%%%%%%%%%%%%%%%%%%%%%%%%%%%%%%%%%%%%%%%%%%%%%%%%%%%%%%%%%%%%%%%%%%%%%%%%%%%%%%%%%%%%%%%%%%%%%%%%%%%%%%%%%%%%%%%%%%%%%%%%%%%%%%%%%%%%%%%%%%%%%%%%%%%%%%%%%%%%%%%%%%%%%%%%%%%%%%%%%%%%%%%%%%%%%%%%%%%%%%%%%%%%%%%%%%%%%%%%%%%%%%%%%%%%%%%%%%%%%%%%%%%%%%%%%%%%%%%%%%%%%%%%%%%%%%%%%%%%%%%%%%%%%%%%%%%%%%%%%%%%%%%%%%%%%%%%%%%%%%%%%%%%%%%%%%%%%%%%%%%%%%%%%%%%%%%%%%%%%%%%%%%%%%%%%%%%%%%%%%%%%%%%%%%%%%%%%%%%%%%%%%%%%%%%%%%%%%%%%%%%%%%%%%%%%%%%%%%%%%%%%%%%%%%%
\subsection{Experiment 2: Estimating the First-Stage Nuisance.}
We next examine the finite-sample performance of the DDML estimator and the variance-selection procedure when $\widehat{\eta}$ is estimated by Nadaraya--Watson (NW), local linear (LL), local quadratic (LQ), or $K$-nearest neighbors (KNN). We retain the two DGPs and conditioning specifications from the first experiment. In each fold, we fit the first-stage learner on the training observations and evaluate
its predictions on the held-out observations. We report the prediction RMSE relative to $\eta_0(S_i)$ alongside the performance of the estimators
of $\theta_0$.

Because of the computational cost of the simulation, we fix each learner's tuning parameters across replications. The notes to the results tables describe these choices. In practice, we recommend choosing tuning parameters by cross-validation within the training sample, as we do for the production-function application in the next section.

For each learner and fold, we use the same first-stage fit $\widehat{\eta}_{\ell}$ across all candidate ISPD kernel bandwidths. The preliminary estimate $\widetilde{\theta}_{\ell}$ and the fitted correction $\check{\vecdel}_{\ell}$ are recomputed at each bandwidth because both depend on the kernel. We use a sample size of $N=200$ and run $10{,}000$ replications for each DGP.

Table \ref{sec06 - table06 - bias} reports the bias, RMSE of $\widehat{\eta}$, RMSE of $\widehat{\theta}$, and coverage probabilities of nominal 90\%, 95\%, and 99\% confidence intervals. Each estimate is obtained by using the fixed reference bandwidth $b_{\textnormal{ref}} = 0.5$. Though the plug-in biases are close to zero and the RMSEs are relatively small to DDML RMSEs in both Panel A and B, the plug-in estimates give undercoverage, meaning invalid inference. In both panel, the correctly specified DDML biases are close to zero and its coverage probabilities are close to the nominal level. Though the performance of DDML estimates using KNN is relatively poor to the others in panel A, but according RMSE of the KNN first stage is relatively larger than the others.

Table~\ref{sec06 - table03 - hausman} reports the empirical rejection probabilities of the specification test at the fixed reference
bandwidth. As in the previous experiment, the rejection probabilities in Panel~A, where the null hypothesis is false, exceed the nominal levels. Power is still modest at $N=200$, but the test rejects more often under the alternative than under the null. The results also suggest that more accurate first-stage predictions can improve power. In Panel~B, where the null hypothesis is true, rejection probabilities are generally at or below their nominal levels. The tendency to under-reject may reflect the small sample size of $N=200$.

We next examine the variance-based bandwidth selection procedure for each first-stage learner. Table~\ref{sec06 - table04 - variance selection} compares the DDML estimates at the fixed bandwidths $b_{\min}=0.1$ and $b_{\max}=2.0$ with the estimates at the variance-selected bandwidth $\widehat{\theta}_{\textnormal{VS}}$. For each learner, the table reports bias, unscaled RMSE, and the empirical coverage probabilities and mean lengths of nominal $90\%$, $95\%$, and $99\%$ confidence intervals. Table~\ref{sec06 - table05 - frequency} reports how often each candidate bandwidth is selected across the simulation replications.

In Panel~A, the DDML estimates under the correctly specified conditioning set exhibit small biases and near-nominal coverage at both the fixed and selected bandwidths. Panel~B shows a similar pattern for both conditioning specifications, $Z^S$ and $Z^1(\eta)$. The near-nominal coverage at the selected bandwidth is consistent with the bandwidth-selection procedure with a fixed finite grid. The modest deviation from nominal coverage may partly reflect finite-sample bias. As shown in the preceding tables, KNN yields relatively less accurate first-stage predictions. The corresponding DDML and estimates at the variance-selected bandwidth also tend to perform less well than those based on the other learners. For correctly specified CMRs, the confidence intervals based on $\widehat{\theta}_{\textnormal{VS}}$ are generally shorter than those at either fixed bandwidth, as expected from a rule that minimizes the estimated asymptotic variance.
\FloatBarrier
\begin{table}[htbp]
\centering
\caption{Biases, RMSEs, and coverage probabilities at the reference bandwidth $b_{\textnormal{ref}}=0.5$}\label{sec06 - table06 - bias}
\begin{footnotesize}
\begin{tabular}{c|c|c|c|c|cc|ccc}
\hline\hline
\multicolumn{10}{l}{\textbf{Panel A: DGP 1 is True}} \\
\hline
\multirow{2}{*}{Estimator} & \multirow{2}{*}{$Z^{c}(\eta)$} & \multirow{2}{*}{Learner} & \multirow{2}{*}{RMSE($\widehat{\eta}$)} & \multirow{2}{*}{Bias} & \multicolumn{2}{c|}{RMSE($\widehat{\theta}_{\textnormal{DDML}}$)} & \multicolumn{3}{c}{Coverage} \\
\cline{6-10}
& & & & & (Unscaled) & (Scaled) & 90\% & 95\% & 99\% \\
\hline
\multirow{8}{*}{Plug-in} & \multirow{4}{*}{$Z^{1}(\eta_0)$} 
    & NW & 0.241 & -0.005 & 0.123 & 1.736  & 0.78 & 0.86 & 0.94 \\
 &  & LL & 0.235 & 0.047 & 0.129 & 1.829  & 0.77 & 0.84 & 0.93 \\
 &  & LQ & 0.198 & 0.017 & 0.115 & 1.626  & 0.80 & 0.87 & 0.95 \\
 &  & KNN & 0.315 & -0.031 & 0.119 & 1.684  & 0.80 & 0.87 & 0.95 \\
\cline{2-10}
 & \multirow{4}{*}{$Z^{S}$} 
    & NW & 0.241 & 0.183 & 0.231 & 3.270  & 0.59 & 0.71 & 0.87 \\
 &  & LL & 0.235 & 0.234 & 0.276 & 3.901  & 0.47 & 0.58 & 0.78 \\
 &  & LQ & 0.198 & 0.200 & 0.246 & 3.478  & 0.55 & 0.67 & 0.84 \\
 &  & KNN & 0.315 & 0.165 & 0.214 & 3.023  & 0.64 & 0.75 & 0.90 \\
\hline
\multirow{8}{*}{DDML} & \multirow{4}{*}{$Z^{1}(\eta_0)$} 
    & NW & 0.241 & 0.026 & 0.119 & 1.688  & 0.91 & 0.95 & 0.99 \\
 &  & LL & 0.235 & 0.018 & 0.114 & 1.606  & 0.93 & 0.97 & 1.00 \\
 &  & LQ & 0.198 & 0.012 & 0.111 & 1.568  & 0.92 & 0.96 & 0.99 \\
 &  & KNN & 0.315 & 0.034 & 0.131 & 1.852  & 0.89 & 0.94 & 0.99 \\
\cline{2-10}
 & \multirow{4}{*}{$Z^{S}$} 
    & NW & 0.241 & 0.183 & 0.230 & 3.257  & 0.66 & 0.77 & 0.93 \\
 &  & LL & 0.235 & 0.183 & 0.231 & 3.265  & 0.68 & 0.80 & 0.94 \\
 &  & LQ & 0.198 & 0.183 & 0.231 & 3.262  & 0.67 & 0.78 & 0.93 \\
 &  & KNN & 0.315 & 0.183 & 0.231 & 3.264  & 0.65 & 0.76 & 0.92 \\
\hline
\multicolumn{10}{l}{\textbf{Panel B: DGP 2 is True}} \\
\hline
\multirow{2}{*}{Estimator} & \multirow{2}{*}{$Z^{c}(\eta)$} & \multirow{2}{*}{Learner} & \multirow{2}{*}{RMSE($\widehat{\eta}$)} & \multirow{2}{*}{Bias} & \multicolumn{2}{c|}{RMSE($\widehat{\theta}_{\textnormal{DDML}}$)} & \multicolumn{3}{c}{Coverage} \\
\cline{6-10}
& & & & & (Unscaled) & (Scaled) & 90\% & 95\% & 99\% \\
\hline
\multirow{8}{*}{Plug-in} & \multirow{4}{*}{$Z^{1}(\eta_0)$} 
    & NW & 0.203 & -0.032 & 0.143 & 2.021  & 0.76 & 0.83 & 0.93 \\
 &  & LL & 0.185 & -0.045 & 0.144 & 2.034  & 0.73 & 0.82 & 0.92 \\
 &  & LQ & 0.177 & -0.036 & 0.144 & 2.034  & 0.74 & 0.83 & 0.93 \\
 &  & KNN & 0.217 & -0.038 & 0.144 & 2.035  & 0.75 & 0.83 & 0.93 \\
\cline{2-10}
 & \multirow{4}{*}{$Z^{S}$} 
    & NW & 0.203 & 0.000 & 0.145 & 2.053  & 0.76 & 0.84 & 0.94 \\
 &  & LL & 0.185 & -0.029 & 0.141 & 1.999  & 0.74 & 0.82 & 0.93 \\
 &  & LQ & 0.177 & -0.023 & 0.142 & 2.013  & 0.75 & 0.83 & 0.93 \\
 &  & KNN & 0.217 & 0.009 & 0.150 & 2.119  & 0.76 & 0.84 & 0.93 \\
\hline
\multirow{8}{*}{DDML} & \multirow{4}{*}{$Z^{1}(\eta_0)$} 
    & NW & 0.203 & 0.020 & 0.166 & 2.352  & 0.87 & 0.93 & 0.98 \\
 &  & LL & 0.185 & 0.027 & 0.168 & 2.381  & 0.87 & 0.93 & 0.98 \\
 &  & LQ & 0.177 & 0.036 & 0.173 & 2.451  & 0.87 & 0.93 & 0.98 \\
 &  & KNN & 0.217 & 0.018 & 0.165 & 2.336  & 0.88 & 0.94 & 0.98 \\
\cline{2-10}
 & \multirow{4}{*}{$Z^{S}$} 
    & NW & 0.203 & 0.031 & 0.161 & 2.272  & 0.90 & 0.95 & 0.99 \\
 &  & LL & 0.185 & 0.025 & 0.157 & 2.215  & 0.88 & 0.94 & 0.99 \\
 &  & LQ & 0.177 & 0.027 & 0.158 & 2.234  & 0.89 & 0.94 & 0.99 \\
 &  & KNN & 0.217 & 0.038 & 0.165 & 2.327  & 0.91 & 0.96 & 0.99 \\
\hline\hline
\end{tabular}
\end{footnotesize}
\par\smallskip\begin{minipage}{.97\textwidth}\footnotesize\textit{Notes:}
This table reports bias, RMSE, the empirical coverage probabilities of nominal $90\%$, $95\%$, and $99\%$ confidence intervals of the plug-in and the DDML estimates using the specified first-stage learner. The number of replications is $10{,}000$, and the sample size is $200$. All estimates are obtained using the reference bandwidth $b_\textnormal{ref}=0.5$. The second column indicates the conditioning specification used for estimation. Panels A and B report results under DGP 1 and DGP 2, respectively. Each learner NW, LL, LQ, and KNN denotes Nadaraya-Watson (smoothing parameter$ = 0.5\times N^{-1/5}$), local linear (smoothing parameter$ = 0.75\times N^{-1/5}$), local quadratic (smoothing parameter$ = 1.5\times N^{-1/5}$), $K$-nearest neighbors($K = 40$), respectively. RMSE($\widehat{\eta}$) measures the prediction error of $\widehat{\eta}_{\ell}(S_i)$ relative to the true $\eta_0(S_i)$.Scaled RMSE is $\sqrt{N}$ times unscaled RMSE. Coverage uses two-sided standard normal critical values.
\end{minipage}
\end{table}

\begin{table}[htbp]
\centering
\caption{Rejection probabilities of the Hausman specification test}\label{sec06 - table07 - hausman}
\begin{footnotesize}
\begin{tabular}{c|ccc}
\hline\hline
\multicolumn{4}{l}{\textbf{Panel A: DGP 1 is True ($H_0$ is False)}} \\
\hline
 & \multicolumn{3}{c}{Rejection Probability} \\
\cline{2-4}
Learner & 10\% & 5\% & 1\% \\
\hline
NW  & 0.58 & 0.42 & 0.17 \\
LL  & 0.64 & 0.48 & 0.19 \\
LQ  & 0.71 & 0.56 & 0.25 \\
KNN & 0.48 & 0.36 & 0.18 \\
\hline
\multicolumn{4}{l}{\textbf{Panel B: DGP 2 is True ($H_0$ is True)}} \\
\hline
 & \multicolumn{3}{c}{Rejection Probability} \\
\cline{2-4}
Learner & 10\% & 5\% & 1\% \\
\hline
NW  & 0.08 & 0.03 & 0.00 \\
LL  & 0.07 & 0.03 & 0.00 \\
LQ  & 0.07 & 0.03 & 0.00 \\
KNN & 0.07 & 0.03 & 0.01 \\
\hline\hline
\end{tabular}
\end{footnotesize}
\par\smallskip\begin{minipage}{.97\textwidth}\footnotesize\textit{Notes:} The fixed bandwidths are $b_{\min}=0.1$, $b_{\max}=2$, and $b_{\mathrm{ref}}=0.5$. The selected-bandwidth test uses the same saved bandwidth for both DDML estimates within each replication. The number of replications is $10{,}000$, and the sample size is $200$. Rejection rates use valid test statistics in each column and $\chi^2_1$ critical values. Each learner NW, LL, LQ, and KNN denotes Nadaraya-Watson (smoothing parameter$ = 0.5\times N^{-1/5}$), local linear (smoothing parameter$ = 0.75\times N^{-1/5}$), local quadratic (smoothing parameter$ = 1.5\times N^{-1/5}$), $K$-nearest neighbors($K = 40$), respectively.
\end{minipage}
\end{table}

\begin{table}[htbp]
\centering
\caption{Selection frequencies for the Gaussian kernel bandwidth.}\label{sec06 - table09 - frequency}
\begin{footnotesize}
\begin{tabular}{c|ccccccccc}
\hline\hline
\multicolumn{10}{l}{\textbf{Panel A: DGP 1 is True}} \\
\hline
Learner & $b=0.1$ & $b=0.2$ & $b=0.3$ & $b=0.4$ & $b=0.5$ & $b=0.75$ & $b=1$ & $b=1.5$ & $b=2$ \\
\hline
NW & 0.06 & 0.10 & 0.10 & 0.05 & 0.05 & 0.18 & 0.42 & 0.06 & 0.00 \\
LL & 0.24 & 0.24 & 0.12 & 0.06 & 0.05 & 0.09 & 0.19 & 0.01 & 0.00 \\
LQ & 0.30 & 0.31 & 0.15 & 0.07 & 0.05 & 0.04 & 0.08 & 0.01 & 0.00 \\
KNN & 0.00 & 0.01 & 0.02 & 0.01 & 0.02 & 0.16 & 0.66 & 0.12 & 0.00 \\
\hline
\multicolumn{10}{l}{\textbf{Panel B: DGP 2 is True}} \\
\hline
Learner & $b=0.1$ & $b=0.2$ & $b=0.3$ & $b=0.4$ & $b=0.5$ & $b=0.75$ & $b=1$ & $b=1.5$ & $b=2$ \\
\hline
NW & 0.01 & 0.07 & 0.19 & 0.19 & 0.28 & 0.23 & 0.02 & 0.00 & 0.00 \\
LL & 0.00 & 0.05 & 0.14 & 0.14 & 0.24 & 0.36 & 0.06 & 0.00 & 0.00 \\
LQ & 0.00 & 0.05 & 0.13 & 0.14 & 0.26 & 0.36 & 0.06 & 0.00 & 0.00 \\
KNN & 0.03 & 0.14 & 0.26 & 0.23 & 0.23 & 0.10 & 0.01 & 0.00 & 0.00 \\
\hline\hline
\end{tabular}
\end{footnotesize}
\par\smallskip\begin{minipage}{.97\textwidth}\footnotesize\textit{Notes:}
This table reports the frequency for the Gaussian kernel bandwidth. The candidate bandwidth grid is $\mathcal{B}=\{0.1,0.2,0.3,0.4,0.5,0.75,1.0,1.5,2.0\}$, with $b_{\min}=0.1$ and $b_{\max}=2.0$. The simulation selects the bandwidth by minimizing the estimated variance of the $Z^1(\hat{\eta})$-DDML estimator and uses that bandwidth for both conditioning specifications.
\end{minipage}
\end{table}

\begin{table}[htbp]
\centering
\caption{Bias, RMSE, and confidence interval for the variance-selected DDML esitmator.}\label{sec06 - table08 - variance selection}
\begin{footnotesize}
\begin{tabular}{c|c|c|cc|cc|cc|cc}
\hline\hline
\multicolumn{11}{l}{\textbf{Panel A: DGP 1 is True}} \\
\hline
\multirow{2}{*}{Estimator} & \multirow{2}{*}{$Z^{c}(\eta)$} & \multirow{2}{*}{Learner} & \multirow{2}{*}{Bias} & \multirow{2}{*}{\shortstack{RMSE\\(unscaled)}} & \multicolumn{2}{c|}{90\% CI} & \multicolumn{2}{c|}{95\% CI} & \multicolumn{2}{c}{99\% CI} \\
\cline{6-11}
& & & & & Coverage & Length & Coverage & Length & Coverage & Length \\
\hline
\multirow{8}{*}{ $\widehat{\theta}_{\textnormal{DDML}}(b_{\min})$ } & \multirow{4}{*}{$Z^{1}(\widehat{\eta})$} 
    & NW & 0.043 & 0.138 & 0.93 & 0.487 & 0.97 & 0.581 & 0.99 & 0.763 \\
 &  & LL & 0.028 & 0.121 & 0.93 & 0.431 & 0.97 & 0.514 & 0.99 & 0.675 \\
 &  & LQ & 0.010 & 0.109 & 0.92 & 0.383 & 0.96 & 0.457 & 0.99 & 0.600 \\
 &  & KNN & -0.012 & 0.171 & 0.92 & 0.590 & 0.96 & 0.703 & 0.99 & 0.923 \\
\cline{2-11}
 & \multirow{4}{*}{$Z^{S}$} 
    & NW & 0.048 & 0.168 & 0.90 & 0.559 & 0.95 & 0.666 & 0.99 & 0.875 \\
 &  & LL & 0.051 & 0.169 & 0.92 & 0.593 & 0.96 & 0.707 & 0.99 & 0.929 \\
 &  & LQ & 0.051 & 0.170 & 0.90 & 0.568 & 0.95 & 0.676 & 0.99 & 0.889 \\
 &  & KNN & 0.051 & 0.169 & 0.90 & 0.560 & 0.95 & 0.667 & 0.99 & 0.877 \\
\hline
\multirow{8}{*}{ $\widehat{\theta}_{\textnormal{DDML}}(b_{\max})$ } & \multirow{4}{*}{$Z^{1}(\widehat{\eta})$} 
    & NW & 0.005 & 0.129 & 0.91 & 0.427 & 0.95 & 0.509 & 0.99 & 0.669 \\
 &  & LL & 0.011 & 0.128 & 0.93 & 0.457 & 0.97 & 0.545 & 0.99 & 0.716 \\
 &  & LQ & 0.012 & 0.128 & 0.92 & 0.436 & 0.96 & 0.519 & 0.99 & 0.682 \\
 &  & KNN & 0.005 & 0.129 & 0.90 & 0.416 & 0.95 & 0.496 & 0.99 & 0.652 \\
\cline{2-11}
 & \multirow{4}{*}{$Z^{S}$} 
    & NW & 0.069 & 0.185 & 0.90 & 0.581 & 0.96 & 0.693 & 0.99 & 0.910 \\
 &  & LL & 0.068 & 0.185 & 0.91 & 0.587 & 0.96 & 0.700 & 0.99 & 0.919 \\
 &  & LQ & 0.069 & 0.185 & 0.91 & 0.586 & 0.96 & 0.698 & 0.99 & 0.918 \\
 &  & KNN & 0.069 & 0.185 & 0.90 & 0.577 & 0.95 & 0.688 & 0.99 & 0.904 \\
\hline
\multirow{8}{*}{ $\widehat{\theta}_{\textnormal{VS}}$ } & \multirow{4}{*}{$Z^{1}(\widehat{\eta})$} 
    & NW & 0.012 & 0.119 & 0.89 & 0.380 & 0.94 & 0.453 & 0.99 & 0.596 \\
 &  & LL & 0.014 & 0.115 & 0.92 & 0.396 & 0.96 & 0.471 & 0.99 & 0.619 \\
 &  & LQ & 0.008 & 0.110 & 0.90 & 0.366 & 0.95 & 0.436 & 0.99 & 0.573 \\
 &  & KNN & 0.011 & 0.119 & 0.89 & 0.378 & 0.94 & 0.451 & 0.99 & 0.592 \\
\cline{2-11}
 & \multirow{4}{*}{$Z^{S}$} 
    & NW & 0.146 & 0.210 & 0.76 & 0.496 & 0.85 & 0.591 & 0.96 & 0.776 \\
 &  & LL & 0.116 & 0.195 & 0.82 & 0.533 & 0.90 & 0.635 & 0.98 & 0.835 \\
 &  & LQ & 0.101 & 0.188 & 0.83 & 0.523 & 0.90 & 0.623 & 0.97 & 0.819 \\
 &  & KNN & 0.160 & 0.220 & 0.73 & 0.495 & 0.84 & 0.590 & 0.96 & 0.775 \\
\hline
\multicolumn{11}{l}{\textbf{Panel B: DGP 2 is True}} \\
\hline
\multirow{2}{*}{Estimator} & \multirow{2}{*}{$Z^{c}(\eta)$} & \multirow{2}{*}{Learner} & \multirow{2}{*}{Bias} & \multirow{2}{*}{\shortstack{RMSE\\(unscaled)}} & \multicolumn{2}{c|}{90\% CI} & \multicolumn{2}{c|}{95\% CI} & \multicolumn{2}{c}{99\% CI} \\
\cline{6-11}
& & & & & Coverage & Length & Coverage & Length & Coverage & Length \\
\hline
\multirow{8}{*}{ $\widehat{\theta}_{\textnormal{DDML}}(b_{\min})$ } & \multirow{4}{*}{$Z^{1}(\widehat{\eta})$} 
    & NW & 0.005 & 0.181 & 0.89 & 0.566 & 0.94 & 0.674 & 0.99 & 0.886 \\
 &  & LL & 0.017 & 0.184 & 0.89 & 0.570 & 0.94 & 0.679 & 0.99 & 0.892 \\
 &  & LQ & 0.028 & 0.187 & 0.90 & 0.581 & 0.95 & 0.693 & 0.99 & 0.910 \\
 &  & KNN & 0.000 & 0.167 & 0.91 & 0.552 & 0.95 & 0.657 & 0.99 & 0.864 \\
\cline{2-11}
 & \multirow{4}{*}{$Z^{S}$} 
    & NW & 0.025 & 0.153 & 0.89 & 0.472 & 0.94 & 0.563 & 0.99 & 0.740 \\
 &  & LL & 0.024 & 0.152 & 0.88 & 0.457 & 0.94 & 0.545 & 0.98 & 0.716 \\
 &  & LQ & 0.030 & 0.154 & 0.88 & 0.463 & 0.94 & 0.552 & 0.99 & 0.725 \\
 &  & KNN & 0.035 & 0.159 & 0.90 & 0.494 & 0.95 & 0.589 & 0.99 & 0.773 \\
\hline
\multirow{8}{*}{ $\widehat{\theta}_{\textnormal{DDML}}(b_{\max})$ } & \multirow{4}{*}{$Z^{1}(\widehat{\eta})$} 
    & NW & 0.085 & 0.265 & 0.93 & 0.862 & 0.97 & 1.027 & 1.00 & 1.349 \\
 &  & LL & 0.075 & 0.252 & 0.92 & 0.804 & 0.97 & 0.958 & 1.00 & 1.259 \\
 &  & LQ & 0.084 & 0.259 & 0.92 & 0.822 & 0.97 & 0.980 & 1.00 & 1.288 \\
 &  & KNN & 0.113 & 0.296 & 0.93 & 0.946 & 0.97 & 1.127 & 1.00 & 1.482 \\
\cline{2-11}
 & \multirow{4}{*}{$Z^{S}$} 
    & NW & 0.101 & 0.331 & 0.96 & 1.251 & 0.98 & 1.491 & 1.00 & 1.960 \\
 &  & LL & 0.079 & 0.309 & 0.96 & 1.180 & 0.98 & 1.407 & 1.00 & 1.849 \\
 &  & LQ & 0.083 & 0.312 & 0.96 & 1.187 & 0.98 & 1.414 & 1.00 & 1.858 \\
 &  & KNN & 0.128 & 0.361 & 0.96 & 1.327 & 0.98 & 1.581 & 1.00 & 2.078 \\
\hline
\multirow{8}{*}{ $\widehat{\theta}_{\textnormal{VS}}$ } & \multirow{4}{*}{$Z^{1}(\widehat{\eta})$}
    & NW & 0.019 & 0.164 & 0.87 & 0.485 & 0.93 & 0.578 & 0.98 & 0.760 \\
 &  & LL & 0.029 & 0.167 & 0.87 & 0.480 & 0.93 & 0.572 & 0.98 & 0.752 \\
 &  & LQ & 0.037 & 0.172 & 0.87 & 0.490 & 0.93 & 0.584 & 0.98 & 0.768 \\
 &  & KNN & 0.013 & 0.160 & 0.89 & 0.495 & 0.94 & 0.590 & 0.99 & 0.776 \\
\cline{2-11}
 & \multirow{4}{*}{$Z^{S}$} 
    & NW & 0.034 & 0.160 & 0.90 & 0.494 & 0.95 & 0.589 & 0.99 & 0.774 \\
 &  & LL & 0.029 & 0.157 & 0.89 & 0.478 & 0.94 & 0.570 & 0.99 & 0.748 \\
 &  & LQ & 0.031 & 0.159 & 0.89 & 0.483 & 0.94 & 0.575 & 0.99 & 0.756 \\
 &  & KNN & 0.038 & 0.163 & 0.90 & 0.508 & 0.95 & 0.605 & 0.99 & 0.795 \\
\hline\hline
\end{tabular}
\end{footnotesize}
\par\smallskip\begin{minipage}{.97\textwidth}\footnotesize\textit{Notes:}
This table reports bias, RMSE, the empirical coverage probabilities of nominal $90\%$, $95\%$, and $99\%$ confidence intervals of the DDML estimators at the smallest, largest bandwidths and at the selected bandwidth $(\widehat{b}_N)$. The simulation selects the bandwidth by minimizing the estimated variance of the $Z^1(\hat{\eta})$-DDML estimator and uses that bandwidth for both conditioning specifications. The number of replications is $10{,}000$, and the sample size is $200$. The second column indicates the conditioning specification used for estimation. Panels A and B report results under DGP 1 and DGP 2, respectively. Each learner NW, LL, LQ, and KNN denotes Nadaraya-Watson (smoothing parameter$ = 0.5\times N^{-1/5}$), local linear (smoothing parameter$ = 0.75\times N^{-1/5}$), local quadratic (smoothing parameter$ = 1.5\times N^{-1/5}$), $K$-nearest neighbors($K = 40$), respectively. RMSE($\widehat{\eta}$) measures the prediction error of $\widehat{\eta}_{\ell}(S_i)$ relative to the true $\eta_0(S_i)$. Coverage uses two-sided standard normal critical values.
\end{minipage}
\end{table} \FloatBarrier
%%%%%%%%%%%%%%%%%%%%%%%%%%%%%%%%%%%%%%%%%%%%%%%%%%%%%%%%%%%%%%%%%%%%%%%%%%%%%%%%%%%%%%%%%%%%%%%%%%%%%%%%%%%%%%%%%%%%%%%%%%%%%%%%%%%%%%%%%%%%%%%%%%%%%%%%%%%%%%%%%%%%%%%%%%%%%%%%%%%%%%%%%%%%%%%%%%%%%%%%%%%%%%%%%%%%%%%%%%%%%%%%%%%%%%%%%%%%%%%%%%%%%%%%%%%%%%%%%%%%%%%%%%%%%%%%%%%%%%%%%%%%%%%%%%%%%%%%%%%%%%%%%%%%%%%%%%%%%%%%%%%%%%%%%%%%%%%%%%%%%%%%%%%%%%%%%%%%%%%%%%%%%%%%%%%%%%%%%%%%%%%%%%%%%%%%%%%%%%%%%%%%%%%%%%%%%%%%%%%%%%%%%%%%%%%%%%%%%%%%%%%%%%%%%%%%%%%%%%%%%%%%%%%%%%%%%%%%%%%%%%%%%%%%%%%%%%%%%%%%%%%%%%%%%%%%%%%%%%%%%%%%%%%%%%%%%%%%%%%%%%%%%%%%%%%%%%%%%%%%%%%%%%%%%%%%%%%%
\section{Empirical illustrations}\label{sec07:empirical}
The first application revisits the production-function example in Section~\ref{sec02}, where finite unconditional moments admit multiple roots. In Section~\ref{sec07:cereal}, we use the cereal scanner data analyzed by \citet{fosgerau_monardo_depalma2024} to illustrate how to combine estimates from two CMRs, as in Corollary~\ref{sec04 - corollary01}, and apply the conditioning-set test discussed in Section~\ref{sec05:test}.

Both applications fit their first stages using local quadratic regression, XGBoost, penalized splines, LASSO, and random forests.\footnote{
    In the production-function application, the local quadratic regression includes an intercept, four linear terms, four squares, and six pairwise interactions, with the regression bandwidth selected from $\{0.3,0.5,0.8,1.2,2,3\}$ in standardized coordinates. The DDML full-sample local-quadratic fit selects $h=3$, with CV RMSE 0.610. XGBoost selects among depths $\{1,2,3\}$, learning rates $\{0.03,0.1\}$, tree counts $\{200,500\}$, and minimum child weights $\{5,20\}$. All 24 settings use squared-error histogram trees, subsample fraction 0.8, L2 penalty 1, and L1 penalty 0. The full-sample fit selects depth 2, learning rate 0.03, 200 trees, and minimum child weight 20, with CV RMSE 0.629. This is a boosted-tree learner of the type considered by \citet{dml_chernozhukov_et_al2018}. Penalized splines use cubic B-spline main effects and pairwise tensor interactions, marginal basis sizes 5 or 7, and second-difference penalties from $\{10^{-6},\allowbreak 10^{-5},\allowbreak 10^{-4},\allowbreak 10^{-3},\allowbreak 10^{-2},\allowbreak 10^{-1},\allowbreak 1,\allowbreak 10,\allowbreak 100\}$. LASSO uses standardized polynomial dictionaries of degrees 2, 3, and 4, and penalty fractions $\{1,\allowbreak .5,\allowbreak .2,\allowbreak .1,\allowbreak .05,\allowbreak .02,\allowbreak .01,\allowbreak .005,\allowbreak .002,\allowbreak .001,\allowbreak .0005,\allowbreak .0001\}$ of the training maximum penalty. Random forests use 500 trees, minimum leaf sizes $\{1,3,5,10,15,30,60\}$, and $\{1,2,4\}$ candidate variables per split.}
We tune each learner within its training sample using three folds of plants for production and five folds of stores for cereal demand.

Both applications use the Gaussian ISPD kernel
\begin{equation}
    K_{b,\ell}(z,z')   = \exp\!\left\{-\frac{\|D_{\ell}^{-1}(z-z')\|^2}{2b^2}\right\},    \qquad
 \mathcal{B}=\{0.1,0.2,0.3,0.4,0.5,0.75,1,1.5,2\},
 \label{sec07:eq:kernel}
\end{equation}
where $D_{\ell}$ is diagonal with training-sample standard deviations of the nonconstant conditioning variables. The fixed grid is common across learners and is distinct from the tuning parameters of the nuisance regressions.

For the production-function application, we apply the bandwidth rule separately to each first-stage learner. The criterion is the mean estimated asymptotic variance of the three input elasticities. For cereal demand, we select a bandwidth for each learner using the estimated variance of the combined price coefficient under the fitted-price ISPD kernel. In the notation of equation~\eqref{sec05:eq:varianceloss}, these criteria are
\begin{equation}\label{sec07:eq:variancecriteria}
 \widehat v_N^{\mathrm{ACF}}(b)=\frac{1}{3}\sum_{j\in\{s,u,k\}}
    \widehat\Omega_{N,jj}(b),
 \qquad
 \widehat v_N^{\mathrm{cereal}}(b)=\widehat\Omega^{1}_{N,\beta\beta}(b).
\end{equation}
The two ISPD kernels for cereal demand are evaluated at the same bandwidth within each comparison. Appendix~\ref{app:varianceselection} reports the production-function variance criterion over the full grid.

\subsection{ACF production-function estimation}\label{sec07:acf}
The model, data, and structural residual are those in Section~\ref{sec02}. The DDML sample contains 1,944 plant-year observations from 401 plants. We use $G=5$ nearly equal observation groups, giving $L=G(2G-1)=45$ pair folds in Definition~\ref{sec04 - definition01 - DML}. Nuisance fits exclude plants represented in each fold's evaluation observations. For every $b\in\mathcal{B}$, the local-quadratic corrections are evaluated at the training pilots and held fixed as in equation~\eqref{sec04 - definition01 - DML - eq01}. We jointly estimate $(\beta_s,\beta_u,\beta_k,\alpha,\gamma)$ by minimizing the squared norm of the pooled corrected score in equation~\eqref{sec04 - definition01 - DML - eq02} over $[-5,5]^3\times[-10,10]\times[0.1,1.2]$. Point estimation and inference use the same observation pairs and the covariance formula in Proposition~\ref{sec04 - proposition02}.

\begin{table}[htbp]\centering\small
\setlength{\tabcolsep}{4pt}
\caption{ACF DDML estimates with variance-selected bandwidths}
\label{sec07:tab:acfselected}
\begin{tabular}{l|r|rrrrr}\hline\hline
First stage & $\widehat{b}_{N}$ & $\beta_s$ & $\beta_u$ & $\beta_k$ & $\alpha$ & $\gamma$\\\hline
Local quadratic & 0.5 & $0.063$ & $0.052$ & $0.087$ & $-0.008$ & $1.002$\\
 & & $(0.098)$ & $(0.082)$ & $(0.059)$ & $(0.117)$ & $(0.010)$\\
XGBoost & 0.3 & $-0.003$ & $-0.053$ & $0.086$ & $0.037$ & $0.998$\\
 & & $(0.158)$ & $(0.133)$ & $(0.097)$ & $(0.136)$ & $(0.011)$\\
Penalized splines & 0.4 & $0.036$ & $0.008$ & $0.126$ & $0.113$ & $0.991$\\
 & & $(0.120)$ & $(0.100)$ & $(0.049)$ & $(0.122)$ & $(0.011)$\\
LASSO & 0.3 & $0.028$ & $0.001$ & $0.156$ & $0.043$ & $0.998$\\
 & & $(0.151)$ & $(0.123)$ & $(0.055)$ & $(0.081)$ & $(0.007)$\\
Random forest & 0.4 & $0.045$ & $0.035$ & $0.000$ & $-0.022$ & $1.003$\\
 & & $(0.104)$ & $(0.087)$ & $(0.101)$ & $(0.147)$ & $(0.012)$\\
\hline\hline
\end{tabular}
\par\smallskip\begin{minipage}{.97\textwidth}\footnotesize\textit{Notes:}
Each row minimizes the estimated mean asymptotic variance of the three input elasticities over all nine bandwidths in equation~\eqref{sec07:eq:kernel}. All five parameters are jointly estimated at that bandwidth. Parentheses report usual standard errors from Proposition~\ref{sec04 - proposition02}, computed with the same observation pairs as point estimation. The sample has 1,944 plant-year observations from 401 plants.
\end{minipage}\end{table}

\begin{figure}[htbp]\centering
    \includegraphics[width=\textwidth]{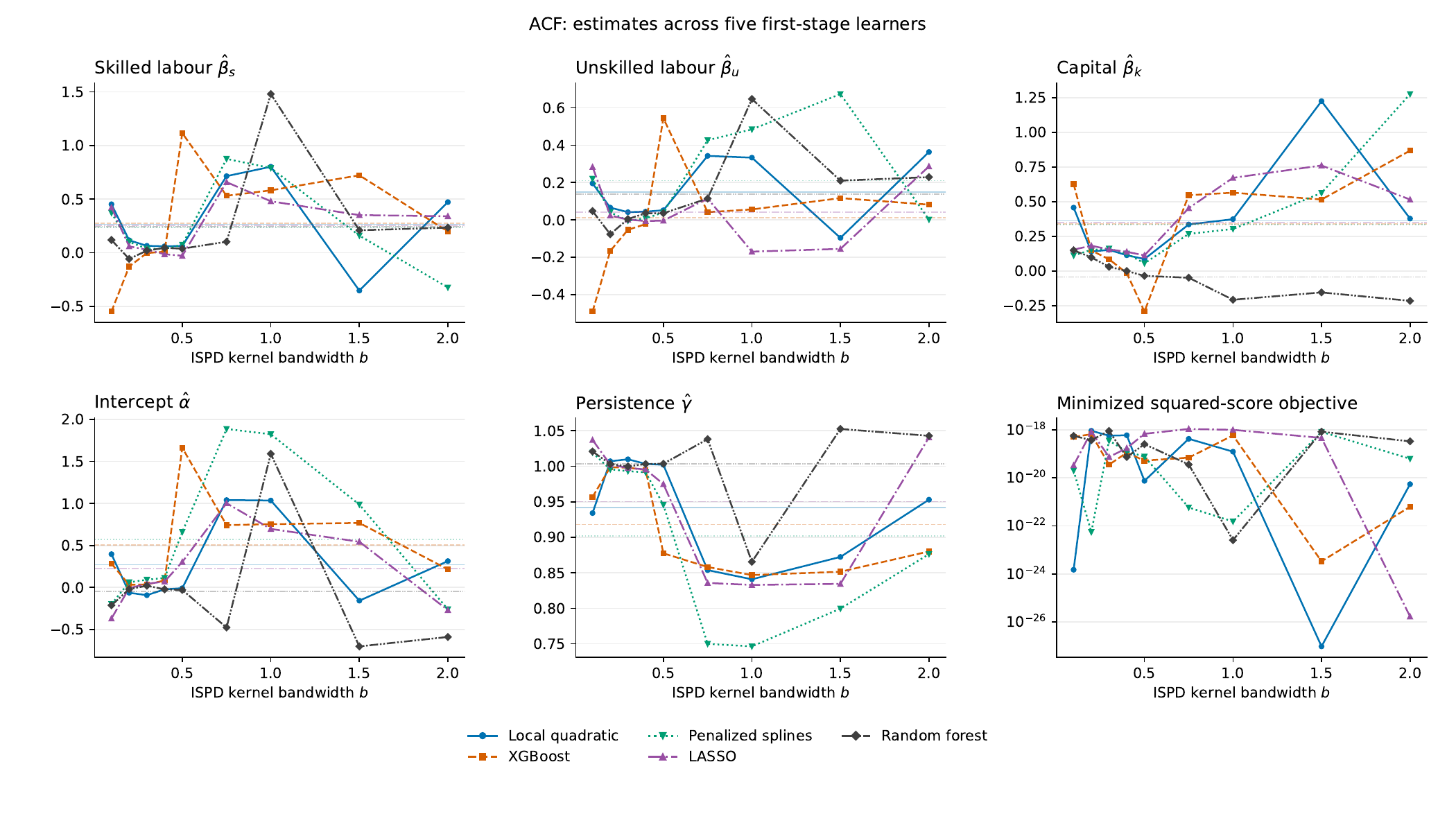}
    \caption{ACF parameter estimates and squared corrected scores across bandwidths} \label{sec07:fig:paths}
    \begin{minipage}{.96\textwidth}\footnotesize\textit{Notes:} Each line corresponds to a first-stage estimator, with the ISPD kernel bandwidth $b$ on the horizontal axis. The faint horizontal lines show the averages of the parameter estimates across the nine candidate bandwidths.\end{minipage}
\end{figure} \FloatBarrier
{ The estimates vary across bandwidths (Figure~\ref{sec07:fig:paths}). At the selected bandwidths, the five learners give similar estimates, except that the random forest yields a lower capital elasticity. For each learner, the individual 95\% confidence intervals for both labor elasticities include zero, and the estimated productivity persistence $\gamma$ is close to one. Appendix~\ref{app:implementation_ACF} gives the implementation details.}

\subsection{Cereal demand with two conditional moment restrictions}\label{sec07:cereal}

\citet{fosgerau_monardo_depalma2024} use Dominick's supermarket scanner data to estimate how consumers substitute among 45 breakfast cereals. The underlying weekly scanner records are available from the \href{https://www.chicagobooth.edu/research/kilts/research-data/dominicks}{Kilts Center for Marketing at the University of Chicago Booth School of Business}. From these records, we construct a sample of 4,363 store-weeks across 86 stores and calculate the shares of Cheerios, Rice Krispies, and Grape-Nuts in their combined unit sales. This simpler setting illustrates how to combine two share restrictions to estimate price sensitivity and how to test whether both restrictions still hold when conditioning directly on acquisition costs.

\paragraph{Retail prices and demand.}
Suppose a box of Cheerios usually sells for \$3.00. Consider two reasons the store might raise its price to \$3.10, holding promotions and the other cereals' prices fixed. First, the store may have to pay more for Cheerios and pass the higher cost on to shoppers. Second, the store may pay its usual cost but raise the price because it expects unusually strong demand for Cheerios. In the second case, Cheerios would have gained market share even at \$3.00. If the researcher cannot observe that demand shift, strong sales at \$3.10 can make shoppers appear less sensitive to price than they are.

To address this endogeneity issue, we use the control-function approach of \citet{petrin_train2010}. We first predict relative retail prices from acquisition costs and promotions. Suppose the higher cost in the first case predicts the full ten-cent increase. The price residual, actual minus predicted price, then stays the same. In the second case, costs and promotions are unchanged, so the fitted price stays the same and the residual rises by ten cents. The demand equation includes the price and its residual and excludes acquisition costs. To estimate price sensitivity, we compare store-weeks with the same promotions and price residuals but different prices predicted by costs. This comparison identifies the price effect if those remaining price differences are unrelated to unobserved demand.

The acquisition cost is the average amount the chain paid for inventory on hand. Earlier sales and restocking decisions can therefore affect the recorded cost. The conditioning-set comparison asks whether the individual acquisition costs still help predict the share errors once actual prices, fitted prices, and promotions are given.

\paragraph{Two share restrictions.}
Take Grape-Nuts as the reference product. For Cheerios ($j=1$) and Rice Krispies ($j=2$), let $p_{ij}$ and $f_{ij}$ be the retail price and promotion indicator relative to Grape-Nuts. Let $A_i$ contain the three acquisition costs. Write $p_i=(p_{i1},p_{i2})^\top$, $f_i=(f_{i1},f_{i2})^\top$, $S_i=(f_i^\top,A_i^\top)^\top$, and $\eta_0(S_i)=\mathbb E[p_i\mid S_i]$. The price control $v_i(\eta)=p_i-\eta(S_i)$ enters each cereal's logit index,
\begin{equation}\label{sec07:eq:cereal-utility}
 u_{i0}=0,\qquad u_{ij}=\alpha_j+\beta p_{ij}+\delta f_{ij}+\gamma v_{ij}(\eta),
 \qquad j=1,2.
\end{equation}
Let $\rho_{ij}(\theta,\eta)=Y_{ij}-P_{ij}(\theta,\eta)$ be the difference between the observed share and its logit probability, where $\theta=(\alpha_1,\alpha_2,\beta,\delta,\gamma)^\top$. The shares sum to one, so the first two share errors determine the third. For each of the two shares, consider
\begin{align}
 \mathbb E[\rho_{ij}(\theta_0,\eta_0)\mid p_i,f_i,\eta_0(S_i)]&=0
    &&\text{(fitted prices)},\label{sec07:eq:cereal-baseline}\\
 \mathbb E[\rho_{ij}(\theta_0,\eta_0)\mid p_i,f_i,A_i]&=0
    &&\text{(acquisition costs)}.\label{sec07:eq:cereal-costs}
\end{align}
Equation~\eqref{sec07:eq:cereal-baseline} is the maintained restriction. Since $\eta_0(S_i)$ is a function of promotions and acquisition costs, equation~\eqref{sec07:eq:cereal-costs} imposes a stronger restriction. Even among store-weeks with the same actual prices, promotions, and predicted prices, differences in acquisition costs must leave both mean share errors at zero. The price control in utility is the same in both equations.

At each bandwidth, the Cheerios equation gives one DDML estimate of all five coefficients and the Rice Krispies equation gives another. We combine the two vectors by minimum distance using their joint covariance, as in Corollary~\ref{sec04 - corollary01}. We carry out these steps first conditioning on fitted prices and then conditioning directly on acquisition costs. Both calculations use the same five groups of stores and the same held-out store-week pairs.

\paragraph{Demand estimates.}
Table~\ref{sec07:tab:cereal-shares} reports the DDML estimates from the two fitted-price restrictions in equation~\eqref{sec07:eq:cereal-baseline}. The first panel uses only the Cheerios share error and the second uses only the Rice Krispies share error. Both cereal-specific intercepts enter each share probability through the logit denominator, so each panel reports all five demand coefficients. We sort the 86 stores by ID and assign consecutive stores to five folds of 18, 17, 17, 17, and 17 stores. All weeks from a store stay in its fold.

\begin{table}[htbp]\centering\small
\caption{DDML estimates from the separate share restrictions, conditioning on fitted prices}
\label{sec07:tab:cereal-shares}
\begin{tabular}{lrrrrrr}
\hline\hline
\multirow{2}{*}{First stage} & \multirow{2}{*}{$\widehat{b}$} & $\alpha_1$ & $\alpha_2$ & $\beta$ & $\delta$ & $\gamma$\\
 & & Cheerios & Rice Krispies & Price & Promotion & Price residual\\
\hline
\multicolumn{7}{l}{\textit{Cheerios share equation}}\\
Local quadratic & $0.5$ & $0.903$ & $1.349$ & $-0.766$ & $0.639$ & $1.018$\\
 & & $(2.476)$ & $(2.943)$ & $(0.578)$ & $(0.292)$ & $(1.444)$\\
XGBoost & $1$ & $2.096$ & $2.624$ & $-0.703$ & $0.597$ & $1.717$\\
 & & $(4.505)$ & $(4.697)$ & $(0.441)$ & $(0.214)$ & $(2.326)$\\
Penalized splines & $0.5$ & $1.113$ & $1.584$ & $-0.759$ & $0.627$ & $1.005$\\
 & & $(2.604)$ & $(3.078)$ & $(0.680)$ & $(0.315)$ & $(1.058)$\\
LASSO & $1.5$ & $0.612$ & $1.003$ & $-0.723$ & $0.646$ & $1.332$\\
 & & $(5.379)$ & $(7.169)$ & $(1.441)$ & $(0.598)$ & $(5.521)$\\
Random forest & $0.75$ & $0.401$ & $0.612$ & $-0.573$ & $0.706$ & $9.483$\\
 & & $(0.570)$ & $(0.609)$ & $(0.480)$ & $(0.275)$ & $(28.295)$\\
\hline
\multicolumn{7}{l}{\textit{Rice Krispies share equation}}\\
Local quadratic & $0.5$ & $-2.292$ & $-0.852$ & $-0.499$ & $0.898$ & $0.743$\\
 & & $(32.553)$ & $(3.883)$ & $(1.649)$ & $(0.434)$ & $(2.827)$\\
XGBoost & $1$ & $6.436$ & $-1.229$ & $-7.640$ & $-10.008$ & $-0.319$\\
 & & $(151.495)$ & $(157.063)$ & $(4.881)$ & $(9.500)$ & $(5.024)$\\
Penalized splines & $0.5$ & $1.515$ & $1.228$ & $-0.954$ & $0.842$ & $-3.918$\\
 & & $(1.828)$ & $(1.625)$ & $(0.545)$ & $(0.363)$ & $(2.339)$\\
LASSO & $1.5$ & $-0.357$ & $-0.278$ & $-0.822$ & $0.817$ & $0.997$\\
 & & $(1.323)$ & $(0.816)$ & $(0.542)$ & $(0.313)$ & $(4.262)$\\
Random forest & $0.75$ & $0.070$ & $0.011$ & $-0.769$ & $0.844$ & $-1.732$\\
 & & $(0.761)$ & $(0.550)$ & $(0.435)$ & $(0.322)$ & $(4.450)$\\
\hline\hline
\end{tabular}
\par\smallskip\begin{minipage}{.92\textwidth}\footnotesize
\textit{Notes.} Both panels use the same 4,363 store-weeks. For each learner, the two share equations use the same ISPD kernel bandwidth $\widehat b$, selected to minimize the estimated variance of the combined price coefficient. Standard errors in parentheses are clustered by store and week.
\end{minipage}
\end{table}

\begin{samepage}
Table~\ref{sec07:tab:cereal-fitted} combines the two coefficient vectors for each learner by minimum distance using their joint covariance. For example, with a local-quadratic first stage, the Cheerios and Rice Krispies equations give price coefficients of $-0.766$ and $-0.499$, with standard errors of $0.578$ and $1.649$. Combining them gives $-0.622$ with a standard error of $0.350$. The table also reports the Petrin--Train benchmark, which fits the two price regressions by OLS on the full sample and estimates the demand coefficients jointly by share-weighted logit.\par
\end{samepage}

\begin{table}[htbp]\centering\small
\caption{Petrin--Train benchmark and combined DDML estimates under fitted-price conditioning}
\label{sec07:tab:cereal-fitted}
\begin{tabular}{lrrrrrr}
\hline\hline
\multirow{2}{*}{First stage} & \multirow{2}{*}{$\widehat{b}$} & $\alpha_1$ & $\alpha_2$ & $\beta$ & $\delta$ & $\gamma$\\
 & & Cheerios & Rice Krispies & Price & Promotion & Price residual\\
\hline
\multicolumn{7}{l}{\textit{Petrin--Train benchmark}}\\
Full-sample OLS &  & $-0.138$ & $-0.073$ & $-0.617$ & $0.731$ & $-0.686$\\
 & & $(0.493)$ & $(0.386)$ & $(0.486)$ & $(0.188)$ & $(0.714)$\\
\hline
\multicolumn{7}{l}{\textit{DDML}}\\
Local quadratic & $0.5$ & $-0.850$ & $-0.726$ & $-0.622$ & $0.706$ & $0.917$\\
 & & $(0.351)$ & $(0.259)$ & $(0.350)$ & $(0.194)$ & $(0.729)$\\
XGBoost & $1$ & $-3.063$ & $-2.763$ & $-0.515$ & $0.639$ & $5.499$\\
 & & $(2.455)$ & $(2.556)$ & $(0.304)$ & $(0.188)$ & $(1.822)$\\
Penalized splines & $0.5$ & $-1.616$ & $-1.781$ & $0.096$ & $1.043$ & $0.001$\\
 & & $(0.478)$ & $(0.402)$ & $(0.386)$ & $(0.220)$ & $(0.881)$\\
LASSO & $1.5$ & $-0.459$ & $-0.359$ & $-0.682$ & $0.599$ & $3.444$\\
 & & $(0.347)$ & $(0.264)$ & $(0.333)$ & $(0.174)$ & $(1.353)$\\
Random forest & $0.75$ & $-0.327$ & $-0.249$ & $-0.806$ & $0.778$ & $-3.697$\\
 & & $(0.319)$ & $(0.244)$ & $(0.297)$ & $(0.217)$ & $(4.013)$\\
\hline\hline
\end{tabular}
\par\smallskip\begin{minipage}{.92\textwidth}\footnotesize
\textit{Notes.} All estimates use the same 4,363 store-weeks. Standard errors in parentheses are clustered by store and week. The benchmark standard errors account for both OLS price regressions. The intercepts are relative to Grape-Nuts. Each DDML row combines the two share-specific coefficient vectors by minimum distance. For each learner, $\widehat b$ minimizes the estimated variance of the combined price coefficient among the eligible bandwidths in equation~\eqref{sec07:eq:kernel}.
\end{minipage}
\end{table}

The benchmark price coefficient is $-0.617$. The DDML price estimates range from $-0.806$ for random forest to $0.096$ for penalized splines. For each of the five learners, the combined price estimate has a smaller standard error than either share-specific estimate.

\paragraph{Testing the conditioning set.}
Table~\ref{sec07:tab:cereal} applies the Hausman test of Section~\ref{sec05:test} to all five combined demand coefficients. Under the null in equation~\eqref{sec07:eq:cereal-costs}, acquisition costs do not help predict either cereal's share error once actual prices, promotions, and fitted prices are given. Four of the five learners reject at $5\%$. LASSO gives $p=0.514$. For the price coefficient alone, only the random-forest difference is significant at $5\%$.

\begin{table}[htbp]\centering\small
\caption{Cereal price estimates and the Hausman test of acquisition-cost conditioning}
\label{sec07:tab:cereal}
\setlength{\tabcolsep}{4pt}
\begin{tabular}{lrrrrrr}
\hline\hline
\multirow{2}{*}{First stage} & \multirow{2}{*}{$\widehat{b}$}& \multicolumn{3}{c}{Price coefficient} & \multicolumn{2}{c}{Joint Hausman test}\\
\cline{3-5}\cline{6-7}
&& Fitted prices & Acquisition costs & Difference (SE) & $\mathcal{T}_N$ & $p$-value\\
\hline
Local quadratic & $0.5$ & $-0.622$ & $-0.535$ & $0.087$ $(0.323)$ & $55.01$ & $<0.001$\\
XGBoost & $1$ & $-0.515$ & $-0.710$ & $-0.195$ $(0.348)$ & $26.74$ & $<0.001$\\
Penalized splines & $0.5$ & $0.096$ & $0.039$ & $-0.057$ $(0.339)$ & $154.32$ & $<0.001$\\
LASSO & $1.5$ & $-0.682$ & $-0.761$ & $-0.079$ $(0.113)$ & $4.25$ & $0.514$\\
Random forest & $0.75$ & $-0.806$ & $-1.928$ & $-1.122$ $(0.255)$ & $196.41$ & $<0.001$\\
\hline\hline
\end{tabular}
\par\smallskip\begin{minipage}{.94\textwidth}\footnotesize
\textit{Notes.} Each row uses the price residuals, store folds, and ISPD kernel bandwidth from Table~\ref{sec07:tab:cereal-fitted}. The difference is the acquisition-cost estimate minus the fitted-price estimate. We cluster the paired influence contributions by store and week for both the standard errors in parentheses and the joint $\chi^2_5$ test.
\end{minipage}
\end{table}
\FloatBarrier

\section{Conclusion}\label{sec08:conclusion}
When an economic model identifies its parameters through a conditional moment restriction (CMR), researchers can estimate those parameters directly from the restriction even when some variables in the conditioning set must be generated in a nonparametric first stage. Building on the identification-preserving kernel approach used by \citet{dml_escanciano_terschuur2026}, we derive an orthogonal score that accounts for first-stage estimation error in both the moment function and the kernel evaluations. Under the stated conditions, the cross-fitted DDML estimator supports root-$N$ inference, including when the bandwidth is selected by minimizing an estimated measure of asymptotic variance over a fixed finite grid.

We examine finite-sample performance in two simulation experiments, first by varying the first-stage convergence rate and then by fitting the first stage with several nonparametric learners. For a correctly specified CMR, the DDML estimator has little bias and its confidence intervals have near-nominal coverage when the first stage converges sufficiently fast. The variance-based bandwidth rule generally yields shorter confidence intervals on average than fixing the bandwidth at either endpoint of the grid, with similar coverage.

The two applications illustrate different decisions practitioners face. In Chilean manufacturing, instruments adapted from \citet{ackerberg_caves_frazer2015} yield four solutions to the same sample moment equations, with conflicting signs for the estimated production elasticities. Estimating directly from the model's CMR avoids choosing a finite set of instruments and deciding which of the resulting roots to use. In cereal demand, replacing predicted retail prices with the acquisition costs used to predict them imposes a stronger conditional restriction. Our Hausman test lets researchers assess that restriction and rejects it for four of the five first-stage learners.

%%%%%%%%%%%%%%%%%%%%%%%%%%%%%%%%%%%%%%%%%%%%%%%%%%%%%%%%%%%%%%%%%%%%%%%%%%%%%%%%%%%%%%%%%%%%%%%%%%%%%%%%%%%%%%%%%%%%%%%%%%%%%%%%%%%%%%%%%%%%%%%%%%%%%%%%%%%%%%%%%%%%%%%%%%%%%%%%%%%%%%%%%%%%%%%%%%%%%%%%%%%%%%%%%%%%%%%%%%%%%%%%%%%%%%%%%%%%%%%%%%%%%%%%%%%%%%%%%%%%%%%%%%%%%%%%%%%%%%%%%%%%%%%%%%%%%%%%%%%%%%%%%%%%%%%%%%%%%%%%%%%%%%%%%%%%%%%%%%%%%%%%%%%%%%%%%%%%%%%%%%%%%%%%%%%%%%%%%%%%%%%%%%%%%%%%%%%%%%%%%%%%%%%%%%%%%%%%%%%%%%%%%%%%%%%%%%%%%%%%%%%%%%%%%%%%%%%%%%%%%%%%%%%%%%%%%%%%%%%%%%%%%%%%%%%%%%%%%%%%%%%%%%%%%%%%%%%%%%%%%%%%%%%%%%%%%%%%%%%%%%%%%%%%%%%%%%%%%%%%%%%%%%%%%%%%%%%%
\bibliographystyle{apalike}  % or plain, abbrv, chicago, etc.
\bibliography{references}    % assumes references.bib is in the same folder
%%%%%%%%%%%%%%%%%%%%%%%%%%%%%%%%%%%%%%%%%%%%%%%%%%%%%%%%%%%%%%%%%%%%%%%%%%%%%%%%%%%%%%%%%%%%%%%%%%%%%%%%%%%%%%%%%%%%%%%%%%%%%%%%%%%%%%%%%%%%%%%%%%%%%%%%%%%%%%%%%%%%%%%%%%%%%%%%%%%%%%%%%%%%%%%%%%%%%%%%%%%%%%%%%%%%%%%%%%%%%%%%%%%%%%%%%%%%%%%%%%%%%%%%%%%%%%%%%%%%%%%%%%%%%%%%%%%%%%%%%%%%%%%%%%%%%%%%%%%%%%%%%%%%%%%%%%%%%%%%%%%%%%%%%%%%%%%%%%%%%%%%%%%%%%%%%%%%%%%%%%%%%%%%%%%%%%%%%%%%%%%%%%%%%%%%%%%%%%%%%%%%%%%%%%%%%%%%%%%%%%%%%%%%%%%%%%%%%%%%%%%%%%%%%%%%%%%%%%%%%%%%%%%%%%%%%%%%%%%%%%%%%%%%%%%%%%%%%%%%%%%%%%%%%%%%%%%%%%%%%%%%%%%%%%%%%%%%%%%%%%%%%%%%%%%%%%%%%%%%%%%%%%%%%%%%%%%%
\clearpage
\appendix
\setcounter{definition}{0}
\setcounter{lemma}{0}
\setcounter{assumption}{0}
\setcounter{theorem}{0}
\setcounter{proposition}{0}
\renewcommand{\theassumption}{A.\arabic{assumption}}
\renewcommand{\thetheorem}{A.\arabic{theorem}}
\renewcommand{\thelemma}{A.\arabic{lemma}}
\renewcommand{\thedefinition}{A.\arabic{definition}}
\renewcommand{\theproposition}{A.\arabic{proposition}}
\renewcommand{\theHassumption}{appendix.\arabic{assumption}}
\renewcommand{\theHtheorem}{appendix.\arabic{theorem}}
\renewcommand{\theHlemma}{appendix.\arabic{lemma}}
\renewcommand{\theHdefinition}{appendix.\arabic{definition}}
\renewcommand{\theHproposition}{appendix.\arabic{proposition}}

\section{Proofs of the main results}

We first introduce five lemmas.

\begin{lemma}\label{appendix - lemma A01}\text{ $\,\,$ }

    \begin{itemize}
        \item[(i)] Under Assumptions~\ref{sec03 - assumption02 - CMR} and \ref{sec03 - assumption03 - kernel}, $\Gamma_{c}$ defined in equation~\eqref{sec04:eq:scorejacobian} is invertible.
        \item[(ii)] Under Assumptions~\ref{sec03 - assumption01 - first stage}, \ref{sec03 - assumption02 - CMR}, \ref{sec03 - assumption03 - kernel}, and \ref{sec04 - assumption04 - linear representation}, $\Sigma_{c}$ defined in equation~\eqref{sec04:eq:scoreprojectioncovariance} is finite and positive definite.
    \end{itemize}
\end{lemma}

\textbf{Proof of Lemma \ref{appendix - lemma A01} : }See Appendix \ref{appendix - lemma A01 - proof}\\

\begin{lemma}\label{appendix - lemma A02}
    Suppose Assumptions \ref{sec03 - assumption02 - CMR} and \ref{sec03 - assumption03 - kernel}. For $\eta, \eta'\in \mathcal{H}_N$, { the following statements hold}:
    \begin{itemize}
        \item[(i)] $\underset{\theta\in\Theta}{\textnormal{sup}} \norm{\rho(W_i, \theta,\eta)-\rho(W_i, \theta,\eta')}\leq\widebar{J}(W_i)\norm{\eta - \eta'}_\infty$.
        \item[(ii)] $\underset{\theta\in\textnormal{int}(\Theta_0)}{\textnormal{sup}} \norm{\nabla_\theta \rho(W_i, \theta,\eta) - \nabla_\theta \rho(W_i, \theta,\eta')}\leq\widebar{J}(W_i)\norm{\eta - \eta'}_\infty$.        
        \item[(iii)] $| K\big(Z_{i}^{c}(\eta), Z_{j}^{c}(\eta) \big) - K\big(Z_{i}^{c}(\eta'), Z_{j}^{c}(\eta') \big)  |\leq2\widebar{K}_1\ind{}{c=1}\norm{\eta - \eta' }_\infty$.
        \item[(iv)] There exists a function $(w,w')\mapsto L_{1}^{c}(w,w')$, independent of $N$, $\theta$ and $\eta\in\mathcal{H}_N$, such that, for any admissible direction $h$,
        \[ \sup_{\theta\in\widebar{\Theta}_{o}, \eta\in \mathcal{H}_N }\norm{ D_{\eta}r^{c}(W_i, W_j, \theta, \eta)[h]} \leq \Big\{ L_{1}^{c}(W_i, W_j) + L_{1}^{c}(W_j, W_i) \Big\}\norm{h}_{\infty},\]
        and $\norm{L_{1}^{c}(W_i, W_j)}_{F_0\otimes F_0, 2} < \infty$.
        \item[(v)] There exists a function $(w,w')\mapsto L_{2}^{c}(w,w')$, independent of $N$, $\theta$ and $\eta\in\mathcal{H}_N$, such that, for any admissible directions $h_1, h_2$,
        \[ \sup_{\theta\in\widebar{\Theta}_{o}, \eta\in \mathcal{H}_N } \Norm{ D_{\eta}^{2}r^{c}(W_i, W_j, \theta, \eta)[h_1, h_2] } \leq \Big\{ L_{2}^{c}(W_i, W_j) + L_{2}^{c}(W_j, W_i) \Big\}\norm{h_1}_\infty \cdot \norm{h_2}_\infty,\]
        and $\norm{L_{2}^{c}(W_i, W_j)}_{F_0\otimes F_0, 2} < \infty$.
    \end{itemize}
\end{lemma}

\textbf{Proof of Lemma \ref{appendix - lemma A02} : }See Appendix \ref{appendix - lemma A02 - proof}\\

\begin{lemma}\label{appendix - lemma A03}
Suppose Assumptions \ref{sec03 - assumption02 - CMR} and \ref{sec03 - assumption03 - kernel}. 
    \begin{itemize}
        \item[(i)] There exists a positive constant $\widebar{L}_{1}^{c} < \infty$, independent of $N$, $\theta$ and $\eta\in\mathcal{H}_N$, such that:
            \[ \sup_{\theta\in \widebar{\Theta}_{o}}\norm{r^{c}(W_i, W_j, \theta, \eta) - r^{c}(W_i, W_j, \theta, \eta_0) }_{F_0\otimes F_0, 2} \leq \widebar{L}_{1}^{c}\cdot \norm{\eta - \eta_0}_{\infty}. \]
        \item[(ii)] For $\eta_1, \eta_2 \in\mathcal{H}_N$ and any admissible direction $h$, there exists a positive constant $\widebar{L}_{2}^{c}<\infty$, independent of $N$, $\theta$ and $\eta\in\mathcal{H}_N$, such that:
            \[ \sup_{\theta\in \widebar{\Theta}_{o}}\norm{\expect{D_\eta r^{c}(W_i,W_j,\theta,\eta_1)[h]-D_\eta r^{c}(W_i,W_j,\theta,\eta_2)[h]}} \leq \widebar{L}_{2}^{c}\norm{\eta_1-\eta_2}_\infty\norm{h}_\infty. \]
        \item[(iii)] There exist positive constants $\widebar{L}_{3}^{c},\widebar{L}_{4}^{c}<\infty$, independent of $N$, $\theta$, and $\eta$, such that, for every $\theta\in\widebar{\Theta}_{o}$ and $\eta\in\mathcal{H}_N$,
        \[  \norm{\nabla_\theta r^{c}(W_i,W_j,\theta,\eta)-\nabla_\theta r^{c}(W_i,W_j,\theta_0,\eta_0)}_{F_0\otimes F_0,2}
                \leq \widebar{L}_{3}^{c}\norm{\eta-\eta_0}_{\infty}+\widebar{L}_{4}^{c}\norm{\theta-\theta_0}.\]
    \end{itemize}
\end{lemma}

\textbf{Proof of Lemma \ref{appendix - lemma A03} : }See Appendix \ref{appendix - lemma A03 - proof}.\\

\begin{lemma}\label{appendix - lemma A04} Maintain the fold-construction and nuisance estimates defined in Definition~\ref{sec04 - definition01 - DML} with fixed $G$. For each $\ell\in[L]$, define a class of functions:
\[  \mathcal{F}_{\ell}^{c} 
        \equiv    \left\{(w,w') \mapsto \psi^{c}\big(w,w',\theta,\widehat{\eta}_{\ell},\check{\vecdel}_{\ell}^{c}\big) : \theta\in\overline{\Theta}_{o}\right\},  \]
where $\overline{\Theta}_0 \subset \textnormal{int}(\theta)$ is the compact subset including $\theta_0$, defined in Assumption~\ref{sec03 - assumption02 - CMR}(e), $w,w'\in\operatorname{supp}(W)$ are generic arguments. Let $\mathcal{W}_{-\ell}$ denote the $\sigma$-field induced by $\{W_i : i\notin J_{\ell}\}$, and $\widebar{\psi}_{\ell}^{c}$ the envelope for $\mathcal{F}_{\ell}^{c}$.

Suppose Assumption~\ref{sec03 - assumption01 - first stage}, \ref{sec03 - assumption02 - CMR}, \ref{sec03 - assumption03 - kernel}, \ref{sec04 - assumption04 - linear representation}, and \ref{sec04 - assumption05 - algorithm stability}. Conditional on $\mathcal W_{-\ell}$ and on the event $\widehat{\eta}_{\ell}\in\mathcal{H}_{N}$, there exist constants $C_0>0$ and $C_1 \geq 1$, independent of $N$, $\ell$, any generic probability measure $P$, and the realized training-sample nuisance estimates, such that:
\begin{equation}\label{sec04 - theorem01 - proof - entropy}
    \sup_{P} \log N\!\left(  \epsilon\|\widebar{\psi}_{\ell}^{c}\|_{P,2},\mathcal{F}_{\ell}^{c},\|\cdot\|_{P,2} \right)
        \leq    C_0 \log\!\left(\frac{C_1}{\epsilon}\right), \qquad 0<\epsilon\leq1.
\end{equation}
\end{lemma}
\textbf{Proof of Lemma \ref{appendix - lemma A04} : }See Appendix \ref{appendix - lemma A04 - proof}.\\

\begin{lemma}\label{appendix - lemma A05}
    Let $\theta \mapsto \varphi_{0,ij}(\theta)$ denote either of the following functions:
    \[\begin{aligned}
        &\rho(W_i, \theta, \eta_0)\rho(W_j, \theta, \eta_0)K\big(Z_{i}^{c}(\eta_0), Z_{j}^{c}(\eta_0)\big) &\textnormal{with $\theta\in\Theta$}&\\
        &r^{c}(W_i, W_j, \theta, \eta_0) &\textnormal{with $\theta \in \widebar{\Theta}_{o}$}.&
    \end{aligned}\]
    Define the first and second Hoeffding-projections by
    \begin{equation}\label{appendix - lemma A05 - Hoeffding Projection}
    \varphi_{1,i}(\theta)
        \equiv        \condexpect{\varphi_{0,ij}(\theta)}{W_i} - \expect{ \varphi_{0,ij}(\theta)}, \qquad
    \varphi_{2,ij}(\theta)
        \equiv        \varphi_{0,ij}(\theta) - \expect{ \varphi_{0,ij}(\theta)} -\varphi_{1,i}(\theta) - \varphi_{1,j}(\theta).        
    \end{equation}
\begin{itemize}
    \item[(i)] Suppose that Assumptions~\ref{sec03 - assumption01 - first stage}, \ref{sec03 - assumption02 - CMR}, and \ref{sec03 - assumption03 - kernel}. Then, { the following statements hold}:
        \[  \sup_{\theta}\Norm{\frac{1}{N}\sum_{i=1}^N \varphi_{1,i}(\theta,\eta)}   = O_P(N^{-1/2}),
            \quad
            \sup_{\theta}\Norm{\binom{N}{2}^{-1}\sum_{i<j} \varphi_{2,ij}(\theta)}  = O_P(N^{-1}),\]
        and therefore:
        \[  \sup_{\theta}\Norm{\binom{N}{2}^{-1}\sum_{i<j} \big\{ \varphi_{0,ij}(\theta) - \expect{ \varphi_{0,ij}(\theta) } \big\} }  = O_P(N^{-1/2}). \]
        Each supremum is taken over $\Theta$ if $\varphi_{0,ij}(\theta) = \rho(W_i, \theta, \eta_0)\rho(W_j, \theta, \eta_0)K\big(Z_{i}^{c}(\eta_0), Z_{j}^{c}(\eta_0)\big)$, $\widebar{\Theta}_{o}$ otherwise.
    \item[(ii)] Maintain the fold-construction and nuisance estimates defined in Definition~\ref{sec04 - definition01 - DML} with fixed $G$. For each $\ell\in[L]$, define:
        \[ \Delta\psi_{\ell, ij}^{c}(\theta) \equiv \psi^{c}(W_i, W_j, \theta, \widehat{\eta}_{\ell}, \check{\vecdel}_{\ell}) -\psi^{c}(W_i, W_j, \theta_0,\widehat{\eta}_{\ell},\check{\vecdel}_{\ell}). \]
        Additionally assume that Assumptions~\ref{sec04 - assumption04 - linear representation} and \ref{sec04 - assumption05 - algorithm stability} hold. Then, for every deterministic sequence $\xi_N\downarrow0$:
        \[  \max_{\ell\in[L]}\sup_{\substack{\theta\in\widebar{\Theta}_{o}\\ \|\theta-\theta_0\|\leq \xi_N}} \Norm{ \frac{1}{|I_{\ell}|}\sum_{(i,j)\in I_{\ell}} \Big\{ \Delta\psi_{\ell,ij}(\theta) - \condexpect{ \Delta\psi_{\ell,ij}(\theta) }{ \mathcal{W}_{-\ell}} \Big\} } = o_{P}(N^{-1/2}). \]
\end{itemize}
\end{lemma}

\textbf{Proof of Lemma \ref{appendix - lemma A05} : }See Appendix \ref{appendix - lemma A05 - proof}.
%%%%%%%%%%%%%%%%%%%%%%%%%%%%%%%%%%%%%%%%%%%%%%%%%%%%%%%%%%%%%%%%%%%%%%%%%%%%%%%%%%%%%%%%%%%%%%%%%%%%%%%%%%%%%%%%%%%%%%%%%%%%%%%%%%%%%%%%%%%%%%%%%%%%%%%%%%%%%%%%%%%%%%%%%%%%%%%%%%%%%%%%%%%%%%%%%%%%%%%%%%%%%%%%%%%%%%%%%%%%%%%%%%%%%%%%%%%%%%%%%%%%%%%%%%%%%%%%%%%%%%%%%%%%%%%%%%%%%%%%%%%%%%%%%%%%%%%%%%%%%%%%%%%%%%%%%%%%%%%%%%%%%%%%%%%%%%%%%%%%%%%%%%%%%%%%%%%%%%%%%%%%%%%%%%%%%%%%%%%%%%%%%%%%%%%%%%%%%%%%%%%%%%%%%%%%%%%%%%%%%%%%%%%%%%%%%%%%%%%%%%%%%%%%%%%%%%%%%%%%%%%%%%%%%%%%%%%%%%%%%%%%%%%%%%%%%%%%%%%%%%%%%%%%%%%%%%%%%%%%%%%%%%%%%%%%%%%%%%%%%%%%%%%%%%%%%%%%%%%%%%%%%%%%%%%%%%%%
\subsection{Proof of Lemma \ref{sec04 - lemma01}}\label{sec04 - lemma01 - proof}
This proof follows the standard consistency argument for extremum estimators \citet{Newey1994}. We fix $c\in\{0,1\}$ and $\eta\in\mathcal{H}_N$. To simplify notation, we write $\rho_i(\theta,\eta)\equiv\rho(W_i,\theta,\eta)$, $d_{i}(\theta, \eta) \equiv \nabla\rho(W_i,\theta,\eta)$, $K_{ij}(\eta)\equiv K(Z_i^{c}(\eta),Z_j^{c}(\eta))$.

\paragraph{Step 1.} We show $\theta_0$ is the unique minimizer of $Q(\theta, \eta_0)$. Let $g(\theta,\eta, z)\equiv\condexpectF{F_0}{\rho(W,\theta,\eta)}{Z^{c}(\eta)=z}$. Conditional Jensen's inequality gives
\[ \int |g(\theta,\eta, z)|^2\nu(dz)\leq  \norm{\rho(W,\theta,\eta)}_2^2<\infty, \]
where $\nu$ denotes the $F_0$-law of $Z^{c}(\eta)$. Thus $g(\theta,\eta,\cdot)\in L^2(\nu)$. By using the law of iterated expectations and $W_{i}\pperp W_{j}$,
\[ Q(\theta, \eta)=\iint g(\theta,\eta, z)g(\theta, \eta, z')K(z,z')\nu(dz)\nu(dz'), \]
Since $K$ is ISPD by Assumption \ref{sec03 - assumption03 - kernel}(a), $Q(\theta, \eta) \geq 0$ and
\begin{equation}\label{sec04 - lemma01 - proof - eq01}
    Q(\theta, \eta)=0\quad\Longleftrightarrow\quad g(\theta,\eta, Z^{c}(\eta))=0\quad F_0\textnormal{-a.s.}
\end{equation}
At $\eta = \eta_0$, the CMR implies $Q(\theta_0, \eta_0)=0$. Conversely, if $Q(\theta, \eta_0)=0$, equation~\eqref{sec04 - lemma01 - proof - eq01} and Assumption \ref{sec03 - assumption02 - CMR}(a) imply $\theta=\theta_0$. 

\paragraph{Step 2.} First, we establish that, for a given $\eta\in\mathcal{H}_{N}$, the map $\theta\mapsto Q(\theta,\eta)$ is uniformly continuous on $\Theta$. Observe that, for any $\theta_1,\theta_2\in\Theta$, adding and subtracting gives:
\begin{align*}
     Q(\theta_1,\eta) - Q(\theta_2, \eta)
        =   &\expectF{F_0}{\rho_{i}(\theta_1, \eta)\{\rho_{j}(\theta_1, \eta)-\rho_{j}(\theta_2, \eta)\}K_{ij}(\eta)}&\\
            &+   \expectF{F_0}{\rho_{j}(\theta_2, \eta)\{\rho_{i}(\theta_1, \eta)-\rho_{i}(\theta_2, \eta)\}K_{ij}(\eta)}.&
\end{align*}
Since $K(\cdot, \cdot)$ is bounded above by Assumption \ref{sec03 - assumption03 - kernel}, and since $W_{i}\pperp W_{j}$, the Cauchy--Schwarz inequality implies:
\[  |Q(\theta_1,\eta)-Q(\theta_2,\eta)| \leq    2\widebar{K}_0\lVert\widebar{\rho}^0\rVert_2 \lVert\rho_i(\theta_1,\eta)-\rho_i(\theta_2,\eta)\rVert_2. \]
For fixed $\eta$, Assumption \ref{sec03 - assumption02 - CMR}(b) gives pointwise continuity in $\theta$. Moreover, the envelope in Assumption \ref{sec03 - assumption02 - CMR}(e) satisfies
\[  |\rho_i(\theta_1,\eta)-\rho_i(\theta_2,\eta)|^2 \leq    4\widebar{\rho}^0(W_i)^2,   \qquad\mathbb E[\widebar{\rho}^0(W_i)^2]<\infty.    \]
Now consider any sequence $\{\theta_k\}_{k\geq1}\subset\Theta$ with $\theta_k\to\theta\in\Theta$. For the fixed $\eta$, continuity in Assumption \ref{sec03 - assumption02 - CMR}(b) implies
\[  \rho_i(\theta_k,\eta) \to \rho_i(\theta,\eta)  \qquad  F_0\text{-almost surely}.    \]
Thus $|\rho_i(\theta_k,\eta)-\rho_i(\theta,\eta)|^2\to0$ almost surely. The envelope bound above supplies the same integrable dominating random variable $4\widebar{\rho}^0(W_i)^2$ for every $k$. Applying the dominated convergence theorem gives
\[  \lVert\rho_i(\theta_k,\eta)-\rho_i(\theta,\eta)\rVert_2^2   =\mathbb E\!\left[|\rho_i(\theta_k,\eta)-\rho_i(\theta,\eta)|^2\right]  \to 0.\]
Substituting into the preceding bound for the criterion difference yields
\[  |Q(\theta_k,\eta)-Q(\theta,\eta)|   \leq    2\widebar{K}_0\lVert\widebar{\rho}^0\rVert_2    \lVert\rho_i(\theta_k,\eta)-\rho_i(\theta,\eta)\rVert_2 \to 0.  \]
Since the sequence and its limit were arbitrary, $\theta\mapsto Q(\theta,\eta)$ is continuous on $\Theta$. Compactness of $\Theta$ then implies uniform continuity for this fixed $\eta$. In particular, for every $\epsilon>0$ there exists $\delta_\eta>0$ such that
\begin{equation}\label{sec04 - lemma01 - proof - eq02}
    \sup_{\substack{\theta_1,\theta_2\in\Theta\\ \lVert\theta_1-\theta_2\rVert<\delta_\eta}}|Q(\theta_1,\eta)-Q(\theta_2,\eta)| < \epsilon.
\end{equation}

Next, we show that replacing $\eta_0$ by a nearby $\eta$ has a uniformly small effect on the criterion over $\Theta$. For any $\theta\in\Theta$, adding and subtracting gives:
\begin{align*}
    |Q(\theta, \eta) - Q(\theta, \eta_0)|
        \leq    &\expect{|\rho_{i}(\theta,\eta)\rho_{j}(\theta,\eta)|\cdot|K_{ij}(\eta)-K_{ij}(\eta_0)|}&\\
                & + \expect{|\rho_{i}(\theta,\eta)\{\rho_{j}(\theta, \eta)-\rho_{j}(\theta, \eta_0)\}|\cdot K_{ij}(\eta_0)}
                  + \expect{|\rho_{j}(\theta,\eta_0)\{\rho_{i}(\theta,\eta)-\rho_{i}(\theta,\eta_0)\}|\cdot K_{ij}(\eta_0)}&\\
        \leq    & \widebar{R}^2 \cdot 2\widebar{K}_1 \ind{}{c=1}  \cdot \norm{\eta - \eta_0}_\infty +  2 \widebar{K}_0 \widebar{R}^2  C_J \cdot \norm{\eta - \eta_0}_\infty  .&
\end{align*}
where the second inequality holds by Lemma \ref{appendix - lemma A02}(i) and (iii), and the third inequality comes from the independence of $W_{i}\pperp W_{j}$, the Cauchy--Schwarz inequality, and Assumptions \ref{sec03 - assumption02 - CMR}(d) and (e). Because $\widebar{R}, \widebar{K}_0, \widebar{K}_1, C_{j}$ do not depend on $\theta$, it follows that:
\[    \sup_{\theta\in\Theta}|Q(\theta, \eta) - Q(\theta, \eta_0)|
        \leq    \big\{  2 \widebar{K}_0 \widebar{R}^2  C_J +  2\widebar{K}_1\widebar{R}^2 \ind{}{c=1}   \big\} \cdot \norm{\eta - \eta_0}_\infty.\]
Let $\theta^*(\eta)$ be the minimizer of $Q(\theta, \eta)$. We have $Q(\theta_0,\eta_0)=0$. Since $\theta^*(\eta)$ minimizes $Q(\theta,\eta)$ over $\theta\in\Theta$, we also have $Q(\theta^*(\eta),\eta)\leq Q(\theta_0,\eta)$. Therefore,
\begin{align}
    Q(\theta^*(\eta), \eta_0)
        \leq    \nonumber
            &|Q( \theta^*(\eta), \eta_0)-Q( \theta^*(\eta), \eta)| + Q( \theta^*(\eta), \eta)&\\
        \leq    \nonumber
            &|Q( \theta^*(\eta), \eta_0)-Q( \theta^*(\eta), \eta)| + Q(\theta_0, \eta) - \underline{Q(\theta_0, \eta_0)}_{= 0}&\\
        \leq    \nonumber
            &2 \sup_{\theta\in\Theta}|Q(\theta, \eta) - Q(\theta, \eta_0)| &\\
        \leq    \label{sec04 - lemma01 - proof - eq03}
            &\big\{  2 \widebar{K}_0 \widebar{R}^2  C_J +  2\widebar{K}_1\widebar{R}^2 \ind{}{c=1}   \big\} \cdot \norm{\eta - \eta_0}_\infty.
\end{align}

Step 1 establishes that $\theta_0$ uniquely minimizes $Q(\theta,\eta_0)$, and the first part of Step 2 establishes continuity on the compact set $\Theta$. Consequently, for every $\epsilon>0$ such that $\Theta_\epsilon=\{\theta\in\Theta:\lVert\theta-\theta_0\rVert\geq\epsilon\}$ is nonempty, we have the separation property
\[  \inf_{\theta\in\Theta_\epsilon}Q(\theta,\eta_0)  >   Q(\theta_0,\eta_0)=0.   \]
For each $\eta\in\mathcal{H}_{N}$, uniform continuity of $\theta\mapsto Q(\theta,\eta)$ on the compact set $\Theta$ ensures the existence of a minimizer $\theta^*(\eta)$. Hence, by equation~\eqref{sec04 - lemma01 - proof - eq03} and the \hyperlink{shrinking-neighborhood-property}{shrinking-neighborhood property} imposed on $\mathcal{H}_{N}$, we have
\[  \sup_{\eta\in\mathcal{H}_{N}}Q(\theta^*(\eta),\eta_0)   \leq C\sup_{\eta\in\mathcal{H}_{N}}\lVert\eta-\eta_0\rVert_\infty \to 0,  \]
where $C$ is the finite constant in equation~\eqref{sec04 - lemma01 - proof - eq03}. Combining this bound with separation gives:
\[  \sup_{\eta\in\mathcal{H}_{N}}\lVert\theta^*(\eta)-\theta_0\rVert\to0.    \]
We note that this convergence is deterministic as $\eta$ is a fixed function ranging over the deterministic class $\mathcal{H}_{N}$, not a random object at this stage of the argument.

\paragraph{Step 3.} In this step, we show that the plug-in estimator is consistent. Formally, we establish that
\[   \sup_{\theta\in\Theta} \left|Q_N(\theta,\widehat\eta)-Q(\theta,\eta_0)\right|=o_P(1).  \]
Together with the minimizing property and separation established in Step 2, this implies consistency. 

We begin with the decomposition
\begin{equation}\label{sec04 - lemma01 - proof - decomposition}
    \sup_{\theta\in\Theta}|Q_N(\theta,\widehat\eta)-Q(\theta,\eta_0)|
    \leq    \sup_{\theta\in\Theta}|Q_N(\theta,\widehat\eta)-Q_N(\theta,\eta_0)|
    +       \sup_{\theta\in\Theta}|Q_N(\theta,\eta_0)-Q(\theta,\eta_0)|.
\end{equation}
We first control the second term on the right-hand side of equation~\eqref{sec04 - lemma01 - proof - decomposition}. By Lemma~\ref{appendix - lemma A05}(i), the second term in equation~\eqref{sec04 - lemma01 - proof - decomposition} satisfies:
\[    \sup_{\theta\in\Theta}|Q_N(\theta,\eta_0)-Q(\theta,\eta_0)| = o_{P}(1). \]
We next control the first term on the right-hand side of equation~\eqref{sec04 - lemma01 - proof - decomposition}. The sample counterpart of Step 2's perturbation bound gives, on the event $\{\widehat\eta\in\mathcal{H}_{N}\}$,
\[  \sup_{\theta\in\Theta}|Q_N(\theta,\widehat\eta)-Q_N(\theta,\eta_0)|
        \leq    2\widebar{R}^2\widebar{K}_0 \cdot \norm{\eta - \eta_0}_{\infty} \]
where the upper bound $2\widebar{R}^2\widebar{K}_0$ comes from equation~\eqref{appendix - lemma A05 - proof - eq05}. Since $\lVert\widehat\eta-\eta_0\rVert_\infty=o_P(1)$ and $\Pr(\widehat\eta\in\mathcal{H}_{N})\to1$ by assumption, the first term is also $o_P(1)$. This inequality holds for every $\eta\in\mathcal{H}_{N}$, so it also applies to $\widehat\eta$ on the event $\{\widehat\eta\in\mathcal{H}_{N}\}$, even when $\widehat\eta$ is estimated using the same sample. Consequently,
\[   \sup_{\theta\in\Theta}|Q_N(\theta,\widehat\eta)-Q(\theta,\eta_0)|=o_P(1).  \]

Since $Q_N(\plugin,\widehat{\eta})\leq Q_N(\theta_0,\widehat{\eta})$ by the minimizing property, we have:
\[ 0\leq Q(\plugin,\eta_0)\leq2\sup_{\theta\in\Theta}|Q_N(\theta,\widehat{\eta})-Q(\theta,\eta_0)|=o_P(1). \]
The separation inequality in Step 2 implies $\plugin\overset{p}{\longrightarrow}\theta_0$. In particular, the minimizer is in $\textnormal{int}(\Theta)$ with probability approaching one.

\paragraph{Step 4.}
Lastly, we derive the convergence rate of the plug-in estimator. Define the sample and population scores by
\[  G_N(\theta,\eta)\equiv\binom N2^{-1}\sum_{i<j}r^{c}(W_i,W_j,\theta,\eta),
    \qquad 
    G(\theta)\equiv\mathbb E[r^{c}(W_i,W_j,\theta,\eta_0)].   \]
Since Step 3 requires only compactness of $\Theta$ and establishes $\plugin \overset{p}{\longrightarrow}\theta_0$ and $\plugin\in\textnormal{int}(\Theta)$ with probability approaching one, it follows that:
\[ \Prob{\plugin\in \widebar{\Theta}_{o} } \longrightarrow 1.\]
On this event the sample first-order condition gives the decomposition
\begin{equation}\label{sec04 - lemma01 - proof - score decomposition}
    \begin{aligned}
    0= G_N(\plugin,\widehat\eta)
        &={}\underbrace{ G_N(\plugin,\widehat\eta)- G_N(\plugin,\eta_0)}_{\text{nuisance perturbation}}\\
        &\quad  +\underbrace{ G_N(\plugin,\eta_0)- G(\plugin)}_{\text{sampling fluctuation}}
                +\underbrace{ G(\plugin)}_{\text{population score}}.
    \end{aligned}
\end{equation}
Put $e_N=\lVert\widehat\eta-\eta_0\rVert_\infty$. We first bound the nuisance perturbation in equation~\eqref{sec04 - lemma01 - proof - score decomposition} by $O_P(e_N)+O_P(e_N^2)$, then control the sampling fluctuation by $O_P(N^{-1/2})$, and finally expand the population score around $\theta_0$.

We begin with the first term in equation~\eqref{sec04 - lemma01 - proof - score decomposition}. Work on the event $\mathcal E_N=\{\widehat\eta\in\mathcal{H}_{N}\}$, whose probability tends to one. Since $\eta_0\in\mathcal{H}_{N}$ and $\mathcal{H}_{N}$ is convex, the path $\eta_\tau=\eta_0+\tau(\widehat\eta-\eta_0)$ for $0\leq\tau\leq1$ lies in $\mathcal{H}_{N}$ on $\mathcal E_N$. For $\theta\in\textnormal{int}(\Theta)$, Taylor expansion along this path gives
\begin{align*}
 r^{c}(W_i,W_j,\theta,\widehat\eta)-r^{c}(W_i,W_j,\theta,\eta_0)
    =   &D_\eta r^{c}(W_i,W_j,\theta,\eta_0)[\widehat\eta-\eta_0]&\\
        &+\int_0^1(1-\tau)D_\eta^2r^{c}(W_i,W_j,\theta,\eta_\tau)
                   [\widehat\eta-\eta_0,\widehat\eta-\eta_0],d\tau.&
\end{align*}
The derivative bound for the integral must hold at every intermediate $\eta_\tau$ (as evident from the second, integral term above).

By Lemma~\ref{appendix - lemma A02}(iv) and (v), there are functions $(w,w')\mapsto L_{1}^{c}(w,w')$ and $L_{2}^{c}(w,w')$, which are independent of $\theta$, $\eta\in\mathcal{H}_N$, and $N$, such that $\norm{ L_{1}^{c}(W_i, W_j) }_{F_0\otimes F_0, 2} < \infty$ and $\norm{ L_{2}^{c}(W_i, W_j) }_{F_0\otimes F_0, 2} < \infty$, and
\begin{align*}
    \lVert D_\eta r^{c}(W_i,W_j,\theta,\eta)[h]\rVert
    &\leq L_{1}^{c}(W_i, W_j)\lVert h\rVert_\infty,\\
    \lVert D_\eta^2 r^{c}(W_i,W_j,\theta,\eta)[h_1,h_2]\rVert
    &\leq L_{2}^{c}(W_i, W_j)\lVert h_1\rVert_\infty\lVert h_2\rVert_\infty.
\end{align*}
Averaging the Taylor expansion over pairs therefore yields, on $\mathcal E_N$,
\begin{equation}\label{sec04 - lemma01 - proof - eq05}
    \sup_{\theta\in\textnormal{int}(\Theta)} \left\lVert\binom N2^{-1}\sum_{i<j} \{r^{c}(W_i,W_j,\theta,\widehat\eta)-r^{c}(W_i,W_j,\theta,\eta_0)\}\right\rVert    \leq     e_N U_{N}^{2}L_1^{c}+\frac{1}{2} e_N^2U_{N}^{2}L_2^{c},
\end{equation}
where $U_{N}^2$ denotes the second-order U-process, defined in \eqref{appendix - lemma A05 - proof - eq00 - Uprocess}.
The law of large numbers for U-statistics gives $U_{N}^{2}L_1^{c}=O_P(1)$ and $U_{N}^{2}L_2^{c}=O_P(1)$. Since $\Pr(\mathcal E_N)\to1$, the score perturbation is $O_P(e_N)+O_P(e_N^2)$. This argument applies even when the nuisance estimator uses the same sample.

We next control the sampling-fluctuation term in equation~\eqref{sec04 - lemma01 - proof - score decomposition}. By Lemma~\ref{appendix - lemma A05}(i), the following holds:
\begin{equation}\label{sec04 - lemma01 - proof - sampling fluctuation}
    \sup_{\theta\in \widebar{\Theta}_{o}}
    \lVert G_N(\theta,\eta_0)- G(\theta)\rVert=O_P(N^{-1/2}).
\end{equation}
Since $\prob{\plugin\in \widebar{\Theta}_{o}}\to1$ by Step 3, this bound applies at the random argument $\theta=\plugin$. Combining it with the nuisance-perturbation bound {in} equation~\eqref{sec04 - lemma01 - proof - eq05} and the sample first-order condition gives:
\begin{equation}\label{sec04 - lemma01 - proof - eq06}
     0= G(\plugin)+O_P(N^{-1/2})+O_P(e_N)+O_P(e_N^2).
\end{equation}
It remains to expand the population score $ G(\plugin)$ around $\theta_0$.

We finally expand the population-score term in equation~\eqref{sec04 - lemma01 - proof - score decomposition}. The CMR and independence imply $ G(\theta_0)=0$. On the compact subset $\widebar{\Theta}_{o}$, Assumption \ref{sec03 - assumption02 - CMR}(b) gives almost-sure continuity of $\nabla_\theta r^{c}(W_i,W_j,\theta,\eta_0)$. The same assumption, independence, and boundedness of $K$ imply:
\[  \sup_{\theta\in \widebar{\Theta}_{o}}\left\lVert\nabla_\theta r^{c}(W_i,W_j,\theta,\eta_0)\right\rVert_2 \leq    4\widebar{K}_0\widebar R^2<\infty,  \]
The bound is uniform over $\theta\in \widebar{\Theta}_{o}$ and implies uniform integrability of the family $\{\nabla_\theta r^{c}(W_i,W_j,\theta,\eta_0):\theta\in\widebar{\Theta}_{o}\}$. For any sequence $\{\theta_k\}\subset \widebar{\Theta}_{o}$ with $\theta_k\to\theta\in \widebar{\Theta}_{o}$, almost-sure continuity, the uniform $L^2$ bound, and Theorem 2.20 and Example 2.21 of \citet{vdv1998} imply:
\[  \mathbb E\!\left[       \left\lVert\nabla_\theta r^{c}(W_i,W_j,\theta_k,\eta_0)
          -\nabla_\theta r^{c}(W_i,W_j,\theta,\eta_0)\right\rVert     \right]\to0.       \]
Hence $\theta\mapsto\mathbb E[\nabla_\theta r^{c}(W_i,W_j,\theta,\eta_0)]$ is continuous on $\widebar{\Theta}_{o}$. These bounds also justify differentiation under the expectation. Consequently, $ G$ is continuously differentiable near $\theta_0$, with:
\[  \nabla_\theta G(\theta_0)    =   \mathbb E[\nabla_\theta r^{c}(W_i,W_j,\theta_0,\eta_0)]    =\Gamma_c. \]
Thus, as $\theta\to\theta_0$,
\[  G(\theta)=\Gamma_c(\theta-\theta_0)+R(\theta),  \qquad  \lVert R(\theta)\rVert=o(\lVert\theta-\theta_0\rVert).   \]
By consistency of $\plugin$, evaluation at the plug-in estimator yields:
\begin{equation}\label{sec04 - lemma01 - proof - population expansion}
    G(\plugin)=\Gamma_c(\plugin-\theta_0) +   o_P(\lVert\plugin-\theta_0\rVert).
\end{equation}

Substituting equation~\eqref{sec04 - lemma01 - proof - population expansion} into equation~\eqref{sec04 - lemma01 - proof - eq06} and using $e_N=o_P(1)$ gives:
\[  \Gamma_c(\plugin-\theta_0)  =O_P(N^{-1/2})+O_P(e_N)  +  o_P(\lVert\plugin-\theta_0\rVert),  \]
since $O_P(e_N^2)$ is absorbed into $O_P(e_N)$. By Lemma \ref{appendix - lemma A01}, the smallest singular value $s_c$ of $\Gamma_c$ is strictly positive. The remainder is bounded by $\tfrac12s_c\lVert\plugin-\theta_0\rVert$ with probability tending to one. Hence it can be moved to the left-hand side, yielding:
\[  \frac{1}{2}s_c\lVert\plugin-\theta_0\rVert  \leq O_P(N^{-1/2})+O_P(e_N).    \]
Therefore,
\[ \lVert\plugin-\theta_0\rVert =O_P(N^{-1/2})+O_P(\lVert\widehat\eta-\eta_0\rVert_\infty), \]
which proves the claimed rate. \qed

%%%%%%%%%%%%%%%%%%%%%%%%%%%%%%%%%%%%%%%%%%%%%%%%%%%%%%%%%%%%%%%%%%%%%%%%%%%%%%%%%%%%%%%%%%%%%%%%%%%%%%%%%%%%%%%%%%%%%%%%%%%%%%%%%%%%%%%%%%%%%%%%%%%%%%%%%%%%%%%%%%%%%%%%%%%%%%%%%%%%%%%%%%%%%%%%%%%%%%%%%%%%%%%%%%%%%%%%%%%%%%%%%%%%%%%%%%%%%%%%%%%%%%%%%%%%%%%%%%%%%%%%%%%%%%%%%%%%%%%%%%%%%%%%%%%%%%%%%%%%%%%%%%%%%%%%%%%%%%%%%%%%%%%%%%%%%%%%%%%%%%%%%%%%%%%%%%%%%%%%%%%%%%%%%%%%%%%%%%%%%%%%%%%%%%%%%%%%%%%%%%%%%%%%%%%%%%%%%%%%%%%%%%%%%%%%%%%%%%%%%%%%%%%%%%%%%%%%%%%%%%%%%%%%%%%%%%%%%%%%%%%%%%%%%%%%%%%%%%%%%%%%%%%%%%%%%%%%%%%%%%%%%%%%%%%%%%%%%%%%%%%%%%%%%%%%%%%%%%%%%%%%%%%%%%%%%%%%
\subsection{Proof of Proposition \ref{sec04 - proposition01}}\label{sec04 - proposition01 - proof}
\paragraph{Step 1.} We first verify that two functions $\vecdel_{0,1}^{c}(\cdot)$ and $\vecdel_{0,2}^{c}(\cdot)$ satisfy {equations}~\eqref{sec04 - assumption04 - a1 - eq01} and \eqref{sec04 - assumption04 - a1 - eq02}. To simplify notation, we write $K_{0,ij}^{c} \equiv  K( Z_{i}^{c}(\eta_0), Z_{j}^{c}(\eta_0) )$ and 
\[  \nabla_1 K_{0,ij} \equiv \frac{\partial}{\partial v}K( (X_{2i}, v), Z_{j}^{c}(\eta_0) )\bigg|_{v = \eta_0(S_i)} 
        \quad\text{and}\quad
    \nabla_2 K_{0,ij} \equiv \frac{\partial}{\partial v}K( Z_{i}^{c}(\eta_0), (X_{2j}, v) )\bigg|_{v = \eta_0(S_j)}. \]
By applying the chain rule and the product rule, we obtain:
\[ \dt{ K( Z_{i}^{c}(\eta_{\tau}), Z_{j}^{c}(\eta_{\tau}) ) }
    =  \ind{}{c=1} \dt{ \Big\{ \nabla_{1}K_{0,ij}\eta_{\tau}(S_i) + \nabla_{2}K_{0,ij}\eta_{\tau}(S_j) \Big\} }. \]
Now we evaluate $\dt{ \expectF{F_0\otimes F_0}{ r^{c}(W_i, W_j, \theta, \eta_{\tau}) } }$. By the chain rule, product rule, and the law of iterated expectations, it follows:
\begin{align}
    &\dt{ \expectF{F_0\otimes F_0}{r^{c}(W_i,W_j, \theta, \eta_{\tau})} }&\nonumber\\
            & \hspace{+1cm} 
                = \dt{\expectF{F_0}{  \condexpectF{F_0\otimes F_0}{\nabla_{v} \nabla_{\theta} \big\{ \rho(W_i, \theta, v) \cdot \rho(W_j, \theta, \eta_0(S_j)) \big\}\big|_{v = \eta_0(S_i)} \cdot K_{0,ij}^{c}}{S_i} \cdot {\eta_{\tau}(S_i)}   }  }&\nonumber\\
            & \hspace{+1.5cm} 
                + \dt{\expectF{F_0}{  \condexpectF{F_0\otimes F_0}{\nabla_{v} \nabla_{\theta} \big\{ \rho(W_i, \theta, \eta_0(S_i)) \cdot \rho(W_j, \theta, v) \big\}\big|_{v = \eta_0(S_j)} \cdot K_{0,ij}^{c}}{S_j} \cdot {\eta_{\tau}(S_j)}   }  }&\nonumber\\
            & \hspace{+1.5cm} 
                + \dt{\expectF{F_0}{ \mathbf{1}\{c = 1\}\condexpectF{F_0\otimes F_0}{\nabla_{\theta}\big\{ \rho(W_i, \theta, \eta_0(S_i)) \cdot \rho(W_j, \theta, \eta_0(S_j)) \big\} \cdot  \nabla_1K_{0,ij}}{S_i}\cdot \eta_{\tau}(S_i)   } }&\nonumber\\
            & \hspace{+1.5cm} 
                + \dt{\expectF{F_0}{ \mathbf{1}\{c = 1\}\condexpectF{F_0\otimes F_0}{\nabla_{\theta}\big\{ \rho(W_i, \theta, \eta_0(S_i)) \cdot \rho(W_j, \theta, \eta_0(S_j)) \big\} \cdot  \nabla_2 K_{0,ij}}{S_j}\cdot\eta_{\tau}(S_j)   } }&\nonumber\\
            & \hspace{+1cm}
                = \dt{ \expectF{F_0\otimes F_0}{ \vecdel_{0,1}^{c}(S_i) \eta_{\tau}(S_i)  + \vecdel_{0,2}^{c}(S_j)\eta_{\tau}(S_j) }  }.&\label{sec04 - proposition01 - proof - eq01}
\end{align}
Therefore, we obtain the condition {in} equation~\eqref{sec04 - assumption04 - a1 - eq01}. 

Now we prove the remaining statement. Fix $\theta\in\textnormal{int}(\Theta)$. Since $\eta_0\in\mathcal{H}_N$, Assumption \ref{sec03 - assumption02 - CMR}(d) gives,
\begin{equation*}
    \max\left\{ \sup_{\theta\in \textnormal{int}(\Theta), \eta\in\mathcal{H}_N}\Norm{ \nabla_v \rho(S_i,\theta,v)\Big|_{v = \eta}}_{2}, \sup_{\theta\in \textnormal{int}(\Theta), \eta\in\mathcal{H}_N}\Norm{ \nabla_v\big(\nabla_\theta \rho(S_i,\theta,v)\big)\Big|_{v = \eta}}_{2}   \right\} < C_{J},
\end{equation*}
and thereby, for $\delta_{0,11}^{c}$, the conditional Jensen's inequality, the triangle inequality, and the bounds on the derivative coefficients and $K$ give:
\[  \norm{\delta_{0,11}^{c}(S_i,\theta)}_2\leq \widebar{K}_0\big\{C_{J} \norm{\rho(W_j,\theta,\eta_0(S_j))}_2 + C_{J}\norm{\nabla_\theta\rho(W_j,\theta,\eta_0(S_j))}_2\big\}. \]
By Assumption \ref{sec03 - assumption02 - CMR}(b), it follows that
\[ \norm{\delta_{0,11}^{c}(S_i,\theta)}_2\leq 2\widebar{K}_0 \cdot \widebar{R}\cdot C_{J}. \]

For $\delta_{0,12}^{c}$, the partial derivatives of $K$ with respect to $v$ are bounded by $\widebar{K}_1$. Conditional Jensen's inequality and the product rule imply
\[  \norm{\delta_{0,12}^{c}(S_i,\theta)}_2\leq \widebar{K}_1\norm{\nabla_\theta\big\{\rho(W_i,\theta,\eta_0(S_i)) \rho(W_j,\theta,\eta_0(S_j)) \big\}}_2. \]
Since $W_i$ and $W_j$ are independent, the $L_2$-norm of each product in the $\theta$-derivative is bounded by the product of the corresponding $L_2$-norms. Thus,
\[  \norm{\nabla_\theta\big\{\rho(W_i,\theta,\eta_0(S_i))\rho(W_j,\theta,\eta_0(S_j))\big\}}_2\leq 2\widebar{R}^2, \]
meaning $\norm{\delta_{0,12}^{c}(S_i,\theta)}_2\leq 2\widebar{K}_1\cdot \widebar{R}^2$.

Combining these bounds with the definitions of $\vecdel_{0,1}^{c}$ yields:
\[  \sup_{\theta\in\textnormal{int}(\Theta)}\norm{\vecdel_{0,1}^{c}(S,\theta)}_2
        \leq    2\widebar{K}_0 \cdot \widebar{R} \cdot C_{J} + 2\ind{}{c=1} \widebar{K}_1 \cdot \widebar{R}^2<\infty. \]
By symmetric structure, $\norm{\vecdel_{0,2}^{c}}_{2}$ has the same upper bound. Consequently, we have:
\[  \sup_{\theta\in\textnormal{int}(\Theta), \eta\in\mathcal{H}_N} \big\{  \norm{\vecdel_{0,1}^{c}(\cdot, \theta)}_{2} + \norm{\vecdel_{0,2}^{c}(\cdot, \theta)}_{2} \big\} < C_{\delta}. \]
if we set $C_{\delta} = 4\widebar{K}_0 \cdot \widebar{R} \cdot C_{J} + 4\ind{}{c=1} \widebar{K}_1 \cdot \widebar{R}^2$. \qed

\paragraph{Step 2.} Next, we establish that $\psi^{c}(W_i, W_j, \theta, \eta(\cdot), \vecdel^{c}(\cdot, \theta))$ is the Neyman orthogonal score. Consider the following functional equation:
\begin{equation}\label{sec04 - proposition01 - proof - eq00}
    \dt{ \expectF{F_0\otimes F_0}{ r^{c}(W_i, W_j, \theta, \eta_{\tau}(\cdot) ) } } = \dt{  \expectF{F_{\tau}\otimes F_{\tau}}{ \phi(W_i, W_j, \theta, \eta_0, \vecdel^{c}_{0}(\cdot, \theta)) }}.
\end{equation}
A function $\phi(\cdot)$, solving the equation~\eqref{sec04 - proposition01 - proof - eq00}, induces the score $\psi^{c}(\cdot) = r^{c}(\cdot) + \phi(\cdot)$ obeying Neyman-orthogonality \citep{dml_chernozhukov_et_al2018, dml_chernozhukov_et_al2022,dml_escanciano_terschuur2026}.
For $\tau\in [0,1]$, the nuisance $\eta_{\tau}(S_i)$ is a solution to the following equation:
\begin{equation*}
    \expectF{F_\tau}{ \varsigma(S_i)( T_i - \eta_{\tau}(S_i)) } = 0 \quad\text{for admissible } \varsigma(\cdot)\in \mathcal{H},
\end{equation*}
meaning, for two measurable test functions $\varsigma_1(\cdot), \varsigma_2(\cdot)\in \mathcal{H}$,
\begin{equation}\label{sec04 - proposition01 - proof - eq02}
    \expectF{F_{\tau} \otimes F_{\tau}}{ \varsigma_1(S_i)( T_i - \eta_{\tau}(S_i)) + \varsigma_2(S_j)( T_j - \eta_{\tau}(S_j)) } = 0
\end{equation}
So, if we choose $\varsigma_1(S_i) \equiv \vecdel_{0,1}^{c}(S_i, \theta)$ and $\varsigma_2(S_j) \equiv \vecdel_{0,2}^{c}(S_j, \theta)$ as the instrument functions, then the first-stage equation~\eqref{sec04 - proposition01 - proof - eq02} gives: for every $\tau \in [0, \widebar{\tau})$, $\widebar{\tau} \in [0, 1]$,
\begin{equation*}
    \expectF{F_{\tau}\otimes F_{\tau} }{ \vecdel_{0,1}^{c}(S_i, \theta) (T_{i} - \eta_{\tau}(S_i) ) + \vecdel_{0,2}^{c}(S_j, \theta)(T_{j} - \eta_{\tau}(S_j) )  } = 0
\end{equation*}
It follows, by the product rule,
\begin{align*}
    \frac{d}{dt}\big( 0 \big)
        =   &\dt{ \ExpectF{F_{\tau} \otimes F_{\tau}}{ \vecdel_{1}^{c}(S_i, \theta) (T_{i} - \eta_{\tau}(S_i) ) + \vecdel_{2}^{c}(S_j, \theta)(T_{j} - \eta_{\tau}(S_j) )  } } &\\
        =   &\dt{ \ExpectF{F_{\tau} \otimes F_{\tau}}{ \vecdel_{1}^{c}(S_i, \theta) (T_{i} - \eta_{0}(S_i) ) + \vecdel_{2}^{c}(S_j, \theta)(T_{j} - \eta_{0}(S_j) )  } }&\\
            & - \dt{ \ExpectF{F_{0}\otimes F_{0}}{ \vecdel_{1}^{c}(S_i, \theta)\eta_{\tau}(S_i) + \vecdel_{2}^{c}(S_j, \theta)\eta_{\tau}(S_j)  } },&
\end{align*}
implying:
\begin{align}
    &\dt{ \ExpectF{F_{0}\otimes F_{0}}{ \vecdel_{0,1}^{c}(S_i, \theta)\eta_{\tau}(S_i) + \vecdel_{0,2}^{c}(S_j, \theta)\eta_{\tau}(S_j)  } }&\nonumber\\
    &\hspace{+2cm} =   \dt{ \ExpectF{F_{\tau} \otimes F_{\tau} }{ \vecdel_{0,1}^{c}(S_i, \theta) (T_{i} - \eta_{0}(S_i) ) + \vecdel_{0,2}^{c}(S_j, \theta)(T_{j} - \eta_{0}(S_j) )  } }.&\label{sec04 - proposition01 - proof - eq03}
\end{align}
By combining {equations}~\eqref{sec04 - proposition01 - proof - eq01} and \eqref{sec04 - proposition01 - proof - eq03} and applying the Riesz representation theorem, we obtain:
\begin{align*}
        \dt{ \expectF{F_{0} \otimes F_{0} }{r^{c}(W_i,W_j, \theta, \eta_{\tau})} }
            =   &\dt{ \ExpectF{ F_{0}\otimes F_{0} }{ \vecdel_{0,1}^{c}(S_i, \theta)\eta_{\tau}(S_i) + \vecdel_{0,2}^{c}(S_j, \theta)\eta_{\tau}(S_j)  } }&\\
            =   &\dt{ \ExpectF{ F_{\tau}\otimes F_{\tau}  }{ \vecdel_{0,1}^{c}(S_i, \theta) (T_{i} - \eta_{0}(S_i) ) + \vecdel_{0,2}^{c}(S_j, \theta)(T_{j} - \eta_{0}(S_j) )  } }.&
\end{align*}
In words, $\phi(W_i, W_j, \theta, \eta, \vecdel_{0}) = \vecdel_{0,1}^{c}(S_i, \theta) (T_{i} - \eta_{0}(S_i) ) + \vecdel_{0,2}^{c}(S_j, \theta)(T_{j} - \eta_{0}(S_j) )  $ satisfies the equation \eqref{sec04 - proposition01 - proof - eq00}. \qed
%%%%%%%%%%%%%%%%%%%%%%%%%%%%%%%%%%%%%%%%%%%%%%%%%%%%%%%%%%%%%%%%%%%%%%%%%%%%%%%%%%%%%%%%%%%%%%%%%%%%%%%%%%%%%%%%%%%%%%%%%%%%%%%%%%%%%%%%%%%%%%%%%%%%%%%%%%%%%%%%%%%%%%%%%%%%%%%%%%%%%%%%%%%%%%%%%%%%%%%%%%%%%%%%%%%%%%%%%%%%%%%%%%%%%%%%%%%%%%%%%%%%%%%%%%%%%%%%%%%%%%%%%%%%%%%%%%%%%%%%%%%%%%%%%%%%%%%%%%%%%%%%%%%%%%%%%%%%%%%%%%%%%%%%%%%%%%%%%%%%%%%%%%%%%%%%%%%%%%%%%%%%%%%%%%%%%%%%%%%%%%%%%%%%%%%%%%%%%%%%%%%%%%%%%%%%%%%%%%%%%%%%%%%%%%%%%%%%%%%%%%%%%%%%%%%%%%%%%%%%%%%%%%%%%%%%%%%%%%%%%%%%%%%%%%%%%%%%%%%%%%%%%%%%%%%%%%%%%%%%%%%%%%%%%%%%%%%%%%%%%%%%%%%%%%%%%%%%%%%%%%%%%%%%%%%%%%%%
\subsection{Proof of Theorem \ref{sec04 - theorem01 - dml}}\label{sec04 - theorem01 - dml - proof}
\newcommand{\foldavg}[1]{ \sum_{\ell=1}^{L}\frac{|I_{\ell}|}{\binom{N}{2}}{#1} }
\newcommand{\avginfold}[1]{ \frac{1}{|I_{\ell}|}\sum_{(i,j)\in I_{\ell}}{#1} }
\newcommand{\calvone}{ {\mathcal{V}_{\ell,ij}^{1}} }
\newcommand{\calvtwo}{ {\mathcal{V}_{\ell,ij}^{2}} }
\newcommand{\calvthree}{ {\mathcal{V}_{\ell,ij}^{3}} }
\newcommand{\foldnorm}[1]{ \| {#1} \|_{-\ell,2} }
\newcommand{\foldNorm}[1]{ \left\| {#1} \right\|_{-\ell,2} }
\newcommand{\deldel}{ \Delta \vecdel_{\ell}  }
\newcommand{\deleta}{ \Delta \eta_{\ell}  }

For each $\ell\in[L]$, let $\mathcal{W}_{-\ell}$ denote the $\sigma$-field induced by $\{W_i : i\notin J_{\ell}\}$, and $\norm{X}_{-\ell,p}\equiv\left\{\condexpect{\norm{X}^p}{\mathcal{W}_{-\ell}}\right\}^{1/p}$. For notational simplicity, let 
\[  \deleta = \norm{\widehat{\eta}_{\ell} - \eta_{0}}_{\infty}, \quad \deldel \equiv \norm{ \check{\vecdel}_{\ell}^{c} - \vecdel_{0}^{c} }_{\infty},  \]
\[  \psi_{0,ij}^{c} \equiv \psi^{c}(W_i, W_j, \theta_0, \eta_0, \vecdel_0^{c}(\cdot,\theta_0) ),\quad \psi_{\ell,ij}^{c}(\theta) \equiv \psi^{c}(W_i, W_j, \theta, \widehat{\eta}_{\ell},\check{\vecdel}_{\ell}^{c}).\]

\paragraph{Proof of (i).} We first establish weak consistency of $\widehat{\theta}_\textnormal{DDML}^{c}$. By Lemma \ref{appendix - lemma A01}, the smallest singular value $s_c$ of $\Gamma_c$ is strictly positive. Assumption \ref{sec03 - assumption02 - CMR}(b), independence of $W_{i}$ and $W_{j}$, and boundedness of the kernel imply that $\theta\mapsto\expect{r^{c}(W_{i},W_{j},\theta,\eta_0)}$ is continuously differentiable near $\theta_0$. This allows us to take $\epsilon>0$ sufficiently small such that
\[
\big\{\theta\in\mathbb R^{d_\theta}:\|\theta-\theta_0\|\leq2\epsilon\big\}
\subset\operatorname{int}(\widebar{\Theta}_{o})
\]
and
\[ \sup_{ \|\theta-\theta_0\|\mathbin{{\leq}} 2\epsilon}\Norm{\expect{\nabla_\theta r^{c}(W_{i},W_{j},\theta,\eta_0)}-\Gamma_c} \leq s_c/2. \]
Since $\expect{r^{c}(W_{i},W_{j},\theta_0,\eta_0)}=0$, integration along the line segment from $\theta_0$ to $\theta$ gives:
\begin{align*}
    \Norm{\expect{r^{c}(W_{i},W_{j},\theta,\eta_0)}}
        =   &\Norm{ \Gamma_c(\theta-\theta_0) + \int_{0}^{1}\big\{\expect{\nabla_\theta r^{c}(W_{i},W_{j},\theta_0+\tau(\theta-\theta_0),\eta_0)}-\Gamma_c\big\}(\theta-\theta_0)d\tau    }&\\
        \geq    &s_c\norm{\theta-\theta_0}-\frac{s_c}{2}\norm{\theta-\theta_0}
        =\frac{s_c}{2}\norm{\theta-\theta_0} > 0,&
\end{align*}
for every $\theta$ satisfying $0<\norm{\theta-\theta_0}\leq2\epsilon$.

Conditional on $\mathcal{W}_{-\ell}$, the nuisance estimates are fixed. The entropy bound in Lemma~\ref{appendix - lemma A04}, together with the moment bounds and Assumption \ref{sec04 - assumption05 - algorithm stability}, therefore permits application of the conditional uniform law of large numbers:
\begin{equation}\label{sec04 - theorem01 - proof - consistency - eq01}
    \sup_{ \|\theta-\theta_0\| \leq2\epsilon}\Norm{\avginfold{\psi_{\ell, ij}^{c}(\theta)}-\condexpect{\psi_{\ell,ij}^{c}(\theta)}{\mathcal{W}_{-\ell}}}=o_P(1).
\end{equation}
By Lemma \ref{appendix - lemma A03}(i) and conditional independence:
\[      \sup_{ \|\theta-\theta_0\| \leq2\epsilon}
            \Norm{  \condexpect{r^{c}(W_{i},W_{j},\theta,\widehat{\eta}_{\ell})}{\mathcal{W}_{-\ell}}-\expect{r^{c}(W_{i},W_{j},\theta,\eta_0)}  }
        \leq    \widebar{L}_1^{c}\deleta. \]
Since the first-stage conditional moment restriction implies
\[ \condexpect{\check{\vecdel}_{\ell,1}^{c}(S_{i})\big\{T_{i}-\eta_0(S_{i})\big\}+\check{\vecdel}_{\ell,2}^{c}(S_{j})\big\{T_{j}-\eta_0(S_{j})\big\}}{\mathcal{W}_{-\ell}}=0, \]
by Proposition \ref{sec04 - proposition01} and Assumption \ref{sec04 - assumption05 - algorithm stability}, it follows that:
\begin{align*}
    \Norm{\condexpect{\phi(W_{i},W_{j},\widehat{\eta}_{\ell},\check{\vecdel}_{\ell}^{c})}{\mathcal{W}_{-\ell}}}
        \leq    &\deleta\Big\{\norm{\check{\vecdel}_{\ell,1}^{c}(S_{i})}_{-\ell,2}+\norm{\check{\vecdel}_{\ell,2}^{c}(S_{j})}_{-\ell,2}\Big\}&\\
        \leq    &\deleta\Big\{\norm{\vecdel_{0,1}^{c}}_2+\norm{\vecdel_{0,2}^{c}}_2+2\deldel\Big\}
        =o_P(1).&
\end{align*}
Therefore, we obtain:
\[ \sup_{ \|\theta-\theta_0\| \leq2\epsilon}\Norm{\condexpect{\psi_{\ell,ij}^{c}(\theta)}{\mathcal{W}_{-\ell}}-\expect{r^{c}(W_{i},W_{j},\theta,\eta_0)}}=o_P(1). \]
Combining this with equation~\eqref{sec04 - theorem01 - proof - consistency - eq01} yields:
\begin{equation}\label{sec04 - theorem01 - proof - consistency - eq02}
    \sup_{ \|\theta-\theta_0\|\leq2\epsilon}\Norm{\binom{N}{2}^{-1}\sum_{\ell=1}^{L}\sum_{(i,j)\in I_{\ell}}\psi_{\ell,ij}^{c}(\theta)-\expect{r^{c}(W_{i},W_{j},\theta,\eta_0)}}=o_P(1).
\end{equation}

Finally, Lemma \ref{sec04 - lemma01} implies $\prob{\norm{\plugin-\theta_0}<\epsilon/2}\to1$. On this event,
\[ \big\{\theta:\norm{\theta-\theta_0}\leq\epsilon/2\big\}
    \subset\big\{\theta\in\textnormal{int}(\Theta):\norm{\theta-\plugin}\leq\epsilon\big\}
    \subset\big\{\theta:\norm{\theta-\theta_0}\leq2\epsilon\big\}. \]
Hence, $\theta_0$ belongs to the set over which the criterion in Definition~\ref{sec04 - definition01 - DML} is minimized, and this entire set lies in the neighborhood where the population separation bound holds. By the minimizing property and equation~\eqref{sec04 - theorem01 - proof - consistency - eq02},
\begin{align*}
    \frac{s_c}{2}\norm{\widehat{\theta}_\textnormal{DDML}^{c}-\theta_0}
        \leq    &\Norm{\expect{r^{c}(W_{i},W_{j},\widehat{\theta}_\textnormal{DDML}^{c},\eta_0)}}&\\
        \leq    &2\sup_{ \|\theta - \theta_0\|\leq2\epsilon}\Norm{\binom{N}{2}^{-1}\sum_{\ell=1}^{L}\sum_{(i,j)\in I_{\ell}}\psi_{\ell,ij}^{c}(\theta)-\expect{r^{c}(W_{i},W_{j},\theta,\eta_0)}} 
        =       o_P(1).&
\end{align*}
Therefore, $\widehat{\theta}_\textnormal{DDML}^{c}\overset{p}{\longrightarrow}\theta_0$.\qed

\paragraph{Proof of (ii).} We establish the $\sqrt{N}$-consistency of the DDML estimator and its limiting distribution. Fix $\epsilon>0$. By adding and subtracting, we have
\begin{equation}\label{sec04 - theorem01 - proof - eq00}
    \sqrt{N}\binom{N}{2}^{-1}\sum_{\ell=1}^{L}\sum_{(i,j)\in I_{\ell}} {\psi}_{\ell,ij}^{c}(\theta_0) 
        =   \sqrt{N}\binom{N}{2}^{-1}\sum_{\ell=1}^{L}\sum_{(i,j)\in I_{\ell}} \psi_{0,ij}^{c} + \sqrt{N}\binom{N}{2}^{-1}\sum_{\ell=1}^{L}\sum_{(i,j)\in I_{\ell}}\big( {\psi}_{\ell,ij}^{c}(\theta_0) - \psi_{0,ij}^{c} \big).
\end{equation}
Since $\sum_{\ell=1}^{L}|I_{\ell}|=\binom{N}{2}$, the triangle inequality and the law of iterated expectations give:
\begin{align}
    \nonumber&\Prob{\Norm{\sqrt{N}\foldavg{\avginfold{ \big\{{\psi}_{\ell,ij}^{c}(\theta_0)-\psi_{0,ij}^{c} \big\} }}}>\epsilon}&\\
    \label{sec04 - theorem01 - proof - eq01}&\hspace{+2cm} \leq L \sup_{\ell\in[L]} \Expect{\condProb{\Norm{\sqrt{N}\avginfold{ \big\{ {\psi}_{\ell,ij}^{c}(\theta_0)-\psi_{0,ij}^{c} \big\} } }>\epsilon}{\mathcal{W}_{-\ell}}}.&
\end{align}
By the conditional Markov inequality \citep{dml_chernozhukov_et_al2018}, we obtain:
\begin{align*}
    &\condProb{\Norm{\sqrt{N}\avginfold{ \big\{ {\psi}_{\ell,ij}^{c}(\theta_0)-\psi_{0,ij}^{c} \big\} }}>\epsilon}{\mathcal{W}_{-\ell}}&\\
        &\hspace{+2cm} \leq
                \frac{N}{\epsilon^2 |I_{\ell}|^2}\condExpect{\Norm{\sum_{(i,j)\in I_{\ell}}({\psi}_{\ell,ij}^{c}(\theta_0)-\psi_{0,ij}^{c})}^2}{\mathcal{W}_{-\ell}}&\\
        &\hspace{+2cm} =
                \frac{N}{\epsilon^2 |I_{\ell}|^2}
                \Bigg\{
                    \sum_{(i,j)\in I_{\ell}}\condexpect{\norm{{\psi}_{\ell,ij}^{c}(\theta_0)-\psi_{0,ij}^{c}}^2}{\mathcal{W}_{-\ell}}&\\
        &\hspace{+3cm}
                    + \sum_{\substack{(i,j),(i',j')\in I_{\ell}\\(i,j)\ne(i',j')}}
                    \condexpect{({\psi}_{\ell,ij}^{c}(\theta_0)-\psi_{0,ij}^{c})^\top({\psi}_{\ell,i'j'}^{c}(\theta_0)-\psi_{0,i'j'}^{c})}{\mathcal{W}_{-\ell}}
                \Bigg\}.&
\end{align*}
For pairs sharing one observation, the absolute cross product expectation is bounded by the diagonal second moment by the conditional Cauchy--Schwarz inequality. For disjoint pairs, it equals $\norm{\condexpect{{\psi}_{\ell,ij}^{c}(\theta_0)-\psi_{0,ij}^{c}}{\mathcal{W}_{-\ell}}}^2$ by conditional independence. Each pair in $I_{\ell}$ shares an observation with at most $2(n_{\ell}-2)$ other pairs. Therefore,
\begin{align}
    \nonumber&\condProb{\Norm{\sqrt{N}\avginfold{ \big\{{\psi}_{\ell,ij}^{c}(\theta_0)-\psi_{0,ij}^{c}\big\} }}>\epsilon}{\mathcal{W}_{-\ell}}&\\
    \nonumber&\hspace{+2cm}\leq
            \frac{N(2n_{\ell}-3)}{\epsilon^2|I_{\ell}|}\foldnorm{{\psi}_{\ell,ij}^{c}(\theta_0)-\psi_{0,ij}^{c}}^2
            +\frac{N}{\epsilon^2}\norm{\condexpect{{\psi}_{\ell,ij}^{c}(\theta_0)-\psi_{0,ij}^{c}}{\mathcal{W}_{-\ell}}}^2&\\
    \label{sec04 - theorem01 - proof - eq02}&\hspace{+2cm}\leq
            \frac{8G+o(1)}{\epsilon^2}\foldnorm{{\psi}_{\ell,ij}^{c}(\theta_0)-\psi_{0,ij}^{c}}^2
            +\frac{N}{\epsilon^2}\norm{\condexpect{{\psi}_{\ell,ij}^{c}(\theta_0)-\psi_{0,ij}^{c}}{\mathcal{W}_{-\ell}}}^2,&
\end{align}
because $G$ is fixed, $n_{\ell}=N/G+O(1)$, and $|I_{\ell}|\geq\lfloor N/(2G)\rfloor^2$ for every $\ell\in[L]$. Therefore, the remainder term of equation~\eqref{sec04 - theorem01 - proof - eq00} vanishes if $\foldnorm{{\psi}_{\ell,ij}^{c}(\theta_0)-\psi_{0,ij}^{c}}=o_P(1)$ and $\sqrt{N}\norm{\condexpect{{\psi}_{\ell,ij}^{c}(\theta_0)-\psi_{0,ij}^{c}}{\mathcal{W}_{-\ell}}}=o_P(1)$.

In step 1, we show $\foldnorm{{\psi}_{\ell,ij}^{c}(\theta_0)-\psi_{0,ij}^{c}} = o_P(1)$.

In step 2, we show $\sqrt{N}\norm{\condexpect{{\psi}_{\ell,ij}^{c}(\theta_0)-\psi_{0,ij}^{c}}{\mathcal{W}_{-\ell}}} = o_P(1)$.

In step 3, we show:
\[ \sqrt{N}\binom{N}{2}^{-1}\sum_{\ell=1}^{L}\sum_{(i,j)\in I_{\ell}}\psi_{0,ij}^{c} \overset{d}{\longrightarrow} \mathcal{N}(0, \Sigma_{c}), \]

In step 4, we finally show the asymptotic distribution of $\sqrt{N}(\widehat{\theta}_{\textnormal{DDML}}^{c} - \theta_0)$.

\subparagraph{Step 1.} By construction of the orthogonal score, ${\psi}_{\ell,ij}^{c}(\theta_0) - \psi_{0,ij}^{c}$ can be decomposed as:
\[ {\psi}_{\ell,ij}^{c}(\theta_0) - \psi_{0,ij}^{c} = \calvone + \calvtwo + \calvthree \]
where
\begin{align*}
    \calvone
        =   & \psi^{c}(W_i, W_j, \theta_0, \widehat{\eta}_{\ell}, {\vecdel}_{0}^{c}) - \psi^{c}(W_i, W_j, \theta_0, {\eta}_0, {\vecdel}_{0}^{c}) &\\
    \calvtwo
        =   & \psi^{c}(W_i, W_j, \theta_0, {\eta}_{0}, \check{\vecdel}_{\ell}^{c}) - \psi^{c}(W_i, W_j, \theta_0, {\eta}_{0}, {\vecdel}_{0}^{c}) &\\
    \calvthree
        =   & \psi^{c}(W_i, W_j, \theta_0, \widehat{\eta}_{\ell}, \check{\vecdel}_{\ell}^{c}) - \psi^{c}(W_i, W_j, \theta_0, \widehat{\eta}_{\ell},   {\vecdel}_{0}^{c}) 
                - \Big\{ \psi^{c}(W_i, W_j, \theta_0, {\eta}_0, \check{\vecdel}_{\ell}^{c}) - \psi^{c}(W_i, W_j, \theta_0, {\eta}_{0}, {\vecdel}_{0}^{c}) \Big\}.&
\end{align*}
First, we evaluate $\foldnorm{\calvone}$ : 
\begin{align*}
    \foldnorm{\calvone}
        \leq    \,\,&\,\, \foldnorm{r^{c}(W_i, W_j, \theta_0, \widehat{\eta}_{\ell}) - r^{c}(W_i, W_j, \theta_0, \eta_0)}&\\
                    &\,\, + \foldnorm{ {\vecdel}_{0}^{c}( S_i, \theta_0)\cdot( \widehat{\eta}_{\ell}(S_i) - \eta_0(S_i) }  + \foldnorm{ {\vecdel}_{0}^{c}( S_i, \theta_0)\cdot( \widehat{\eta}_{\ell}(S_j) - \eta_0(S_j) } &\\
        \leq    \,\,&\,\, \foldnorm{r^{c}(W_i, W_j, \theta_0, \widehat{\eta}_{\ell}) - r^{c}(W_i, W_j, \theta_0, \eta_0)} + \big\{ \foldnorm{ {\vecdel}_{0}^{c}( S_i, \theta_0) } + \foldnorm{ {\vecdel}_{0}^{c}( S_i, \theta_0) } \big\} \cdot \deleta.&
\end{align*}

By Lemma \ref{appendix - lemma A03}(i) and the conditional independence,
\[ \foldnorm{r^{c}(W_i, W_j, \theta_0, \widehat{\eta}_{\ell}) - r^{c}(W_i, W_j, \theta_0, \eta_0)} = \norm{r^{c}(W_i, W_j, \theta_0, \widehat{\eta}_{\ell}) - r^{c}(W_i, W_j, \theta_0, \eta_0)}_{2} \leq \widebar{L}_{1}^{c}\cdot \deleta.\] 
Combining this with Assumption \ref{sec04 - assumption04 - linear representation}, it follows:
\[     \foldnorm{\calvone}
        \leq    \,\, \widebar{L}_{1}^{c} \cdot \deleta  + \big\{\foldnorm{ \vecdel_{0,1}^{c}(S_i, \theta_0) }  + \foldnorm{\vecdel_{0,2}^{c}(S_i, \theta_0) } \big\} \cdot \deleta
        \leq    \,\, \big\{ \widebar{L}_{1}^{c} + C_{\delta} \big\} \cdot \deleta. \]
Second, we compute the upper bound of $\foldnorm{\calvtwo}$:
\begin{align*}
    \foldnorm{\calvtwo}
        \leq    \,\,&\,\, \foldNorm{ \big( \check{\vecdel}_{\ell,1}^{c}(S_i, \widetilde{\theta}_{\ell}) - \vecdel_{0,1}^{c}(S_i,\theta_0 ) \big)(T_i - \eta_0(S_i) ) }
                + \foldNorm{ \big( \check{\vecdel}_{\ell,2}^{c}(S_j, \widetilde{\theta}_{\ell}) - \vecdel_{0,2}^{c}(S_j, \theta_0 ) \big)(T_j - \eta_0(S_j) ) } &\\
        \leq    \,\,&\,\, \foldnorm{ T_i - \eta_0(S_i) } \cdot \big( \norm{ \check{\vecdel}_{\ell,1}^{c} - \vecdel_{0,1}^{c}}_{\infty} + \norm{ \check{\vecdel}_{\ell,2}^{c} - \vecdel_{0,2}^{c} }_{\infty} \big)&\\
        \leq    \,\,&\,\, 2 C_{T} \cdot \deldel,&
\end{align*}
because $\foldnorm{T_{i} - \eta_0(S_i)} = \norm{ T_i - \eta_0(S_i) }_{2} < C_{T}$ is imposed by Assumption \ref{sec03 - assumption01 - first stage} and conditional independence. Lastly,
\[ \foldnorm{ \calvthree }
        \leq    \,\, \deleta\cdot \big( \norm{ \check{\vecdel}_{\ell,1}^{c} - \vecdel_{0,1}^{c} }_{\infty} + \norm{ \check{\vecdel}_{\ell,2}^{c} - \vecdel_{0,2}^{c} }_{\infty} \big)
        \leq    \,\, 2\deleta\cdot \deldel. \]
Combining three results give:
\begin{align*}
    \foldnorm{{\psi}_{\ell,ij}^{c}(\theta_0) - \psi_{0,ij}^{c}}
        \,\,\leq\,\,    & \big( \foldnorm{\calvone} + \foldnorm{\calvtwo} + \foldnorm{\calvthree} \big) &\\
        \,\,\leq\,\,    & \big( \widebar{L}_{1}^{c} + C_{\delta,2} \big)\cdot \deleta  + 2C_{T} \cdot \deldel + 2\deleta \cdot \deldel&\\
        \,\,=\,\,       &o_{P}\left( a_N + b_N + a_N \cdot b_N\right).&
\end{align*}
By Assumption \ref{sec04 - assumption05 - algorithm stability}, this term is $o_P(1)$.

\subparagraph{Step 2.} For each $\ell\in[L]$ and $\tau\in[0,1]$, let $\eta_{\tau} \equiv \eta_0 + \tau (\widehat{\eta} - \eta_0)$. Taylor expansion of $\psi^{c}$ yields:
\begin{align*}
    \condexpect{{\psi}_{\ell,ij}^{c}(\theta_0) - \psi_{0,ij}^{c}}{\mathcal{W}_{-\ell}}
        =   &\condexpect{D_\eta\psi^{c}(W_i,W_j,\theta_0,\eta_0, \vecdel_{0}^{c} )[ \widehat{\eta} - \eta_0 ]}{\mathcal{W}_{-\ell}}&\\
            & + \condexpect{D_\eta\psi^{c}(W_i,W_j,\theta_0,\eta_0,\check{\vecdel}_{\ell}^{c})[ \widehat{\eta} - \eta_0 ]}{\mathcal{W}_{-\ell}}
                - \condexpect{D_\eta\psi^{c}(W_i,W_j,\theta_0,\eta_0, {\vecdel}_{0}^{c} )[ \widehat{\eta} - \eta_0 ]}{\mathcal{W}_{-\ell}}&\\
            & +\int_{0}^{1}\condexpect{D_{\eta} \psi^{c}(W_i,W_j,\theta_0,\eta_{\tau},\check{\vecdel}_{\ell}^{c})[ \widehat{\eta} - \eta_0 ]- D_{\eta} \psi^{c}(W_i,W_j, \theta_0, \eta_0, \check{\vecdel}_{\ell}^{c})[ \widehat{\eta} - \eta_0 ] }{ \mathcal{W}_{-\ell} }d\tau.&
\end{align*}
By orthogonality, the first term in the second equality is zero. The difference between the next two terms is bounded in norm by $\deleta\cdot \deldel$, since the bias-correction term is linear in $\vecdel$.

On the event $\widehat{\eta}_{\ell}\in\mathcal{H}_N$, convexity gives $\eta_{\tau}\in\mathcal{H}_N$ for every $\tau\in[0,1]$. By Lemma \ref{appendix - lemma A03}(ii),
\[ \Norm{\int_{0}^{1}\condexpect{D_{\eta} \psi^{c}(W_i,W_j,\theta_0,\eta_{\tau},\check{\vecdel}_{\ell}^{c})[ \widehat{\eta} - \eta_0 ]- D_{\eta} \psi^{c}(W_i,W_j, \theta_0, \eta_0, \check{\vecdel}_{\ell}^{c})[ \widehat{\eta} - \eta_0 ] }{ \mathcal{W}_{-\ell} }d\tau} \leq \int_{0}^{1} \big\{\widebar{L}_{2}^{c} \cdot \deleta^2 \big\} \tau d\tau, \]
and therefore:
\begin{equation*}
    \norm{\condexpect{{\psi}_{\ell,ij}^{c}(\theta_0)-\psi_{0,ij}^{c}}{\mathcal{W}_{-\ell}}}
        \leq    2 \deleta \cdot \deldel +\frac{\widebar{L}_{2}^{c}}{2}\deleta^2 = o_P(a_Nb_N+a_N^2).    
\end{equation*}
Since $\sqrt{N}a_N^2=O(1)$ and $\sqrt{N}a_N b_N = O(1)$ by Assumption \ref{sec04 - assumption05 - algorithm stability}(b), we have:
\[ \sqrt{N}\norm{\condexpect{{\psi}_{\ell,ij}^{c}(\theta_0)-\psi_{0,ij}^{c}}{\mathcal{W}_{-\ell}}}=o_P(1). \]

\subparagraph{Step 3.} In this step, we show the limiting distribution of $\sqrt{N}\binom{N}{2}^{-1}\sum_{\ell=1}^{L}\sum_{(i,j)\in I_{\ell}} \psi_{0,ij}^{c}$. Fix $\theta\in \widebar{\Theta}_{o}$ and let $\Psi_0^{c}(\theta) \equiv \expect{ \psi^{c}(W_i,W_j,\theta, \eta_0, \vecdel_0)}$. Define the Hoeffding-projections:
\[ \varphi_1^{c}(W_i,\theta) \equiv \condexpect{ \psi^{c}(W_i,W_j,\theta, \eta_0, \vecdel_0 )}{W_i} - \Psi_0^{c}(\theta), \]
\[ \varphi_2^{c}(W_i,W_j,\theta) \equiv \psi^{c}(W_i,W_j,\theta, \eta_0, \vecdel_0) - \Psi_0^{c}(\theta) - \varphi_1^{c}(W_i,\theta) - \varphi_1^{c}(W_j,\theta). \]
By the law of iterated expectations, $\expect{\varphi_1^{c}(W_i,\theta)}=0$. Moreover, symmetry and independence imply:
\begin{align*}
    \condexpect{\varphi_2^{c}(W_i,W_j,\theta)}{W_i}
        =   &\condexpect{\psi^{c}(W_i,W_j,\theta, \eta_0, \vecdel_0^{c})}{W_i} - \Psi_0^{c}(\theta) - \varphi_1^{c}(W_i,\theta) - \expect{\varphi_1^{c}(W_j,\theta)} = 0,&\\
    \condexpect{\varphi_2^{c}(W_i,W_j,\theta)}{W_j}
        =   &\condexpect{\psi^{c}(W_i,W_j,\theta, \eta_0, \vecdel_0^{c})}{W_j} - \Psi_0^{c}(\theta) - \expect{\varphi_1^{c}(W_i,\theta)} - \varphi_1^{c}(W_j,\theta) = 0,&
\end{align*}
meaning $\varphi_2^{c}(\cdot,\cdot,\theta)$ is degenerate of order two.

Since $I_1,\ldots,I_L$ partition the set of all pairs, and since $\Psi_0^{c}(\theta_0)=0$, the Hoeffding decomposition \citep{serfling1980, vdv1998} gives:
\begin{align*}
    \sqrt{N}\binom{N}{2}^{-1}\sum_{\ell=1}^{L}\sum_{(i,j)\in I_{\ell}} \psi^{c}(W_i,W_j,\theta, \eta_0, \vecdel_0)
        =   &\sqrt{N}\binom{N}{2}^{-1}\sum_{i<j}\psi^{c}(W_i,W_j,\theta, \eta_0, \vecdel_0)&\\
        =   &\Psi_{0}^{c}(\theta) + \frac{2}{\sqrt{N}}\sum_{i=1}^{N}\varphi_1^{c}(W_i,\theta) + \sqrt{N}U_{N}^{2}\varphi_2^{c}(\cdot,\cdot,\theta).&
\end{align*}
Because $\varphi_2^{c}(\cdot,\cdot,\theta)$ is degenerate,
\[ \Expect{\Norm{\sqrt{N}U_{N}^{2}\varphi_2^{c}(\cdot,\cdot,\theta)}^2}
    = {N}\binom{N}{2}^{-1}\expect{\Norm{\varphi_2^{c}(W_i,W_j,\theta)}^2}
    = O(N^{-1}). \]
Because $\Sigma_{c}$ is finite and positive definite matrix by Lemma~\ref{appendix - lemma A01}(ii), therefore, the central limit theorem and Slutsky's lemma give:
\[ \sqrt{N}\binom{N}{2}^{-1}\sum_{\ell=1}^{L}\sum_{(i,j)\in I_{\ell}} \psi_{0,ij}^{c} 
    \overset{d}{\longrightarrow}
    \mathcal{N}\big(0,\expect{4\varphi_1^{c}(W_i,\theta_0)\varphi_1^{c}(W_i,\theta_0)^\top}\big)
    \overset{d}{=} \mathcal{N}(0,\Sigma_c). \]

\subparagraph{Step 4.} Lastly, we derive the limit distribution of $\sqrt{N}(\widehat{\theta}_{\textnormal{DDML}}^{c}-\theta_0)$. Define
\[  \theta_N^* \equiv \argmin{\substack{\theta\in\widebar{\Theta}_{o},\,\|\theta-\plugin\|\leq\epsilon}}{\norm{\Psi_{N}(\theta)}}
    \quad\text{where}\quad
    \Psi_{N}(\theta) \equiv \Gamma_c(\theta-\theta_0)+\binom{N}{2}^{-1}\sum_{\ell=1}^{L}\sum_{(i,j)\in I_{\ell}}\psi_{\ell,ij}^{c}(\theta_0). \]
By adding and subtracting $\theta_N^*$, we have
\[  \sqrt{N}(\widehat{\theta}_{\textnormal{DDML}}^{c}-\theta_0)
        =   \sqrt{N}(\theta_N^*-\theta_0)+\sqrt{N}(\widehat{\theta}_{\textnormal{DDML}}^{c}-\theta_N^*).\]

First, we show that $\theta_N^*-\theta_0=O_P(N^{-1/2})$ and $\sqrt{N}(\theta_N^*-\theta_0)\overset{d}{\longrightarrow}\mathcal{N}(0,\Omega_c)$. By Lemma~\ref{appendix - lemma A01}, $\Gamma_c$ is invertible. Hence, $\Psi_{N}(\theta)$ is zero at:
\[ \theta_0-(\Gamma_c^\top\Gamma_c)^{-1}\Gamma_c^\top \binom{N}{2}^{-1}\sum_{\ell=1}^{L}\sum_{(i,j)\in I_{\ell}}\psi_{\ell,ij}^{c}(\theta_0). \]
By the preceding score expansion at $\theta_0$, the fitted score average differs from the oracle score average by $o_P(N^{-1/2})$, and the latter is $O_P(N^{-1/2})$, meaning the solution of $\Psi_{N}(\theta) = 0$ is $\theta_0+O_P(N^{-1/2})$. Since $\widebar{\Theta}_{o}$ contains a neighborhood of $\theta_0$, $\plugin\overset{p}{\longrightarrow}\theta_0$, and $\epsilon>0$ is fixed, this zero is feasible with probability approaching one. On this event, it equals $\theta_N^*$ and $\Psi_{N}(\theta_{N}^{*})=0$. Therefore,
\[  \begin{aligned}
        \theta_N^*-\theta_0
            =   &-(\Gamma_c^\top\Gamma_c)^{-1}\Gamma_c^\top \binom{N}{2}^{-1}\sum_{\ell=1}^{L}\sum_{(i,j)\in I_{\ell}}  \Big\{\psi_{0,ij}^{c}+\big(\psi_{\ell,ij}^{c}(\theta_0)-\psi_{0,ij}^{c}\big)\Big\}  +   o_P(N^{-1/2})&\\
            =   &O_P(N^{-1/2}).&
\end{aligned}   \]
Since the folds partition all unordered pairs, the limit distribution of the oracle score gives
\[  \begin{aligned}
        \sqrt{N}(\theta_N^*-\theta_0)
            =   &-(\Gamma_c^\top\Gamma_c)^{-1}\Gamma_c^\top \sqrt{N}\binom{N}{2}^{-1}\sum_{i<j}\psi_{0,ij}^{c}+o_P(1)&\\
            \overset{d}{\longrightarrow}    & \mathcal{N}(0, \Omega_c).&
\end{aligned}   \]

We next show that $\norm{\widehat{\theta}_{\textnormal{DDML}}^{c}-\theta_N^*}=o_P(N^{-1/2})$. On the event $\Psi_{N}(\theta_{N}^{*})=0$, the definition of $\Psi_N(\cdot)$ gives
\[  \norm{\Gamma_c(\widehat{\theta}_{\textnormal{DDML}}^{c}-\theta_N^*)}
        =  \norm{\Psi_{N}(\widehat{\theta}_{\textnormal{DDML}}^{c})-\Psi_{N}(\theta_{N}^{*})}
        =  \norm{\Psi_{N}(\widehat{\theta}_{\textnormal{DDML}}^{c})}.   \]
By the triangle inequality and the minimizing property of $\widehat{\theta}_{\textnormal{DDML}}^{c}$,
\[  \begin{aligned}
        \norm{\Psi_{N}(\widehat{\theta}_{\textnormal{DDML}}^{c})}
        \leq   &\Norm{\binom{N}{2}^{-1}\sum_{\ell=1}^{L}\sum_{(i,j)\in I_{\ell}}\psi_{\ell,ij}^{c}(\widehat{\theta}_{\textnormal{DDML}}^{c})}
                + \Norm{\Psi_{N}(\widehat{\theta}_{\textnormal{DDML}}^{c})-\binom{N}{2}^{-1}\sum_{\ell=1}^{L}\sum_{(i,j)\in I_{\ell}}\psi_{\ell,ij}^{c}(\widehat{\theta}_{\textnormal{DDML}}^{c})}&\\
        \leq  &\Norm{\binom{N}{2}^{-1}\sum_{\ell=1}^{L}\sum_{(i,j)\in I_{\ell}}\psi_{\ell,ij}^{c}(\theta_N^*)}
                + \Norm{\Psi_{N}(\widehat{\theta}_{\textnormal{DDML}}^{c})-\binom{N}{2}^{-1}\sum_{\ell=1}^{L}\sum_{(i,j)\in I_{\ell}}\psi_{\ell,ij}^{c}(\widehat{\theta}_{\textnormal{DDML}}^{c})}.&
\end{aligned}\]

To bound the two terms, fix $\theta\in\widebar{\Theta}_{o}$. Adding and subtracting the conditional mean of $\Delta\psi_{\ell,ij}^{c}(\theta)$ gives
\[\begin{aligned}
    \Psi_{N}(\theta)-\binom{N}{2}^{-1}\sum_{\ell=1}^{L}\sum_{(i,j)\in I_{\ell}}\psi_{\ell,ij}^{c}(\theta)
        =   &\Gamma_c(\theta-\theta_0)  -\sum_{\ell=1}^{L}\frac{|I_{\ell}|}{\binom{N}{2}} \condexpect{\Delta\psi_{\ell,ij}^{c}(\theta)}{\mathcal W_{-\ell}}&\\
            & -\binom{N}{2}^{-1}\sum_{\ell=1}^{L}\sum_{(i,j)\in I_{\ell}} \Big\{\Delta\psi_{\ell,ij}^{c}(\theta)-\condexpect{\Delta\psi_{\ell,ij}^{c}(\theta)}{\mathcal W_{-\ell}}\Big\}.&
\end{aligned} \]
Let $\theta_\tau=\theta_0+\tau(\theta-\theta_0)$ for $\theta$ in a neighborhood of $\theta_0$. Conditional on $\mathcal W_{-\ell}$, the fitted nuisance $\widehat{\eta}_{\ell}$ is fixed and $(W_i,W_j)$ has law $F_0\otimes F_0$ for $(i,j)\in I_{\ell}$. Hence, on the event $\widehat{\eta}_{\ell}\in\mathcal{H}_{N}$, Lemma~\ref{appendix - lemma A03}(iii) applied at $(\theta_\tau,\widehat{\eta}_{\ell})$ gives
\[ \Norm{\Gamma_c-\condexpect{\nabla_\theta r^{c}(W_i,W_j,\theta_\tau,\widehat{\eta}_{\ell})}{\mathcal W_{-\ell}}}
 \leq \widebar L_3^{c}\norm{\widehat{\eta}_{\ell}-\eta_0}_\infty
 +\tau\widebar L_4^{c}\norm{\theta-\theta_0}. \]
Since the fitted correction is fixed in $\theta$, we have:
\[\begin{aligned}
    \Norm{\Gamma_c(\theta-\theta_0) - \sum_{\ell=1}^{L}\frac{|I_{\ell}|}{\binom{N}{2}}\condexpect{\Delta\psi_{\ell,ij}^{c}(\theta)}{\mathcal W_{-\ell}}}
        \leq    &\max_{\ell\in[L]}\int_{0}^{1} \Norm{\Gamma_c-\condexpect{\nabla_\theta r^{c}(W_i,W_j,\theta_\tau,\widehat{\eta}_{\ell})}{\mathcal W_{-\ell}}} d\tau\,\norm{\theta-\theta_0}&\\
        \leq    &\left\{\widebar{L}_{3}^{c}\max_{\ell\in[L]}\norm{\widehat{\eta}_{\ell}-\eta_0}_\infty +\frac{1}{2}\widebar{L}_{4}^{c}\norm{\theta-\theta_0}\right\}\norm{\theta-\theta_0}.&
\end{aligned} \]
By Lemma~\ref{appendix - lemma A05}(ii) and $\sum_{\ell=1}^{L}|I_{\ell}|/\binom{N}{2}=1$, the centered term satisfies, for every deterministic sequence $\xi_N\downarrow0$,
\[  \sup_{\substack{\theta\in\widebar{\Theta}_{o}\\\norm{\theta-\theta_0}\leq\xi_N}}    \Norm{\binom{N}{2}^{-1}\sum_{\ell=1}^{L}\sum_{(i,j)\in I_{\ell}}    \Big\{\Delta\psi_{\ell,ij}^{c}(\theta)-\condexpect{\Delta\psi_{\ell,ij}^{c}(\theta)}{\mathcal W_{-\ell}}\Big\}} =   o_P(N^{-1/2}).
\]

By consistency of $\widehat{\theta}_{\textnormal{DDML}}^{c}$ and $\theta_N^*$, choose $\xi_N\downarrow0$ sufficiently slowly that both lie within $\xi_N$ of $\theta_0$ with probability approaching one. Since $\max_{\ell\in[L]}\norm{\widehat{\eta}_{\ell}-\eta_0}_\infty=o_P(1)$, the coefficient multiplying $\norm{\theta-\theta_0}$ in the preceding bound is $o_P(1)$ uniformly on this neighborhood. Evaluating the two bounds at $\theta_N^*$ and using $\Psi_{N}(\theta_{N}^{*})=0$ with probability approaching one and $\theta_N^*-\theta_0=O_P(N^{-1/2})$, we obtain
\[ \Norm{\binom{N}{2}^{-1}\sum_{\ell=1}^{L}\sum_{(i,j)\in I_{\ell}}\psi_{\ell,ij}^{c}(\theta_N^*)}  =o_P(N^{-1/2}).\]
Evaluating the same bounds at $\widehat{\theta}_{\textnormal{DDML}}^{c}$ and applying the minimizing inequality above therefore gives
\[  \begin{aligned}
        s_c\norm{\widehat{\theta}_{\textnormal{DDML}}^{c}-\theta_N^*}
            \leq    &\norm{\Psi_{N}(\widehat{\theta}_{\textnormal{DDML}}^{c})-\Psi_{N}(\theta_{N}^{*})}&\\
            \leq    &o_P(N^{-1/2})+o_P(1)\norm{\widehat{\theta}_{\textnormal{DDML}}^{c}-\theta_0}&\\
            \leq    &o_P(N^{-1/2})+o_P(1)\norm{\widehat{\theta}_{\textnormal{DDML}}^{c}-\theta_N^*},&
\end{aligned}\]
where the last inequality uses $\norm{\theta_N^*-\theta_0}=O_P(N^{-1/2})$. Since $s_c>0$ by Lemma~\ref{appendix - lemma A01}, absorbing the last term gives $\norm{\widehat{\theta}_{\textnormal{DDML}}^{c}-\theta_N^*}=o_P(N^{-1/2})$. It follows that $\sqrt{N}(\widehat{\theta}_{\textnormal{DDML}}^{c}-\theta_0)\overset{d}{\longrightarrow}\mathcal{N}(0,\Omega_c)$.\qed
%%%%%%%%%%%%%%%%%%%%%%%%%%%%%%%%%%%%%%%%%%%%%%%%%%%%%%%%%%%%%%%%%%%%%%%%%%%%%%%%%%%%%%%%%%%%%%%%%%%%%%%%%%%%%%%%%%%%%%%%%%%%%%%%%%%%%%%%%%%%%%%%%%%%%%%%%%%%%%%%%%%%%%%%%%%%%%%%%%%%%%%%%%%%%%%%%%%%%%%%%%%%%%%%%%%%%%%%%%%%%%%%%%%%%%%%%%%%%%%%%%%%%%%%%%%%%%%%%%%%%%%%%%%%%%%%%%%%%%%%%%%%%%%%%%%%%%%%%%%%%%%%%%%%%%%%%%%%%%%%%%%%%%%%%%%%%%%%%%%%%%%%%%%%%%%%%%%%%%%%%%%%%%%%%%%%%%%%%%%%%%%%%%%%%%%%%%%%%%%%%%%%%%%%%%%%%%%%%%%%%%%%%%%%%%%%%%%%%%%%%%%%%%%%%%%%%%%%%%%%%%%%%%%%%%%%%%%%%%%%%%%%%%%%%%%%%%%%%%%%%%%%%%%%%%%%%%%%%%%%%%%%%%%%%%%%%%%%%%%%%%%%%%%%%%%%%%%%%%%%%%%%%%%%%%%%%%%%
\subsection{Proof of Proposition \ref{sec04 - proposition02}}\label{sec04 - proposition02 - proof}
Since the correction term does not depend on $\theta$, $\nabla_\theta\psi^{c}=\nabla_\theta r^{c}$. We decompose $\widehat{\Gamma}_c-\Gamma_c$ as:
\begin{align*}
    \widehat{\Gamma}_c-\Gamma_c
        =   &\foldavg{ \left\{ \avginfold{\nabla_\theta r^{c}(W_i,W_j,\widehat{\theta}_\textnormal{DDML}^{c},\widehat{\eta}_{\ell})}-
        \condexpect{\nabla_\theta r^{c}(W_i,W_j,\widehat{\theta}_\textnormal{DDML}^{c},\widehat{\eta}_{\ell})}{\mathcal{W}_{-\ell}} \right\}}&\\
            &+\foldavg{ \left\{ \condexpect{\nabla_\theta r^{c}(W_i,W_j,\widehat{\theta}_\textnormal{DDML}^{c},\widehat{\eta}_{\ell})}{\mathcal{W}_{-\ell}} - \Gamma_c \right\}}.&
\end{align*}
Here, the conditional expectation is taken for each fixed $\theta$ and then evaluated at $\widehat{\theta}_\textnormal{DDML}^{c}$. The first term is $o_P(1)$ by consistency of $\widehat{\theta}_\textnormal{DDML}^{c}$ and the conditional uniform law of large numbers for $\nabla_\theta r^{c}$. The second term is $o_P(1)$ by consistency of $\widehat{\theta}_\textnormal{DDML}^{c}$, the first-stage rate conditions, and Lemma \ref{appendix - lemma A03}(iii). Thus, $\widehat{\Gamma}_c=\Gamma_c+o_P(1)$.

For each $\ell\in[L]$ and $i\in J_{\ell}$, define
\[ \widehat{\varphi}_{\ell,i}^{c}\equiv\frac{1}{n_{\ell}-1}\sum_{j\in J_{\ell}\backslash\{i\}}\psi^{c}(W_i,W_j,\widehat{\theta}_\textnormal{DDML}^{c},\widehat{\eta}_{\ell},\check{\vecdel}_{\ell}^{c}),\qquad \varphi_{0,i}^{c}\equiv\condexpect{\psi_{0,ij}^{c}}{W_i}. \]
By construction of the sample partition, every $i\in[N]$ appears in $2G-1$ sets among $J_1,\ldots,J_L$. It follows that:
\[    \widehat{\Sigma}_c-\Sigma_c
        =   \frac{4}{N(2G-1)}\sum_{\ell=1}^{L}\sum_{i\in J_{\ell}}\big(\widehat{\varphi}_{\ell,i}^{c}\widehat{\varphi}_{\ell,i}^{c\top}-\varphi_{0,i}^{c}\varphi_{0,i}^{c\top}\big) +\frac{4}{N}\sum_{i=1}^{N}\big\{\varphi_{0,i}^{c}\varphi_{0,i}^{c\top}-\expect{\varphi_{0,i}^{c}\varphi_{0,i}^{c\top}}\big\}. \]
The second term is $o_P(1)$ by the law of large numbers and square integrability of $\psi_{0,ij}^{c}$.

For the first term, Jensen's inequality gives:
\begin{align*}
    \frac{1}{n_{\ell}}\sum_{i\in J_{\ell}}\norm{\widehat{\varphi}_{\ell,i}^{c}-\varphi_{0,i}^{c}}^2
        \leq    &\frac{2}{n_{\ell}(n_{\ell}-1)}\sum_{i\in J_{\ell}}\sum_{j\in J_{\ell}\backslash\{i\}}\norm{\psi^{c}(W_i,W_j,\widehat{\theta}_\textnormal{DDML}^{c},\widehat{\eta}_{\ell},\check{\vecdel}_{\ell}^{c})-\psi_{0,ij}^{c}}^2&\\
                &+\frac{2}{n_{\ell}}\sum_{i\in J_{\ell}}\Norm{\frac{1}{n_{\ell}-1}\sum_{j\in J_{\ell}\backslash\{i\}}\big\{\psi_{0,ij}^{c}-\varphi_{0,i}^{c}\big\}}^2.&
\end{align*}
For the first term on the right-hand side, adding and subtracting $\widehat{\psi}_{\ell,ij}^{c}$ gives:
\begin{align*}
    &\frac{1}{n_{\ell}(n_{\ell}-1)}\sum_{i\in J_{\ell}}\sum_{j\in J_{\ell}\backslash\{i\}}\norm{\psi^{c}(W_i, W_j, \widehat{\theta}_\textnormal{DDML}^{c}, \widehat{\eta}_{\ell}, \check{\vecdel}_{\ell}^{c})-\psi_{0,ij}^{c}}^2&\\
    &\hspace{+1cm}\leq\frac{2}{n_{\ell}(n_{\ell}-1)}\sum_{i\in J_{\ell}}\sum_{j\in J_{\ell}\backslash\{i\}}\norm{\widehat{\psi}_{\ell,ij}^{c}-\psi_{0,ij}^{c}}^2&\\
    &\hspace{+1cm}+\frac{2}{n_{\ell}(n_{\ell}-1)}\sum_{i\in J_{\ell}}\sum_{j\in J_{\ell}\backslash\{i\}}\norm{r^{c}(W_i,W_j,\widehat{\theta}_\textnormal{DDML}^{c},\widehat{\eta}_{\ell})-r^{c}(W_i,W_j,\theta_0,\widehat{\eta}_{\ell})}^2.&
\end{align*}
The first term is $o_P(1)$ by the conditional Markov inequality and the conditional $L^2$ bound established in Step 1 of the proof of Theorem \ref{sec04 - theorem01 - dml}. The second term is $o_P(1)$ by consistency of $\widehat{\theta}_\textnormal{DDML}^{c}$, the entropy condition, and the derivative bounds.

For the remaining term, conditional on $W_i$, the random variables $\psi_{0,ij}^{c}-\varphi_{0,i}^{c}$ are independent and have mean zero. Hence,
\[ \Expect{\frac{1}{n_{\ell}}\sum_{i\in J_{\ell}}\Norm{\frac{1}{n_{\ell}-1}\sum_{j\in J_{\ell}\backslash\{i\}}\big\{\psi_{0,ij}^{c}-\varphi_{0,i}^{c}\big\}}^2}
    =\frac{1}{n_{\ell}-1}\Expect{\norm{\psi_{0,ij}^{c}-\varphi_{0,i}^{c}}^2}=O(n_{\ell}^{-1}). \]
Markov's inequality therefore implies:
\[ \frac{1}{n_{\ell}}\sum_{i\in J_{\ell}}\norm{\widehat{\varphi}_{\ell,i}^{c}-\varphi_{0,i}^{c}}^2=o_P(1). \]
Using $\norm{\pmb{u}\pmb{u}^\top-\pmb{v}\pmb{v}^\top}\leq\norm{\pmb{u}-\pmb{v}}\big(\norm{\pmb{u}}+\norm{\pmb{v}}\big)$ and the Cauchy--Schwarz inequality gives:
\begin{align*}
    \Norm{\frac{1}{n_{\ell}}\sum_{i\in J_{\ell}}\big(\widehat{\varphi}_{\ell,i}^{c}\widehat{\varphi}_{\ell,i}^{c\top}-\varphi_{0,i}^{c}\varphi_{0,i}^{c\top}\big)}
        \leq    &\left(\frac{1}{n_{\ell}}\sum_{i\in J_{\ell}}\norm{\widehat{\varphi}_{\ell,i}^{c}-\varphi_{0,i}^{c}}^2\right)^{1/2}&\\
                &\quad\times\left\{\left(\frac{1}{n_{\ell}}\sum_{i\in J_{\ell}}\norm{\widehat{\varphi}_{\ell,i}^{c}}^2\right)^{1/2}+\left(\frac{1}{n_{\ell}}\sum_{i\in J_{\ell}}\norm{\varphi_{0,i}^{c}}^2\right)^{1/2}\right\}&\\
        =       &o_P(1),&
\end{align*}
where the expression in braces is $O_P(1)$ by the preceding bound and the law of large numbers. Since $L$ is fixed and $n_{\ell}=O(N)$, it follows that $\widehat{\Sigma}_c=\Sigma_c+o_P(1)$.

Finally, Lemma \ref{appendix - lemma A01} and the continuous mapping theorem establish consistency of the variance estimator. \qed

%%%%%%%%%%%%%%%%%%%%%%%%%%%%%%%%%%%%%%%%%%%%%%%%%%%%%%%%%%%%%%%%%%%%%%%%%%%%%%%%%%%%%%%%%%%%%%%%%%%%%%%%%%%%%%%%%%%%%%%%%%%%%%%%%%%%%%%%%%%%%%%%%%%%%%%%%%%%%%%%%%%%%%%%%%%%%%%%%%%%%%%%%%%%%%%%%%%%%%%%%%%%%%%%%%%%%%%%%%%%%%%%%%%%%%%%%%%%%%%%%%%%%%%%%%%%%%%%%%%%%%%%%%%%%%%%%%%%%%%%%%%%%%%%%%%%%%%%%%%%%%%%%%%%%%%%%%%%%%%%%%%%%%%%%%%%%%%%%%%%%%%%%%%%%%%%%%%%%%%%%%%%%%%%%%%%%%%%%%%%%%%%%%%%%%%%%%%%%%%%%%%%%%%%%%%%%%%%%%%%%%%%%%%%%%%%%%%%%%%%%%%%%%%%%%%%%%%%%%%%%%%%%%%%%%%%%%%%%%%%%%%%%%%%%%%%%%%%%%%%%%%%%%%%%%%%%%%%%%%%%%%%%%%%%%%%%%%%%%%%%%%%%%%%%%%%%%%%%%%%%%%%%%%%%%%%%%%%
\subsection{Proof of Corollary \ref{sec04 - corollary01}}\label{sec04 - corollary01 - proof}
For each $t\in[T]$, the asymptotic linear representation and Hoeffding decomposition established in the proof of Theorem \ref{sec04 - theorem01 - dml} give:
\[ \sqrt{N}(\widehat{\theta}_{N,t}^{c}-\theta_0)=-(\Gamma_{c,t}^\top\Gamma_{c,t})^{-1}\Gamma_{c,t}^\top\frac{2}{\sqrt{N}}\sum_{i=1}^{N}\condexpect{\psi_t^{c}(W_i,W_j,\theta_0,\eta_0,\vecdel_{0,t}^{c})}{W_i}+o_P(1). \]
The summands have mean zero and finite second moments. Since $T$ is fixed, stacking these representations and applying the Cramer-Wold device yields $\sqrt{N}\big(\widehat{\pmb{\theta}}_N^{c}-\pmb{A}\theta_0\big)\overset{d}{\longrightarrow}\mathcal{N}(0,\pmb{\Omega}_c)$. Since $\pmb{A} = \pmb{1}_T \otimes \mathbb{I}_{d_\theta}$ has full column rank, $\pmb{\Omega}_c$ is positive definite, and $\widehat{\pmb{\Omega}}_c\overset{p}{\longrightarrow}\pmb{\Omega}_c$, the unconstrained minimum-distance solution satisfies
\[  (\pmb{A}^\top\widehat{\pmb{\Omega}}_c^{-1}\pmb{A})^{-1} \pmb{A}^\top\widehat{\pmb{\Omega}}_c^{-1} \widehat{\pmb{\theta}}_N^{c} = \theta_0 + O_{P}(N^{-1/2}). \]
Because $\theta_0\in\textnormal{int}(\Theta)$, this solution belongs to $\Theta$ with probability approaching one. Consequently,
\[     \sqrt{N}(\widehat{\theta}_{\mathrm{MD}}^{c}-\theta_0)
        =   (\pmb{A}^\top\widehat{\pmb{\Omega}}_c^{-1}\pmb{A})^{-1}\pmb{A}^\top\widehat{\pmb{\Omega}}_c^{-1}\sqrt{N}\big(\widehat{\pmb{\theta}}_N^{c}-\pmb{A}\theta_0\big)+o_P(1)
        \overset{d}{\longrightarrow}   \mathcal{N}\left(0,\,(\pmb{A}^\top\pmb{\Omega}_c^{-1}\pmb{A})^{-1}\right), \]
by Slutsky's lemma. Consistency of the asymptotic variance estimator follows from the continuous mapping theorem. \qed

%%%%%%%%%%%%%%%%%%%%%%%%%%%%%%%%%%%%%%%%%%%%%%%%%%%%%%%%%%%%%%%%%%%%%%%%%%%%%%%%%%%%%%%%%%%%%%%%%%%%%%%%%%%%%%%%%%%%%%%%%%%%%%%%%%%%%%%%%%%%%%%%%%%%%%%%%%%%%%%%%%%%%%%%%%%%%%%%%%%%%%%%%%%%%%%%%%%%%%%%%%%%%%%%%%%%%%%%%%%%%%%%%%%%%%%%%%%%%%%%%%%%%%%%%%%%%%%%%%%%%%%%%%%%%%%%%%%%%%%%%%%%%%%%%%%%%%%%%%%%%%%%%%%%%%%%%%%%%%%%%%%%%%%%%%%%%%%%%%%%%%%%%%%%%%%%%%%%%%%%%%%%%%%%%%%%%%%%%%%%%%%%%%%%%%%%%%%%%%%%%%%%%%%%%%%%%%%%%%%%%%%%%%%%%%%%%%%%%%%%%%%%%%%%%%%%%%%%%%%%%%%%%%%%%%%%%%%%%%%%%%%%%%%%%%%%%%%%%%%%%%%%%%%%%%%%%%%%%%%%%%%%%%%%%%%%%%%%%%%%%%%%%%%%%%%%%%%%%%%%%%%%%%%%%%%%%%%%
\subsection{Proof of Proposition \ref{sec05 - proposition03 - test}}\label{sec05 - proposition03 - proof}

We first show that the $Z^{S}$-CMR has no solution under the alternative hypothesis {in} equation~\eqref{sec05 - eq01 - hypothesis - alt}. Suppose that there exists $\theta^*\in\Theta$ such that $\condexpect{\rho(W,\theta^*,\eta_0)}{Z^{S}}=0$ almost surely. Since $\sigma(Z^{1}(\eta_0))\subseteq\sigma(Z^{S})$, the tower property gives:
\[ \condexpect{\rho(W,\theta^*,\eta_0)}{Z^{1}(\eta_0)}
        = \condexpect{\condexpect{\rho(W,\theta^*,\eta_0)}{Z^{S}}}{Z^{1}(\eta_0)} = 0. \]
Assumption \ref{sec05 - assumption06}(a) therefore implies $\theta^*=\theta_0$, contradicting the alternative hypothesis.

Under the null hypothesis, both models identify $\theta_0$. Applying the asymptotic linear representation in the proof of Theorem \ref{sec04 - theorem01 - dml} to each estimator gives:
\begin{align*}
    \sqrt{N}\big(\widehat{\theta}_{\textnormal{DDML}}^{1}-\widehat{\theta}_{\textnormal{DDML}}^{S}\big)
        =   &-\big(\Gamma_{1}^\top\Gamma_{1}\big)^{-1}\Gamma_{1}^\top\sqrt{N}\binom{N}{2}^{-1}\sum_{\ell=1}^{L}\sum_{(i,j)\in I_{\ell}}\psi^{1}(W_i,W_j,\theta_0,\eta_0,\vecdel_0^{1})&\\
            &+\big(\Gamma_{S}^\top\Gamma_{S}\big)^{-1}\Gamma_{S}^\top\sqrt{N}\binom{N}{2}^{-1}\sum_{\ell=1}^{L}\sum_{(i,j)\in I_{\ell}}\psi^{S}(W_i,W_j,\theta_0,\eta_0,\vecdel_0^{S})+o_P(1).&
\end{align*}
The first-order Hoeffding projections of these two U-statistics are sums over the same observations. The multivariate central limit theorem therefore yields:
\[ \sqrt{N}\big(\widehat{\theta}_{\textnormal{DDML}}^{1}-\widehat{\theta}_{\textnormal{DDML}}^{S}\big)
        \overset{d}{\longrightarrow}\mathcal{N}\big(0\,,\, {\Omega}_{11}+{\Omega}_{SS}-{\Omega}_{1S}-{\Omega}_{S1}\big). \]
By Assumption \ref{sec05 - assumption06}(c) and (d), the covariance matrix is positive definite and its estimator is consistent. Then Slutsky's lemma gives $\mathcal{T}_N\overset{d}{\longrightarrow}\chi^2_{d_\theta}$.

Under the alternative hypothesis, consistency of the estimators and Assumption \ref{sec05 - assumption06}(d) give:
\[ \widehat{\theta}_{\textnormal{DDML}}^{1}-\widehat{\theta}_{\textnormal{DDML}}^{S}\overset{p}{\longrightarrow}\theta_0-\theta^S\ne0. \]
Consequently, Assumption \ref{sec05 - assumption06}(c) and (d) imply:
\[    \frac{\mathcal{T}_N}{N}
        \overset{p}{\longrightarrow} (\theta_0-\theta^S)^\top\big({\Omega}_{11}+{\Omega}_{SS}-{\Omega}_{1S}-{\Omega}_{S1}\big)^{-1}(\theta_0-\theta^S)>0.\]
Thus, $\mathcal{T}_N\overset{p}{\longrightarrow}+\infty$, completing the proof. \qed

%%%%%%%%%%%%%%%%%%%%%%%%%%%%%%%%%%%%%%%%%%%%%%%%%%%%%%%%%%%%%%%%%%%%%%%%%%%%%%%%%%%%%%%%%%%%%%%%%%%%%%%%%%%%%%%%%%%%%%%%%%%%%%%%%%%%%%%%%%%%%%%%%%%%%%%%%%%%%%%%%%%%%%%%%%%%%%%%%%%%%%%%%%%%%%%%%%%%%%%%%%%%%%%%%%%%%%%%%%%%%%%%%%%%%%%%%%%%%%%%%%%%%%%%%%%%%%%%%%%%%%%%%%%%%%%%%%%%%%%%%%%%%%%%%%%%%%%%%%%%%%%%%%%%%%%%%%%%%%%%%%%%%%%%%%%%%%%%%%%%%%%%%%%%%%%%%%%%%%%%%%%%%%%%%%%%%%%%%%%%%%%%%%%%%%%%%%%%%%%%%%%%%%%%%%%%%%%%%%%%%%%%%%%%%%%%%%%%%%%%%%%%%%%%%%%%%%%%%%%%%%%%%%%%%%%%%%%%%%%%%%%%%%%%%%%%%%%%%%%%%%%%%%%%%%%%%%%%%%%%%%%%%%%%%%%%%%%%%%%%%%%%%%%%%%%%%%%%%%%%%%%%%%%%%%%%%%%%
\subsection{Proof of Proposition~\ref{sec05:prop:bandwidth}}
\label{app:proof:bandwidth}
For brevity, write $\widehat{\theta}_{\textnormal{DDML}}^{c}(b)=\widehat{\theta}_{\textnormal{DDML}}^{c}(b)$, and suppress the superscript $c$. If $\mathcal{B}$ contains a single bandwidth, all conclusions follow directly from the assumptions at that bandwidth. Henceforth suppose $|\mathcal{B}|>1$.

\paragraph{Step 1.} We first establish uniform consistency of the variance criterion. For each $b\in\mathcal{B}$, consistency of $\widehat{\Omega}_{N}(b)$ and $\widehat{\pmb{W}}_{N}$ implies their joint convergence in probability to $(\Omega_{c}(b),\pmb{W})$. Joint convergence in probability and the continuous mapping theorem yield $\widehat{v}_{N}(b)\overset{p}{\longrightarrow} v(b)$, since the trace and matrix products are continuous. For every $\varepsilon>0$, the union bound therefore yields
\[  \Prob{ \max_{b\in\mathcal{B}}|\widehat{v}_{N}(b)-v(b)|>\varepsilon }
    \leq    \sum_{b\in\mathcal{B}}  \Prob{ |\widehat{v}_{N}(b)-v(b)|>\varepsilon }\longrightarrow0, \]
because the grid is fixed and finite.

\paragraph{Step 2.} We next show that the selected optimal grid point $\widehat{b}_N$ converges in probability to $b_*$. By hypothesis, $b_*$ is the unique minimizer of $v(b)$ over $b\in\mathcal{B}$. By construction, the grid $\mathcal{B}$ is finite. Let
\[  \Delta=\min_{b\in\mathcal{B}\setminus\{b_*\}}   \{v(b)-v(b_*)\}.    \]
Together, uniqueness and finiteness imply that the constant satisfies $\Delta>0$. Now define the event
\[  E_N=\left\{\max_{b\in\mathcal{B}}|\widehat{v}_{N}(b)-v(b)|<\Delta/3\right\}.    \]
On $E_N$, every estimated criterion is within $\Delta/3$ of its population value.   Fix any $b\in\mathcal{B}\setminus\{b_*\}$. Then
\[  \widehat{v}_{N}(b)>v(b)-\Delta/3,
    \qquad
    \widehat{v}_{N}(b_*)<v(b_*)+\Delta/3.   \]
Subtracting these inequalities gives
\[  \widehat{v}_{N}(b)-\widehat{v}_{N}(b_*) > v(b)-v(b_*)-2\Delta/3.    \]
By the definition of $\Delta$, $v(b)-v(b_*)\geq\Delta$. Consequently,
\[  \widehat{v}_{N}(b)-\widehat{v}_{N}(b_*)>\Delta/3>0. \]
Since this holds for every $b\ne b_*$, the estimated criterion is uniquely  minimized at $b_*$ on $E_N$. Hence
\[  E_N\subseteq\{\widehat{b}_{N}=b_*\},
    \qquad
    \prob{\widehat{b}_{N}=b_*} \geq \prob{E_N}. \]
Step 1 implies $\Pr(E_N)\to1$, because the maximum estimation error converges to zero in probability and $\Delta/3$ is a fixed positive constant. It follows that $\Pr(\widehat{b}_{N}=b_*)\to1$.

\paragraph{Step 3.} Lastly, we show the $\sqrt{N}$-consistency of $\widehat{\theta}_{VS}$, its limiting distribution, and the consistency of the variance estimator. The selected estimator equals $\widehat{\theta}_{\textnormal{DDML}}^{c}(b_*)$ whenever $\widehat{b}_{N}=b_*$. Consequently, for every $\varepsilon>0$,
\[  \Prob{ \sqrt{N}\|\widehat{\theta}_{\textnormal{DDML}}^{c}(\widehat{b}_{N}) -\widehat{\theta}_{\textnormal{DDML}}^{c}(b_*)\|>\varepsilon }
        \leq    \prob{ \widehat{b}_{N}\ne b_*} \longrightarrow 0. \]
It follows that
\[  \sqrt{N}\big(\widehat{\theta}_{\textnormal{VS}}-\theta_0\big) = \sqrt{N}\big(\widehat{\theta}_{\textnormal{DDML}}^{c}(b_*)-\theta_0\big) + o_P(1)
    \overset{d}{\longrightarrow}\mathcal{N}(0,\Omega(b_*)).  \]
Similarly, using any fixed matrix norm,
\[    \Prob{ \norm{\widehat{\Omega}_{N}(\widehat{b}_{N})-\Omega(b_*)}>\varepsilon } \leq  \Prob{\widehat{b}_{N}\ne b_*}
    + \Prob{ \norm{\widehat{\Omega}_{N}(b_*)-\Omega(b_*)}>\varepsilon } \longrightarrow0. \]
We have therefore established consistent selection of $b_*$, asymptotic normality of $\widehat{\theta}_{\textnormal{VS}}$, and consistency of its estimated asymptotic covariance matrix, proving all assertions of the proposition.
\qed

%%%%%%%%%%%%%%%%%%%%%%%%%%%%%%%%%%%%%%%%%%%%%%%%%%%%%%%%%%%%%%%%%%%%%%%%%%%%%%%%%%%%%%%%%%%%%%%%%%%%%%%%%%%%%%%%%%%%%%%%%%%%%%%%%%%%%%%%%%%%%%%%%%%%%%%%%%%%%%%%%%%%%%%%%%%%%%%%%%%%%%%%%%%%%%%%%%%%%%%%%%%%%%%%%%%%%%%%%%%%%%%%%%%%%%%%%%%%%%%%%%%%%%%%%%%%%%%%%%%%%%%%%%%%%%%%%%%%%%%%%%%%%%%%%%%%%%%%%%%%%%%%%%%%%%%%%%%%%%%%%%%%%%%%%%%%%%%%%%%%%%%%%%%%%%%%%%%%%%%%%%%%%%%%%%%%%%%%%%%%%%%%%%%%%%%%%%%%%%%%%%%%%%%%%%%%%%%%%%%%%%%%%%%%%%%%%%%%%%%%%%%%%%%%%%%%%%%%%%%%%%%%%%%%%%%%%%%%%%%%%%%%%%%%%%%%%%%%%%%%%%%%%%%%%%%%%%%%%%%%%%%%%%%%%%%%%%%%%%%%%%%%%%%%%%%%%%%%%%%%%%%%%%%%%%%%%%%%
\subsection{Proof of Lemma \ref{appendix - lemma A01}}\label{appendix - lemma A01 - proof}
Fix a vector $\pmb{a} \in \mathbb{R}^{d_{\theta}}\backslash\{0\}$. To simplify notation, suppress the superscript $c$, and let $\rho_{0,i} \equiv \rho(W_i, \theta_0, \eta_0)$, $d_{0,i} \equiv \nabla_{\theta}\rho(W_i, \theta_0, \eta_0)$, $h_{0,i} \equiv \nabla_{\theta}^{2}\rho(W_i,\theta_0, \eta_0)$, $K_{0,ij} \equiv K\big( Z_{i}^{c}(\eta_0), Z_{j}^{c}(\eta_0) \big)$. 

\paragraph{Proof of (i).}The product rule gives:
\[  \Gamma_{c}  =   \expect{ \big\{h_{0,i}\rho_{0,j} + \rho_{0,i} h_{0,j} + 2 d_{0,i}d_{0,j}^\top \big\} K_{0,ij} }. \]
The law of iterated expectations and Assumption \ref{sec03 - assumption02 - CMR}(a) yield:
\[    \Gamma_{c}
        =    \expect{ \big\{h_{0,i}\condexpect{\rho_{0,j}}{Z_{j}^{c}} + \condexpect{\rho_{0,i}}{Z_{i}^{c}} h_{0,j} + 2 d_{0,i}d_{0,j}^\top \big\} K_{0,ij} }
        =    2\expect{ d_{0,i}d_{0,j}^\top K_{0,ij} }.  \]

Lastly, we show that $\Gamma_{c}$ is invertible. Let $g(z) \equiv \condexpect{ d_{0,i} }{ Z_{i}^{c} = z } \in \mathbb{R}^{d_{\theta}}$. Since $\expect{g(Z)g(Z)^\top}$ is assumed to be invertible by Assumption \ref{sec03 - assumption02 - CMR}(c), we have:
\begin{equation}\label{appendix - lemma A01 - proof - eq01}
    \Expect{ \big( \pmb{a}^\top g(Z) \big)^{2} } = \pmb{a}^\top \Expect{ g(Z)g(Z)^\top }\pmb{a} \geq \lambda_{\min}\big( \expect{ g(Z)g(Z)^\top} \big)\norm{\pmb{a}}^2 > 0,    
\end{equation}
where $\lambda_{\min}(\pmb{A})$ denotes the smallest eigenvalue of a generic square matrix $\pmb{A}$. Then, we have:
\begin{align}
    \pmb{a}^\top \Gamma_{c} \pmb{a}
        =   & 2 \times \pmb{a}^\top\iint g(z)g(z')^\top K(z,z')P_{Z}(dz)P_{Z}(dz') \pmb{a}&\nonumber\\
        =   & 2 \times \iint \big( \pmb{a}^\top g(z) \big) \big( \pmb{a}^\top g(z') \big) K(z,z')P_{Z}(dz)P_{Z}(dz').&\label{appendix - lemma A01 - proof - eq02}
\end{align}
Because of Assumption \ref{sec03 - assumption03 - kernel} and equation~\eqref{appendix - lemma A01 - proof - eq01}, equation~\eqref{appendix - lemma A01 - proof - eq02} is strictly positive. Therefore, $\Gamma_{c}$ is invertible.\qed

\paragraph{Proof of (ii).} Let $\psi_{0,ij}$ denote the corrected score at the truth and $\varphi_{0,1i} \equiv \condexpect{\psi_{0,ij}}{W_i}$ its first-order Hoeffding projection. By Assumption~\ref{sec03 - assumption02 - CMR}(a) and independence $W_i \pperp W_j$, $\condexpect{\rho_{0,j}K_{0,ij}}{W_i}=0$. By Assumption~\ref{sec03 - assumption01 - first stage}(b), $\condexpect{T_j-\eta_0(S_j)}{S_j}=0$. Hence,
\[  \varphi_{0,1i}  = \rho_{0,i}\condexpect{d_{0,j}K_{0,ij}}{W_i}   + \vecdel_{0,1}(S_i,\theta_0)\{T_i-\eta_0(S_i)\}.   \]
Since $\psi_{0,ij}$ is $L_2(F_0\otimes F_0)$-integrable by the Assumptions~\ref{sec03 - assumption02 - CMR}(b), \ref{sec03 - assumption02 - CMR}(e), \ref{sec03 - assumption03 - kernel}(b), and \ref{sec04 - assumption04 - linear representation}, conditional Jensen's inequality gives
\[ \norm{\varphi_{0,1i}}_{F_0\otimes F_0, 2} \leq \norm{\psi_{0,ij}}_{F_0\otimes F_0, 2} < \infty.\]
Thus $\Sigma_c=4\expect{\varphi_{0,1i}\varphi_{0,1i}^{\top}}$ is finite.

We next establish positive definiteness. By independence of $W_i$ and $W_j$, $\condexpect{d_{0,j}K_{0,ij}}{W_i}$ depends on $W_i$ only through $Z_i^{c}(\eta_0)$. Conditional on $Z_i^{c}(\eta_0)$, $S_i$, and $T_i$, both $\condexpect{d_{0,j}K_{0,ij}}{W_i}$ and the bias-correction term are therefore fixed. By the law of total variance,
\[  \begin{aligned}
    \frac{1}{4}\pmb{a}^\top\Sigma_c \pmb{a}
        =   &\Expect{\big\{\pmb{a}^\top\condexpect{d_{0,j}K_{0,ij}}{W_i}\big\}^{2} \textnormal{\textbf{Var}}\big(\rho_{0,i}\mid Z_i^{c}(\eta_0),S_i,T_i\big)}\\
            &   + \textnormal{\textbf{Var}}\left(  \pmb{a}^\top\condexpect{d_{0,j}K_{0,ij}}{W_i}\condexpect{\rho_{0,i}}{Z_i^{c}(\eta_0),S_i,T_i}
                + \pmb{a}^\top\vecdel_{0,1}(S_i,\theta_0)\{T_i-\eta_0(S_i)\} \right).
\end{aligned} \]
Then, it suffices to show that the first term is strictly positive. By the second condition of Assumption~\ref{sec03 - assumption02 - CMR}(c), $\expect{\big\{ \pmb{a}^\top\condexpect{d_{0,i}}{Z_i^{c}(\eta_0)}\big\}^{2}}>0$. Applying the ISPD condition in Assumption~\ref{sec03 - assumption03 - kernel}(a) to the function $z\mapsto \pmb{a}^\top\condexpect{d_{0,i}}{Z_i^{c}(\eta_0)=z}$ gives
\[  \Expect{\big\{\pmb{a}^\top \condexpect{d_{0,i}}{Z_i^{c}(\eta_0)}\big\} \big\{\pmb{a}^\top \condexpect{d_{0,j}K_{0,ij}}{W_i}\big\}}
        =   \Expect{\big\{ \pmb{a}^\top\condexpect{d_{0,i}}{Z_i^{c}(\eta_0)}\big\} K_{0,ij} \big\{ \pmb{a}^\top\condexpect{d_{0,j}}{Z_j^{c}(\eta_0)}\big\}} > 0. \]
It follows that $\prob{\pmb{a}^\top\condexpect{d_{0,j}K_{0,ij}}{W_i}\ne 0}>0$. Since $\textnormal{\textbf{Var}}\big(\rho_{0,i}\mid Z_i^{c}(\eta_0),S_i,T_i\big)>0$ almost surely by the first condition of Assumption~\ref{sec03 - assumption02 - CMR}(c), the first term is strictly positive. This complete the proof.\qed 
%%%%%%%%%%%%%%%%%%%%%%%%%%%%%%%%%%%%%%%%%%%%%%%%%%%%%%%%%%%%%%%%%%%%%%%%%%%%%%%%%%%%%%%%%%%%%%%%%%%%%%%%%%%%%%%%%%%%%%%%%%%%%%%%%%%%%%%%%%%%%%%%%%%%%%%%%%%%%%%%%%%%%%%%%%%%%%%%%%%%%%%%%%%%%%%%%%%%%%%%%%%%%%%%%%%%%%%%%%%%%%%%%%%%%%%%%%%%%%%%%%%%%%%%%%%%%%%%%%%%%%%%%%%%%%%%%%%%%%%%%%%%%%%%%%%%%%%%%%%%%%%%%%%%%%%%%%%%%%%%%%%%%%%%%%%%%%%%%%%%%%%%%%%%%%%%%%%%%%%%%%%%%%%%%%%%%%%%%%%%%%%%%%%%%%%%%%%%%%%%%%%%%%%%%%%%%%%%%%%%%%%%%%%%%%%%%%%%%%%%%%%%%%%%%%%%%%%%%%%%%%%%%%%%%%%%%%%%%%%%%%%%%%%%%%%%%%%%%%%%%%%%%%%%%%%%%%%%%%%%%%%%%%%%%%%%%%%%%%%%%%%%%%%%%%%%%%%%%%%%%%%%%%%%%%%%%%%%
\subsection{Proof of Lemma \ref{appendix - lemma A02}}\label{appendix - lemma A02 - proof}

Fix $c\in\{0,1\}$. To simplify notation, let $\rho_i(\theta, \eta) \equiv \rho(W_i, \theta, \eta)$, $d_i(\theta, \eta) \equiv \nabla_{\theta} \rho(W_i, \theta, \eta)$, $K_{ij}(\eta) \equiv K\big( Z_{i}(\eta), Z_{j}(\eta) \big)$, and:
\[ r_{ij}(\theta, \eta) \equiv \big\{ \rho_{i}(\theta,\eta) \cdot d_{j}(\theta,\eta) +  d_{i}(\theta, \eta)\cdot \rho_{j}(\theta, \eta) \big\} \cdot K_{ij}(\eta). \]

\paragraph{Proof of (i) - (iii).}Fix $\theta\in\textnormal{int}(\Theta)$ and $\eta, \eta'\in \mathcal{H}_N$. For $\tau\in[0,1]$, let $\eta_{\tau} \equiv \eta' + \tau\cdot h$ where $h = \eta - \eta'$. By Assumption \ref{sec03 - assumption02 - CMR}(d), Taylor's theorem gives:
\begin{align*}
    \norm{ \rho_i(\theta, \eta) - \rho_i(\theta, \eta') }
        =       &\Norm{ \int_{0}^{1} D_{\eta}\rho_i(\theta, \eta_{\tau})[h] d\tau }
        \leq     \int_{0}^{1}\norm{ D_{\eta}\rho_i(\theta, \eta_{\tau})[h] }d\tau 
        \leq     \widebar{J}(W_i)\cdot \norm{\eta - \eta'}_{\infty},&\\
    \norm{ d_i(\theta, \eta) - d_i(\theta, \eta') }
        =       &\Norm{ \int_{0}^{1} D_{\eta} d_i(\theta, \eta_{\tau})[h] d\tau }
        \leq     \int_{0}^{1}\norm{ D_{\eta}d_i(\theta, \eta_{\tau})[h] } d\tau 
        \leq     \widebar{J}(W_i)\cdot \norm{\eta - \eta'}_{\infty}.&
\end{align*}
Analogously, by Assumption \ref{sec03 - assumption03 - kernel}, Taylor's theorem gives:
\begin{align*}
    \norm{ K_{ij}(\eta) - K_{ij}(\eta')}
        =       &\ind{}{c=1}\Norm{ \int_{0}^{1} \big\{ D_{\eta}K\big( Z_{i}^{c}(\eta_\tau), Z_{j}^{c}(\eta_0) \big)[h] + D_{\eta}K\big( Z_{i}^{c}(\eta_0), Z_{j}^{c}(\eta_\tau) \big)[h] \big\} d\tau }&\\
        \leq    &\ind{}{c=1}\left\{ \int_{0}^{1} \norm{ D_{\eta}K\big( Z_{i}^{c}(\eta_\tau), Z_{j}^{c}(\eta_0) \big)[h] }d\tau  + \int_{0}^{1} \norm{ D_{\eta}K\big( Z_{i}^{c}(\eta_0), Z_{j}^{c}(\eta_\tau) \big)[h] }d\tau\right\}&\\
        \leq    &2\widebar{K}_{1}\cdot \ind{}{c=1} \cdot  \norm{\eta - \eta'}_{\infty}.\qed&
\end{align*}

\paragraph{Proof of (iv).} Fix $\theta\in \widebar{\Theta}_{o}$ and $\eta \in \mathcal{H}_N$. By the product rule, we have:
\begin{align}
    D_{\eta}r_{ij}(\theta, \eta)[h]
        =   &  \Big\{ D_{\eta}\rho_{i}(\theta,\eta)[h]\cdot d_{j}(\theta,\eta) + \rho_{i}(\theta,\eta)\cdot D_{\eta}d_{j}(\theta,\eta) [h] \Big\}    \times K_{ij}(\eta)&\nonumber\\ 
            &+ \Big\{ d_{i}(\theta,\eta)\cdot D_{\eta}\rho_{j}(\theta,\eta) [h] + D_{\eta}d_{i}(\theta,\eta)[h]\cdot \rho_{j}(\theta,\eta) \Big\}     \times K_{ij}(\eta)&\nonumber\\
            &+ \big\{ \rho_{i}(\theta,\eta)\cdot d_{j}(\theta,\eta)  + d_{i}(\theta,\eta)\cdot \rho_{j}(\theta,\eta) \big\} \times D_{\eta}K_{ij}(\eta)[h].&\label{appendix - lemma A02 - proof - eq02}
\end{align}
Define a function $(w, w') \mapsto L_{1}^{c}(w,w')$ as:
\begin{equation}\label{appendix - lemma A02 - proof - eq03}
    L_{1}^{c}(w,w')
        \equiv  \widebar{K}_0\Big(  \widebar{J}(w)\widebar{\rho}^{1}(w') + \widebar{\rho}^{0}(w)\widebar{J}(w') \Big) + 2\widebar{K}_1 \widebar{\rho}^{0}(w)\widebar{\rho}^{1}(w') \ind{}{c=1}.
\end{equation}
Then, using Lemma \ref{appendix - lemma A02}(i)-(iii), Assumption \ref{sec03 - assumption02 - CMR}(e), and equations~\eqref{appendix - lemma A02 - proof - eq02}, \eqref{appendix - lemma A02 - proof - eq03} yields:
\begin{align}
    \Norm{ D_{\eta}r_{ij}(\theta, \eta)[h]}
        \leq&  \Big\{ \widebar{J}(W_i)\widebar{\rho}^{1}(W_j) + \widebar{\rho}^{0}(W_i)\widebar{J}(W_j) \Big\} \cdot \widebar{K}_0 \cdot \norm{h}_{\infty} &\nonumber\\
            & +\Big\{ \widebar{\rho}^{1}(W_i)\widebar{J}(W_j) + \widebar{J}(W_i)\widebar{\rho}^{0}(W_j) \Big\} \cdot \widebar{K}_0 \cdot \norm{h}_{\infty}&\nonumber\\
            & + \ind{}{c=1} \cdot \Big\{ \widebar{\rho}^{0}(W_i)\widebar{\rho}^{1}(W_j) + \widebar{\rho}^{1}(W_i)\widebar{\rho}^{0}(W_j) \Big\} \cdot 2\widebar{K}_{1} \cdot \norm{h}_{\infty}&\nonumber\\
        =   &   \Big\{ L_{1}^{c}(W_i, W_j) + L_{1}^{c}(W_j, W_i) \Big\}\norm{h}_{\infty}.&\label{appendix - lemma A02 - proof - eq04}
\end{align}
By Assumptions \ref{sec03 - assumption02 - CMR} and \ref{sec03 - assumption03 - kernel}, the function $L_{1}^{c}(w,w')$ does not rely on $\theta$ and $\eta\in\mathcal{H}_N$. In addition, $\norm{ L_{1}^{c}(W_i, W_j) }_{F_0\otimes F_0, 2}$ is bounded by:
\begin{align*}
    \norm{ L_{1}^{c}(W_i, W_j) }_{F_0\otimes F_0, 2}
        \leq    &\widebar{K}_0\norm{\widebar{J}(W_i)\widebar{\rho}^{1}(W_j) + \widebar{\rho}^{0}(W_j)\widebar{J}(W_i)}_{F_0\otimes F_0, 2} + 2\widebar{K}_1 \norm{\widebar{\rho}^{0}(W_i)\widebar{\rho}^{1}(W_j)}_{F_0\otimes F_0, 2} \ind{}{c=1}&\\
        \leq    &\widebar{K}_0\norm{\widebar{J}}_{F_0,2}\big( \norm{\widebar{\rho}^{1}}_{F_0,2} + \norm{\widebar{\rho}^{0}}_{F_0, 2} \big) + 2\widebar{K}_1 \norm{\widebar{\rho}^{0}}_{F_0,2}\norm{\widebar{\rho}^{1}}_{F_0, 2} \ind{}{c=1},&
\end{align*}
where the second inequality comes from $W_{i}\pperp W_{j}$. By Assumption \ref{sec03 - assumption02 - CMR}(d), (e) and Assumption \ref{sec03 - assumption03 - kernel}, therefore, $\norm{ L_{1}^{c}(W_i, W_j) }_{F_0\otimes F_0, 2}$ is finite.\qed

\paragraph{Proof of (v).} Fix $\theta\in \widebar{\Theta}_{o}$ and $\eta \in \mathcal{H}_N$. By the product rule, we have:
\begin{align}
    D_{\eta}^{2}r_{ij}(\theta, \eta)[h_1, h_2]
        =   &  \Big\{ D_{\eta}^{2}\rho_{i}(\theta,\eta)[h_1, h_2]\cdot d_{j}(\theta,\eta) + \rho_{i}(\theta,\eta)\cdot D_{\eta}^{2}d_{j}(\theta,\eta) [h_1, h_2] \Big\}    \times K_{ij}(\eta)&\nonumber\\ 
            &+ \Big\{ d_{i}(\theta,\eta)\cdot D_{\eta}^{2}\rho_{j}(\theta,\eta) [h_1, h_2] + D_{\eta}^{2}d_{i}(\theta,\eta)[h_1, h_2]\cdot \rho_{j}(\theta,\eta) \Big\}     \times K_{ij}(\eta)&\nonumber\\
            &+ \big\{ \rho_{i}(\theta,\eta)\cdot d_{j}(\theta,\eta)  + d_{i}(\theta,\eta)\cdot \rho_{j}(\theta,\eta) \big\} \times D_{\eta}^{2}K_{ij}(\eta)[h_1, h_2].&\label{appendix - lemma A03 - proof - eq02}
\end{align}
We note that Assumption \ref{sec03 - assumption03 - kernel}(b) gives:
\begin{equation}\label{appendix - lemma A03 - proof - eq03}
\norm{D_\eta^2K_{ij}(\eta)[h_1,h_2]}
        \leq    4\widebar{K}_2\ind{}{c=1}\norm{h_1}_\infty\cdot \norm{h_2}_\infty. 
\end{equation}
Define a function $(w,w')\mapsto L_{2}^{c}(w,w')$ as:
\begin{equation}\label{appendix - lemma A03 - proof - eq04}
    L_{2}^{c}(w,w')
        \equiv  \Big( \widebar{H}(w)\widebar{\rho}^{1}(w') + \widebar{\rho}^{0}(w)\widebar{H}(w') \Big) \widebar{K}_0 + 4\widebar{\rho}^{0}(w)\widebar{\rho}^{1}(w')\widebar{K}_2 \ind{}{c=1}.
\end{equation}
Then, using Lemma \ref{appendix - lemma A02}(i)-(iii), Assumption \ref{sec03 - assumption02 - CMR}(e) and equations~\eqref{appendix - lemma A03 - proof - eq02}, \eqref{appendix - lemma A03 - proof - eq03} and \eqref{appendix - lemma A03 - proof - eq04} give:
\begin{align*}
    \Norm{ D_{\eta}^{2}r_{ij}(\theta, \eta_\tau)[h_1, h_2] }
        \leq&  \Big\{ \widebar{H}(W_i)\widebar{\rho}^{1}(W_j) + \widebar{\rho}^0(W_i)\widebar{H}(W_j) \Big\} \widebar{K}_{0}\times \norm{h_1}_\infty\cdot \norm{h_2}_\infty&\\ 
            &+ \Big\{ \widebar{\rho}^{1}(W_i)\widebar{H}(W_j) + \widebar{H}(W_i)\widebar{\rho}^0(W_j) \Big\} \widebar{K}_{0}\times \norm{h_1}_\infty\cdot \norm{h_2}_\infty&\\
            &+ 4\cdot \ind{}{c=1} \big\{ \widebar{\rho}^0(W_i)\widebar{\rho}^1(W_j) + \widebar{\rho}^1(W_i)\widebar{\rho}^0(W_j) \big\} \cdot \widebar{K}_2 \times \norm{h_1}_\infty \cdot \norm{h_2}_\infty&\\
        =   & \Big\{ L_{2}^{c}(W_i, W_j) + L_{2}^{c}(W_j, W_i) \Big\} \times \norm{h_1}_\infty \cdot \norm{h_2}_\infty.&
\end{align*}
By Assumptions \ref{sec03 - assumption02 - CMR} and \ref{sec03 - assumption03 - kernel}, $(w,w')\mapsto L_{2}^{c}(w,w')$ does not rely on $\theta$ and $\eta\in\mathcal{H}_N$. In addition, $\norm{ L_{2}^{c}(W_i, W_j) }_{F_0\otimes F_0, 2}$ is bounded by:
\begin{align*}
    \norm{L_{2}^{c}(W_i, W_j)}_{F_0\otimes F_0, 2}
        \leq    &  \widebar{K}_0 \norm{ \widebar{H}(W_i)\widebar{\rho}^{1}(W_j) + \widebar{\rho}^{0}(W_i)\widebar{H}(W_j)}_{F_0\otimes F_0, 2}  + 4\widebar{K}_2 \norm{\widebar{\rho}^{0}(W_i)\widebar{\rho}^{1}(W_j)}_{F_0\otimes F_0, 2}\ind{}{c=1},&\\
        \leq    &  \widebar{K}_0 \norm{ \widebar{H}}_{F_0,2} \Big( \norm{\widebar{\rho}^{0}}_{F_0, 2} + \norm{\widebar{\rho}^{1}}_{F_0,2}\Big)  + 4\widebar{K}_2 \norm{\widebar{\rho}^{0}}_{F_0,2}\norm{\widebar{\rho}^{1}}_{F_0, 2}\ind{}{c=1},&
\end{align*}
where the second inequality comes from $W_{i}\pperp W_{j}$. By Assumption \ref{sec03 - assumption02 - CMR}(d), (e) and Assumption \ref{sec03 - assumption03 - kernel}, $\norm{ L_{2}^{c}(W_i, W_j) }_{F_0\otimes F_0, 2}$ is finite.\qed
%%%%%%%%%%%%%%%%%%%%%%%%%%%%%%%%%%%%%%%%%%%%%%%%%%%%%%%%%%%%%%%%%%%%%%%%%%%%%%%%%%%%%%%%%%%%%%%%%%%%%%%%%%%%%%%%%%%%%%%%%%%%%%%%%%%%%%%%%%%%%%%%%%%%%%%%%%%%%%%%%%%%%%%%%%%%%%%%%%%%%%%%%%%%%%%%%%%%%%%%%%%%%%%%%%%%%%%%%%%%%%%%%%%%%%%%%%%%%%%%%%%%%%%%%%%%%%%%%%%%%%%%%%%%%%%%%%%%%%%%%%%%%%%%%%%%%%%%%%%%%%%%%%%%%%%%%%%%%%%%%%%%%%%%%%%%%%%%%%%%%%%%%%%%%%%%%%%%%%%%%%%%%%%%%%%%%%%%%%%%%%%%%%%%%%%%%%%%%%%%%%%%%%%%%%%%%%%%%%%%%%%%%%%%%%%%%%%%%%%%%%%%%%%%%%%%%%%%%%%%%%%%%%%%%%%%%%%%%%%%%%%%%%%%%%%%%%%%%%%%%%%%%%%%%%%%%%%%%%%%%%%%%%%%%%%%%%%%%%%%%%%%%%%%%%%%%%%%%%%%%%%%%%%%%%%%%%%%
\subsection{Proof of Lemma \ref{appendix - lemma A03}}\label{appendix - lemma A03 - proof}
Fix $c\in\{0,1\}$, $\theta\in \widebar{\Theta}_{o}$. To simplify notation, let $\rho_i(\theta, \eta) \equiv \rho(W_i, \theta, \eta)$, $d_i(\theta, \eta) \equiv \nabla_{\theta} \rho(W_i, \theta, \eta)$, $K_{ij}(\eta) \equiv K\big( Z_{i}(\theta, \eta), Z_{j}(\theta, \eta) \big)$, and:
\[ r_{ij}(\theta, \eta) \equiv \big\{ \rho_{i}(\theta,\eta) \cdot d_{j}(\theta,\eta) +  d_{i}(\theta, \eta)\cdot \rho_{j}(\theta, \eta) \big\} \cdot K_{ij}(\eta). \]

\paragraph{Proof of (i).} Fix $\eta, \eta'\in \mathcal{H}_N$. For $\tau\in[0,1]$, let $\eta_{\tau} \equiv \eta_0 + \tau\cdot h$ where $h = \eta - \eta_0$ is an admissible direction with $\eta\in\mathcal{H}_N$. Convexity gives $\eta_\tau \in \mathcal{H}_N$. By Assumption \ref{sec03 - assumption02 - CMR}(d), Taylor's theorem gives:
\begin{align}
    \norm{ r_{ij}(\theta, \eta) - r_{ij}(\theta, \eta_0) }_{F_0\otimes F_0, 2}
        =       &\Norm{ \int_{0}^{1}  D_{\eta}r_{ij}(\theta, \eta_{\tau})[h] d\tau }_{F_0\otimes F_0, 2}&\nonumber\\
        \leq    &\int_{0}^{1}  \Norm{ D_{\eta}r_{ij}(\theta, \eta_{\tau})[h] }_{F_0\otimes F_0, 2} d\tau.&\label{appendix - lemma A02 - proof - eq01}
\end{align}
Let $\widebar{L}_{1}^{c} \equiv 2 \norm{ L_{1}^{c}(W_i, W_j) }_{F_0\otimes F_0, 2}$, { which does not depend on} $\theta$, $\eta\in\mathcal{H}_N$, and $N$ by  Lemma~\ref{appendix - lemma A02}(iv). Then, combining the result of Lemma~\ref{appendix - lemma A02}(iv) with \eqref{appendix - lemma A02 - proof - eq01} gives:
\begin{align*}
    \norm{ r_{ij}(\theta, \eta) - r_{ij}(\theta, \eta_0) }_{F_0\otimes F_0, 2}
        \leq    &\int_{0}^{1}  \Norm{ D_{\eta}r_{ij}(\theta, \eta_{\tau})[h] }_{F_0\otimes F_0, 2} d\tau &\\
        \leq    &\int_{0}^{1} \big\{ \norm{L_{1}^{c}(W_i, W_j)}_{F_0\otimes F_0, 2} + \norm{L_{1}^{c}(W_j, W_i)}_{F_0\otimes F_0, 2} \big\} \cdot \norm{h}_{\infty} d\tau&\\
        \leq    &  \widebar{L}_{1}^{c} \cdot \norm{h}_{\infty}.&
\end{align*}
Since $\norm{ L_{1}^{c}(W_i, W_j) }_{F_0\otimes F_0, 2}$ is finite by Lemma~\ref{appendix - lemma A02}(iv), $\widebar{L}_{1}^{c} < \infty$.\qed

\paragraph{Proof of (ii).} Fix $\eta_1, \eta_2\in \mathcal{H}_N$, and let $h_1 = \eta_1 - \eta_0$, $h_2 = \eta_2 - \eta_0$ be an admissible direction. Let $\eta_\tau=\eta_2+\tau(\eta_1-\eta_2)$ for $\tau\in[0,1]$. Then, convexity gives $\eta_\tau\in\mathcal{H}_N$. By Assumption \ref{sec03 - assumption02 - CMR}(d), Taylor's theorem gives:
\begin{align}
    \Norm{ \Expect{ D_\eta r_{ij}(\theta,\eta_1)[h] - D_\eta r_{ij}(\theta,\eta_2)[h] }}
        =       &\Norm{ \Expect{ \int_{0}^{1} D_\eta^2 r_{ij}(\theta,\eta_\tau)[\eta_1-\eta_2,h]d\tau  } }&\nonumber\\
        \leq    &\int_{0}^{1}  \Norm{D_\eta^2 r_{ij}(\theta,\eta_\tau)[\eta_1-\eta_2,h]}_{F_0\otimes F_0, 2} d\tau.&\label{appendix - lemma A03 - proof - eq01}
\end{align}
Let $\widebar{L}_{2}^{c} \equiv 2 \norm{ L_{2}^{c}(W_i, W_j) }_{F_0\otimes F_0, 2}$, { which does not depend on} $\theta$, $\eta\in\mathcal{H}_N$, and $N$ by Lemma~\ref{appendix - lemma A02}(v). Combining the result of Lemma~\ref{appendix - lemma A02}(v) with equation~\eqref{appendix - lemma A03 - proof - eq01} yields:
\begin{align*}
    \Norm{ \Expect{ D_\eta r_{ij}(\theta,\eta_1)[h] - D_\eta r_{ij}(\theta,\eta_2)[h] }}
        \leq&\int_{0}^{1}  \Norm{D_\eta^2 r_{ij}(\theta,\eta_\tau)[\eta_1-\eta_2,h]}_{F_0\otimes F_0, 2} d\tau&\\
        \leq&\int_{0}^{1}  \big\{ \norm{L_{2}^{c}(W_i, W_j)}_{F_0\otimes F_0, 2} + \norm{L_{2}^{c}(W_j, W_i)}_{F_0\otimes F_0, 2} \big\} d \tau \cdot \norm{\eta_1 - \eta_2}_\infty \cdot \norm{h}_\infty&\\
        \leq& \widebar{L}_{2}^{c} \norm{\eta_1 - \eta_2}_\infty \cdot \norm{h}_\infty.& 
\end{align*}
Since $\norm{ L_{2}^{c}(W_i, W_j) }_{F_0\otimes F_0, 2}$ is finite by Lemma~\ref{appendix - lemma A02}(v), $\widebar{L}_{2}^{c}$ is finite as well.\qed

\paragraph{Proof of (iii).} Fix $\theta\in\widebar{\Theta}_{o}$ and $\eta\in\mathcal{H}_N$. Let
\[ \begin{aligned}
        \widebar{L}_{3}^{c} \equiv &2\widebar{K}_{0}\norm{\widebar{J}}_{F_0, 2}\big\{ \norm{\widebar{\rho}^0}_{F_0, 2} + \norm{\widebar{\rho}^{1}}_{F_0, 2}\big\} +  4\widebar{K}_{1}\norm{\widebar{\rho}^{0}}_{F_0,2}\norm{\widebar{\rho}^{1}}_{F_0,2} \ind{}{c=1} \\
        \widebar{L}_{4}^{c} \equiv &2\widebar{K}_0\norm{ \widebar{H} }_{F_0,2}\norm{ \widebar{\rho}^{0} }_{F_0,2}.
\end{aligned} \]
By Assumptions \ref{sec03 - assumption02 - CMR} and \ref{sec03 - assumption03 - kernel}, two constants $\widebar{L}_{3}^{c}$ and $\widebar{L}_{4}^{c}$ are finite, and do not rely on $N$, $\theta$, and $\eta\in\mathcal{H}_N$. 

By adding and subtracting $\nabla_{\theta}r_{ij}(\theta, \eta_0)$, the triangle inequality gives:
\[     \norm{\nabla_{\theta}r_{ij}(\theta, \eta) - \eta_{\theta}r_{ij}(\theta_0, \eta_0)}_{F_0\otimes F_0, 2}
        \leq    \norm{\nabla_{\theta}r_{ij}(\theta, \eta) - \nabla_{\theta}r_{ij}(\theta, \eta_0)}_{F_0\otimes F_0, 2} + \norm{\nabla_{\theta}r_{ij}(\theta, \eta_0) - \nabla_{\theta}r_{ij}(\theta_0, \eta_0)}_{F_0\otimes F_0, 2}. \]
Since $\nabla_\theta r_{ij}(\theta, \eta) = d_{i}(\theta,\eta)\rho_{j}(\theta,\eta)K_{ij}(\eta) + \rho_{i}(\theta,\eta)d_{j}(\theta,\eta)K_{ij}(\eta)$ by construction, $\norm{\nabla_\theta r_{ij}(\theta, \eta) - \nabla_\theta r_{ij}(\theta, \eta_0)}_{F_0\otimes F_0,2}$ is bounded by:
\begin{equation}\label{appendix - lemma A03 - proof - eq05}
    \begin{aligned}
    \norm{\nabla_\theta r_{ij}(\theta, \eta) - \nabla_\theta r_{ij}(\theta, \eta_0)}_{F_0\otimes F_0,2}
        \leq    &2 \norm{ \{d_{i}(\theta,\eta) - d_{i}(\theta,\eta_0)\}\rho_{j}(\theta,\eta)K_{ij}(\eta) }_{F_0\otimes F_0,2}&\\
                &+ 2 \norm{ d_{i}(\theta,\eta_0)\{\rho_{j}(\theta,\eta) - \rho_{j}(\theta,\eta_0)\}K_{ij}(\eta) }_{F_0\otimes F_0,2}&\\
                &+ 2 \norm{ d_{i}(\theta,\eta_0)\rho_{j}(\theta,\eta_0)\{K_{ij}(\eta) - K_{ij}(\eta_0)\} }_{F_0\otimes F_0,2}.&
    \end{aligned}
\end{equation}
Using Lemma~\ref{appendix - lemma A02}(i), (ii), (iii) gives:
\begin{equation}\label{appendix - lemma A03 - proof - eq06}
    \begin{aligned}
    \norm{ \{d_{i}(\theta,\eta) - d_{i}(\theta,\eta_0)\}\rho_{j}(\theta,\eta)K_{ij}(\eta) }_{F_0\otimes F_0,2}
        \leq    &\widebar{K}_{0}\norm{\widebar{J}(W_i)}_{F_0, 2}\norm{\widebar{\rho}^0(W_j)}_{F_0, 2} \times \norm{\eta - \eta_0}_{\infty}&\\
    \norm{ d_{i}(\theta,\eta_0)\{\rho_{j}(\theta,\eta) - \rho_{j}(\theta,\eta_0)\}K_{ij}(\eta) }_{F_0\otimes F_0,2}
        \leq    &\widebar{K}_{0}\norm{\widebar{\rho}^{1}(W_i)}_{F_0, 2} \norm{\widebar{J}(W_j)}_{F_0, 2} \times \norm{\eta - \eta_0}_{\infty}&\\
    \norm{ d_{i}(\theta,\eta_0)\rho_{j}(\theta,\eta_0)\{K_{ij}(\eta) - K_{ij}(\eta_0)\} }_{F_0\otimes F_0,2}
        \leq    &2\widebar{K}_{1}\norm{\widebar{\rho}^{1}(W_i)}_{F_0,2}\norm{\widebar{\rho}^{0}(W_j)}_{F_0,2} \ind{}{c=1} \times \norm{\eta - \eta_0}_{\infty}.&
    \end{aligned}
\end{equation}
Then combining equation~\eqref{appendix - lemma A03 - proof - eq06} with equation~\eqref{appendix - lemma A03 - proof - eq05} yields:
\begin{equation}\label{appendix - lemma A03 - proof - eq07}
    \norm{\nabla_\theta r_{ij}(\theta, \eta) - \nabla_\theta r_{ij}(\theta, \eta_0)}_{F_0\otimes F_0,2} \leq \widebar{L}_{3}^{c}\norm{\eta - \eta_0}_{\infty}.
\end{equation}

We next derive the upper bound of the second term $\norm{\nabla_{\theta}r_{ij}(\theta, \eta_0) - \nabla_{\theta}r_{ij}(\theta_0, \eta_0)}_{F_0\otimes F_0, 2}$. For $\tau\in[0,1]$, let $\theta_\tau = \theta_0 + \tau(\theta - \theta_0)$. By the Taylor Theorem, we have:
\begin{align}
    \norm{\nabla_{\theta}r_{ij}(\theta, \eta_0) - \nabla_{\theta}r_{ij}(\theta_0, \eta_0)}_{F_0\otimes F_0, 2}
        \leq    & \int_{0}^{1} \norm{ \nabla_{\theta}^{2} r_{ij}(\theta_\tau, \eta_0)}_{F_0\otimes F_0, 2} d\tau \times \norm{\theta - \theta_0} &\nonumber\\
        \leq    & \int_{0}^{1} 2\norm{ \nabla_\theta^2 \rho_{i}(\theta_\tau, \eta_0)\rho_{j}(\theta_\tau, \eta_0) K_{ij}(\eta_0)  }_{F_0\otimes F_0, 2} d\tau \times \norm{\theta - \theta_0} &\nonumber\\
        \leq    & \int_{0}^{1} 2\widebar{K}_0\norm{ \widebar{H}(W_i) }_{F_0,2}\norm{ \widebar{\rho}^{0}(W_j) }_{F_0,2}  d\tau \times \norm{\theta - \theta_0}&\nonumber\\
        \leq    & \widebar{L}_{4}^{c} \norm{\theta - \theta_0},&\label{appendix - lemma A03 - proof - eq08}
\end{align}
where the third inequality comes from Assumptions~\ref{sec03 - assumption02 - CMR} and \ref{sec03 - assumption03 - kernel}. In consequence, equations~\eqref{appendix - lemma A03 - proof - eq07} and \eqref{appendix - lemma A03 - proof - eq08} together imply:
\[ \norm{\nabla_{\theta}r_{ij}(\theta, \eta) - \nabla_{\theta}r_{ij}(\theta_0, \eta_0)}_{F_0\otimes F_0, 2}
        \leq    \widebar{L}_{3}^{c}\norm{\eta - \eta_0}_{\infty} + \widebar{L}_{4}^{c} \norm{\theta - \theta_0}, \]
which complete the proof.\qed
%%%%%%%%%%%%%%%%%%%%%%%%%%%%%%%%%%%%%%%%%%%%%%%%%%%%%%%%%%%%%%%%%%%%%%%%%%%%%%%%%%%%%%%%%%%%%%%%%%%%%%%%%%%%%%%%%%%%%%%%%%%%%%%%%%%%%%%%%%%%%%%%%%%%%%%%%%%%%%%%%%%%%%%%%%%%%%%%%%%%%%%%%%%%%%%%%%%%%%%%%%%%%%%%%%%%%%%%%%%%%%%%%%%%%%%%%%%%%%%%%%%%%%%%%%%%%%%%%%%%%%%%%%%%%%%%%%%%%%%%%%%%%%%%%%%%%%%%%%%%%%%%%%%%%%%%%%%%%%%%%%%%%%%%%%%%%%%%%%%%%%%%%%%%%%%%%%%%%%%%%%%%%%%%%%%%%%%%%%%%%%%%%%%%%%%%%%%%%%%%%%%%%%%%%%%%%%%%%%%%%%%%%%%%%%%%%%%%%%%%%%%%%%%%%%%%%%%%%%%%%%%%%%%%%%%%%%%%%%%%%%%%%%%%%%%%%%%%%%%%%%%%%%%%%%%%%%%%%%%%%%%%%%%%%%%%%%%%%%%%%%%%%%%%%%%%%%%%%%%%%%%%%%%%%%%%%%%%
\subsection{Proof of Lemma \ref{appendix - lemma A04}}\label{appendix - lemma A04 - proof}
The class $\mathcal{F}_{\ell}^{c}$ is random through the training-sample nuisance estimates. Conditional on $\mathcal{W}_{-\ell}$, those estimates, and hence the class, are fixed. Its functions are then evaluated at the held-out observation pairs $(W_i,W_j)$. By construction,
\[  \psi^{c}\big(w,w',\theta,\widehat{\eta}_{\ell},\check{\vecdel}_{\ell}^{c}\big)
        =   r^{c}(w,w',\theta,\widehat{\eta}_{\ell})    +   \phi(w,w',\widehat{\eta}_{\ell},\check{\vecdel}_{\ell}^{c}), \]
where the last term does not depend on $\theta$, and therefore it does not affect covering numbers at a fixed radius. Conditional on $\mathcal W_{-\ell}$, work on the event $\widehat{\eta}_{\ell}\in\mathcal{H}_{N}$. Assumption~\ref{sec03 - assumption02 - CMR}(e) supplies measurable envelopes satisfying, simultaneously for all $\theta\in\widebar{\Theta}_{o}$ and for $F_0$-almost every $w$,
\[  |\rho(w,\theta,\widehat{\eta}_{\ell})| \leq \widebar{\rho}^0(w),    \qquad
    \norm{ \nabla_\theta\rho(w,\theta,\widehat{\eta}_{\ell}) }  \leq    \widebar{\rho}^1(w). \]
By Assumption~\ref{sec03 - assumption03 - kernel}(b), the kernel is bounded by $\widebar{K}_0$. The product rule for the raw score and the triangle inequality therefore give, for $(F_0\otimes F_0)$-almost every $(w,w')$,
\[  \sup_{\theta\in\widebar{\Theta}_{o}}   \norm{r^{c}(w,w',\theta,\widehat{\eta}_{\ell})}
        \leq    \widebar{K}_0\big\{\widebar{\rho}^1(w)\widebar{\rho}^0(w') + \widebar{\rho}^0(w)\widebar{\rho}^1(w')\big\}.  \]
Applying the triangle inequality once more yields the pointwise envelope bound
\[  \sup_{\theta\in\widebar{\Theta}_{o}}
    \norm{\psi^{c}(w,w',\theta,\widehat{\eta}_{\ell}, \check{\vecdel}_{\ell}^{c}) }  \leq    \widebar{\psi}_{\ell}^{c}(w,w'),  \]
where the envelope is given by:
\[  \widebar{\psi}_{\ell}^{c}(w,w') 
        \equiv  \widebar{K}_0 \big\{\widebar{\rho}^1(w)\widebar{\rho}^0(w') + \widebar{\rho}^0(w)\widebar{\rho}^1(w') \big\} + \norm{\phi(w,w',\widehat{\eta}_{\ell},\check{\vecdel}_{\ell}^{c}) }. \]
Since the fitted correction-term $\phi(w,w',\widehat{\eta}_{\ell},\check{\vecdel}_{\ell}^{c})$ does not depend on $\theta$, this bound is uniform over $\theta$ and pointwise in the observation arguments. Conditional on the training sample, a held-out pair $(W_i,W_j)$ has law $F_0\otimes F_0$. Thus, evaluating the functions at $(W_i,W_j)$ gives the same inequality almost surely on the event $\widehat{\eta}_{\ell}\in\mathcal{H}_{N}$.

Conditional on $\mathcal W_{-\ell}$ and on the event $\widehat{\eta}_{\ell}\in\mathcal{H}_{N}$, preservation of Euclidean classes under products and finite sums gives
\begin{equation}\label{sec04 - theorem01 - proof - entropy - second}
    \sup_{Q}\log N\!\left(  \epsilon\|\widebar{\psi}_{\ell}^{c}\|_{Q,2},\mathcal{F}_{\ell}^{c},\|\cdot\|_{Q,2}  \right)
    \leq C_0\log\!\left(\frac{C_1}{\epsilon}\right),    \qquad 0 < \epsilon \leq 1,
\end{equation}
where $C_0>0$ and $C_1\geq1$ are independent of $N$, $\ell$, $Q$, and the realized training-sample nuisance estimates; see, for example, \citet[Lemma~2.14(i)--(ii)]{pakespollard1989} and \citet[Theorem~2.10.20]{vw2023}.\qed
%%%%%%%%%%%%%%%%%%%%%%%%%%%%%%%%%%%%%%%%%%%%%%%%%%%%%%%%%%%%%%%%%%%%%%%%%%%%%%%%%%%%%%%%%%%%%%%%%%%%%%%%%%%%%%%%%%%%%%%%%%%%%%%%%%%%%%%%%%%%%%%%%%%%%%%%%%%%%%%%%%%%%%%%%%%%%%%%%%%%%%%%%%%%%%%%%%%%%%%%%%%%%%%%%%%%%%%%%%%%%%%%%%%%%%%%%%%%%%%%%%%%%%%%%%%%%%%%%%%%%%%%%%%%%%%%%%%%%%%%%%%%%%%%%%%%%%%%%%%%%%%%%%%%%%%%%%%%%%%%%%%%%%%%%%%%%%%%%%%%%%%%%%%%%%%%%%%%%%%%%%%%%%%%%%%%%%%%%%%%%%%%%%%%%%%%%%%%%%%%%%%%%%%%%%%%%%%%%%%%%%%%%%%%%%%%%%%%%%%%%%%%%%%%%%%%%%%%%%%%%%%%%%%%%%%%%%%%%%%%%%%%%%%%%%%%%%%%%%%%%%%%%%%%%%%%%%%%%%%%%%%%%%%%%%%%%%%%%%%%%%%%%%%%%%%%%%%%%%%%%%%%%%%%%%%%%%%%
\subsection{Proof of Lemma \ref{appendix - lemma A05}}\label{appendix - lemma A05 - proof}

For notational simplicity, let $q_{ij}(\theta)\equiv\rho(W_i,\theta,\eta_0)\rho(W_j,\theta,\eta_0)K\big(Z_i^{c}(\eta_0),Z_j^{c}(\eta_0)\big)$, $r_{ij}(\theta)\equiv r^{c}(W_i,W_j,\theta,\eta_0)$, and $\psi_{ij}(\theta)\equiv\psi^{c}(W_i,W_j,\theta,\eta_0,\vecdel_0^{c}(\cdot,\theta_0))$.

\paragraph{Proof of (i).} For a generic measurable function $(w_1,\ldots,w_k)\mapsto f(w_1,\ldots,w_k)$, define $U_{n}^{k} f$ as follows:
\begin{equation}\label{appendix - lemma A05 - proof - eq00 - Uprocess}
    U_{n}^{k} f \equiv \frac{1}{n(n-1)\cdots(n-k+1)} \sum_{\sigma_k} f(W_{i_1},\ldots,W_{i_k}),
\end{equation}
where $\sigma_k=(i_1,\ldots,i_k)$ ranges over all ordered $k$-tuples of distinct indices from $[n]$.

We begin with the Hoeffding decomposition for the generic kernel $\varphi_{0,ij}(\theta)$. By the law of iterated expectations,
\begin{equation}\label{appendix - lemma A05 - proof - eq01}
    \expect{\varphi_{1,i}(\theta)} = \Expect{\condexpect{\varphi_{0,ij}(\theta)}{W_i} - \expect{\varphi_{0,ij}(\theta)}} = 0.
\end{equation}
Since $W_i\pperp W_j$ and $\varphi_{0,ij}(\theta)=\varphi_{0,ji}(\theta)$, we have
\begin{equation}\label{appendix - lemma A05 - proof - eq02}
    \begin{aligned}
        \condexpect{\varphi_{2,ij}(\theta)}{W_i}
            &= \condexpect{\varphi_{0,ij}(\theta)}{W_i} - \expect{\varphi_{0,ij}(\theta)} - \varphi_{1,i}(\theta) - \expect{\varphi_{1,j}(\theta)} = 0,\\
        \condexpect{\varphi_{2,ij}(\theta)}{W_j}
            &= \condexpect{\varphi_{0,ij}(\theta)}{W_j} - \expect{\varphi_{0,ij}(\theta)} - \varphi_{1,j}(\theta) - \expect{\varphi_{1,i}(\theta)} = 0.
    \end{aligned}
\end{equation}
Thus $\varphi_{2,ij}(\theta)$ is degenerate of order two.

We next define the classes associated with the kernel and its two projections. If $\varphi_{0,ij}(\theta)=q_{ij}(\theta)$, define
\[  \mathcal{F}_{0} \equiv \{\varphi_{0,ij}(\theta):\theta\in\Theta\},\quad
    \mathcal{F}_{1} \equiv \{\varphi_{1,i}(\theta):\theta\in\Theta\},\quad\text{and}\quad
    \mathcal{F}_{2} \equiv \{\varphi_{2,ij}(\theta):\theta\in\Theta\}.  \]
Otherwise, define
\[  \mathcal{F}_{0} \equiv \{\varphi_{0,ij}(\theta):\theta\in\widebar{\Theta}_{o}\},\quad
    \mathcal{F}_{1} \equiv \{\varphi_{1,i}(\theta):\theta\in\widebar{\Theta}_{o}\},\quad\text{and}\quad
    \mathcal{F}_{2} \equiv \{\varphi_{2,ij}(\theta):\theta\in\widebar{\Theta}_{o}\}.  \]

Suppose first that $\mathcal{F}_{0}$ is Euclidean \citep{pakespollard1989} with an envelope $\widebar{\varphi}_{0}\in L_2(F_0\otimes F_0)$, coordinatewise for a vector-valued kernel.

The first projection has mean zero by \eqref{appendix - lemma A05 - proof - eq01}, and the second is degenerate conditional on either observation by \eqref{appendix - lemma A05 - proof - eq02}. By Lemmas~5--6 of \citet{sherman1994}, $\mathcal{F}_{1}$ and $\mathcal{F}_{2}$ inherit Euclidean covering bounds and square-integrable envelopes. Applying Corollary~4 of that paper to the first- and second-order processes, coordinatewise, gives
\begin{equation}\label{appendix - lemma A05 - proof - eq03}
    \sup_{\theta}\Norm{\frac{1}{N}\sum_{i=1}^N\varphi_{1,i}(\theta)} = O_P(N^{-1/2}),   \qquad
    \sup_{\theta}\Norm{U_{N}^{2}\varphi_{2,..}(\theta)} = O_P(N^{-1}),
\end{equation}
where each supremum is taken over $\Theta$ for $q_{ij}(\theta)$ and over $\widebar{\Theta}_{o}$ otherwise.

The parameter dimension is fixed, so the coordinatewise bounds give the same orders in Euclidean norm. The Hoeffding decomposition \citep{serfling1980,vdv1998} yields
\begin{equation}\label{appendix - lemma A05 - proof - eq04}
    \binom{N}{2}^{-1}\sum_{i<j}\varphi_{0,ij}(\theta)   = \expect{\varphi_{0,ij}(\theta)} + \frac{2}{N}\sum_{i=1}^N\varphi_{1,i}(\theta) + U_{N}^{2}\varphi_{2,..}(\theta).
\end{equation}
Combining \eqref{appendix - lemma A05 - proof - eq03} and \eqref{appendix - lemma A05 - proof - eq04} gives
\[    \sup_{\theta}\Norm{\binom{N}{2}^{-1}\sum_{i<j}\big\{\varphi_{0,ij}(\theta)-\expect{\varphi_{0,ij}(\theta)}\big\}} = O_P(N^{-1/2}).  \]

It remains to verify that $\mathcal{F}_{0}$ is Euclidean with a square-integrable envelope for $q_{ij}(\theta)$ and $r_{ij}(\theta)$. By the product and finite-sum covering bounds in \citet{pakespollard1989}, Assumption~\ref{sec03 - assumption02 - CMR}(e) implies that $\{q_{ij}(\theta):\theta\in\Theta\}$ and each coordinate of $\{r_{ij}(\theta):\theta\in\widebar{\Theta}_{o}\}$ are Euclidean, since the kernel is fixed at $\eta_0$. Under Assumptions~\ref{sec03 - assumption02 - CMR}(e) and~\ref{sec03 - assumption03 - kernel}, the corresponding pointwise envelope bounds are
\begin{equation}\label{appendix - lemma A05 - proof - eq05}
    \begin{aligned}
        \norm{q_{ij}(\theta)}
            &\leq \widebar{K}_0\widebar{\rho}^{0}(W_i)\widebar{\rho}^{0}(W_j),\\
        \norm{r_{ij}(\theta)}
            &\leq \widebar{K}_0\big\{\widebar{\rho}^{1}(W_i)\widebar{\rho}^{0}(W_j) + \widebar{\rho}^{0}(W_i)\widebar{\rho}^{1}(W_j)\big\}.
    \end{aligned}
\end{equation}
Since $W_i\pperp W_j$ and $\widebar{\rho}^{0},\widebar{\rho}^{1}\in L_2(F_0)$, both envelopes belong to $L_2(F_0\otimes F_0)$. This completes the proof of (i).\qed

\paragraph{Proof of (ii).} Fix $\ell\in[L]$ and a deterministic sequence $\xi_N\downarrow0$. Conditional on $\mathcal W_{-\ell}$, work on the event $\widehat{\eta}_{\ell}\in\mathcal{H}_{N}$. Lemma~\ref{appendix - lemma A04} establishes that the class $\mathcal{F}_{\ell}$ is Euclidean with $L_2(F_0\otimes F_0)$-envelope $\widebar{\psi}_{\ell}^{c}$. Since subtracting the same function from each member of a Euclidean class preserves its covering numbers, the following class is Euclidean:
\[   \mathcal{D}_{\ell} \equiv  \left\{
        (w,w')\mapsto\psi^{c}\big(w,w',\theta,\widehat{\eta}_{\ell},\check{\vecdel}_{\ell}^{c}\big)- \psi^{c}\big(w,w',\theta_0,\widehat{\eta}_{\ell},\check{\vecdel}_{\ell}^{c}\big)
        : \theta\in\widebar{\Theta}_{o}  \right\},   \]
with square-integrable envelope $2\widebar{\psi}_{\ell}^{c}$. Define its subclass
\[  \mathcal{D}_{\ell,\xi_N}    \equiv  \left\{
        (w,w')\mapsto\psi^{c}\big(w,w',\theta,\widehat{\eta}_{\ell},\check{\vecdel}_{\ell}^{c}\big) - \psi^{c}\big(w,w',\theta_0,\widehat{\eta}_{\ell},\check{\vecdel}_{\ell}^{c}\big)
        : \theta\in\widebar{\Theta}_{o}\textnormal{ such that }\norm{\theta-\theta_0} \leq \xi_N    \right\}.\]

Let $\Delta\psi_{1,\ell,i}^{c}(\theta)$ and $\Delta\psi_{2,\ell,ij}^{c}(\theta)$ denote the first and second Hoeffding projections of $\Delta\psi_{\ell,ij}^{c}(\theta)$, respectively, conditional on $\mathcal W_{-\ell}$. As in the proof of Lemma~\ref{appendix - lemma A05}(i), we have
\[  \condexpect{\Delta\psi_{1,\ell,i}^{c}(\theta)}{\mathcal W_{-\ell}}=0,   \qquad
    \condexpect{\Delta\psi_{2,\ell,ij}^{c}(\theta)}{W_i,\mathcal W_{-\ell}} =   \condexpect{\Delta\psi_{2,\ell,ij}^{c}(\theta)}{W_j,\mathcal W_{-\ell}}=0,  \]
so $\Delta\psi_{2,\ell,ij}^{c}(\theta)$ is degenerate of order two.

Let $\mathcal{D}_{1,\ell,\xi_N}$ and $\mathcal{D}_{2,\ell,\xi_N}$ denote the classes of functions $\Delta\psi_{1,\ell,i}^{c}(\theta)$ and $\Delta\psi_{2,\ell,ij}^{c}(\theta)$, respectively, indexed by $\theta\in\widebar{\Theta}_{o}$ such that $\norm{\theta-\theta_0}\leq\xi_N$. Both classes inherit Euclidean covering bounds and square-integrable envelopes by Lemmas~5--6 of \citet{sherman1994}. Since the fitted correction cancels in the increment, Lemma~\ref{appendix - lemma A03}(i), consistency of $\widehat{\eta}_{\ell}$, and $L_2$-continuity of the raw score at $(\theta_0,\eta_0)$ in $\theta$ imply that the $L_2$ radius of $\mathcal{D}_{\ell,\xi_N}$ is $o_P(1)$. The same holds for $\mathcal{D}_{1,\ell,\xi_N}$ by contraction. Applying Lemma~2.2 of \citet{vdv_wellner2007} to the first projection and Corollary~4 of \citet{sherman1994} to the second projection, coordinatewise, gives
\begin{equation}\label{appendix - lemma A05 - proof - eq03 - second}
    \begin{aligned}
    &\sup_{\substack{\theta\in\widebar{\Theta}_{o}\\ \|\theta-\theta_0 \|\leq\xi_N}}    \Norm{\frac{1}{|I_{\ell}|}\sum_{(i,j)\in I_{\ell}}  \big\{\Delta\psi_{1,\ell,i}^{c}(\theta)+\Delta\psi_{1,\ell,j}^{c}(\theta)\big\}}     =  o_P(n_{\ell}^{-1/2}),\\
    &\sup_{\substack{\theta\in\widebar{\Theta}_{o}\\ \|\theta-\theta_0 \|\leq\xi_N}}    \Norm{\frac{1}{|I_{\ell}|}\sum_{(i,j)\in I_{\ell}}\Delta\psi_{2,\ell,ij}^{c}(\theta)}   =   O_P(n_{\ell}^{-1}).
    \end{aligned}
\end{equation}
Here $n_{\ell}=|J_{\ell}|$. For a within-group fold, the first sum after normalization equals $(2/n_{\ell})\sum_{i\in J_{\ell}}\Delta\psi_{1,\ell,i}^{c}(\theta)$. For a cross-half fold, it equals the sum of the two half-sample averages. The second sum over a cross-half fold is the complete-pair sum over the union minus the complete-pair sums within the two halves, each of size proportional to $n_{\ell}$.

Therefore, the Hoeffding decomposition \citep{serfling1980,vdv1998} yields
\begin{equation}\label{appendix - lemma A05 - proof - eq04 - second}
    \begin{aligned}
        \frac{1}{|I_{\ell}|}\sum_{(i,j)\in I_{\ell}}\Delta\psi_{\ell,ij}^{c}(\theta) - \condexpect{\Delta\psi_{\ell,ij}^{c}(\theta)}{\mathcal W_{-\ell}}
            =     \frac{1}{|I_{\ell}|}\sum_{(i,j)\in I_{\ell}}\big\{\Delta\psi_{1,\ell,i}^{c}(\theta)+\Delta\psi_{1,\ell,j}^{c}(\theta)\big\}
                + \frac{1}{|I_{\ell}|}\sum_{(i,j)\in I_{\ell}}\Delta\psi_{2,\ell,ij}^{c}(\theta),
    \end{aligned}
\end{equation}
and hence
\[  \sup_{\substack{\theta\in\widebar{\Theta}_{o}\\ \|\theta-\theta_0 \|\leq\xi_N}} \Norm{\frac{1}{|I_{\ell}|}\sum_{(i,j)\in I_{\ell}}  \big\{\Delta\psi_{\ell,ij}^{c}(\theta) -\condexpect{\Delta\psi_{\ell,ij}^{c}(\theta)}{\mathcal W_{-\ell}}\big\}}  = o_P(n_{\ell}^{-1/2}).   \]
Since $G$ is fixed, $n_{\ell}\asymp N$ and $L$ is fixed. Taking the maximum over $\ell\in[L]$ gives (iii). \qed
%%%%%%%%%%%%%%%%%%%%%%%%%%%%%%%%%%%%%%%%%%%%%%%%%%%%%%%%%%%%%%%%%%%%%%%%%%%%%%%%%%%%%%%%%%%%%%%%%%%%%%%%%%%%%%%%%%%%%%%%%%%%%%%%%%%%%%%%%%%%%%%%%%%%%%%%%%%%%%%%%%%%%%%%%%%%%%%%%%%%%%%%%%%%%%%%%%%%%%%%%%%%%%%%%%%%%%%%%%%%%%%%%%%%%%%%%%%%%%%%%%%%%%%%%%%%%%%%%%%%%%%%%%%%%%%%%%%%%%%%%%%%%%%%%%%%%%%%%%%%%%%%%%%%%%%%%%%%%%%%%%%%%%%%%%%%%%%%%%%%%%%%%%%%%%%%%%%%%%%%%%%%%%%%%%%%%%%%%%%%%%%%%%%%%%%%%%%%%%%%%%%%%%%%%%%%%%%%%%%%%%%%%%%%%%%%%%%%%%%%%%%%%%%%%%%%%%%%%%%%%%%%%%%%%%%%%%%%%%%%%%%%%%%%%%%%%%%%%%%%%%%%%%%%%%%%%%%%%%%%%%%%%%%%%%%%%%%%%%%%%%%%%%%%%%%%%%%%%%%%%%%%%%%%%%%%%%%%
\pagebreak\newpage
\section{Numerical details for the empirical examples}\label{app:empiricaldetails}

\subsection{Finite-instrument GMM in Section~\ref{sec02}}
The baseline first stage uses a full quadratic in the four input and proxy variables: one intercept, four linear terms, four squares, and six pairwise interactions. The regression is fitted on the same 1,944 current-period observations that enter the second-stage estimation sample. Its fitted function is evaluated at the union of current and lagged inputs, comprising 2,397 plant-year rows. Lag-only outcomes are excluded from fitting and tuning. Fitted values and all five instruments remain fixed throughout the GMM search.

All five structural parameters are optimized jointly over $[-3,3]^3\times[-2,2]\times[0.1,0.99]$. Starting values form the Cartesian product of nine equally spaced points for each beta, seven for alpha, and thirteen for gamma, including endpoints: $9^3\cdot7\cdot13=66,339$ starts. Each start is refined by continuous bounded optimization with analytic derivatives. A numerical root requires $Q_n<10^{-16}$ and maximum absolute direct sample moment below $10^{-8}$. Solutions are grouped using maximum-coordinate distance $10^{-5}$. All 66,339 baseline runs reach one of the four interior roots in Table~\ref{sec02:tab:roots}, with no optimizer errors or boundary fits.

An independent polynomial calculation checks the zero-moment solutions. With all sums normalized by $n$, define
\[  A=\frac1n\sum\widehat Z_{i,t-1}X_{it}^\top,
    \quad B=\frac1n\sum\widehat Z_{i,t-1}X_{i,t-1}^\top,
    \quad a=\frac1n\sum\widehat Z_{i,t-1},  \]
\[  c_0=\frac1n\sum\widehat Z_{i,t-1}\widehat\eta_{it},
    \qquad c_1=\frac1n\sum\widehat Z_{i,t-1}\widehat\eta_{i,t-1},
    \qquad\widehat Z_{i,t-1}=Z_{i,t-1}(\widehat\eta).   \]
The moment equations become
\begin{equation}\label{app:eq:linearsystem}
    (A-\gamma B)\beta+a\alpha=c_0-\gamma c_1.
\end{equation}
A necessary condition for a solution is $\det\big[A-\gamma B,\ a,\ c_0-\gamma c_1\big]=0$. Multilinearity makes this a polynomial of degree at most four in $\gamma$. We numerically enumerate its real roots, solve equation~\eqref{app:eq:linearsystem} where the coefficient matrix has rank four, and check feasibility and the original moments directly. This recovers the four baseline roots. Every displayed root has a rank-five moment Jacobian and maximum absolute direct moment below $1.7\times10^{-13}$. The computations provide numerical checks rather than interval-arithmetic certification.

For local linear and random forests, tuning uses three folds grouped by plant. Each selected model is refitted on the same 1,944 observations. Local linear selects bandwidth $h=1.2$ in standardized coordinates; the forest selects 500 trees, minimum leaf size 10, and one candidate variable per split. Both use the same second-stage box, instruments, grid of starts, and polynomial check as the quadratic baseline. Their feasible interior root counts are four and three, respectively.

We also repeat the exercise using the full sample of 2,544 plant-year observations from 497 plants to fit $\eta_0$. The quadratic regression is refitted on this larger sample; local linear is retuned by three-fold cross-validation grouped by plant and refitted on all observations, again selecting $h=1.2$. The second-stage sample remains the same 1,944 plant-year observations from 401 plants, and the parameter box and instrument definitions are unchanged. Both the quadratic and local-linear first stages yield four distinct interior roots. In each case, all 66,339 optimization runs converge to a numerical root, with no boundary fits or optimizer errors, and the independent polynomial calculation recovers the same four roots. Thus, the multiplicity persists when the first stage uses the full available sample as well as when the two stages use matched sample sizes. Related numerical sensitivity of ACF implementations is documented by \citet{rovigatti_mollisi2018}.

The GMM weight is the inverse instrument second-moment matrix in equation~\eqref{sec02:eq:weight}. It has rank five, with smallest eigenvalue approximately 0.003149 in the quadratic specification; the corresponding values for local linear and random forest are approximately 0.003137 and 0.003151. For any fixed positive-definite weight,
\begin{equation} \label{app:eq:weightinvariance}
    \widehat{g}_N(\theta)^\top W\widehat{g}_N(\theta)=0 \quad\Longleftrightarrow\quad   \widehat{g}_N(\theta)=0.
\end{equation} This also applies to a fixed positive-definite second-step weight estimated after an identity-weight first step: all exact roots of the same moment vector remain zero-objective minimizers. Weights can change nonzero minima and numerical attraction basins, but cannot distinguish the existing exact roots. This invariance follows algebraically and does not require a separate optimizer comparison.

For the local-quadratic alternative reported in Section~\ref{sec02}, we replace the local-linear eta regression by a degree-two local polynomial, including all squares and pairwise interactions. We retain three-fold CV grouped by plant with the same GMM seed and bandwidth grid. The matched-sample fit selects $h=2$; fitting on all 2,544 observations selects $h=3$. Both versions produce four interior GMM roots with rank-five moment Jacobians. In each case, all 66,339 starts converge to one of the roots, with no boundary fits or errors, and the independent polynomial calculation agrees. The 1,944 second-stage observations, five instruments, and compact GMM parameter box are unchanged.
%%%%%%%%%%%%%%%%%%%%%%%%%%%%%%%%%%%%%%%%%%%%%%%%%%%%%%%%%%%%%%%%%%%%%%%%%%%%%%%%%%%%%%%%%%%%%%%%%%%%%%%%%%%%%%%%%%%%%%%%%%%%%%%%%%%%%%%%%%%%%%%%%%%%%%%%%%%%%%%%%%%%%%%%%%%%%%%%%%%%%%%%%%%%%%%%%%%%%%%%%%%%%%%%%%%%%%%%%%%%%%%%%%%%%%%%%%%%%%%%%%%%%%%%%%%%%%%%%%%%%%%%%%%%%%%%%%%%%%%%%%%%%%%%%%%%%%%%%%%%%%%%%%%%%%%%%%%%%%%%%%%%%%%%%%%%%%%%%%%%%%%%%%%%%%%%%%%%%%%%%%%%%%%%%%%%%%%%%%%%%%%%%%%%%%%%%%%%%%%%%%%%%%%%%%%%%%%%%%%%%%%%%%%%%%%%%%%%%%%%%%%%%%%%%%%%%%%%%%%%%%%%%%%%%%%%%%%%%%%%%%%%%%%%%%%%%%%%%%%%%%%%%%%%%%%%%%%%%%%%%%%%%%%%%%%%%%%%%%%%%%%%%%%%%%%%%%%%%%%%%%%%%%%%%%%%%%%%
\subsection{Implementation details for the production function application}\label{app:implementation_ACF}

This subsection describes the implementation of the estimation procedure for the production function application in Subsection~\ref{sec07:acf}.

\paragraph{Model and nuisance functions.}
The model, data, and structural residual are those in Section~\ref{sec02}. For each $t$, the first-stage conditional mean is $\eta_0(S_{i,t})=\mathbb E[Y_{i,t}\mid S_{i,t}]$, where $S_{i,t}=(\ell_{i,t}^s,\ell_{i,t}^u,k_{i,t},M_{i,t})^\top$ and $Z_{i,t}(\eta)=(\ell_{i,t-1}^s,\ell_{i,t-1}^u,k_{i,t-1}, \eta(S_{i,t-1}))^\top$. The generalized residual in equation~\eqref{sec02:eq:residual} and its gradient with respect to $\theta$ are
\[  \rho_{i,t}(\theta,\eta) = \eta(S_{i,t})-\beta_s\ell_{i,t}^s-\beta_u \ell_{i,t}^u-\beta_k k_{i,t}-\alpha - \gamma\big(\eta(S_{i,t-1})-\beta_s\ell_{i,t-1}^s-\beta_u\ell_{i,t-1}^u - \beta_k k_{i,t-1}\big). \]
\[  d_{i,t}(\theta,\eta) \equiv \nabla_{\theta} \rho_{i,t}(\theta,\eta) =
    \begin{bmatrix}
    -\ell_{i,t}^s+\gamma\ell_{i,t-1}^s\\
    -\ell_{i,t}^u+\gamma\ell_{i,t-1}^u\\
    -k_{i,t}+\gamma k_{i,t-1}\\
    -1\\
    -\eta(S_{i,t-1}) + \beta_s\ell_{i,t-1}^s+\beta_u\ell_{i,t-1}^u+\beta_k k_{i,t-1}
    \end{bmatrix}. \]

\paragraph{Bias-correction term.}
Fix a kernel bandwidth $b$. Write $\eta_{i,t}=\eta(S_{i,t})$ and $\eta_{i,t-1}=\eta(S_{i,t-1})$. Let $K_{b,ij,t}(\eta) = K_{b,\ell}(Z_{i,t}(\eta),Z_{j,t}(\eta))$ and $\pmb{e}_5=(0,0,0,0,1)^\top$. The partial derivatives with respect to the two evaluations of $\eta$ are
\[  \begin{aligned}
    \frac{\partial}{\partial\eta_{i,t}}\rho_{i,t}(\theta,\eta)
        =   & 1, &
    \frac{\partial}{\partial\eta_{i,t}}d_{i,t}(\theta,\eta)
        =   & \pmb{0}_5, &\\
    \frac{\partial}{\partial\eta_{i,t-1}}\rho_{i,t}(\theta,\eta)
        =   & -\gamma, &
    \frac{\partial}{\partial\eta_{i,t-1}}d_{i,t}(\theta,\eta)
        =   & -\pmb{e}_5. &
\end{aligned}   \]
For the Gaussian kernel in equation~\eqref{sec07:eq:kernel}, holding the coordinate scales in $\pmb{D}_{\ell}$ fixed,
\[  \frac{\partial}{\partial\eta_{i,t}}K_{b,ij,t}(\eta) = 0,   \qquad
    \frac{\partial}{\partial\eta_{i,t-1}}K_{b,ij,t}(\eta)
        =   -\frac{\eta_{i,t-1}-\eta_{j,t-1}}{b^2(D_{\ell})_{44}^{2}}K_{b,ij,t}(\eta). \]
Let $r_{b,ij,t}(\theta,\eta) = \big\{d_{i,t}(\theta,\eta)\rho_{j,t}(\theta,\eta) + \rho_{i,t}(\theta,\eta)d_{j,t}(\theta,\eta)\big\}K_{b,ij,t}(\eta)$. Then,
\[\begin{aligned}
    \lambda^{1}_{b,ij}(\theta,\eta)
        \equiv  & \frac{\partial}{\partial\eta_{i,t}}r_{b,ij,t}(\theta,\eta)
        =       d_{j,t}(\theta,\eta)K_{b,ij,t}(\eta), &\\
    \lambda^{2}_{b,ij}(\theta,\eta)
        \equiv  & \frac{\partial}{\partial\eta_{i,t-1}}r_{b,ij,t}(\theta,\eta)
        =       \big\{-\pmb{e}_5\rho_{j,t}(\theta,\eta)-\gamma d_{j,t}(\theta,\eta)\big\}K_{b,ij,t}(\eta) 
              - r_{b,ij,t}(\theta,\eta)\frac{\eta_{i,t-1}-\eta_{j,t-1}}{b^2(D_{\ell})_{44}^{2}}. &
\end{aligned}\]
Following equations~\eqref{sec04 - def02 - delta} and~\eqref{sec04 - def02 - delta2}, the correction coefficients for observation $i$ are
\[  \begin{aligned}
    \vecdel_{0,1}^{1}(S_{i,t},\theta)
        =   & \condExpectF{F_0\otimes F_0}{ \lambda^{1}_{b,ij}(\theta,\eta_0) }{S_{i,t}}, &\\
    \vecdel_{0,1}^{2}(S_{i,t-1},\theta)
        =   & \condExpectF{F_0\otimes F_0}{ \lambda^{2}_{b,ij}(\theta,\eta_0) }{S_{i,t-1}}. &
\end{aligned}   \]
By symmetry, the correction coefficients for observation $j$ are obtained by exchanging $i$ and $j$.

\paragraph{Pair folds and first-stage estimation.}
We set $G=5$, and partition $\binom{1944}{2}=1,888,596$ distinct unordered observation pairs into $L=G(2G-1)=45$ folds $I_{1},...,I_{45}$ according to Definition~\ref{sec04 - definition01 - DML}. Repeat the following procedure across $\ell\in [L]$. For each $\ell\in[L]$, we construct two training sets. One is $I_{-\ell} \equiv \{(i,j)\in [N]^2 : i < j,\; i,j\in J_{-\ell}\}$ where $J_{\ell} \equiv \{i\in [N] : i$ appears in some pair in $I_{\ell}\}$ and $n_{\ell} \equiv |J_{\ell}|$. Another one is $J_{-\ell} = \{i\in [N] : \text{the plant of }i\text{ does not appear in }J_{\ell}\}$. 

First, using training observations in $J_{-\ell}$, we estimate $\widehat{\eta}_{\ell}$ by regressing $Y_{i,t}$ on $S_{i,t}$ using the specified learners explained in Subsection~~\ref{sec07:acf}.

Second, we obtain the training pilot $\widetilde{\theta}_{\ell}$ at each bandwidth $b\in\mathcal{B}$ by minimizing
\[  \frac{1}{|J_{-\ell}|^2}\sum_{i\in J_{-\ell}}\sum_{j\in J_{-\ell}} \rho_{i,t}(\theta,\widehat{\eta}_{\ell})\rho_{j,t}(\theta,\widehat{\eta}_{\ell}) K_{b,ij,t}(\widehat{\eta}_{\ell}),   \]
over $\Theta_{\mathrm{emp}}=[-5,5]^3\times[-10,10]\times[0.1,1.2].$ This pilot initializes the final numerical search and serves as the reference for choosing among feasible roots.

Third, we obtain $\lambda^{1}_{b,ij}( \widetilde{\theta}_{\ell}, \widehat{\eta}_{\ell} )$ and $\lambda^{2}_{b,ij}( \widetilde{\theta}_{\ell}, \widehat{\eta}_{\ell} )$ at each bandwidth $b\in\mathcal{B}$ for both orderings of each pair in $I_{-\ell}$. Then regressing $\lambda^{1}_{b,ij}$ on $S_{i,t}$ and $\lambda^{2}_{b,ij}$ on $S_{i,t-1}$ using local quadratic regression gives $\check{\vecdel}^{1}_{b,\ell}(S_{i,t})$ and $\check{\vecdel}^{2}_{b,\ell}(S_{i,t-1})$, respectively.

Using pairs of observations in the test data $I_{\ell}$, we construct the orthogonal score
\[  \begin{aligned}
    \psi_{b,\ell,ij,t}(\theta)
        =   & \big\{ d_{i,t}(\theta,\widehat{\eta}_{\ell})\rho_{j,t}(\theta,\widehat{\eta}_{\ell}) + \rho_{i,t}(\theta,\widehat{\eta}_{\ell})d_{j,t}(\theta,\widehat{\eta}_{\ell}) \big\}K_{b,ij,t}(\widehat{\eta}_{\ell}) &\\
            & + \check{\vecdel}^{1}_{b,\ell}(S_{i,t})\big(Y_{i,t} - \widehat{\eta}_{\ell}(S_{i,t})\big) + \check{\vecdel}^{2}_{b,\ell}(S_{i,t-1})\big(Y_{i,t-1} - \widehat{\eta}_{\ell}(S_{i,t-1})\big) &\\
            & + \check{\vecdel}^{1}_{b,\ell}(S_{j,t})\big(Y_{j,t} - \widehat{\eta}_{\ell}(S_{j,t})\big) + \check{\vecdel}^{2}_{b,\ell}(S_{j,t-1})\big(Y_{j,t-1} - \widehat{\eta}_{\ell}(S_{j,t-1})\big).&
\end{aligned} \]

\paragraph{DDML estimate and its variance.} 

For each bandwidth $b$, the DDML estimator minimizes the squared norm of the pooled corrected score over the compact parameter box:
\[  \widehat{\theta}_b\in\argmin{\theta\in\Theta_{\mathrm{emp}}}{ \Norm{ {\binom{N}{2}}^{-1}\sum_{\ell=1}^{L}\sum_{(i,j)\in I_{\ell}} \psi_{b,\ell,ij,t}(\theta) }^2 } \]
We first start the numerical search at the full-sample plug-in estimate. Further starts comprise the six original controls, training-fold pilots, and independent profile candidates.

For each bandwidth with a full-rank Jacobian, we evaluate the scores and derivatives at $\widehat{\theta}_b$. We calculate $\widehat\Omega_b = (\widehat{\Gamma}_b^\top\widehat{\Gamma}_b)^{-1}\widehat{\Gamma}_b^\top \widehat{\Sigma}_b \widehat{\Gamma}_b (\widehat{\Gamma}_b^\top\widehat{\Gamma}_b)^{-1}$ where each component is obtained by:
\[  \begin{aligned}
    \widehat{\Gamma}_b
        =   &\binom{N}{2}^{-1}\sum_{\ell=1}^{L}\sum_{(i,j)\in I_{\ell}}\nabla_\theta\psi_{b,\ell,ij,t}(\widehat{\theta}_b), &\\
    \widehat{\Sigma}_b
        =   &\frac{4}{N(2G-1)}\sum_{\ell=1}^{L}\sum_{i\in J_{\ell}}\widebar{\psi}_{b,\ell,i,t}\widebar{\psi}_{b,\ell,i,t}^\top,&
\end{aligned}\]
where, as in equation~\eqref{sec04 - eq - variance},
\[  \widebar{\psi}_{b,\ell,i,t}
        =   \frac{1}{n_{\ell}-1}\sum_{j\in J_{\ell}\backslash\{i\}}\psi_{b,\ell,ij,t}(\widehat{\theta}_b). \]
%%%%%%%%%%%%%%%%%%%%%%%%%%%%%%%%%%%%%%%%%%%%%%%%%%%%%%%%%%%%%%%%%%%%%%%%%%%%%%%%%%%%%%%%%%%%%%%%%%%%%%%%%%%%%%%%%%%%%%%%%%%%%%%%%%%%%%%%%%%%%%%%%%%%%%%%%%%%%%%%%%%%%%%%%%%%%%%%%%%%%%%%%%%%%%%%%%%%%%%%%%%%%%%%%%%%%%%%%%%%%%%%%%%%%%%%%%%%%%%%%%%%%%%%%%%%%%%%%%%%%%%%%%%%%%%%%%%%%%%%%%%%%%%%%%%%%%%%%%%%%%%%%%%%%%%%%%%%%%%%%%%%%%%%%%%%%%%%%%%%%%%%%%%%%%%%%%%%%%%%%%%%%%%%%%%%%%%%%%%%%%%%%%%%%%%%%%%%%%%%%%%%%%%%%%%%%%%%%%%%%%%%%%%%%%%%%%%%%%%%%%%%%%%%%%%%%%%%%%%%%%%%%%%%%%%%%%%%%%%%%%%%%%%%%%%%%%%%%%%%%%%%%%%%%%%%%%%%%%%%%%%%%%%%%%%%%%%%%%%%%%%%%%%%%%%%%%%%%%%%%%%%%%%%%%%%%%%%
\subsection{Cereal-demand implementation}\label{app:implementation_cereal}

\paragraph{Sample construction.}
We select one product per manufacturer using a year of sales before the estimation period. Among ready-to-eat cereals in 10- to 32-ounce packages, the top sellers by units in Dominick's weeks 69--120 were 10-ounce Cheerios (General Mills), 15-ounce Rice Krispies (Kellogg's), and 24-ounce Grape-Nuts (Post). In weeks 121--172, each cereal's share is its units sold divided by total units sold for these three products in the same store-week.

We divide the recorded retail price by the number of packages in the price quote. The acquisition cost per package is this price times one minus the gross-margin fraction. We code a promotion when a sale code is present.\footnote{The \href{https://www.chicagobooth.edu/-/media/enterprise/centers/kilts/datasets/dominicks-dataset/dominicks-manual-and-codebook_kiltscenter}{Dominick's manual and codebook} describes these fields and notes that sale codes were recorded intermittently.} We retain store-weeks in which all three products have valid scanner records, nonnegative unit sales, and positive prices and costs. The resulting 4,363 store-weeks receive equal weight. The 86 stores contribute between 17 and 52 weeks each.

\paragraph{Petrin--Train benchmark.}
Let $p_i$ and $f_i$ be the two retail price and promotion differences relative to Grape-Nuts, and let $A_i$ contain the three acquisition costs in levels. We regress each component of $p_i$ on $w_i=(1,f_i^\top,A_i^\top)^\top$ by full-sample OLS. Both the observed relative price and its OLS residual enter the utility index in equation~\eqref{sec07:eq:cereal-utility}. With $\widehat\eta_{\rm OLS}(S_i)=\widehat B^\top w_i$, the share-weighted logit estimator is
\[
 \widehat\theta_{\rm PT}=\arg\max_\theta\frac{1}{N}\sum_{i=1}^N\sum_{j=0}^2
 Y_{ij}\log P_{ij}(\theta,\widehat\eta_{\rm OLS}).
\]
For its standard errors in Table~\ref{sec07:tab:cereal-fitted}, we stack the two OLS estimating equations with the logit score and cluster by store and week.

\paragraph{Cross-fitted price predictions.}
We sort stores by ID and divide them into five consecutive folds of 18, 17, 17, 17, and 17 stores. Each fold in turn serves as the evaluation sample, with all weeks from its stores held out together. On the other four folds, each learner predicts the two relative prices from $S_i=(f_i^\top,A_i^\top)^\top$. We use the learner specifications and tuning grids in Section~\ref{sec07:empirical}. We tune each price regression by minimizing squared prediction error over five inner folds, assigning training stores at random to these folds. Both share equations and both conditioning sets use the same held-out predictions.

Table~\ref{app:tab:cereal-first-stage} shows how well each learner predicts prices at held-out stores. In the random-forest row, 2,543 store-weeks have a counterpart at another store with the same observed relative prices and promotion differences, similar fitted prices, and different acquisition costs.

\begin{table}[htbp]\centering\small
\caption{Relative-price prediction errors and matches across stores}
\label{app:tab:cereal-first-stage}
\begin{tabular}{lrrr}
\hline\hline
 & \multicolumn{2}{c}{Price RMSE (\$)} & Matched\\
\cline{2-3}
First stage & Cheerios & Rice Krispies & store-weeks\\
\hline
Local quadratic & $0.080$ & $0.102$ & 1,771\\
XGBoost & $0.036$ & $0.060$ & 2,311\\
Penalized splines & $0.073$ & $0.090$ & 2,171\\
LASSO & $0.078$ & $0.100$ & 1,838\\
Random forest & $0.028$ & $0.052$ & 2,543\\
\hline\hline
\end{tabular}
\par\smallskip\begin{minipage}{.94\textwidth}\footnotesize
\textit{Notes.} RMSE is the root mean squared prediction error for the price relative to Grape-Nuts. A store-week counts once if it has a match at another store in its held-out fold. Both observed relative prices and promotion differences must agree, both fitted prices must be within 2.5 cents, and at least one acquisition cost must differ by at least half its training-sample standard deviation.
\end{minipage}
\end{table}

\paragraph{DDML score and first-stage correction.}
The five evaluation folds contain 920, 869, 857, 877, and 840 store-weeks. Let $I_{\ell}$ contain all unordered pairs of store-weeks in evaluation fold $\ell$, giving $M=1{,}903{,}188$ pairs in total. For each share $j$ and conditioning set $c$, we fit the training pilot by minimizing the plug-in quadratic criterion including its diagonal. We use $c=1$ for fitted prices and $c=S$ for acquisition costs.

Fix a kernel bandwidth $b$. Write $\eta_i=\eta(S_i)$ and $d_{ij}=\nabla_\theta\rho_{ij}$. From equation~\eqref{sec07:eq:cereal-utility}, for $j,h=1,2$,
\[  \begin{aligned}
    &d_{ij}
        =     -P_{ij}\sum_{h=1}^{2}\big(\mathbf 1\{j=h\}-P_{ih}\big)
              \big(\mathbf 1\{h=1\},\mathbf 1\{h=2\},p_{ih},f_{ih},v_{ih}(\eta)\big)^\top, &\\
    &\frac{\partial}{\partial\eta_{ih}}\rho_{ij}
        =    \gamma P_{ij}\big(\mathbf 1\{j=h\}-P_{ih}\big),
    \qquad
     \frac{\partial}{\partial\eta_{ih}}d_{ij}
        =    \big(\pmb{e}_5 P_{ij}-\gamma d_{ij}\big)\big(\mathbf 1\{j=h\}-P_{ih}\big) + \gamma P_{ij}d_{ih}. &
\end{aligned}   \]
For the Gaussian kernel in equation~\eqref{sec07:eq:kernel}, let $D_{\ell,\eta}$ denote the diagonal matrix of training-sample standard deviations of the two fitted prices. Holding the coordinate scales fixed,
\[\begin{aligned}
        &\frac{\partial}{\partial\eta_{ih}}K_{ik}^{1}
        =   -\frac{\eta_{ih}-\eta_{kh}}{b^2(D_{\ell,\eta})_{hh}^{2}}K_{ik}^{1},&
        &\frac{\partial}{\partial\eta_{ih}}K_{ik}^{S} = 0&
\end{aligned} \]
Therefore, by following equations~\eqref{sec04 - def02 - delta} and~\eqref{sec04 - def02 - delta2}, the analytic forms of the correction matrices for share $j$ are
\[  \begin{aligned}
    &\vecdel_{0,1}^{c,j}(S_i,\theta)
        =   \condExpectF{F_0\otimes F_0}{ \begin{bmatrix}\lambda_{b,ik,1}^{c,j}(\theta,\eta_0) & \lambda_{b,ik,2}^{c,j}(\theta,\eta_0)\end{bmatrix} }{S_i} &\\
    &\vecdel_{0,2}^{c,j}(S_k,\theta)
        =   \condExpectF{F_0\otimes F_0}{ \begin{bmatrix}\lambda_{b,ki,1}^{c,j}(\theta,\eta_0) & \lambda_{b,ki,2}^{c,j}(\theta,\eta_0)\end{bmatrix} }{S_k}, &
\end{aligned}   \]
where
\[\begin{aligned}
    \lambda_{b,ik,h}^{c,j}(\theta,\eta)
        \equiv  &   \big\{ \big(\pmb{e}_5 P_{ij}-\gamma d_{ij}\big)\big(\mathbf 1\{j=h\}-P_{ih}\big) + \gamma P_{ij}d_{ih} \big\}\rho_{kj}K_{ik}^{c} &\\
                & + \gamma P_{ij}\big(\mathbf 1\{j=h\}-P_{ih}\big)d_{kj}K_{ik}^{c}
                  - \ind{}{c=1}\big\{ d_{ij}\rho_{kj}+\rho_{ij}d_{kj} \big\}\frac{\eta_{ih}-\eta_{kh}}{b^2(D_{\ell,\eta})_{hh}^{2}}K_{ik}^{c}. &
\end{aligned}\]

At each bandwidth $b\in\mathcal{B}$, we evaluate $\lambda_{b,ik,h}^{c,j}(\theta,\eta)$ at the training pilot and $\widehat{\eta}_{\ell}$ for $h=1,2$. For each training store-week $i$, we average each column over all other training store-weeks $k$. Regressing the resulting $5\times2$ matrix on $S_i$ by local quadratic regression gives $\check{\vecdel}_{b,\ell}^{c,j}(S_i)$. For each price column, five-fold cross-validation over training stores selects a smoothing bandwidth from $\{0.3,0.5,0.8,1.2,2,3,5\}$ using the mean squared prediction error standardized for each of its five entries.

Using pairs of observations in the test data $I_{\ell}$, we construct the orthogonal score
\[  \begin{aligned}
    \psi_{b,\ell,ik}^{c,j}(\theta)
        =   & \big\{ d_{ij}(\theta,\widehat{\eta}_{\ell})\rho_{kj}(\theta,\widehat{\eta}_{\ell}) + \rho_{ij}(\theta,\widehat{\eta}_{\ell})d_{kj}(\theta,\widehat{\eta}_{\ell}) \big\}K_{ik}^{c}(\widehat{\eta}_{\ell}) &\\
            & + \check{\vecdel}_{b,\ell}^{c,j}(S_i)\big(p_i-\widehat{\eta}_{\ell}(S_i)\big)
              + \check{\vecdel}_{b,\ell}^{c,j}(S_k)\big(p_k-\widehat{\eta}_{\ell}(S_k)\big). &
\end{aligned} \]
These corrections stay fixed as we solve for $\theta$.

\paragraph{Numerical solution.}
For each share and conditioning set, we search the box $[-40,40]^5$ in scaled coefficients. We scale the price and promotion coefficients by the root mean squares of their regressors, and the price-control coefficient by the root mean square price residual from the full-sample linear first stage. Starting values include the full-sample plug-in pilot, the joint logit estimate, and solutions from a 12-week-per-store subsample or nearby ISPD kernel bandwidths.

We report an interior root when the normalized score norm is below $10^{-7}$, the Newton step has norm below $10^{-4}$, the ratio of the smallest to largest Jacobian singular value exceeds $10^{-10}$, and every coordinate lies more than $10^{-5}$ from the boundary. A failed check leaves open whether the squared score has a local minimum near the pilot, as in Definition~\ref{sec04 - definition01 - DML}.

\paragraph{Combining the share estimates.}
For one share equation, let $s_i$ be the sum of the corrected scores over all evaluation pairs containing store-week $i$, and let $\overline s=N^{-1}\sum_i s_i$. With $\widehat\Gamma$ the mean score Jacobian over the $M$ pairs, the contribution of store-week $i$ to the linearization is $-(N/M)\widehat\Gamma^{-1}(s_i-\overline s)$. Using the sum retains the actual number of partners for each store-week in its fold.

Stack the two shares' contributions into the ten-dimensional vector $a_i^c$. For a clustering dimension $d$ with $G_d$ groups, define
\[
 \widehat V_d(a^c)=\frac{G_d}{G_d-1}\frac{1}{N^2}
   \sum_{g=1}^{G_d}\left(\sum_{i\in g}a_i^c\right)\left(\sum_{i\in g}a_i^c\right)^\top .
\]
We use $\widehat V_c=\widehat V_{\rm store}(a^c)+\widehat V_{\rm week}(a^c)-\widehat V_{\rm store,week}(a^c)$ to allow for dependence across weeks within a store and across stores within a week, following \citet{cameron_gelbach_miller2011}. Section~\ref{sec04 - DDML construction} gives the theory for iid observations.

Let $\widehat t_c$ stack the Cheerios and Rice Krispies coefficient vectors, and let $B=(I_5,I_5)^\top$ stack two identity matrices. The minimum-distance weights and combined estimate are
\[
 L_c=(B^\top\widehat V_c^{-1}B)^{-1}B^\top\widehat V_c^{-1},
 \qquad \widehat\theta^c_{\rm MD}=L_c\widehat t_c .
\]
We combine the estimates when both shares yield verified roots and their joint covariance in scaled coordinates is positive definite with condition number at most $10^{12}$. Table~\ref{app:tab:cereal-vectors} shows the calculation for the local-quadratic first stage.

\begin{table}[htbp]\centering\small
\caption{Local-quadratic DDML estimates by share and conditioning set at $b=0.5$}
\label{app:tab:cereal-vectors}
\begin{tabular}{llrrrrr}
\hline\hline
Conditioning variable & Estimate & $\alpha_1$ & $\alpha_2$ & $\beta$ & $\delta$ & $\gamma$\\
\hline
Fitted prices & Cheerios & $0.903$ & $1.349$ & $-0.766$ & $0.639$ & $1.018$\\
 & Rice Krispies & $-2.292$ & $-0.852$ & $-0.499$ & $0.898$ & $0.743$\\
 & Combined & $-0.850$ & $-0.726$ & $-0.622$ & $0.706$ & $0.917$\\
Acquisition costs & Cheerios & $1.432$ & $1.942$ & $-0.797$ & $0.536$ & $0.221$\\
 & Rice Krispies & $0.467$ & $0.317$ & $-0.904$ & $0.756$ & $-2.318$\\
 & Combined & $-1.366$ & $-1.193$ & $-0.535$ & $0.522$ & $0.908$\\
\hline\hline
\end{tabular}
\par\smallskip\begin{minipage}{.94\textwidth}\footnotesize
\textit{Notes.} The intercepts are relative to Grape-Nuts. The slopes multiply relative price, relative promotion, and the price residual. Within each conditioning set, the combined row uses the joint covariance of the two preceding share estimates.
\end{minipage}
\end{table}

\FloatBarrier

\subsection{Bandwidth selection in the production-function application}\label{app:varianceselection}

\begin{figure}[htbp]\centering
 \includegraphics[width=.80\textwidth]{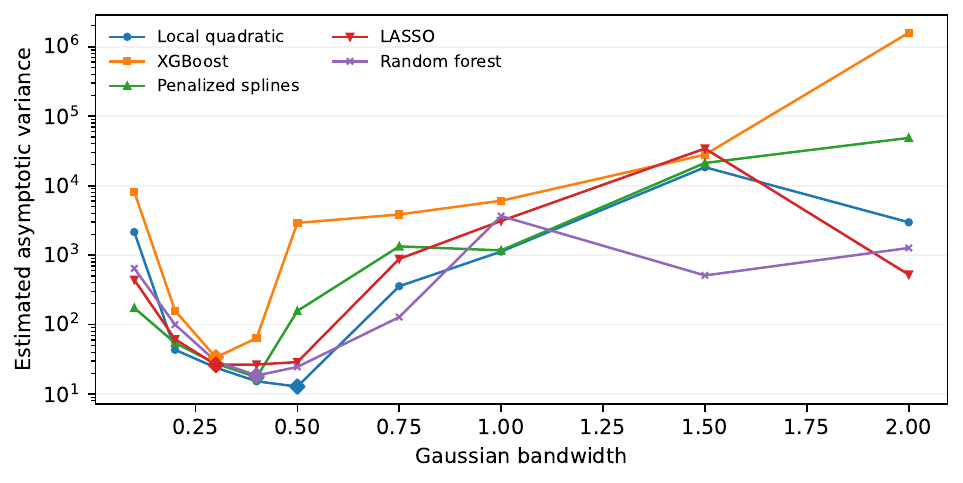}
 \caption{Production-function variance criteria over the bandwidth grid}
 \label{app:fig:variancepaths}
 \begin{minipage}{.97\textwidth}\footnotesize
 \textit{Notes.} The vertical axis is logarithmic. Diamonds mark the minimum of each curve.
 \end{minipage}
\end{figure}

Figure~\ref{app:fig:variancepaths} reports the estimated mean asymptotic variance of the three input elasticities over the nine bandwidths in equation~\eqref{sec07:eq:kernel}. The selected bandwidths are $0.5$ for local quadratic, $0.3$ for XGBoost, $0.4$ for penalized splines, $0.3$ for LASSO, and $0.4$ for random forests. The LASSO criteria at $0.3$ and $0.4$ differ by about 1\%.

\FloatBarrier

\end{document}